\documentclass[12pt]{article}
\usepackage{jheppub} 
\usepackage{amsfonts,amsmath,amssymb,epsf,mathtools}
\usepackage{amsmath,braket}
\usepackage{graphicx,hyperref,color}
\usepackage{pgfplots}
\usepackage{comment}
\hypersetup{
    bookmarks=true,         
    unicode=false,          
    pdftoolbar=true,        
    pdfmenubar=true,        
    linktocpage=true,       
    pdffitwindow=false,     
    pdfstartview={FitH},    
    pdfnewwindow=true,      
    colorlinks=true,       
    linkcolor=blue,          
    citecolor=blue,        
    filecolor=blue,      
    urlcolor=blue           
}

\numberwithin{equation}{section}									

\def\d#1{\,{\rm d}#1}

\newcommand{\de}{\partial}
\newcommand{\be}{\begin{equation}}
\newcommand{\ba}{\begin{eqnarray}}
\newcommand{\ea}{\end{eqnarray}}
\newcommand{\ee}{\end{equation}}

\newcommand{\s}{\sqrt}
\newcommand{\vp}{\varphi}

\newcommand{\ti}{\tilde}

\newcommand{\no}{\nonumber \\}
\newcommand{\la}{\langle}
\newcommand{\lb}{\rangle}
\newcommand{\bea}{\begin{eqnarray}}
\newcommand{\eea}{\end{eqnarray}}
\newcommand{\bes}{\begin{equation*}}
\newcommand{\beas}{\begin{eqnarray*}}
\newcommand{\eeas}{\end{eqnarray*}}
\newcommand{\bas}{\begin{array*}}
\newcommand{\eas}{\end{array*}}
\newcommand{\ees}{\end{equation*}}

\newcommand{\ep}{\epsilon}

\def\dd{\mathrm{d}}

\def\calT{\mathcal{T}}

\newcommand{\Tr}{\textrm{Tr}}

 \let\b=\beta  \let\d=\delta \let\e=\epsilon       
  \let\r=\rho 
\let\t=\tau   \let\vp=\varphi \let\x=\xi  \let\z=\zeta

\def\pa{\partial}

\renewcommand{\Re}{\operatorname{Re}}
\renewcommand{\Im}{\operatorname{Im}}
\newcommand{\Log}{\operatorname{Log}}
\newcommand{\Arg}{\operatorname{Arg}}

\newcommand{\JQ}{\mathrm{JQ}}

\subheader{\today}
\title{\boldmath Pseudo entropy under joining local quenches}

\author[a]{Kotaro Shinmyo}
\author[a,b,c]{Tadashi Takayanagi}
\author[a]{Kenya Tasuki}

\affiliation[a]{Center for Gravitational Physics and Quantum Information, Yukawa Institute for Theoretical Physics, Kyoto University,\\
    Kitashirakawa Oiwakecho, Sakyo-ku, Kyoto 606-8502, Japan}
\affiliation[b]{Inamori Research Institute for Science,\\
    620 Suiginya-cho, Shimogyo-ku, Kyoto 600-8411, Japan}
\affiliation[c]{Kavli Institute for the Physics and Mathematics of the Universe (WPI),\\
    University of Tokyo, Kashiwa, Chiba 277-8582, Japan}

\emailAdd{kotaro.shinmyo@yukawa.kyoto-u.ac.jp}
\emailAdd{takayana@yukawa.kyoto-u.ac.jp}
\emailAdd{kenya.tasuki@yukawa.kyoto-u.ac.jp}

\abstract{We compute the pseudo entropy in two-dimensional holographic and free Dirac fermion CFTs for excited states under joining local quenches. Our analysis reveals two of its characteristic properties that are missing in the conventional entanglement entropy. 
One is that, under time evolution, the pseudo entropy exhibits a dip behavior as the excitations propagate from the joined point to the boundaries of the subsystem. 
The other is that the excess of pseudo entropy over entanglement entropy can be positive in holographic CFTs, whereas it is always non-positive in free Dirac fermion CFTs. 
We argue that the entropy excess can serve as a measure of multi-partite entanglement.
Its positivity implies that the vacuum state in holographic CFTs possesses multi-partite entanglement, in contrast to free Dirac fermion CFTs.}

\begin{document} 

\begin{flushright}
YITP-23-132\\
\end{flushright}
\maketitle
\flushbottom

\section{Introduction and summary}
The entanglement entropy has played essential roles in studying the dynamical properties of quantum field theories, especially conformal field theories (CFTs) \cite{Bombelli:1986rw, Srednicki:1993im, Holzhey:1994we, Calabrese:2004eu, Casini:2009sr, Calabrese:2009qy}. 
Its significance stems from its ability to quantify the degrees of freedom and provide a non-equilibrium and local extension of thermal entropy, particularly in scenarios involving non-trivial time evolution, such as quantum quenches \cite{Calabrese:2005in, Calabrese:2007mtj, Nozaki:2014hna}. 
Moreover, the entanglement entropy has a geometric description in terms of AdS/CFT correspondence \cite{Maldacena:1997re, Gubser:1998bc, Witten:1998qj}, referred to as the holographic entanglement entropy \cite{Ryu:2006bv, Ryu:2006ef, Hubeny:2007xt}.

Recently, a novel generalization of entanglement entropy called pseudo entropy was introduced in \cite{Nakata:2021ubr}. 
As opposed to the entanglement entropy, this quantity is defined for a pair of quantum states \(\ket{\psi}\) and \(\ket{\vp}\) as follows. 
First, we introduce the transition matrix:
\begin{equation}
    \calT^{\psi | \vp} \coloneqq \frac{|\psi\lb \la\vp|}{\la \vp|\psi\lb}.
\end{equation}
The reduced transition matrix for subsystem \(A\) is given by tracing out the degrees of freedom in \(B\), the complement of \(A\), from the transition matrix:
\begin{equation}
    \calT_A^{\psi|\vp} \coloneqq \mbox{Tr}_B \left[\calT^{\psi|\vp}\right].
\end{equation}
The \(n\)-th pseudo R\'{e}nyi entropy \(S^{(n)}(\calT_A^{\psi|\vp})\) is defined as:
\begin{equation}
    S^{(n)}(\calT_A^{\psi|\vp}) \coloneqq \frac{1}{1-n}\log\mbox{Tr}[(\calT_A^{\psi|\vp})^{n}].
\end{equation}
We are interested in the \(n\to1\) limit, the pseudo entropy \cite{Nakata:2021ubr}:
\begin{equation}
    S(\calT_A^{\psi|\vp}) \coloneqq \lim_{n\to1} S^{(n)}(\calT_A^{\psi|\vp}) = -\mbox{Tr}[\calT_A^{\psi|\vp}\log{\calT_A^{\psi|\vp}}].
\end{equation}
For cases in which \(\ket{\psi} = \ket{\vp}\), the pseudo entropy is equal to the entanglement entropy:
\begin{equation}
    S(\calT_A^{\psi | \psi}) = S(\rho_A^{\psi}), \quad \rho_A^{\psi} \coloneqq \Tr_B\left[ \rho^{\psi} \right] = \Tr_B\left[ | \psi \lb \la \psi | \right].
\end{equation}
In this sense, the pseudo entropy is a generalization of the entanglement entropy.

A remarkable feature of pseudo entropy is its gravity dual description \cite{Nakata:2021ubr} via AdS/CFT, which can be viewed as a natural extension of holographic entanglement entropy. 
The minimal area in a Euclidean time-dependent asymptotically AdS space, divided by \(4G_N\), coincides with the pseudo entropy.
Furthermore, this quantity has a peculiar property that it takes complex values. 
This is because transition matrices are not hermitian in general as opposed to density matrices. 
This property is vital in holography for de Sitter spaces, or dS/CFT \cite{Strominger:2001pn,Maldacena:2002vr}.
The holographic entanglement entropy for dS$_3/$CFT$_2$ \cite{Hikida:2021ese, Hikida:2022ltr} takes complex values, as there are no space-like geodesics but only time-like ones that connect two boundary points. 
This complex valued entropy should be regarded as a noteworthy example of pseudo entropy \cite{Doi:2022iyj, Narayan:2022afv, Narayan:2023ebn, Doi:2023zaf, Jiang:2023loq, Kawamoto:2023nki, Chen:2023eic}. 
We also encounter such complex-valued pseudo entropy when considering time-like extensions of entanglement entropy \cite{Doi:2022iyj, Liu:2022ugc, Li:2022tsv, Doi:2023zaf, Chu:2023zah, He:2023ubi}. 
Motivated by pseudo entropy, a modified quantity called SVD entropy was introduced in \cite{Parzygnat:2023avh}.
SVD entropy always takes non-negative real values and has an intriguing interpretation in quantum information theory. 
However, it lacks its direct gravitational dual.

Moreover, pseudo entropy can be used as an order parameter of quantum phase transition \cite{Mollabashi:2020yie, Mollabashi:2021xsd, Nishioka:2021cxe, Kanda:2023zse}. 
In particular, the excess of the pseudo entropy above the average of the entanglement entropy, the entropy excess \(\varDelta{S}_A\), plays an important role:
\begin{equation}
    \varDelta{S}_A \coloneqq \mbox{Re}\left[S(\calT_A^{\psi | \vp})\right]-\frac{1}{2}S(\rho_A^{\psi})-\frac{1}{2}S(\rho_A^{\vp}).\label{dPE}
\end{equation}
This quantity is always non-positive when \(\ket{\psi}\) and \(\ket{\vp}\) belong to the same quantum phase. 
However, it can take positive values when the states are in different quantum phases. 
For other properties of pseudo entropy, refer to, e.g., \cite{Camilo:2021dtt,Goto:2021kln,Miyaji:2021lcq,Murciano:2021dga,Akal:2021dqt, Guijosa:2022jdo,Akal:2022qei,Mori:2022xec,Mukherjee:2022jac,Guo:2022sfl,Ishiyama:2022odv,Miyaji:2022dna,Bhattacharya:2022wlp,Guo:2022jzs,Chen:2023gnh,He:2023eap,Chandra:2023rhx,Carignano:2023xbz,He:2023wko,Guo:2023aio}.

Despite these advancements in understanding pseudo entropy in quantum field theories and its importance in AdS/CFT, its precise physical interpretation still needs to be clarified.
In this paper, we aim to explore this fundamental question by studying its dynamical properties, focusing on a class of excited states in two-dimensional CFTs generated through joining local quenches. 
Local quenches, in general, are used to describe local excitations (for classification, see \cite{Shimaji:2018czt}). 
A joining local quench is triggered by joining two semi-infinite systems, each hosting a two-dimensional CFT \cite{Calabrese:2007mtj}. 
The gravity dual of joining quenches was developed in \cite{Ugajin:2013xxa} and later examined in detail in \cite{Shimaji:2018czt}. 
Joining local quenches have a crucial technical advantage that allows universal computation of the time evolution of physical quantities using conformal mapping techniques.
This advantageous feature enables us to calculate the pseudo entropy for both the free Dirac fermion CFT and the holographic CFT (i.e., a maximally strongly coupled CFT) in this paper (refer to Table~\ref{tab:def}).
We represent the quantum state created by joining two CFTs at \(x=x_1\) as \(\ket{\JQ(x_1,0)}\) and its time evolution, referred to as the joining quenched state, as \(\ket{\JQ(x_1,t)}=e^{-iHt}\ket{\JQ(x_1,0)}\), where \(H\) denotes the CFT Hamiltonian.
We will investigate two distinct setups for pseudo entropy: 
\begin{itemize}
    \item[(i)] Select \(\ket{\psi}=\ket{\JQ(0,t)}\) and \(\ket{\vp}=\ket{\Omega}\), where \(\ket{\Omega}\) is the CFT vacuum state, described by a single-slit geometry in Euclidean path integral (see Fig.~\ref{fig:single-slit_geometry}). 
    \item[(ii)] Select \(\ket{\psi}=\ket{\JQ(-x_1,t)}\) and \(\ket{\vp}=\ket{\JQ(x_1,t)}\) described by a double-slit geometry in Euclidean path integral (see Fig.~\ref{fig:double-slit_geometry}).
\end{itemize}
Our primary focus will be on the entropy excess \eqref{dPE} and a comparative analysis of its behavior in the two distinct CFTs.
It is worth noting that the evolution of pseudo entropy for operator local quenches \cite{Nozaki:2014hna,Nozaki:2013wia,Nozaki:2014uaa,Asplund:2014,He:2014mwa,Asplund:2014coa,Caputa:2014vaa,Kusuki:2019avm} has been computed for integrable CFTs in \cite{Guo:2022sfl,He:2023eap,He:2023wko}.

\begin{figure}[t]
    \centering
    \includegraphics[width=8cm]{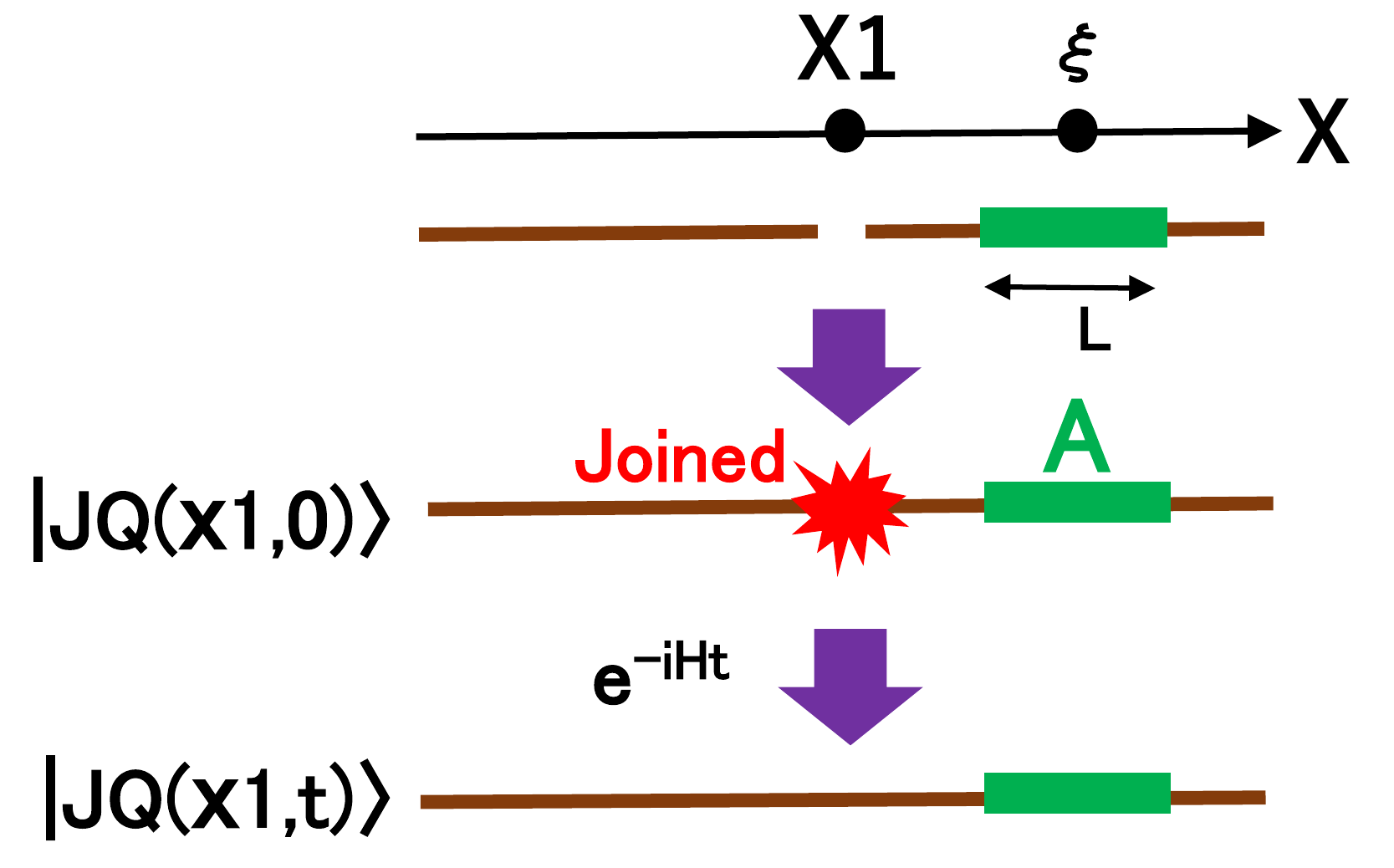}
    \caption{A sketch of the quantum state created by a joining local quench and subsystem \(A\), where we measure the pseudo entropy.}
    \label{fig:Join}
\end{figure}

\begin{table}[t]
    \caption{Three setups for entanglement entropy (EE) and pseudo entropy (PE) calculations. 
    We consider the pseudo entropy \(S(\calT_A^{\psi|\vp})\) defined from the transition matrix for the two different states, \(\ket{\psi}\) and \(\ket{\vp}\), in each case.}
    \label{tab:def}
    \vspace{5mm}
    \centering
    \begin{tabular}{|c|c|c|c|}
        \hline
        & EE in Joining Quench & (i) PE in Single-slit & (ii) PE in Double-slit \\
        \hline
        \(\ket{\psi}\) & \(\ket{\JQ(0,t)}\) & \(\ket{\JQ(0,t)}\) & \(\ket{\JQ(-x_1,t)}\)
        \\
        \hline
        \(\ket{\vp}\) & \(\ket{\JQ(0,t)}\) & \(\ket{\Omega}\) & \(\ket{\JQ(x_1,t)}\) \\
        \hline
        \end{tabular}
\end{table}

This paper is organized as follows.
Section \ref{sec:4qubit} presents a toy analysis of four-qubit models that mimic the joining local quenches. 
In section \ref{sec:methods}, we review the details of the field-theoretic analysis of joining local quenches and the calculation of entanglement entropy. 
In section \ref{sec:ee}, we examine the behavior of entanglement entropy in joining local quenches, including a review of earlier works and introducing our new findings.
In section \ref{sec:pejqgs}, we present the calculation of the time evolution of pseudo entropy for (i) the single-slit setup. 
In section \ref{sec:pejqjq}, we compute the time evolution of pseudo entropy for (ii) the double-slit setup. 
In section \ref{sec:discussions}, we discuss the final results and future problems. 

\subsection{Summary of main results in this paper}\label{ssec:PSummary}
For the reader's convenience, we summarize the key findings of this paper. 
Our primary quantity of interest is the pseudo entropy  \(S(\calT_A^{\psi|\vp})\) for (i) single-slit setup and (ii) double-slit setup. 
Furthermore, we revisit the analysis of entanglement entropy of a joining quenched state. 
Refer to Table \ref{tab:def} for the definitions of the respective transition matrices. 

We consider a two-dimensional CFT on a flat space with the time-like and space-like coordinates \(t\) and \(x\). 
Our analysis concentrates on the subsystem \(A(\x)\), which takes the form of an interval defined as \(\x - L/2 \leq x \leq \x + L/2\) with a fixed length \(L\). 
We are particularly interested in the dependence upon the center of mass \(\x\), and the time evolution of the pseudo entropy.
We focus on the difference \(S(\calT_A^{\psi|\vp})-S_A^{(0)}\), where \(S_A^{(0)}\) denotes the entanglement entropy of the ground state \(\ket{\Omega}\).
For an interval \(A\), we have the standard result \cite{Holzhey:1994we}:
\begin{equation}
    S_A^{(0)} \coloneqq S(\r_A^{(0)}) = \frac{c}{3} \log{\frac{L}{\epsilon}}, \quad \rho_A^{(0)} \coloneqq \mbox{Tr}_B\left[ \ket{\Omega}\bra{\Omega} \right].
\end{equation}
We also examine whether the entropy excess \(\varDelta{S}_A\) defined in \eqref{dPE} can take positive values. 
We can explicitly and analytically compute the pseudo entropy in these setups in both the free massless Dirac fermion CFT and the holographic CFT. 
In the Dirac fermion CFT, we can express the twist operator in the replica method explicitly via the bosonization method, which enables us to evaluate the pseudo entropy exactly. 
In the holographic CFT, the two-point function of the twist operators can be computed from simple Wick contractions by combining the mirror image method and generalized free field properties.
For a detailed description of these analytical methods, refer to, e.g., \cite{Shimaji:2018czt} and the references therein. 
The plots of pseudo entropy presented in this paper are summarized in Table \ref{tab:plots}.

\begin{table}
    \centering
    \caption{A summary of entanglement/pseudo entropy plots in this paper. 
    Here, Fig.~A/Fig.~B means that the real/imaginary part of the entropy is described in Fig.~A/Fig.~B.}
    \vspace{5mm}
    \begin{tabular}{|c|c|c|c|}
        \hline
         & Entanglement entropy & Pseudo entropy & Pseudo entropy \\
         & \(S_A^{\JQ}\) & \(S_A^{\JQ|\Omega}\) & \(S_A^{\JQ_1|\JQ_2}\) \\
        \hline
        \(S_{A(\xi)}(t)\) & Fig.~\ref{fig:EE} & Fig.~\ref{fig:rePE_JQGS}/Fig.~\ref{fig:imPE_JQGS} & Fig.~\ref{fig:rePE_JQJQ}/Fig.~\ref{fig:imPE_JQJQ} \\
        \hline
        \(\varDelta{S}_{A(\xi)}(t)\) & 0 & Fig.~\ref{fig:deltaS_JQGS} & Fig.~\ref{fig:deltaS_JQJQ} \\
        \hline
    \end{tabular}
    \label{tab:plots}
\end{table}

\begin{figure}
    \centering
    \includegraphics[width=6cm]{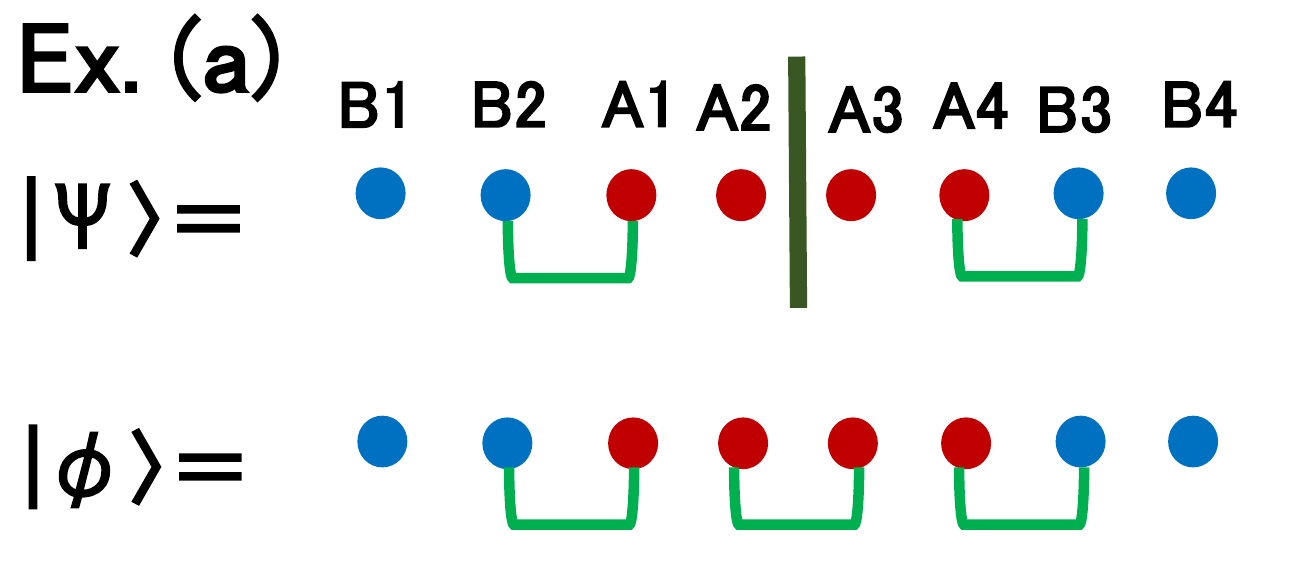}
    \hspace{1cm}
    \includegraphics[width=6cm]{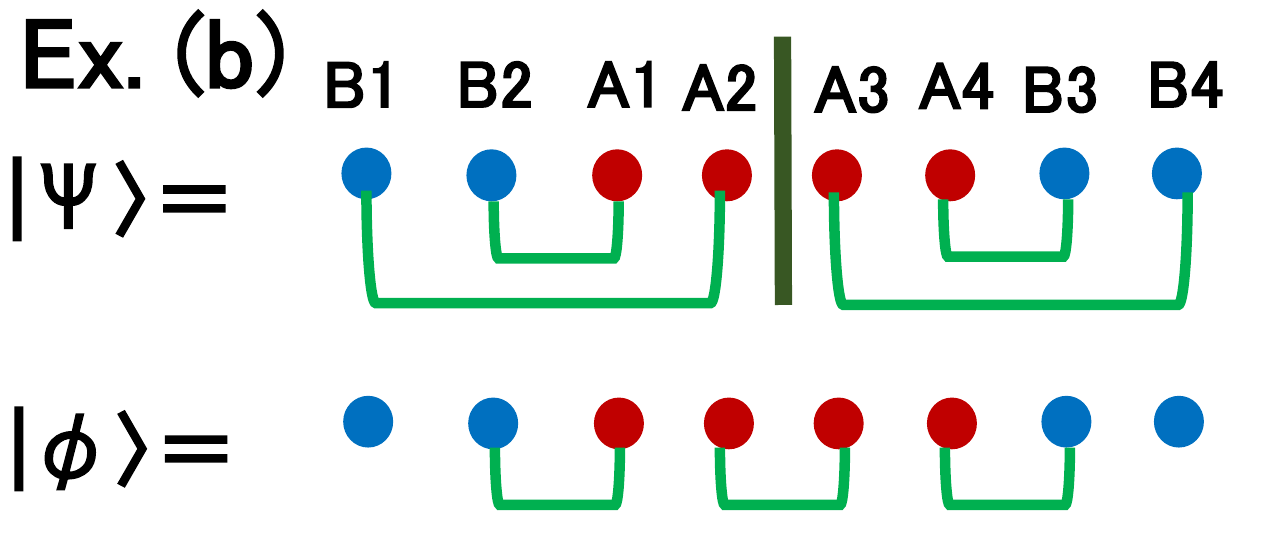}
    \caption{A sketch of examples (a) and (b), that are states in an eight-qubit system with only bipartite quantum entanglement. 
    The red/blue dots and green links represent the qubits (in A/B) and quantum entanglement between two qubits.
    The state \(\ket{\psi}\) represents the original state, in which \(A_2\) and \(A_3\) are disentangled.  
    On the other hand, \(\ket{\vp}\) is the one after the joining quench which connects \(A_2\) and \(A_3\). 
    We find \(\varDelta{S}_A=0\) for (a) and \(\varDelta{S}_A<0\) for (b).}
    \label{fig:DScase}
\end{figure}

As a warm-up, we consider the entropy excess \(\varDelta{S}_A\) in a quantum spin system with only bipartite quantum entanglement.
This preliminary investigation serves as a foundation for our later exploration of the single-slit setup in CFTs.
We choose subsystem \(A\) so that the joining procedure is performed in it.
We find that the entropy excess is non-positive for such simple structures of quantum entanglement. 
For example, in the examples shown in Fig.~\ref{fig:DScase}, it is straightforward to find 
\(\varDelta{S}_A=0\) for (a) as \(S(\calT_A^{\psi|\vp})=S(\rho_A^{\psi})=S(\rho_A^{\vp})\).
On the other hand, we have \(\varDelta{S}_A<0\) for (b) as \(S(\calT_A^{\psi|\vp})=S(\rho_A^{\vp})<S(\rho_A^{\psi})\).
The non-positive nature arises because the pseudo entropy gets decreased when we flip the (bipartite) quantum entanglement, which follows from the fundamental property that the pseudo entropy vanishes when either \(\ket{\psi}\) or \(\ket{\vp}\) has no entanglement between \(A\) and \(B\). 
Therefore, \(\varDelta{S}_A>0\) implies the presence of non-trivial multi-partite quantum entanglement in the given quantum states, as we will provide qubit examples in section \ref{sec:4qubit}.

Now, we consider joining local quenches in two-dimensional CFTs, where we join two semi-infinite systems at \(x=0\).
In (i) the single-slit setup, where the transition matrix is defined by the joining quenched state and the CFT vacuum, it turns out that the entropy excess \(\varDelta{S}_A\) is always non-positive in the Dirac fermion CFT, while it becomes positive for a large enough subsystem \(A\) in the holographic CFT, as can be seen from Fig.~\ref{fig:deltaS_JQGS}. 
Note that \(\varDelta{S}_A>0\) occurs in the holographic CFT when the joining point \(x=0\) is included in the subsystem \(A\), which is consistent with the result mentioned above in the qubit model. 
We also found an intriguing upper bound \(\varDelta{S}_A(t=0) \leq (c/6) \log{2}\).

Moreover, we find \(\varDelta{S}_A>0\) can occur in (ii) the double-slit setup only for the holographic CFT from Fig.~\ref{fig:deltaS_JQJQ} when we take the size of \(A\) to be large enough compared with the displacement \(x_1\). 
We argue that the positive entropy excess is due to the multi-partite entanglement in holographic CFT, which is missing in the free Dirac fermion CFT. 

Next, we turn to the time evolution of pseudo entropy. 
The pseudo entropy generally becomes complex-valued except for the initial time \(t=0\). 
First, let's focus on (i) the single-slit setup. 
Similar to the evolution of entanglement entropy under local quenches depicted in Fig.~\ref{fig:EE}, the real part of the pseudo entropy for the single-slit setup grows in the time interval when the excitation created at \(x=0\) by the joining quench propagates to the subsystem \(A\), as in Fig.~\ref{fig:rePE_JQGS}. 
Since this excitation consists of a pair of entangled modes, each located at \(x=t\) and \(x=-t\), propagating at the speed of light, they contribute to the entanglement during the time interval when one of them is within the interval \(A\). 
However, at the edges of this time interval, the real part of pseudo entropy decreases, unlike the entanglement entropy under the joining quenches. 
These qualitative profiles of pseudo entropy are common to both the free Dirac fermion and holographic CFTs. 
We expect this dip behavior is caused by the fact mentioned above that the pseudo entropy decreases when the quantum entanglement is flipped. 
Indeed, the value at the peak is also much smaller than that of the joining quenches. 
Another way to understand this behavior is to look at the expectation value of energy density. 
This is because the first law relation of entanglement entropy \cite{Bhattacharya:2012mi,Blanco:2013joa}, which relates energy to entanglement entropy, can also be applied to pseudo entropy \cite{Mollabashi:2020yie}. 
The energy density in the joining quench and the single-slit setup is obtained in Fig.~\ref{fig:EStensor_JQJQ} and Fig.~\ref{fig:EStensor_JQGS}.
These also show that the dip appears only in the latter case.

We also compute the evolution of pseudo entropy in (ii) the double-slit setup.
Notice that when \(x_1=0\), the pseudo entropy coincides with the entanglement entropy for the joining quenches.
In the limit of \(x_1\to\infty\), it reduces to the single-slit result. 
At \(t=0\), we observe that the real part of pseudo entropy exhibits a dip, particularly when either of the two endpoints of the interval \(A\) approaches the joining points, as shown in Fig.~\ref{fig:rePE_JQJQ}. 
The standard property can explain this behavior: pseudo entropy decreases when quantum entanglement is flipped. 
The time evolution of pseudo entropy can be explained by the excitations created at two joining points \(x=\pm x_1\) propagating at the speed of light, which leads to two peaks as in Fig.~\ref{fig:rePE_JQJQ}. 
Also, we again find a dip when these excitations reach the boundaries of the interval \(A\), similar to (i) the single-slit setup.

\section{Simple qubit models}\label{sec:4qubit}
Before delving into the analysis of pseudo entropy in two-dimensional CFTs, we will begin by examining simpler qubit examples. 
In the context of the spin model setup in Fig.~\ref{fig:DScase}, it is noteworthy that qubit system states with solely bipartite entanglement between two spins cannot yield positive entropy excess \(\varDelta{S}_A>0\).
This limitation arises from the fact that if the spins at \(A_2\) and \(A_3\) get entangled as a pure state \(\ket{\vp}_{A_2A_3}\) within the state \(\ket{\vp}\) after the quench, then the reduced transition matrix factorizes as \(\calT_A^{\psi|\vp}=\calT_{A_2A_3}\otimes \calT_{A_1A_4}\).
Here, \(\calT_{A_2A_3}\) takes a `pure state form,' \(\calT_{A_2A_3} \propto |\ti{\psi}\lb\la\vp|_{A_2A_3}\).
Therefore the contribution from \(\calT_{A_2A_3}\) to the pseudo entropy \(S(\calT_A^{\psi | \vp})\) vanishes and we find \(\varDelta{S}_A\leq 0\).
Below are examples of four-qubit models with multi-partite entanglement, where we observe positive entropy excess.

\subsection{Example 1: spins with four-body entanglement}
We consider a four-qubit system with the spins denoted as \(B_1, A_1, A_2, \) and \(B_2\), arranged from left to right.
We choose a pair of states that emulate the pre- and post-quench states in CFTs.
The quench couples two spins, \(A_1\) and \(A_2\), and creates quantum entanglement between them.
The state before the quench is represented as:
\begin{equation}
    \ket{\psi} = (\cos{\theta'}\ket{00}+\sin{\theta'}\ket{11})_{B_1A_1} \otimes (\cos{\theta'}\ket{00}+\sin{\theta'}\ket{11})_{A_2B_2}.\label{fLQSa}
\end{equation}
This initial state exhibits no entanglement between \(A_1\) and \(A_2\).
Next, we consider a state with four-body entanglement as the one after the joining quench:
\begin{equation}
    \ket{\vp}  = \cos{\theta}\ket{0000}_{B_1A_1A_2B_2} + \sin{\theta}\ket{1111}_{B_1A_1A_2B_2}.\label{fLQSb}
\end{equation}
In the cases where \(\theta=\pi/4\), this state corresponds to the GHZ state. 

\begin{figure}[t]
    \centering
    \includegraphics[width=0.45\linewidth]{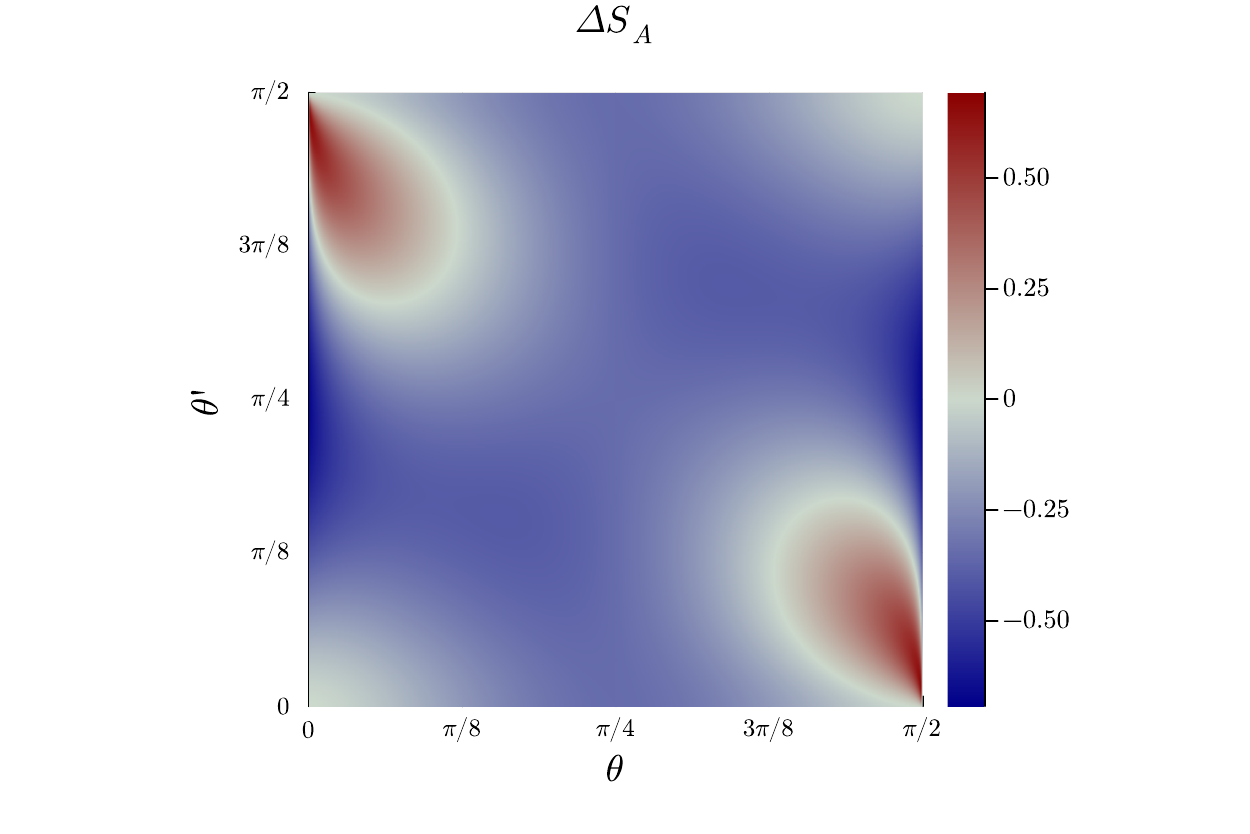}
    \caption{A plot of entropy excess \(\varDelta{S}_A\) for the four-qubit states \eqref{fLQSa} and \eqref{fLQSb}. 
    The horizontal and vertical coordinates are \(\theta\) and \(\theta'\), in the range \(0\leq \theta, \theta' \leq \pi/2\).}
    \label{fig:4qubitGHZ}
\end{figure}

The calculation of entropy excess is carried out straightforwardly.
As depicted in Fig.~\ref{fig:4qubitGHZ}, positive values of \(\varDelta{S}_A\) are detected within regions where \(\theta\) is small (or large) and \(\theta'\) is large (or small). 
These conditions are satisfied when the two states, \(\ket{\psi}\) and \(\ket{\vp}\), exhibit significant dissimilarity.
This is consistent with the observation in quantum spin chains \cite{Mollabashi:2020yie,Mollabashi:2021xsd} that positive entropy excess is realized only when the two states are in different quantum phases. 

\subsection{Example 2: Ising spin chain model}
We consider a spin-1/2 Ising chain with both transverse and longitudinal fields.
The Hamiltonian is 
\begin{equation}
    H = \sum_{i = 1}^{L - 1} J \sigma_i^{(z)} \sigma_{i+1}^{(z)} + \sum_{i=1}^{L} h \sigma_i^{(z)} + \sum_{i=1}^L g \sigma_i^{(x)},
    \label{spinHamiltonian}
\end{equation}
where \(\sigma_i^{(x)}\) and \(\sigma_i^{(z)}\) are the Pauli matrices of the spin at site \(i\).
This model was introduced in \cite{kim_ballistic_2013, mezei_entanglement_2017}.
We focus on the ground state of a four-qubit system \(B_1A_1A_2B_2\):
\begin{equation}
    \ket{\varphi} = \ket{\Omega}_{B_1A_1A_2B_2} \label{spinchain1},
\end{equation}
and the product state composed of the ground states of two-qubit systems \(B_1A_1\) and \(A_2B_2\):
\begin{equation}
    \ket{\psi} = \ket{\Omega}_{B_1A_1} \otimes \ket{\Omega}_{A_2B_2}. \label{spinchain2}
\end{equation}

\begin{figure}[t]
   \centering
   \includegraphics[width=0.45\linewidth]{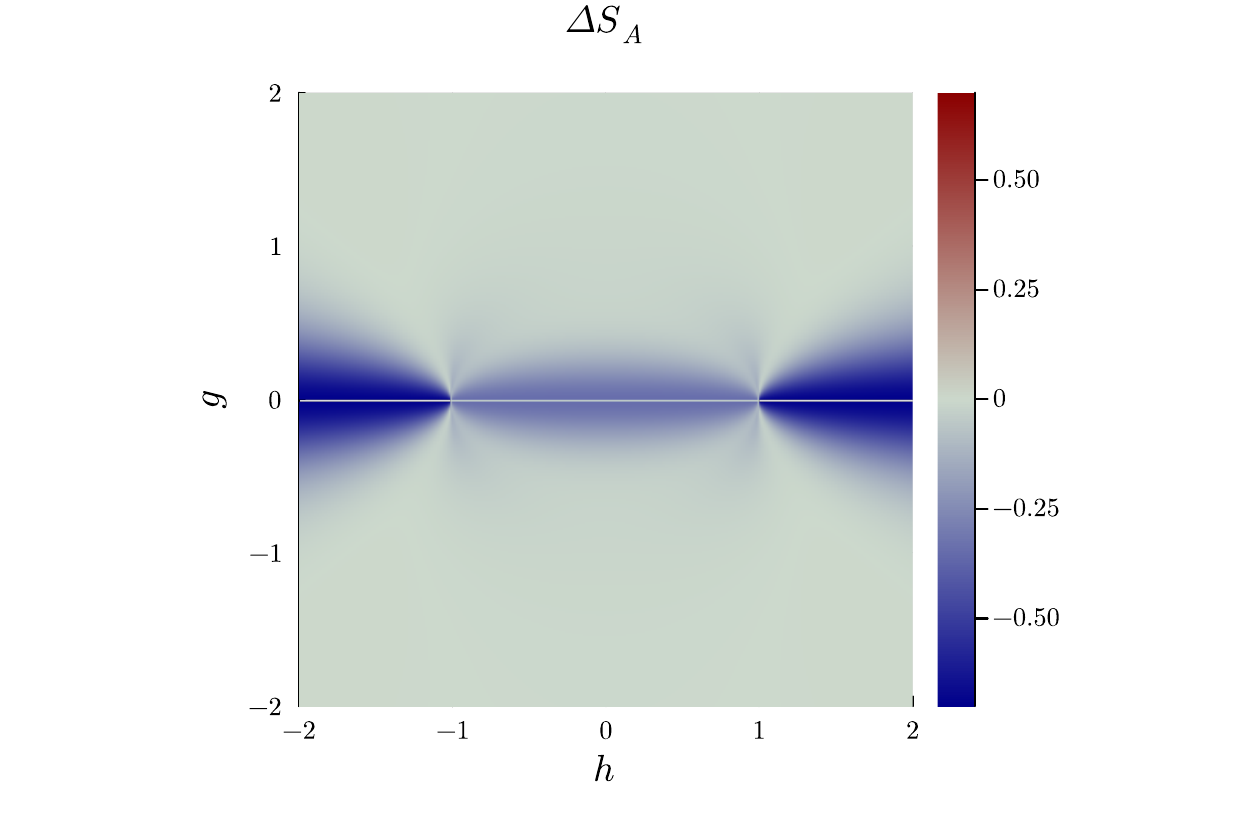}
   \includegraphics[width=0.45\linewidth]{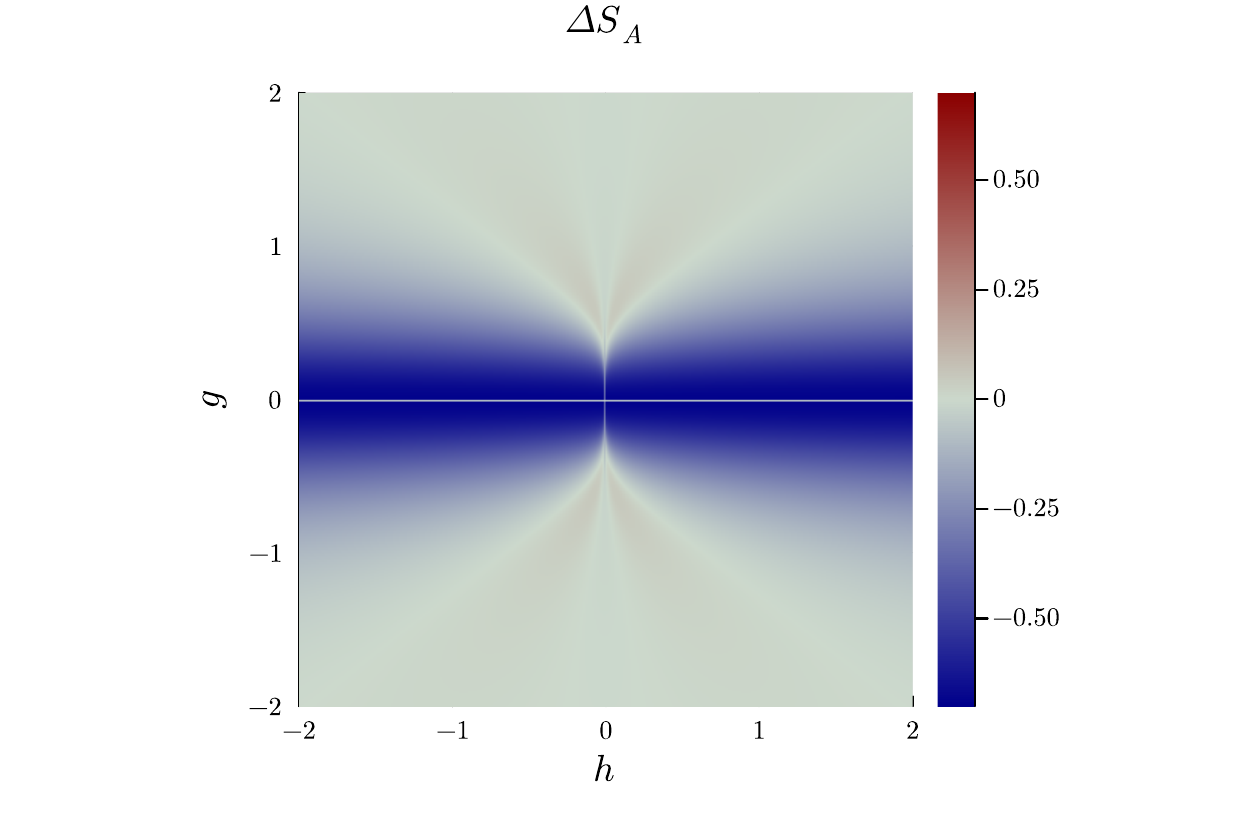}
   \caption{Plots of entropy excess \(\varDelta{S_A}\) for the four-qubit states \eqref{spinchain1} and  \eqref{spinchain2}. 
   The horizontal and vertical coordinates are \(h\) and \(g\) in the Hamiltonian \eqref{spinHamiltonian} with \(J = 1\) (left) and \(J = -1\) (right).}
   \label{fig:4qubitchaotic}
\end{figure}

Entropy excess for these two states is plotted in Fig.~\ref{fig:4qubitchaotic}.
Due to the presence of \(\mathbb{Z}_2\) symmetry, \(\varDelta{S}_A\) exhibits symmetry about \(h=0\) and \(g=0\). 
For \(J=-1\), the entropy excess \(\varDelta{S}_A\) can take positive values in the region \(|h| < |g|, h \neq 0\) . 
Also, \(\varDelta{S}_A>0\) is possible only when \(h\neq 0\), implying that the chaotic property is a prerequisite for this phenomenon. 

\section{General calculation methods in conformal field theory}\label{sec:methods}
We consider two-dimensional CFTs and choose a pair of quantum states represented as \(\ket{\psi}\) and \(\ket{\vp}\).
Pseudo entropy is calculated employing the replica method, which involves a path integral on a plane with a geometric shape corresponding to the selected pair of quantum states.
The replica method for pseudo entropy of general states is discussed in section \ref{ssec:generalcalc}. 
The geometries corresponding to each pair of states are described in detail in section \ref{ssec:geometries}.
The energy stress tensor is also calculated and discussed in section \ref{ssec:estensor}.

\subsection{Pseudo entropy calculation for general setups}\label{ssec:generalcalc}
The complex coordinate \(w=\tau+ix\) describes the Euclidean spacetime of the two-dimensional CFT with the Euclidean time \(\tau\) and the spatial coordinate \(x\).
The Lorentzian time evolution is realized through Wick rotation \(\t=it\).
Note that the complex conjugate of \(w\) after Wick rotation is represented as \(\bar{w}=it-ix\), leaving the sign of \(it\) unchanged.

A joint subsystem \(A\) at a fixed Euclidean time \(\t = \t_0\) is represented as the interval \(A \coloneqq \{(\t_0, x) | x_1\leq x\leq x_2\}\).
We also write the \(x\)-coordinate of the center of such \(A\) as \(\x\), and the length as \(L\), i.e., \(x_1 = \x - L/2\) and \(x_2 = \x + L/2\).
Subsystems consisting of a general number of intervals are similarly represented.

Pseudo entropy is obtained by taking the limit of the \(n\)-th R\'{e}nyi entropy, a process known as the replica method:
\begin{equation}
    S^{(n)}(\calT_A^{\psi|\vp}) \coloneqq \frac{1}{1-n} \log\mbox{Tr}[(\calT_A^{\psi|\vp})^n] \xrightarrow{n\to1} S(\mathcal{T}_A^{\psi|\varphi}).
    \label{PEFGHT}
\end{equation}
For a subsystem \(A\) consisting of \(k\) intervals, the trace is given by the \(2k\)-point function of twist operators \(\sigma_n\) situated at the edges of \(A\) within the original geometry.
To facilitate this computation, we employ a conformal map \(z = f(w)\) that maps the original geometry into the upper half-plane.
The explicit form of this map is provided in subsection \ref{ssec:geometries}. 
Using such a conformal map, the trace is represented using the \(k\)-point function in the upper half-plane:
\begin{align}
   \mbox{Tr} [(\calT_A^{\psi|\vp})^n] &= \la\sigma_n(w_1) \bar{\sigma}_n(w_2) \cdots \sigma_n(w_{2k-1}) \bar{\sigma}_n(w_{2k}) \lb_{\mathrm{original}} \no
   &= \left|\frac{dz_1}{dw_1}\right|^{2h_n}\left|\frac{dz_2}{dw_2}\right|^{2h_n} \cdots \left|\frac{dz_{2k-1}}{dw_{2k-1}}\right|^{2h_n}\left|\frac{dz_{2k}}{dw_{2k}}\right|^{2h_n} \no 
   &\qquad \times \la\sigma_n(z_1) \bar{\sigma}_n(z_2) \cdots \sigma_n(z_{2k-1}) \bar{\sigma}_n(z_{2k}) \lb_{\mathrm{UHP}}, \label{trhon}
\end{align}
where \(z_i = f(w_i)\), and \(h_n = (c/24) (n - 1/n)\) denotes the chiral dimension of the twist operator \(\sigma_n\).

The explicit form of the \(2k\)-point functions depends on the type of CFT under consideration.
For simplicity, we consider single intervals (\(k=1\)) and present the explicit form of pseudo entropy in the holographic CFT and free Dirac fermion CFT, respectively.

Since the pseudo entropy is complex-valued in general, we need to deal with the branch cut when we take the logarithm in \eqref{PEFGHT}. 
We consistently employ a configuration in which the pseudo entropy assumes a non-negative real value at \(t=0\), avoiding the branch cut issue at the initial time.
As we progress into the subsequent time evolution \(t>0\), we can uniquely select the branch that smoothly connects with the \(t=0\) result.

\subsubsection{Pseudo entropy in holographic CFTs}
In the holographic CFTs, the two-point functions on the upper half-plane can be approximated by either the connected or disconnected contributions:
\begin{align}
    \begin{split}
        \mathrm{Connected}:& \quad \la \sigma_n(z_1) \bar{\sigma}_n(z_2) \lb_{\mathrm{UHP}} = \frac{1}{|z_1-z_2|^{4h_n}} \\
       \mathrm{Disconnected}:& \quad \la \sigma_n(z_1) \bar{\sigma}_n(z_2) \lb_{\mathrm{UHP}} = \frac{1}{|z_1 - \bar{z}_1|^{2h_n}|z_2 - \bar{z}_2|^{2h_n}}
    \end{split}.
    \label{con_and_discon}
\end{align}
Note that these results are derived by considering the mirror images of the two twist operators and applying the saddle point approximation. 
This prescription follows from the AdS/BCFT formulation \cite{Takayanagi:2011zk, Fujita:2011fp, Shimaji:2018czt}. 
The AdS/BCFT provides a gravity dual of a boundary conformal field theory (BCFT), where a CFT is defined on a manifold with a boundary (or boundaries). 
In AdS/BCFT, the spacetime of the gravity dual is given by an asymptotically AdS spacetime surrounded by both the AdS boundary and the end-of-the-world brane (EOW brane). 
This EOW brane is a bulk extension of the boundary of the BCFT.

By substituting these into \eqref{trhon}, we obtain the pseudo R\'{e}nyi entropy \(S^{(n)}_A\).
In the limit of \(n\to1\), we find the connected contribution
\begin{align}
    \label{S_con}
    S^{\mathrm{con}}(\calT_A^{\psi|\varphi}) &= \frac{c}{6} \log{\frac{\left|z_1 - z_2\right|^2}{\tilde{\epsilon}_1 \tilde{\epsilon}_2}} \no
    &= \frac{c}{6} \log\left[ \frac{\left|f(w_1) - f(w_2)\right|^2}{|f'(w_1)| |f'(w_2)| \epsilon^2} \right],
\end{align}
and the disconnected contribution
\begin{align}
    \label{S_dis}
    S^{\mathrm{dis}}(\calT_A^{\psi|\varphi}) &= \frac{c}{6} \log{\frac{z_1 - \bar{z}_1}{\tilde{\epsilon}_1}} + \frac{c}{6} \log{\frac{z_2 - \bar{z}_2}{\tilde{\epsilon}_2}} + 2S_\mathrm{bdy} \no
    &= \frac{c}{6}\log\left[\frac{|f(w_1)-\overline{f(w_1)}| |f(w_2)-\overline{f(w_2)}|}{|f'(w_1)||f'(w_2)|\ep^2}\right] + 2S_{\mathrm{bdy}},
\end{align}
where \(\ep\) is the UV cut-off (or lattice constant) in the original geometry, and \(S_\mathrm{bdy}\) is the boundary entropy. 
Pseudo entropy for an interval \(A\) is approximated by the smaller of these two contributions:
\begin{equation}
    \label{S_hol}
    S(\calT_A^{\psi|\varphi}) = \min \left\{S^{\mathrm{con}}(\calT_A^{\psi|\varphi}), \, S^{\mathrm{dis}}(\calT_A^{\psi|\varphi}) \right\}.
\end{equation}

This holographic CFT calculation reproduces the gravity dual calculation, which is obtained by combining the holographic pseudo entropy computation \cite{Nakata:2021ubr} with the holographic entanglement entropy calculation in the AdS/BCFT \cite{Takayanagi:2011zk,Fujita:2011fp}.
This is expressed as 
\begin{equation}
    \label{PE_AdS}
    S(\calT_A^{\psi|\varphi}) = \min \left\{ \frac{A(\Gamma^{\mathrm{con}}_A)}{4G_N}, \, \frac{A(\Gamma^{\mathrm{dis}}_A)}{4G_N} \right\}.
\end{equation}
In this equation, \(\Gamma^{\mathrm{con}}_A\) represents the length of the connected geodesic that links two endpoints of subsystem \(A\) through the bulk AdS space. 
On the other hand, \(\Gamma^{\mathrm{dis}}_A\) consists of two disconnected geodesics that connect each endpoint of \(A\) to a point on the EOW brane, which is dual to the boundary of the BCFT.
As our calculations for the two-dimensional BCFT can provide all the required results using the field-theoretical approach described in this section, there is no need to rely on the gravity dual picture.
Therefore, we will forgo providing the details of the gravity dual calculation.

\subsubsection{Pseudo entropy in free Dirac fermion CFTs}
In the free Dirac fermion CFTs, the two-point functions on the upper half-plane can be approximated as
\begin{equation}
    \la \sigma_n(z_1) \bar{\sigma}_n(z_2) \lb_{\mathrm{UHP}} = \tilde{d}_n \left( 
\frac{|z_1 - \bar{z}_2|^2}{|z_1 - \bar{z}_1| |z_2 - \bar{z}_2| |z_1 - z_2|^2} \right)^{2h_n},
\end{equation}
where \(\tilde{d}_n\) is the normalization factor of the two-point function on the upper half-plane.
Since \(c=1\) for the free Dirac fermion CFTs, pseudo entropy is given as follows:
\begin{equation}
    \label{S_Dir}
    S(\calT_A^{\psi|\varphi}) = \frac{1}{6} \log\left[ \frac{|z_1 - \bar{z}_1| |z_2 - \bar{z}_2| |z_1 - z_2|^2}{|z_1 - \bar{z}_2|^2 |f'(w_1)| |f'(w_2)| \ep^2} \right] + \text{const.}.
\end{equation}

\subsection{Geometries of path integral}\label{ssec:geometries}
To calculate pseudo entropy in a given setup, it is necessary to substitute the explicit form of the corresponding conformal map \(f\). 
In the following, we will provide the expressions for conformal maps applicable to the opposing-slit, single-slit, and double-slit geometries.

\subsubsection{Opposing-slit geometry: entanglement entropy of JQ state}
\begin{figure}[t]
    \centering
    \centering
    \begin{tikzpicture}
        \fill[lightgray](-2,0.85)--(-3,0.85)--(-3,-3)--(3,-3)--(3,0.85)--(2,0.85)--(2,1.15)--(3,1.15)--(3,3)--(-3,3)--(-3,1.15)--(-2,1.15)--cycle;
        
        \draw [->, >=stealth, very thick](-3,0)--(3,0) node [below left]{$\Re{w}$};
        \draw [->, >=stealth, very thick](0,-3)--(0,3) node [below right]{$\Im{w}$};
        
        \draw [brown, very thick] (-2,1.15)--(-3,1.15);
        \draw [brown, very thick] (-2,0.85)--(-3,0.85);
        \draw [brown, very thick] (-2,0.85)arc(-90:90:0.15);
        \draw [brown, very thick] (2,1.15)--(3,1.15);
        \draw [brown, very thick] (2,0.85)--(3,0.85);
        \draw [brown, very thick] (2,1.15)arc(90:270:0.15);
        
        \filldraw [fill=yellow, draw=black] (-2,1) circle(0.125);
        \draw (-2,0.9) node [below]{$- \delta + i x^*$};
        
        \filldraw [fill=red, draw=black] (2,1) circle(0.125);
        \draw (2,0.9) node [below]{$\delta + i x^*$};
        
        \draw [blue, dashed, very thick] (-3,0.85)--(-3,-3)--(3,-3)--(3,0.85);
        
        \draw [red, very thick] (0.5,2)--(0.5,-1);
        \draw (0.6,0.5) node [right]{\textcolor{red}{$A$}};
        
        \filldraw [fill=black, draw=black] (0.5,-1) circle(0.125);
        \draw (0.6,-1) node [right]{$\sigma_n(w_1)$};
        \filldraw [fill=black, draw=black] (0.5,2) circle(0.125);
        \draw (0.6,2) node [right]{$\sigma_n(w_2)$};
    \end{tikzpicture}
    \begin{tikzpicture}
        \draw [white] (0,3)--(0,-3);
        \draw [->, >=stealth, very thick] (-1,0.2)--(1,0.2);
        \draw [->, >=stealth, very thick] (1,-0.2)--(-1,-0.2);
        \draw (0,0.2) node [above]{$f(w)$};
        \draw (0,-0.2) node [below]{$f^{-1}(z)$};
    \end{tikzpicture}
    \begin{tikzpicture}
        \fill[lightgray](3,0)--(3,3)--(-3,3)--(-3,0)--cycle;
        
        \draw [->, >=stealth, very thick](-3,0)--(3,0) node [below left]{$\Re{z}$};
        \draw [->, >=stealth, very thick](0,-3)--(0,3) node [below right]{$\Im{z}$};
        
        \draw [brown, very thick] (3,0.01)--(-3,0.01);
        
        \filldraw [fill=blue, draw=black] (0,0) circle (0.125);
        \draw (0,0) node [below right]{$0$};
        \filldraw [fill=red, draw=black] (1.5,0) circle (0.125);
        \draw (1.5,0) node [below]{$b$};
        \filldraw [fill=yellow, draw=black] (-1,0) circle (0.125);
        \draw (-1,0) node [below]{$-a$};
        
        \filldraw [fill=black, draw=black] (2,2) circle(0.125);
        \draw (2,2.1) node [above]{$\sigma_n(z_1)$};
        \filldraw [fill=black, draw=black] (-1,1) circle(0.125);
        \draw (-1,1.1) node [above]{$\sigma_n(z_2)$};
    \end{tikzpicture}
    \caption{The imaginary time-space region with opposing-slit geometry for the density matrix under a local quench at \(x=x^*\) (left) is mapped to the upper half-plane (right) by \(z = f(w)\).
    This geometry corresponds to the path integral calculation of the entanglement entropy of \(\ket{\JQ(x^*, t)}\).}
    \label{fig:opposing-slit_geometry}
\end{figure}
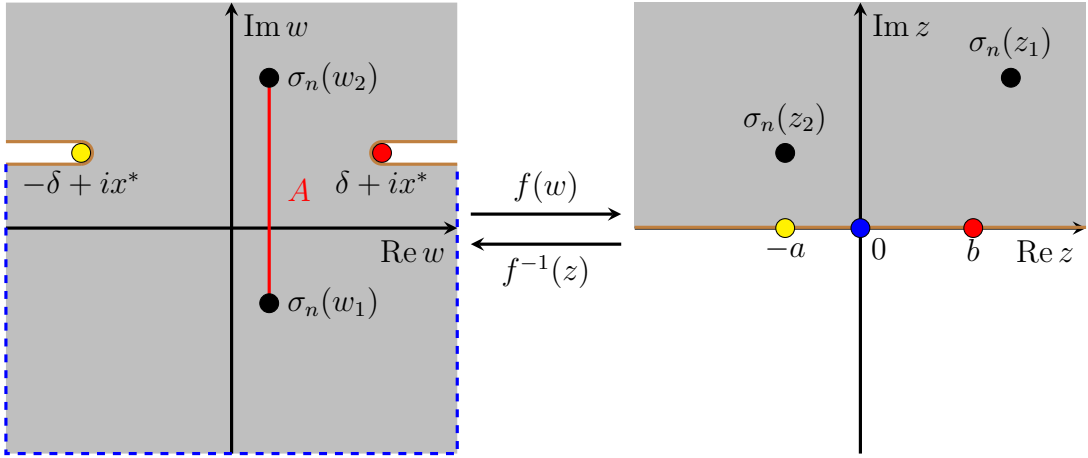

Firstly, we choose \(\ket{\psi} = \ket{\vp} = \ket{\JQ(x^*, t)}\) and consider the entanglement entropy under joining quenches.
The corresponding geometry in the path integral formalism is the \emph{opposing-slit} geometry shown in Fig.~\ref{fig:opposing-slit_geometry}. 
This is identical to the local quench setup considered in the pioneering work of \cite{Calabrese:2007mtj}.

By the Schwarz--Christoffel formula, we obtain the inverse map \(f^{-1}(z)\) as
\begin{equation}
    f^{-1}(z) = z + \frac{\d^2}{4z} + ix^*,
\end{equation}
where \(\d\) is the regularization parameter of the joining local quench such that \(\ket{\psi} = e^{-\d\cdot  H_{\mathrm{CFT}}}\ket{\JQ(x^*, t)}\). 
We can analytically solve for \(z\) and get
\begin{equation}
    f(w) = \frac{w -ix^* \pm \sqrt{(w - ix^*)^2 - \d^2}}{2}.
\end{equation}
We have to choose the sign to make the imaginary part of \(f(w)\) positive.

\subsubsection{Single-slit geometry: pseudo entropy of JQ state and ground state}
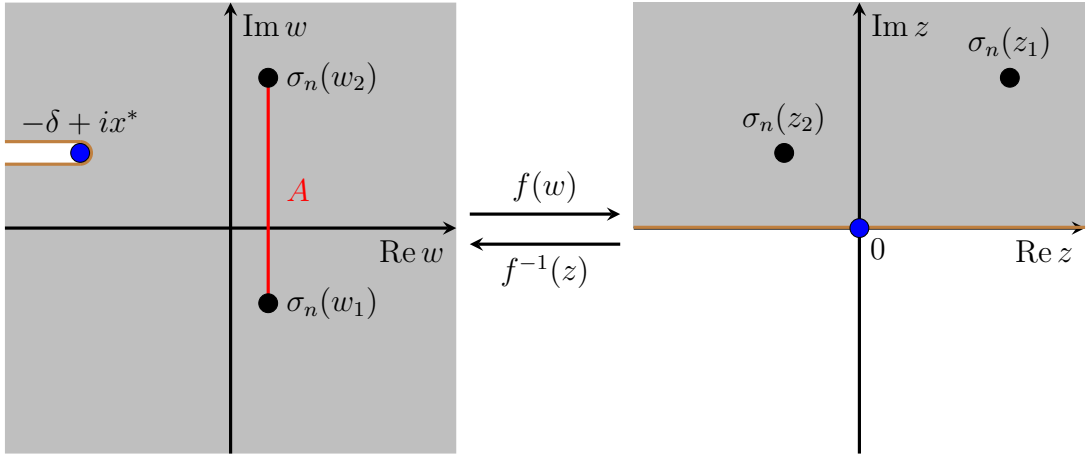
\begin{figure}[t]
    \centering
    \begin{tikzpicture}
        \fill[lightgray](-2,0.85)--(-3,0.85)--(-3,-3)--(3,-3)--(3,3)--(-3,3)--(-3,1.15)--(-2,1.15)--cycle;
        
        \draw [->, >=stealth, very thick](-3,0)--(3,0) node [below left]{$\Re{w}$};
        \draw [->, >=stealth, very thick](0,-3)--(0,3) node [below right]{$\Im{w}$};
        
        \draw [brown, very thick] (-2,1.15)--(-3,1.15);
        \draw [brown, very thick] (-2,0.85)--(-3,0.85);
        \draw [brown, very thick] (-2,0.85)arc(-90:90:0.15);
        
        \filldraw [fill=blue, draw=black] (-2,1) circle(0.125);
        \draw (-2,1.1) node [above]{$- \delta + i x^*$};
        
        \draw [red, very thick] (0.5,2)--(0.5,-1);
        \draw (0.6,0.5) node [right]{\textcolor{red}{$A$}};
        
        \filldraw [fill=black, draw=black] (0.5,-1) circle(0.125);
        \draw (0.6,-1) node [right]{$\sigma_n(w_1)$};
        
        \filldraw [fill=black, draw=black] (0.5,2) circle(0.125);
        \draw (0.6,2) node [right]{$\sigma_n(w_2)$};
    \end{tikzpicture}
    \begin{tikzpicture}
    \draw [white] (0,3)--(0,-3);
    \draw [->, >=stealth, very thick] (-1,0.2)--(1,0.2);
    \draw [->, >=stealth, very thick] (1,-0.2)--(-1,-0.2);
    \draw (0,0.2) node [above]{$f(w)$};
    \draw (0,-0.2) node [below]{$f^{-1}(z)$};
    \end{tikzpicture}
    \begin{tikzpicture}
        \fill[lightgray](3,0)--(3,3)--(-3,3)--(-3,0)--cycle;
        
        \draw [->, >=stealth, very thick](-3,0)--(3,0) node [below left]{$\Re{z}$};
        \draw [->, >=stealth, very thick](0,-3)--(0,3) node [below right]{$\Im{z}$};
        
        \draw [brown, very thick] (3,0.01)--(-3,0.01);
        
        \filldraw [fill=blue, draw=black] (0,0) circle (0.125);
        \draw (0,0) node [below right]{$0$};
        
        \filldraw [fill=black, draw=black] (2,2) circle(0.125);
        \draw (2,2.1) node [above]{$\sigma_n(z_1)$};
        
        \filldraw [fill=black, draw=black] (-1,1) circle(0.125);
        \draw (-1,1.1) node [above]{$\sigma_n(z_2)$};
    \end{tikzpicture}
    \caption{The imaginary time-space region with single-slit geometry for the transition matrix under a local quench at \(x=x^*\) (left) is mapped to the upper half-plane (right) by \(z = f(w)\).
    This geometry corresponds to the path integral calculation of pseudo entropy of \(\ket{\JQ(x^*, t)}\) and \(\ket{\Omega}\).}
    \label{fig:single-slit_geometry}
\end{figure}

Secondly, we choose \(\ket{\psi}\) as the joining quenched state \(\ket{\JQ(x^*, t)}\), while \(\ket{\vp}\) as the CFT vacuum \(\ket{\Omega}\). 
In the path integral formalism, the corresponding geometry is the \emph{single-slit} geometry shown in the left picture of Fig.~\ref{fig:single-slit_geometry}. 

This geometry is conformal mapped into the upper half-plane depicted in the right of Fig.~\ref{fig:single-slit_geometry} by the map
\begin{equation}
    f(w) = i\s{w + \delta - i x^*}.\label{comapsin}
\end{equation}

\subsubsection{Double-slit geometry: pseudo entropy of different JQ states}
\begin{figure}[t]
   \centering
   \begin{tikzpicture}
      \fill[lightgray](-2,-1.15)--(-3,-1.15)--(-3,-3)--(3,-3)--(3,0.85)--(2,0.85)--(2,1.15)--(3,1.15)--(3,3)--(-3,3)--(-3,-0.85)--(-2,-0.85)--cycle;

      \draw [->, >=stealth, very thick](-3,0)--(3,0) node [below left]{$\Re{w}$};
      \draw [->, >=stealth, very thick](0,-3)--(0,3) node [below right]{$\Im{w}$};

      \draw [brown, very thick] (-2,-0.85)--(-3,-0.85);
      \draw [brown, very thick] (-2,-1.15)--(-3,-1.15);
      \draw [brown, very thick] (-2,-1.15)arc(-90:90:0.15);
      \draw [brown, very thick] (2,0.85)--(3,0.85);
      \draw [brown, very thick] (2,1.15)--(3,1.15);
      \draw [brown, very thick] (2,1.15)arc(90:270:0.15);

      \filldraw [fill=yellow, draw=black] (-2,-1) circle(0.125);
      \draw (-2,-1.1) node [below]{$- \delta + i x_1$};

      \filldraw [fill=red, draw=black] (2,1) circle(0.125);
      \draw (2,1.1) node [above]{$\delta + i x_2$};

      \draw [blue, dashed, very thick] (-3,-1.15)--(-3,-3)--(3,-3)--(3,0.85);

      \draw [red, very thick] (0.5,2)--(0.5,-1);
      \draw (0.6,0.5) node [right]{\textcolor{red}{$A$}};

      \filldraw [fill=black, draw=black] (0.5,-1) circle(0.125);
      \draw (0.6,-1) node [right]{$\sigma_n(w_1)$};
      \filldraw [fill=black, draw=black] (0.5,2) circle(0.125);
      \draw (0.6,2) node [right]{$\sigma_n(w_2)$};
   \end{tikzpicture}
   \begin{tikzpicture}
      \draw [white] (0,3)--(0,-3);
      \draw [->, >=stealth, very thick] (-1,0.2)--(1,0.2);
      \draw [->, >=stealth, very thick] (1,-0.2)--(-1,-0.2);
      \draw (0,0.2) node [above]{$f(w)$};
      \draw (0,-0.2) node [below]{$f^{-1}(z)$};
   \end{tikzpicture}
   \begin{tikzpicture}
      \fill[lightgray](3,0)--(3,3)--(-3,3)--(-3,0)--cycle;

      \draw [->, >=stealth, very thick](-3,0)--(3,0) node [below left]{$\Re{z}$};
      \draw [->, >=stealth, very thick](0,-3)--(0,3) node [below right]{$\Im{z}$};

      \draw [brown, very thick] (3,0.01)--(-3,0.01);

      \filldraw [fill=blue, draw=black] (0,0) circle (0.125);
      \draw (0,0) node [below right]{$0$};
      \filldraw [fill=red, draw=black] (1.5,0) circle (0.125);
      \draw (1.5,0) node [below]{$b$};
      \filldraw [fill=yellow, draw=black] (-1,0) circle (0.125);
      \draw (-1,0) node [below]{$-a$};

      \filldraw [fill=black, draw=black] (2,2) circle(0.125);
      \draw (2,2.1) node [above]{$\sigma_n(z_1)$};
      \filldraw [fill=black, draw=black] (-1,1) circle(0.125);
      \draw (-1,1.1) node [above]{$\sigma_n(z_2)$};
   \end{tikzpicture}
   \caption{The imaginary time-space region with double-slit geometry for the transition matrix for the local quench (left) is mapped to the upper half-plane (right) by \(z = f(w)\).
   This geometry corresponds to the path integral calculation of pseudo entropy of \(\ket{\JQ(x_1, t)}\) and \(\ket{\JQ(x_2, t)}\).}
   \label{fig:double-slit_geometry}
\end{figure}
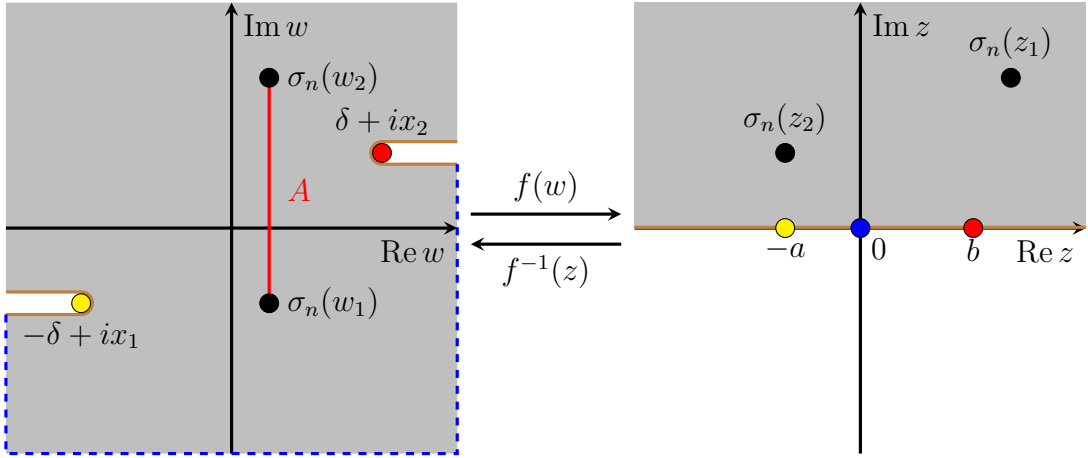

Thirdly, we choose \(\ket{\psi} = \ket{\JQ(x_1, t)}\) and \(\ket{\vp} = \ket{\JQ(x_2, t)}\).
In the path integral formalism, the corresponding geometry is
the \emph{double-slit} geometry shown in the left picture of Fig.~\ref{fig:double-slit_geometry}.
If \(x_1 = x_2\), this geometry reduces to the opposing-slit geometry, and then the corresponding pseudo entropy equals the entanglement entropy of \(\ket{\psi}\).

By the Schwarz--Christoffel formula, we obtain the inverse map as
(see appendix \ref{ap:formulae})
\begin{equation}
    f^{-1}(z) = z - A \Log{z} + \frac{a^*(a^*+A)}{z} + \left( 2a^* + A + A\log{a^*} - \delta + i x_2 \right),
\end{equation}
where \(A \coloneqq (x_2 - x_1) / \pi\) and \(a=a^*>0\) is the unique solution of
\begin{equation}
    4a + 2A - A \log{\left(1 + \frac{A}{a}\right)} = 2\delta.
\end{equation}
Since it is generally not possible to find analytical solutions for \(z\), we will employ the numerically obtained results in the following chapters.

\subsection{Energy stress tensor}\label{ssec:estensor}
Before we move on to the evaluation of pseudo entropy, here we calculate the energy stress tensor, which is relatively more accessible and will aid in our subsequent discussions concerning pseudo entropy.
The expectation value of the energy stress tensor can be computed via the conformal map from the original setup (with coordinates \(w\)) to the upper half-plane (with coordinates \(z\)):
\begin{equation}
    T(w)=\left(\frac{dw}{dz}\right)^{-2}\left[T(z)-\frac{c}{12}\{w:z\}\right],
\end{equation}
where we defined the Schwarzian derivative:
\begin{equation}
    \{w:z\}=\frac{\de^3_z w}{\de_z w}-\frac{3}{2}\left(\frac{\de^2_z w}{\de_z w}\right)^2.
\end{equation}
The anti-chiral component \(\bar{T}(\bar{w})\) is given similarly. 
Since the energy stress tensor on the upper half-plane vanishes, we can set \(T(z)=0\). 

\begin{figure}
    \centering
    \includegraphics[width=0.45\linewidth]{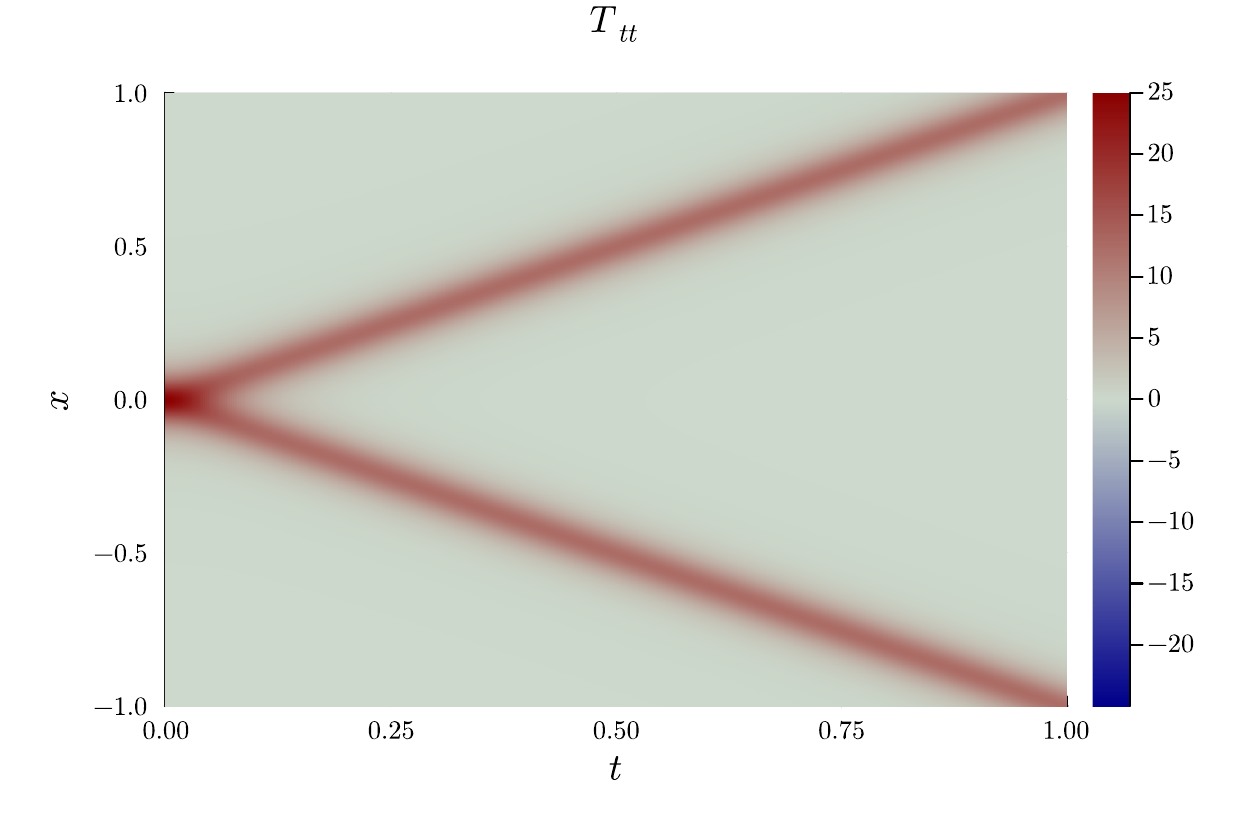}
    \includegraphics[width=0.45\linewidth]{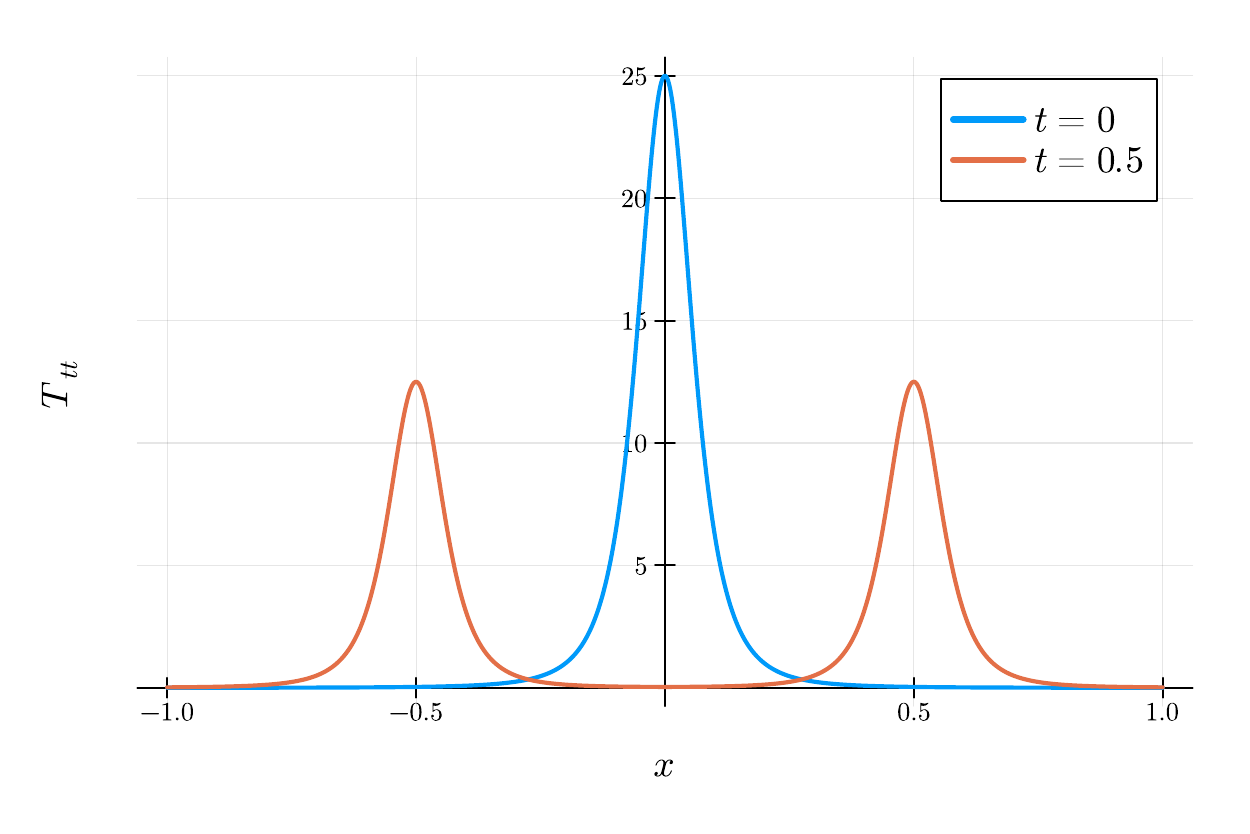}
    \includegraphics[width=0.45\linewidth]{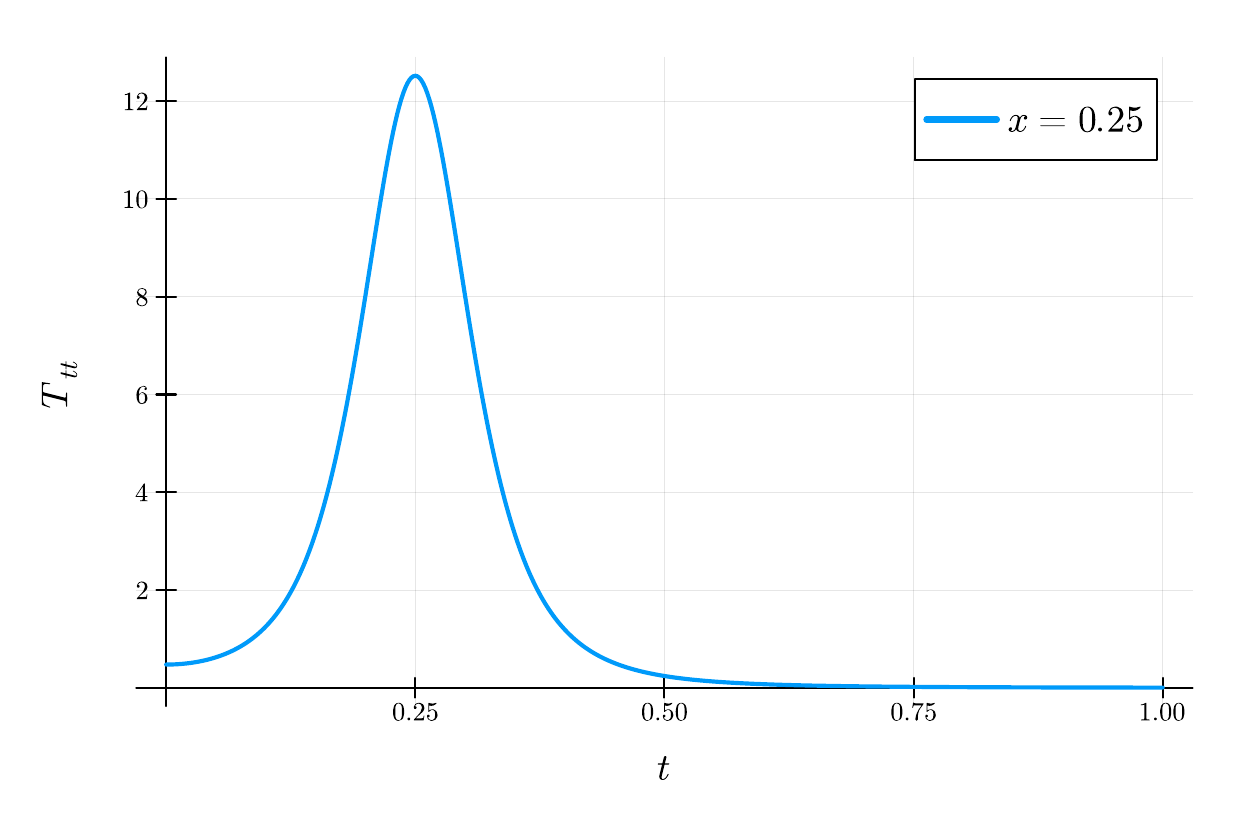}
    \caption{Plots of the energy stress tensor \(T_{tt}(x, t)\) under a joining quench.
    The left/right/bottom row represents the spatiotemporal/spatial/temporal dependence.
    We chose \(\delta=0.1\) and \(c=1\).}
    \label{fig:EStensor_JQJQ}
\end{figure}

For the joining quenched state, the conformal map reads \(w=z+\delta^2 / (4z)\), we obtain
\begin{align}
    T(w) &= \frac{c}{8} \cdot \frac{\delta^2 z^4}{\left(z^2-\delta^2 / 4\right)^4},\notag\\
    \bar{T}(\bar{w}) &= \frac{c}{8} \cdot \frac{\delta^2 \bar{z}^4}{\left(\bar{z}^2-\delta^2 / 4\right)^4}.
\end{align}
It is straightforward to confirm that \(T_{tt}=T(w)+\bar{T}(\bar{w})\) is always real-valued and localized at \(x=\pm t\). 
This profile is depicted in Fig.~\ref{fig:EStensor_JQJQ}.

\begin{figure}
    \centering
    \includegraphics[width=0.45\linewidth]{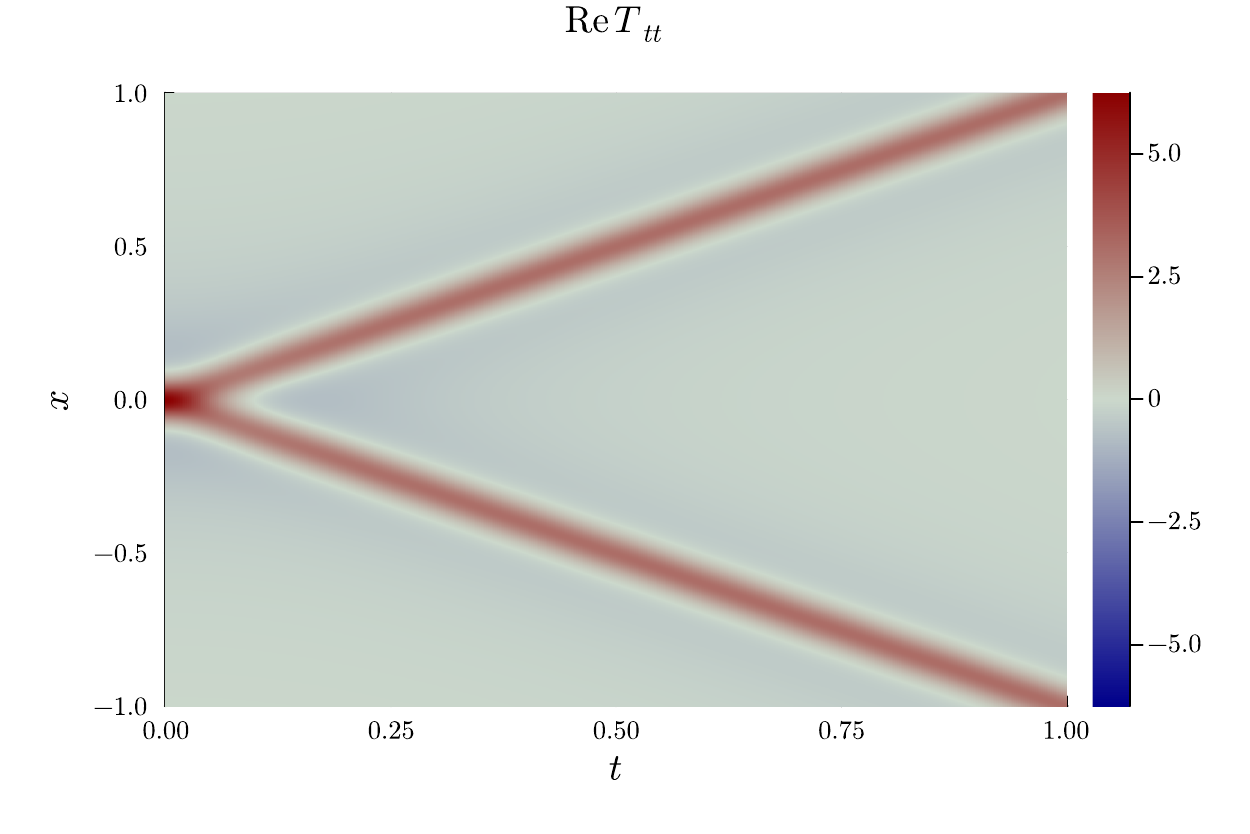}
    \includegraphics[width=0.45\linewidth]{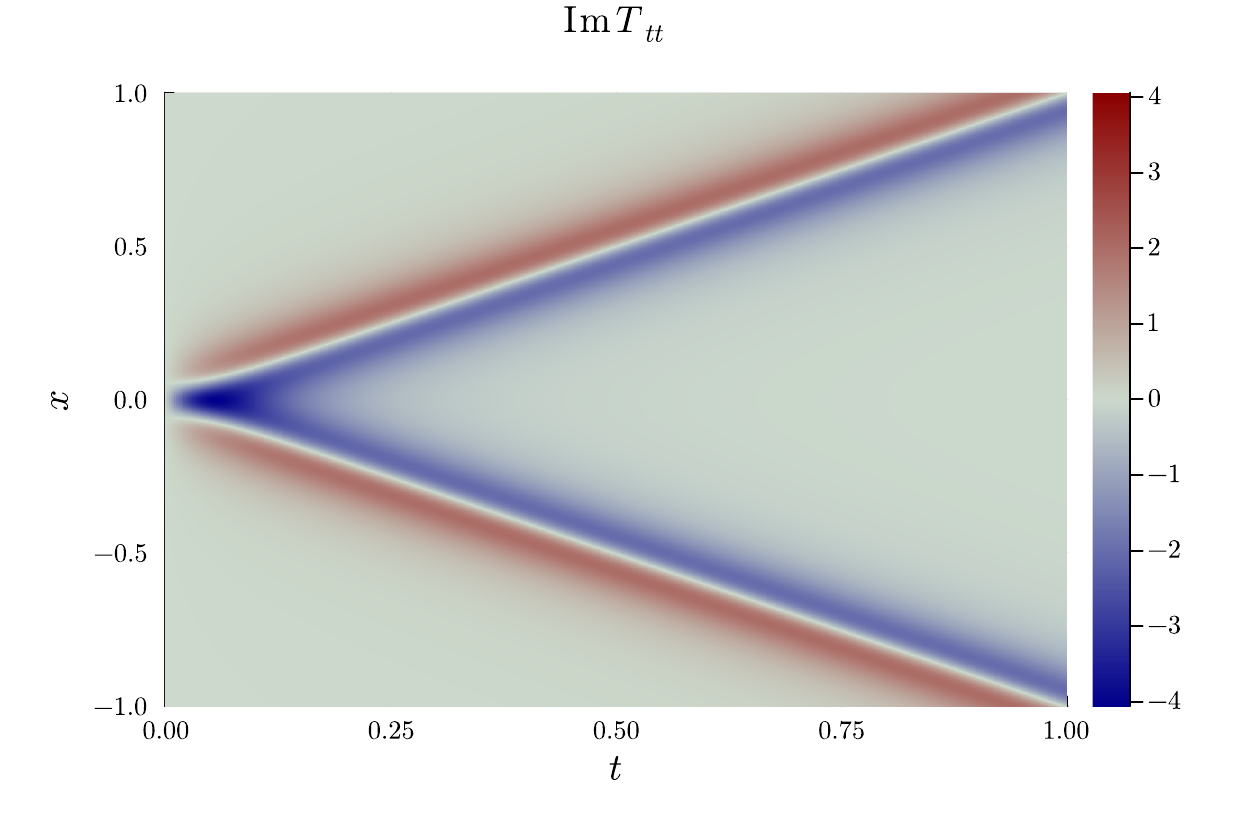}
    \includegraphics[width=0.45\linewidth]{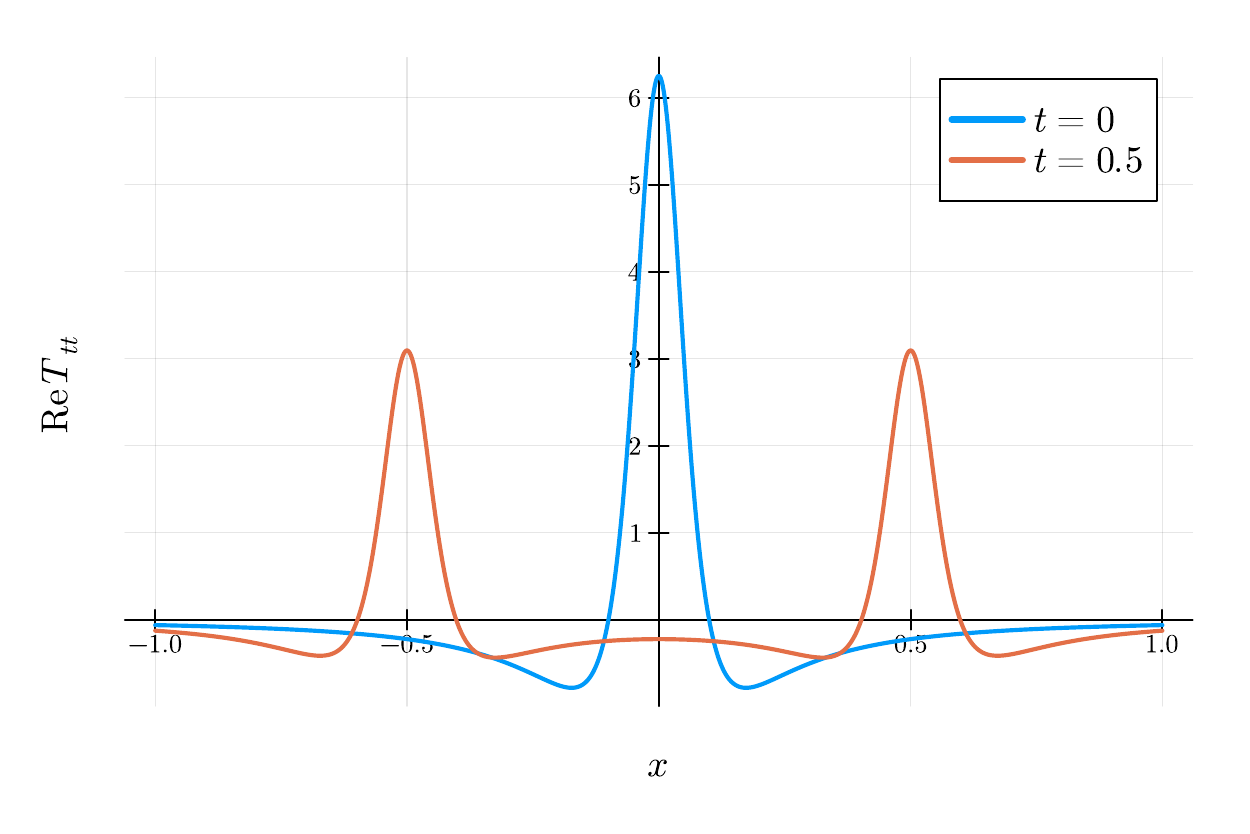}
    \includegraphics[width=0.45\linewidth]{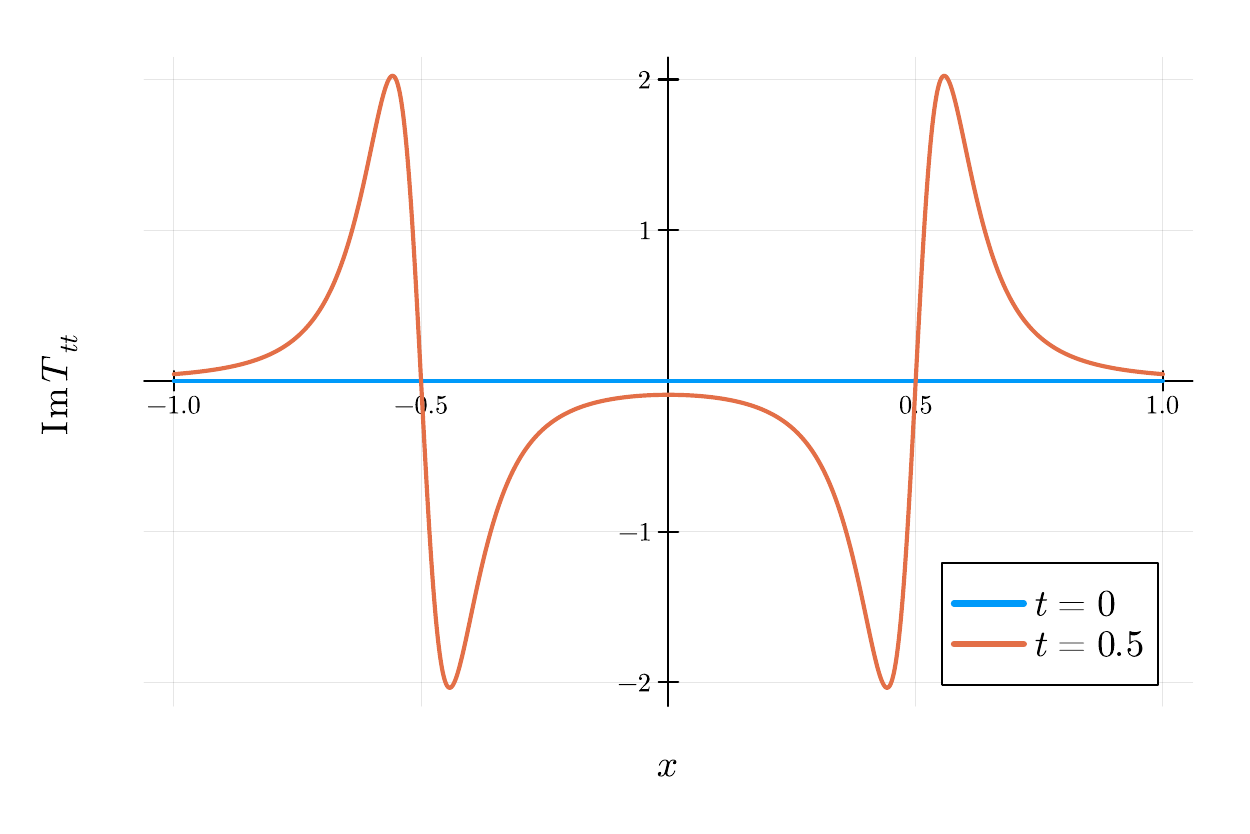}
    \includegraphics[width=0.45\linewidth]{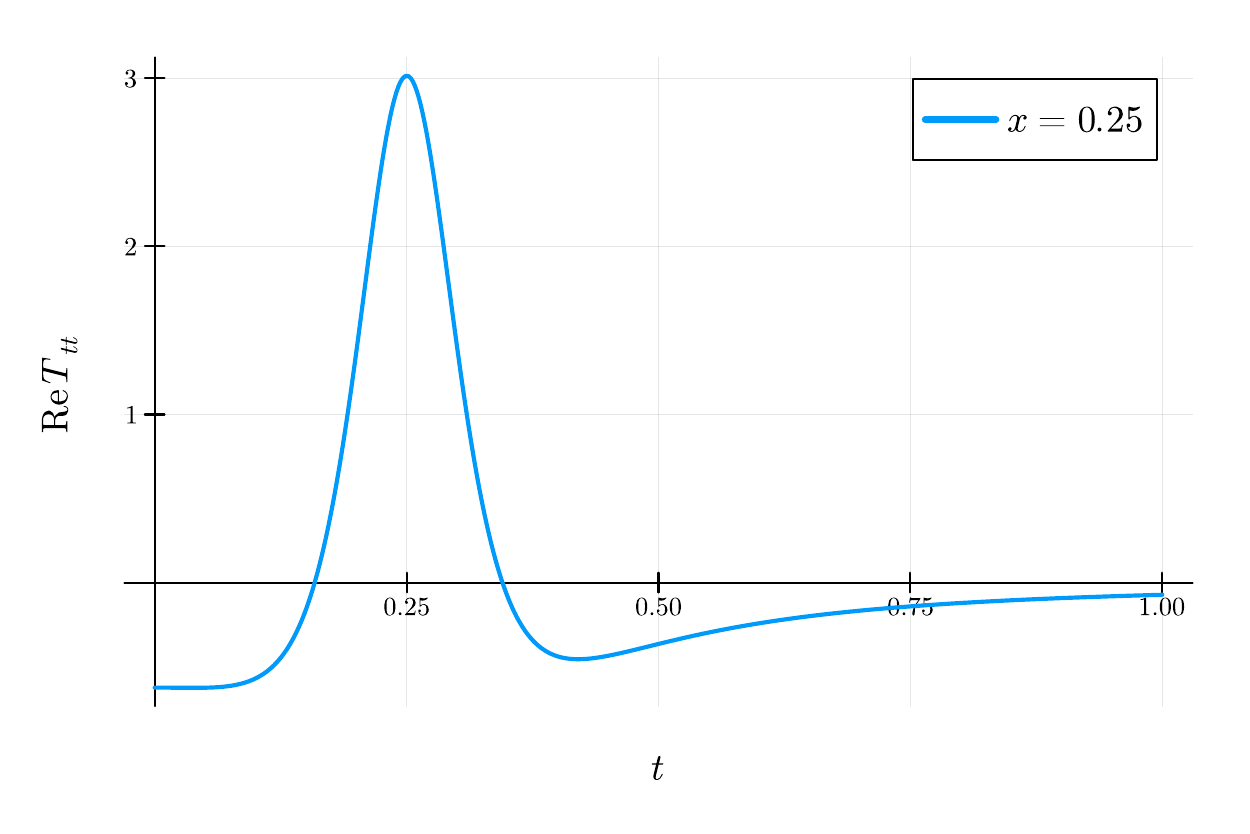}
    \includegraphics[width=0.45\linewidth]{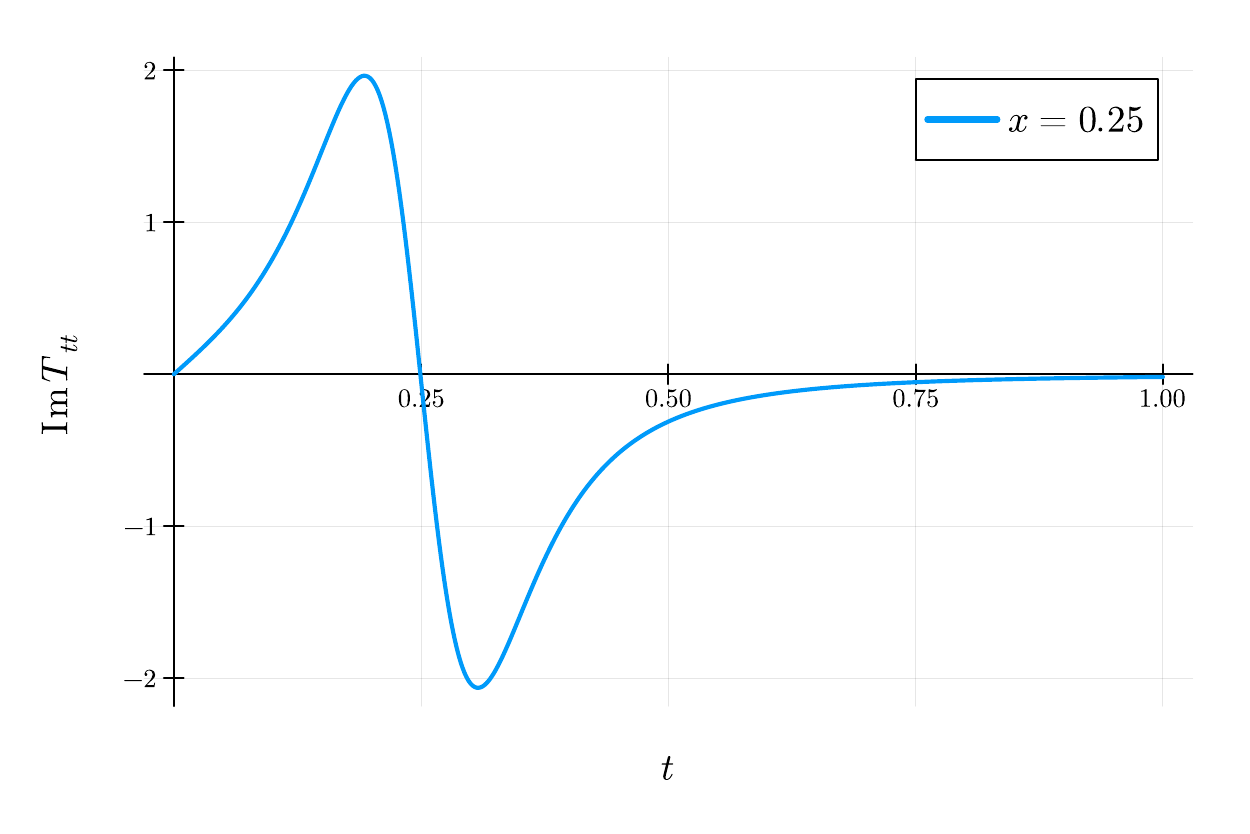}
    \caption{The real/imaginary part of energy stress tensor \(T_{tt}(x, t)\) for the single-slit setup (i.e., for the joining quenched state and the ground state) is shown in the left/right column.
    The top/middle/bottom row represents the spatiotemporal/spatial/temporal dependence.
    We chose \(\delta=0.1\) and \(c=1\).}
    \label{fig:EStensor_JQGS}
\end{figure}

For the single-slit geometry, where the conformal map reads \(z^2=-w-\delta\), we obtain
\begin{align}
    T(w) &= \frac{c}{32} \cdot \frac{1}{(w + \delta)^2}, \notag\\
    \bar{T}(\bar{w}) &= \frac{c}{32} \cdot \frac{1}{(\bar{w} + \delta)^2}.
\end{align}
Under the time evolution, we set \(w=it+ix\) and \(\bar{w}=it-ix\). 
Thus the energy density \(T_{tt}\) reads
\begin{equation}
    T_{tt} = T(w)+\bar{T}(\bar{w}) = -\frac{c}{32} \cdot \left[\frac{1}{(t+x-i\delta)^2} + \frac{1}{(t-x-i\delta)^2} \right].
\end{equation}
This profile is depicted in Fig.~\ref{fig:EStensor_JQGS}. 
It shows that the real part has a dip, a negative contribution, in addition to the peak at \(x=|t|\). 
The propagation at the speed of light looks analogous to what we found in the time evolution of pseudo entropy. 

\clearpage
\section{Entanglement entropy under joining quenches}\label{sec:ee}
In this section, we take the opposing-slit geometry, \(\ket{\psi} = \ket{\vp} = \ket{\JQ(x=0)}\), and consider the entanglement entropy of joining quenches at \(x=0\).
Many authors have studied this type of joining local quenches since the pioneering work in \cite{Calabrese:2007mtj}.
Hence, in the following, we reanalyze certain aspects of entanglement entropy that are helpful for comparison with our subsequent pseudo entropy calculations. 
We will also study the behavior of mutual information to manifest the difference between the holographic CFT and the free Dirac fermion CFT (see also \cite{Asplund:2014} for a relevant analysis).

\subsection{Entanglement entropy for joint subsystems}
\begin{figure}[ttt]
    \centering
    \includegraphics[width=0.45\linewidth]{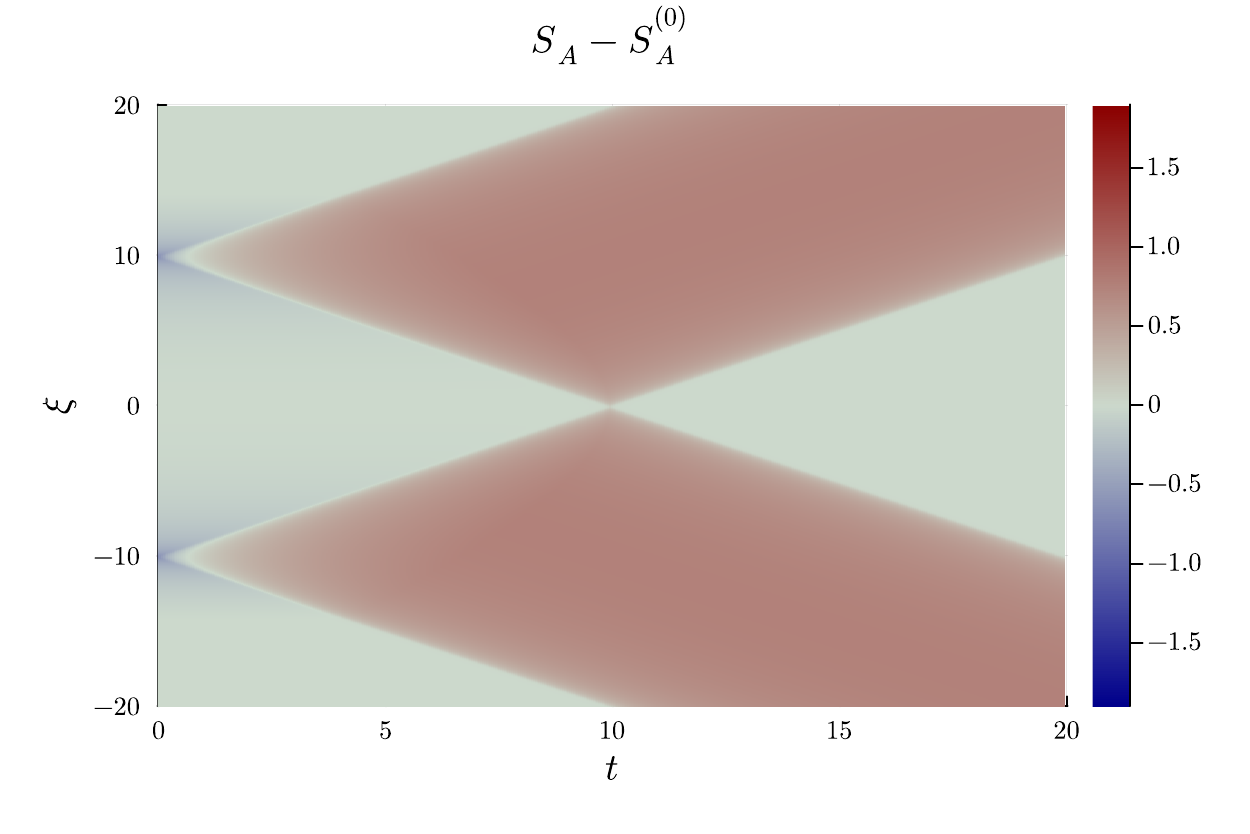}
    \includegraphics[width=0.45\linewidth]{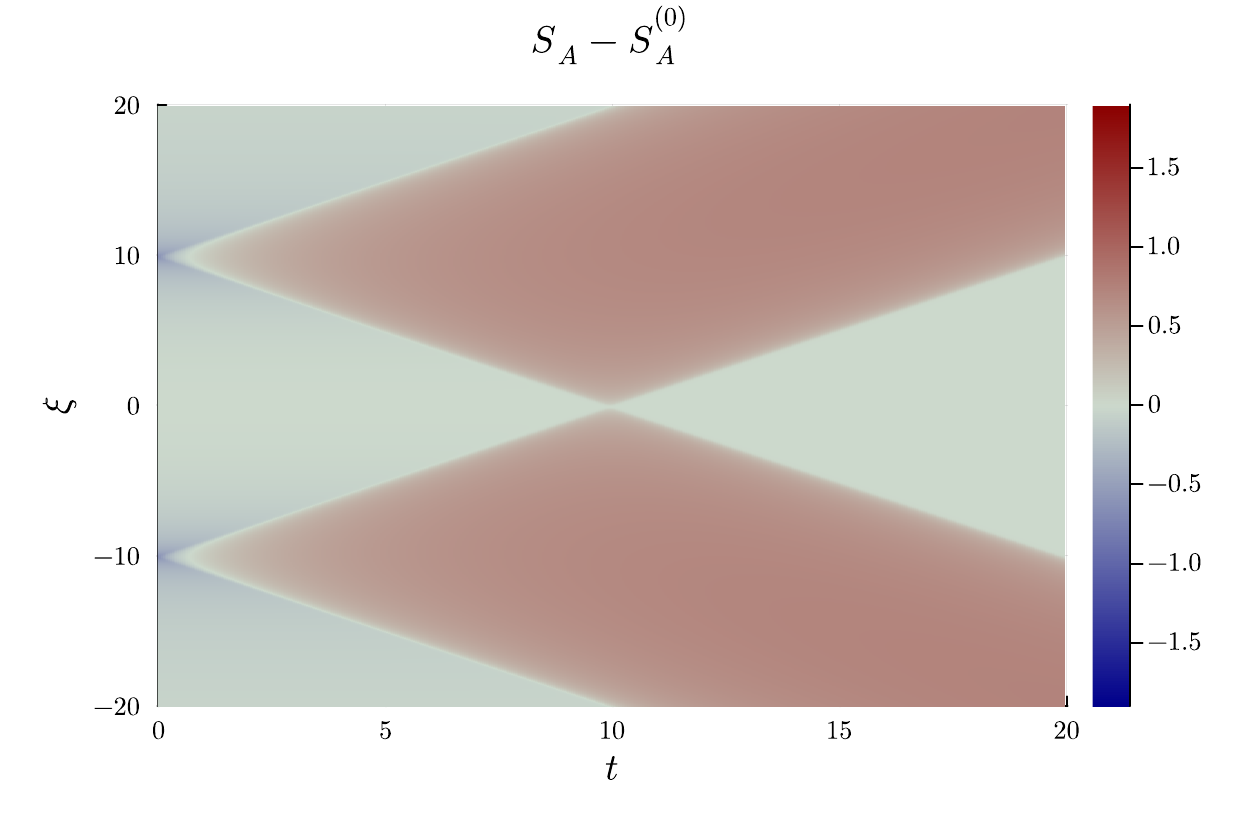}
    \includegraphics[width=0.45\linewidth]{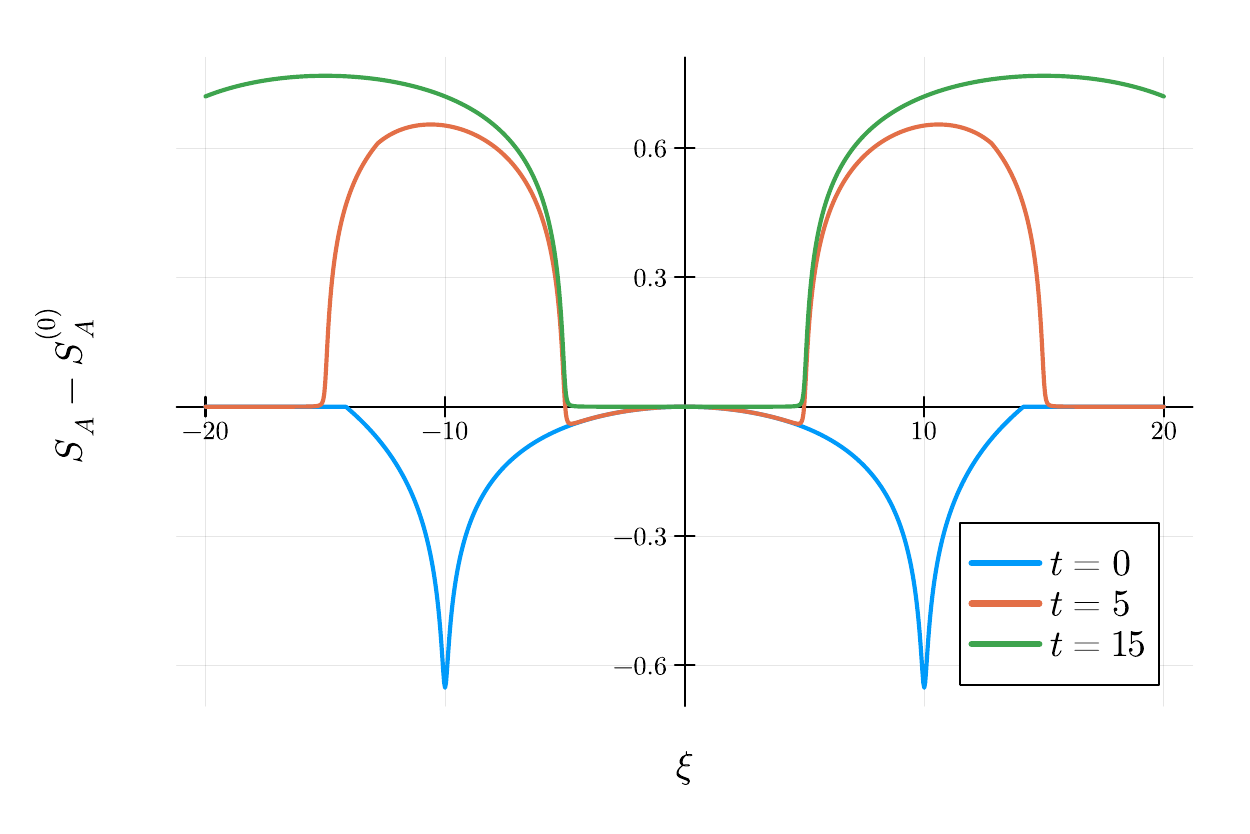}
    \includegraphics[width=0.45\linewidth]{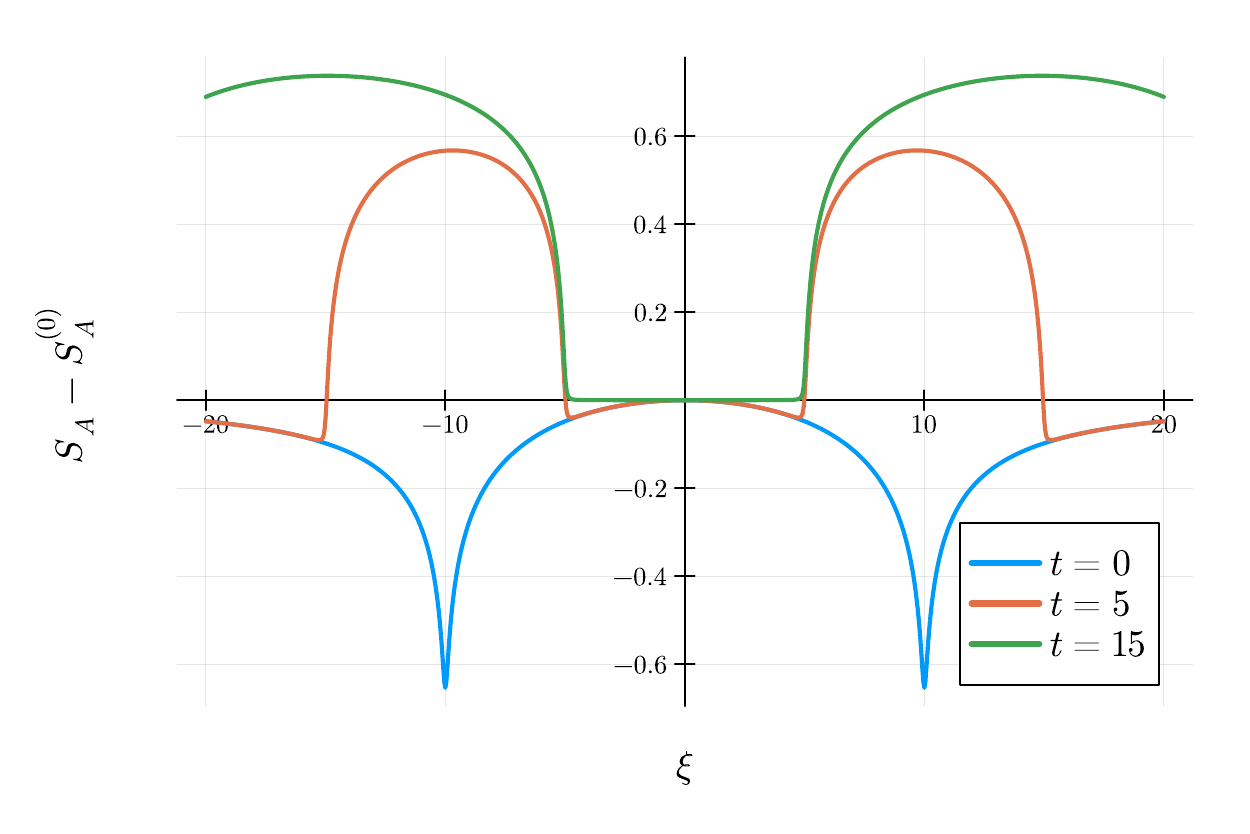}
    \includegraphics[width=0.45\linewidth]{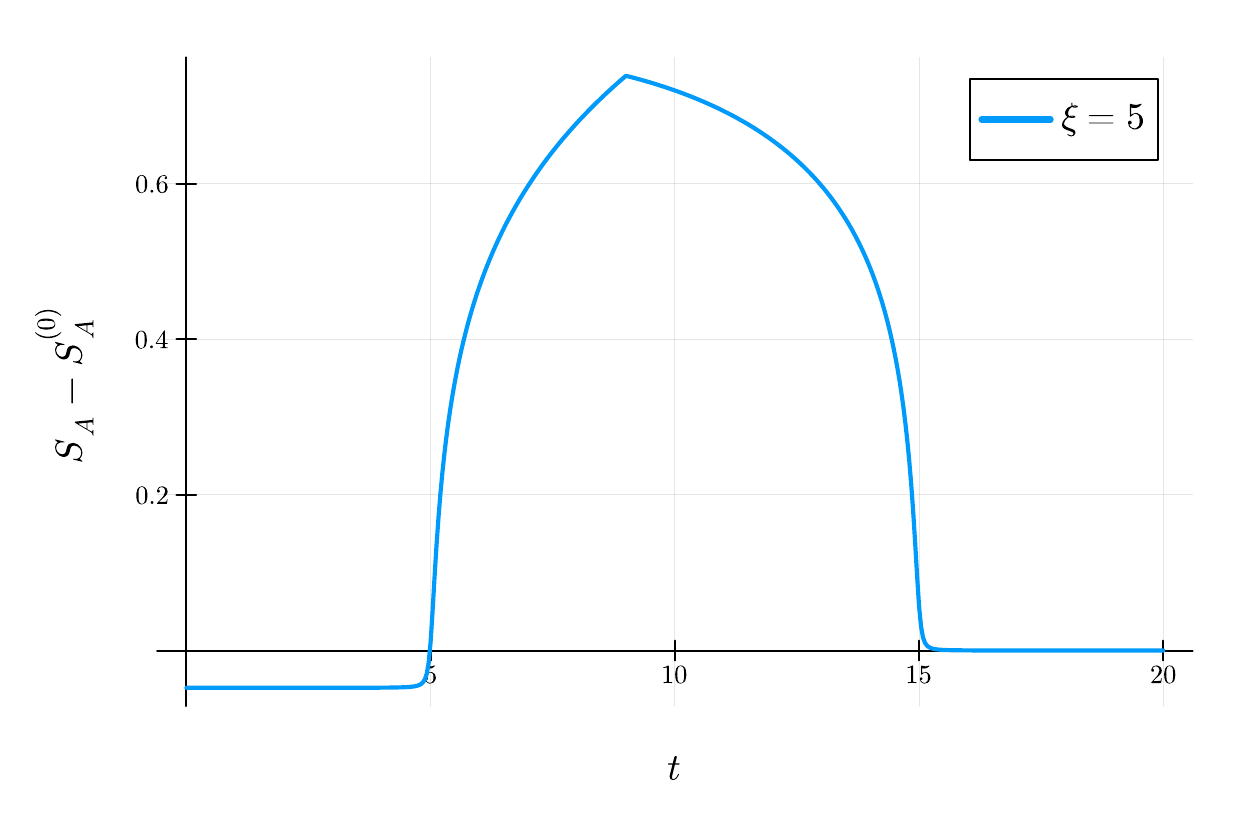}
    \includegraphics[width=0.45\linewidth]{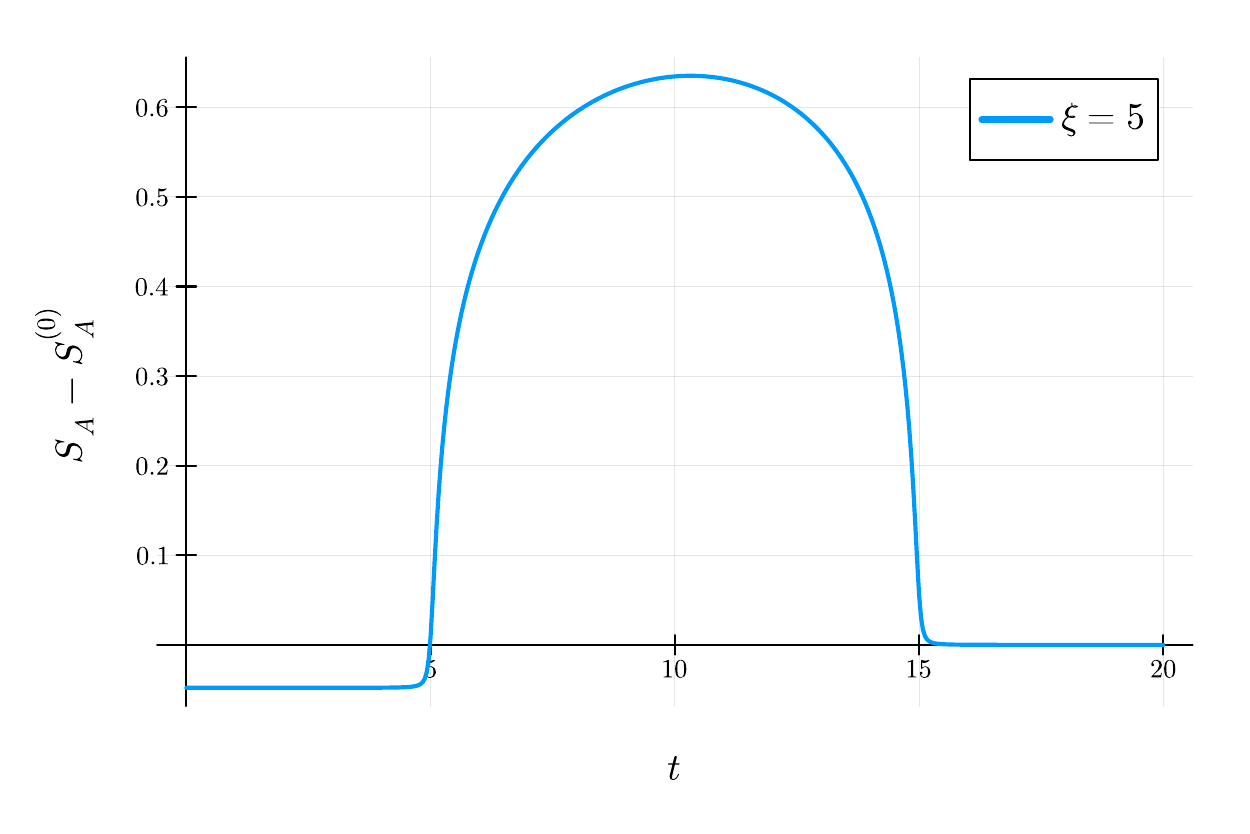}
    \caption{The entanglement entropy \(S_{A(\x)}(t) - S_{A(\x)}^{(0)}\) under joining quenches in the holographic CFT/the free Dirac fermion CFT for \(A(\x) = [\x - 10, \x + 10]\) is shown in the left/right column.
    The top/middle/bottom row represents the spatiotemporal/spatial/temporal dependence.
    We chose \(c=1\), \(\ep=1\), \(\d=0.1\), and \(S_{\mathrm{bdy}} = 0\).}
    \label{fig:EE}
\end{figure}

We take the subsystem \(A\) to be the interval
\begin{equation}
    A(\x) = [\x - L / 2, \x + L / 2]
\end{equation}
and analyze entanglement entropy for \(A\) as a function of \(\xi\), \(L\) and time \(t\). 
We obtain the entanglement entropy in the holographic CFT by selecting the smaller contribution, as indicated by the formula \eqref{S_hol}. 
For plots of each contribution, refer to appendix \ref{ap:condis}.
Together with the free Dirac fermion CFT result, we plotted the result of entanglement entropy in Fig.~\ref{fig:EE}.

The qualitative behavior depicted in Fig.~\ref{fig:EE} highlights a fundamental feature of the picture: relativistic propagation of entangled pairs, which is consistent in both holographic and free Dirac fermion CFTs. 
Initially, the joining quench creates entangled pairs localized at \(x=0\).
One of the quasiparticles consisting of the pair propagates at the speed of light from the left to the right while the other moves opposite.
The entanglement entropy increases when only one member of these pairs enters subsystem \(A\).
The derivative of the entanglement entropy is discontinuous in the holographic CFT, which arises from the phase transition between the connected and disconnected contributions.

Explaining this result from the gravity dual via the AdS/BCFT is also helpful. 
The behavior of the connected contribution \(S_{A(\x)}^{\mathrm{con}}(t)\) can be understood from the shock wave geometry, which is triggered by the gravitational quench at \(t=x=0\). 
As the gravitational shock waves propagate outward, the backreaction to the metric is suppressed, reducing the minimal area.
The disconnected contribution \(S_{A(\x)}^{\mathrm{dis}}(t)\) diverges in the limit of \(|\x| \to \infty\).
In the context of AdS/BCFT, this divergence of \(S_{A(\x)}^{\mathrm{dis}}(t)\) is explained by that of the minimum geodesic length connecting the endpoints of \(A(\x)\) to the EOW brane \cite{Ugajin:2013xxa,Shimaji:2018czt}.
The divergence of \(S_{A(\x)}^{\mathrm{dis}}(t)\) in the limit of \(t \to \infty\) is also explained in the context of AdS/BCFT.
The EOW brane is heavy, and as time passes, it falls toward \(z\to\infty\) in the extra dimension, almost at the speed of light. 
Consequently, also in this case, the minimum geodesic length connecting the endpoints of \(A(\x)\) to the EOW brane diverges.

\subsection{Entanglement entropy and mutual information for disjoint subsystems}
We take a disjoint subsystem \(A(\x) \cup B(\x)\), where 
\begin{equation}
    A(\x) = [\x - D/2 - L, \x - D/2], \quad B(\x) = [\x + D/2, \x + D/2 + L].
\end{equation}

\begin{figure}
    \centering
    \includegraphics[width=0.45\linewidth]{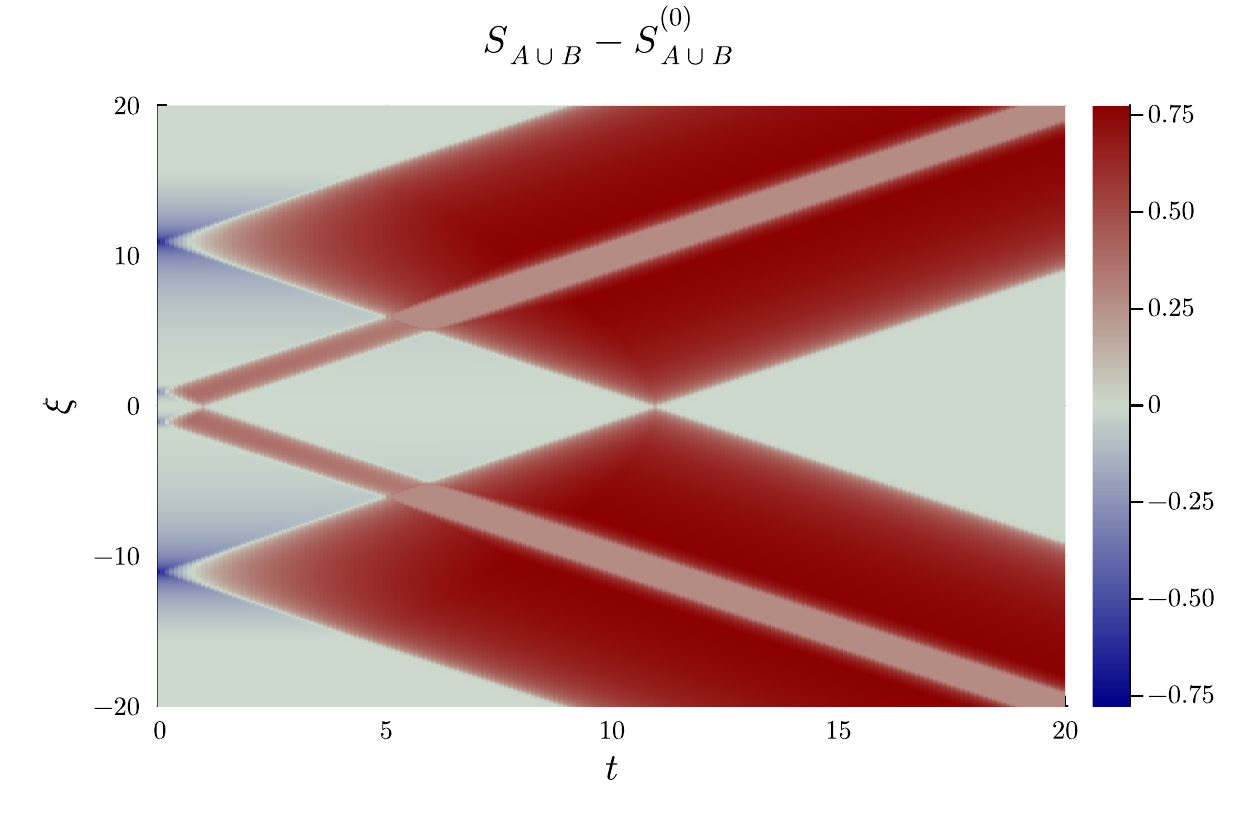}
    \includegraphics[width=0.45\linewidth]{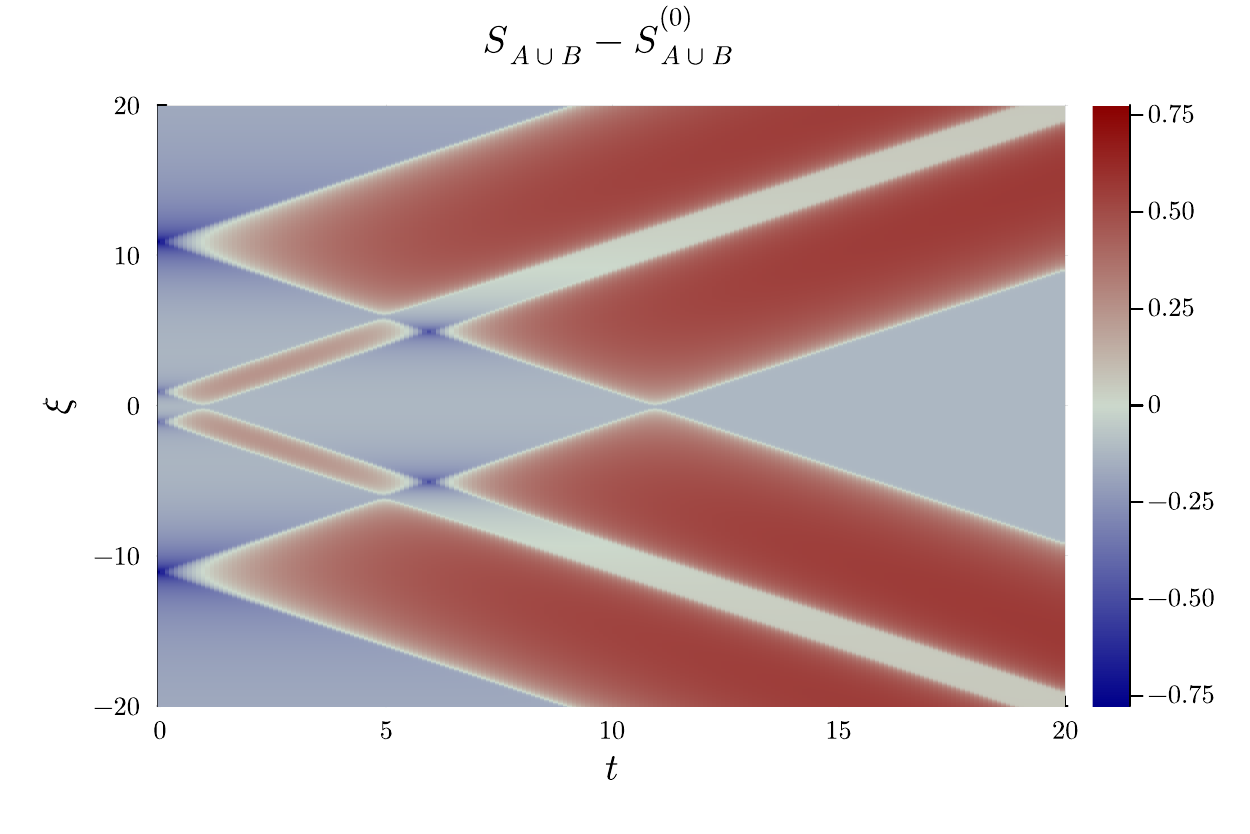}
    \includegraphics[width=0.45\linewidth]{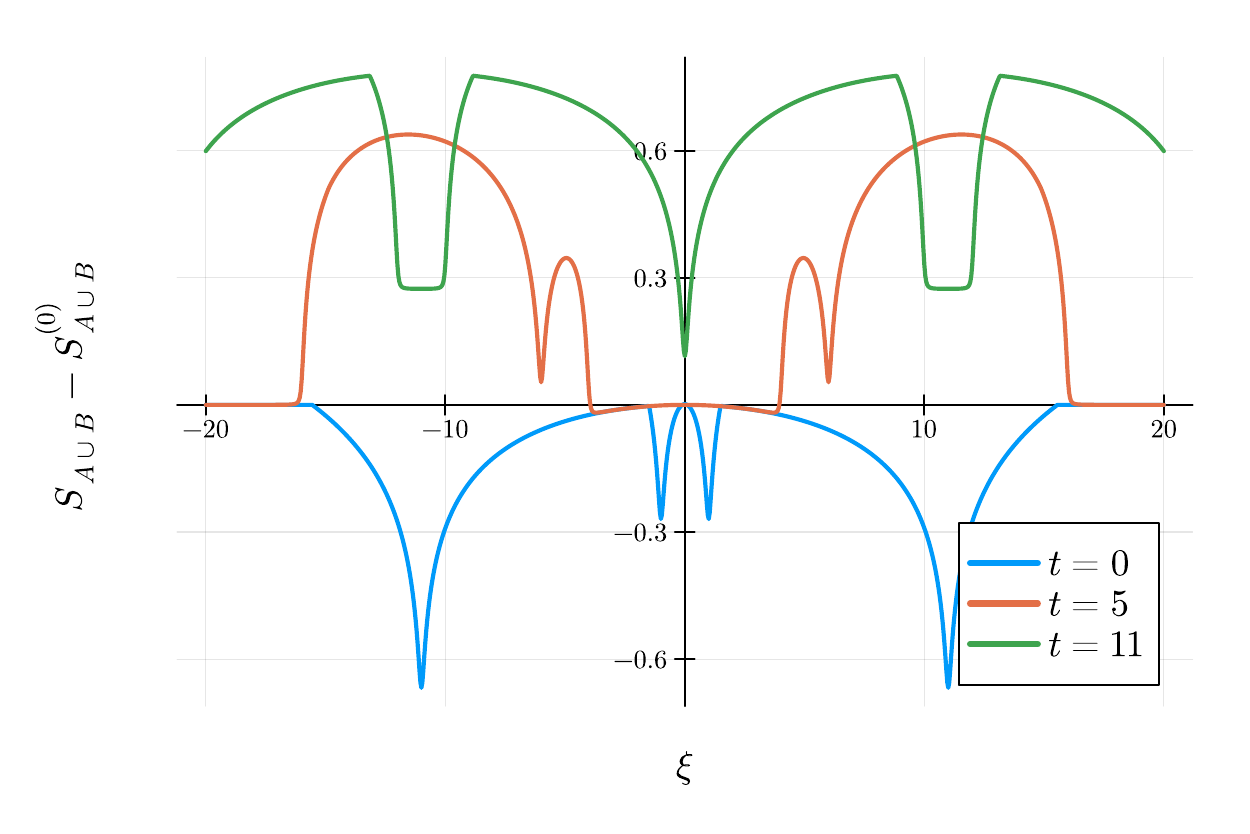}
    \includegraphics[width=0.45\linewidth]{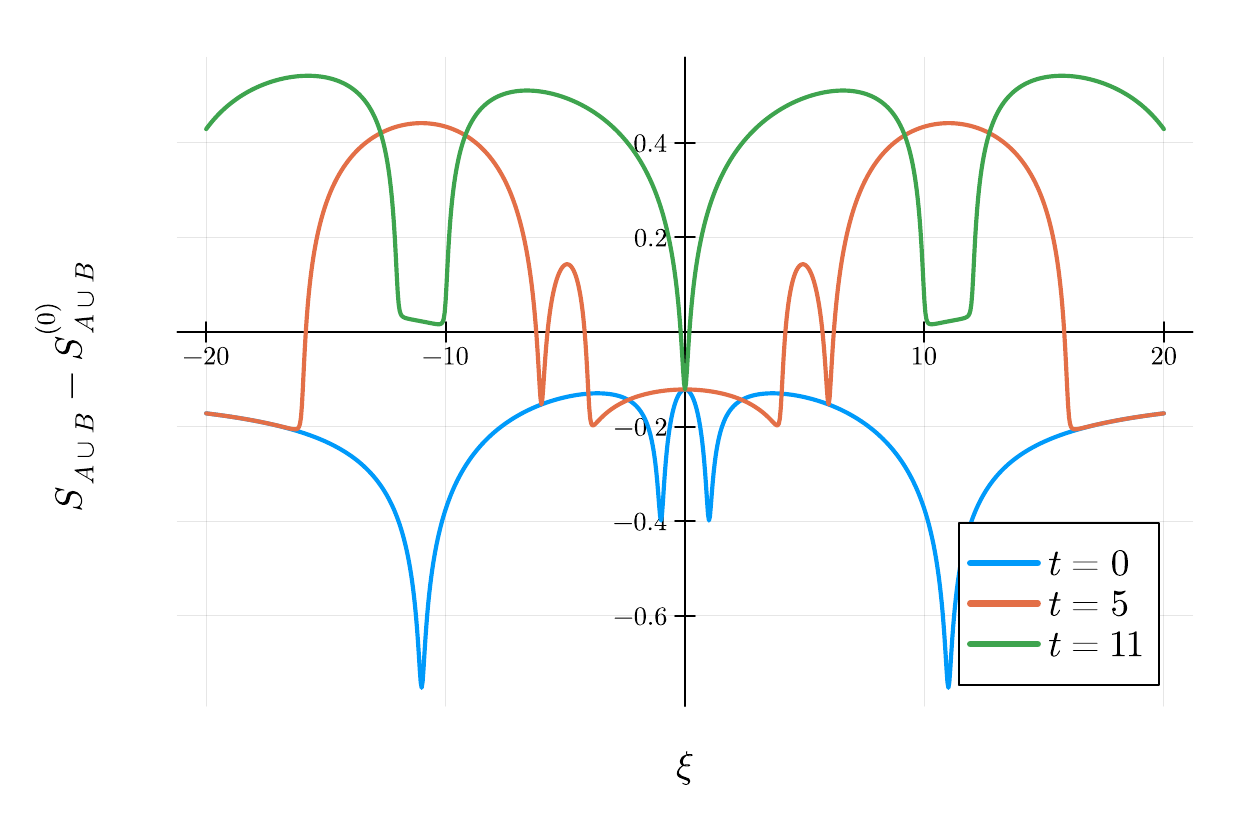}
    \includegraphics[width=0.45\linewidth]{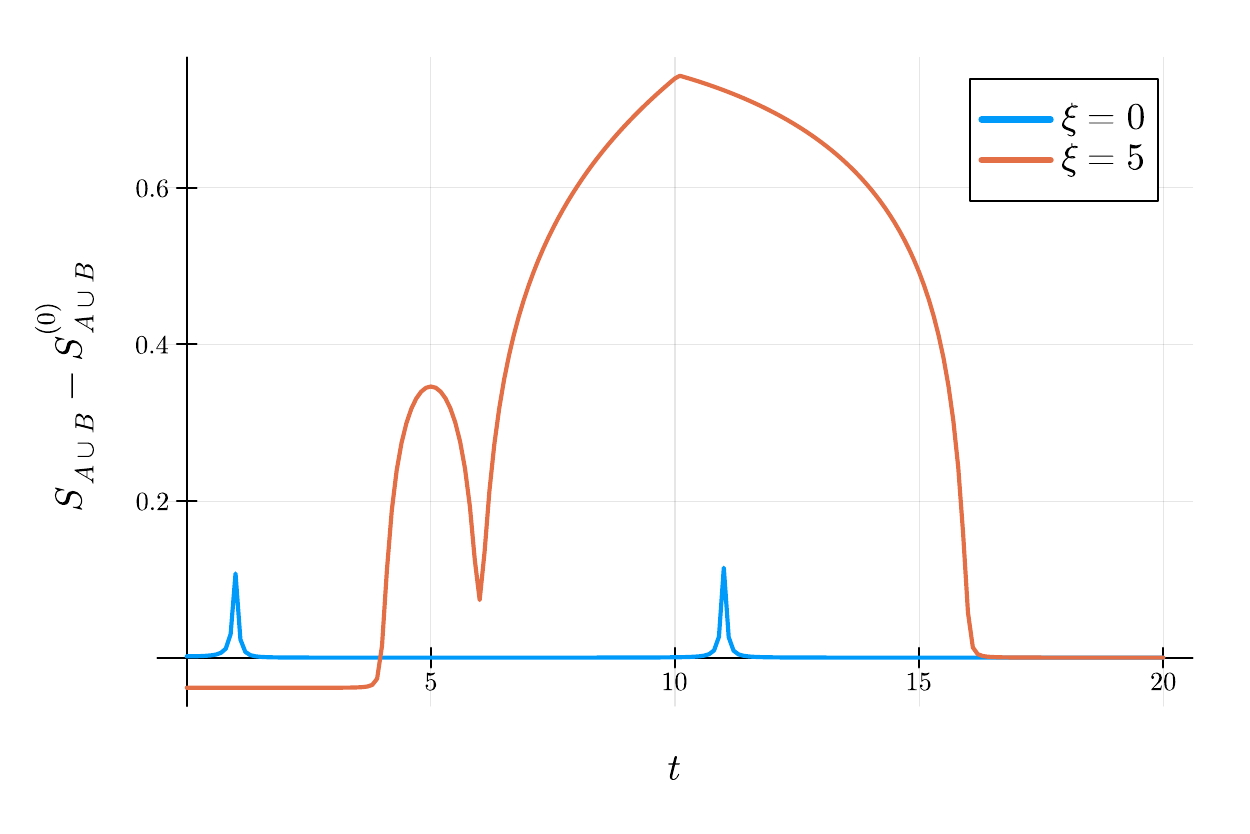}
    \includegraphics[width=0.45\linewidth]{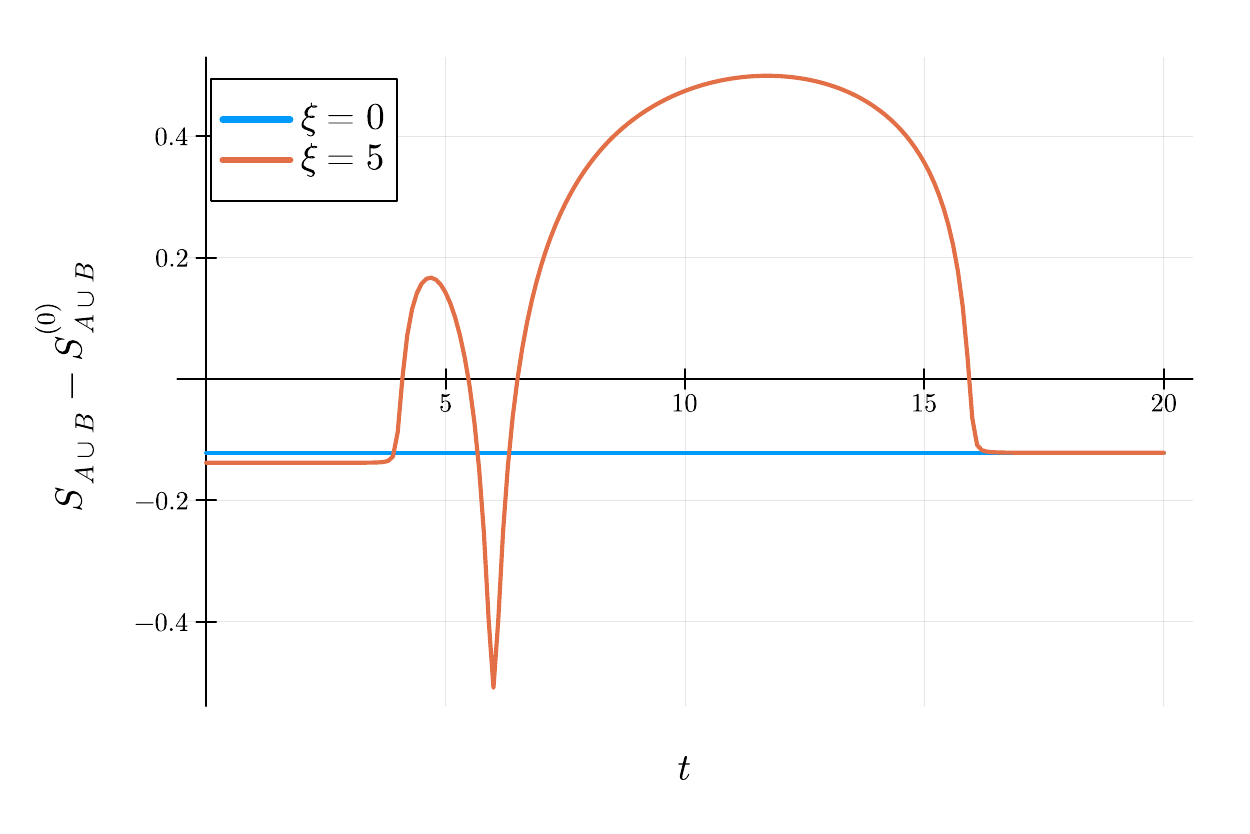}
    \caption{The entanglement entropy \(S_{A(\x) \cup B(\x)}(t) - S_{A(\x) \cup B(\x)}^{(0)}\) in the holographic CFT/the free Dirac fermion CFT for \(A(\x) = [\x - 11, \x - 1], B(\x) = [\x + 1, \x + 11]\) is shown in the left/right column.
    The top/middle/bottom row represents the spatiotemporal/spatial/temporal dependence.
    We chose \(c=1\), \(\ep=1\), \(\d=0.1\), and \(S_{\text{bdy}}=0\).}
    \label{fig:disjEE_L10}
\end{figure}

\begin{figure}
    \centering
    \includegraphics[width=0.45\linewidth]{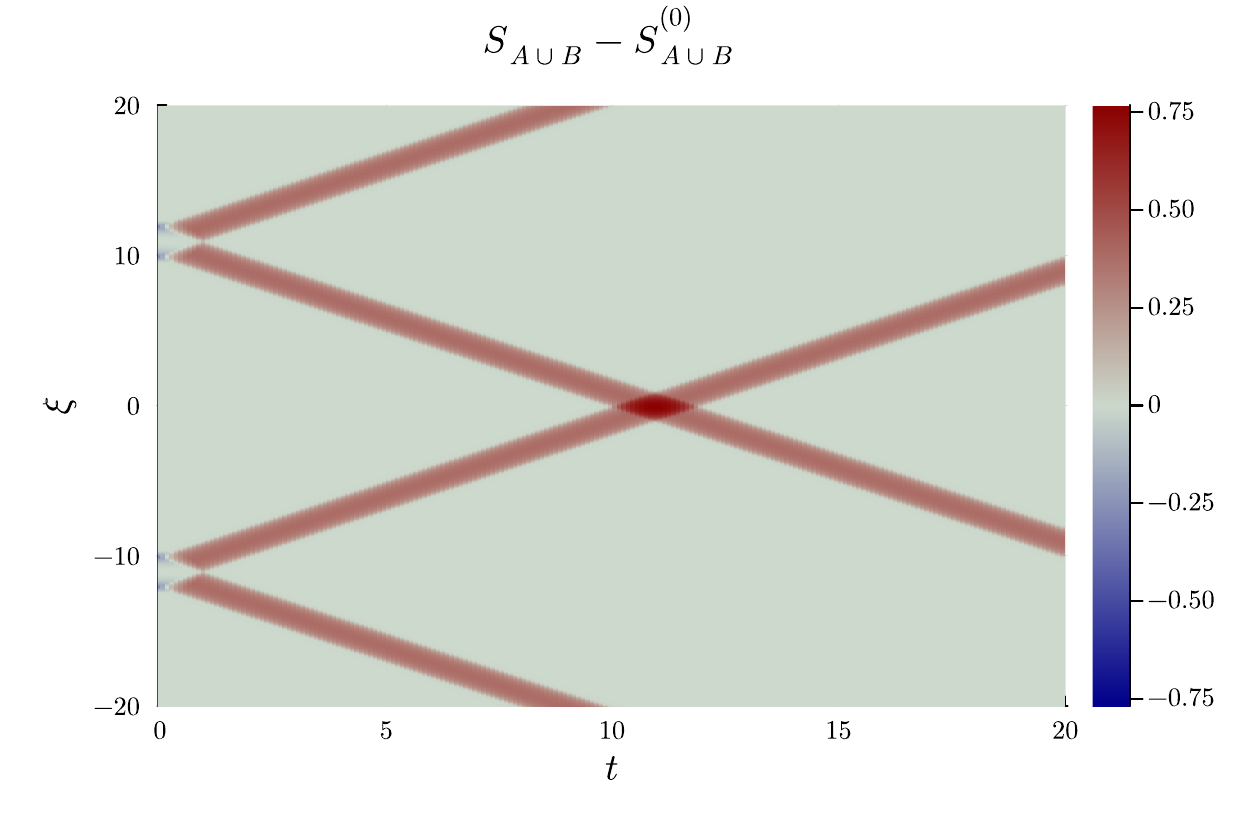}
    \includegraphics[width=0.45\linewidth]{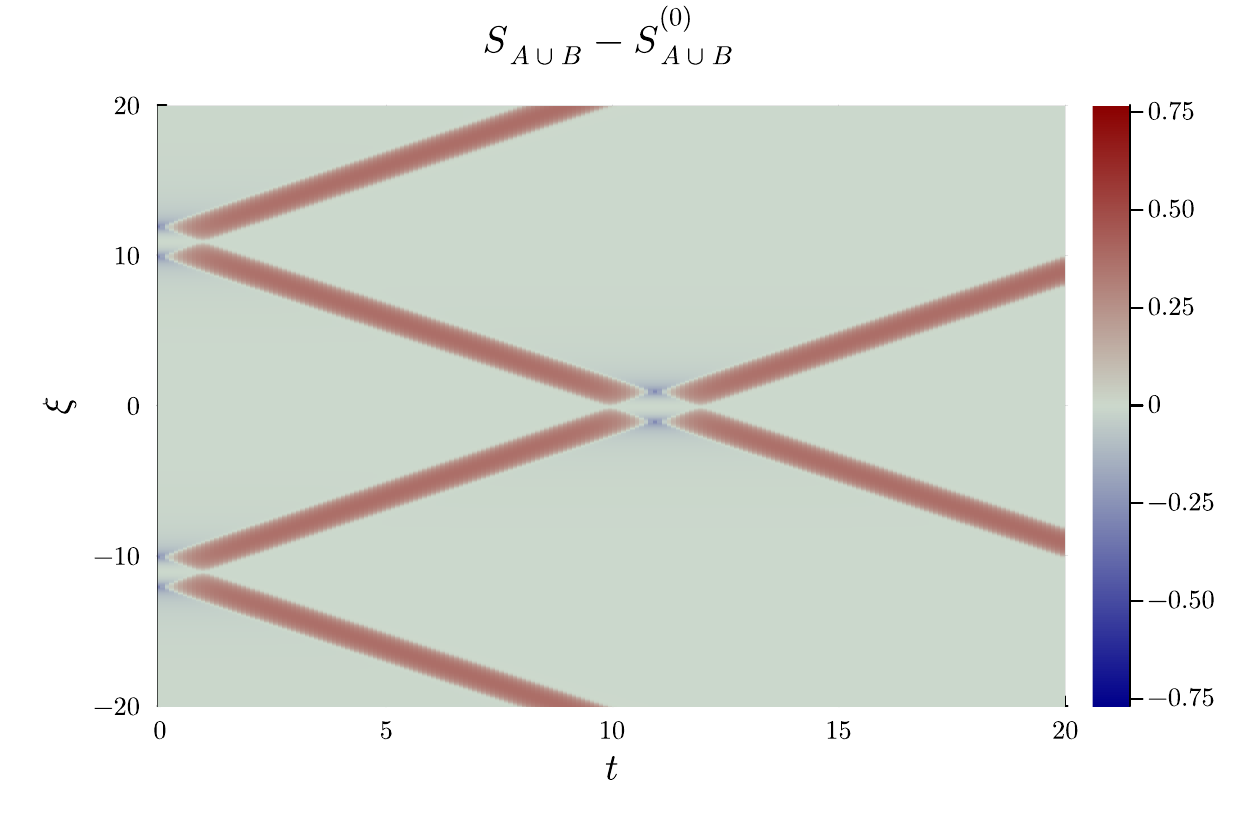}
    \includegraphics[width=0.45\linewidth]{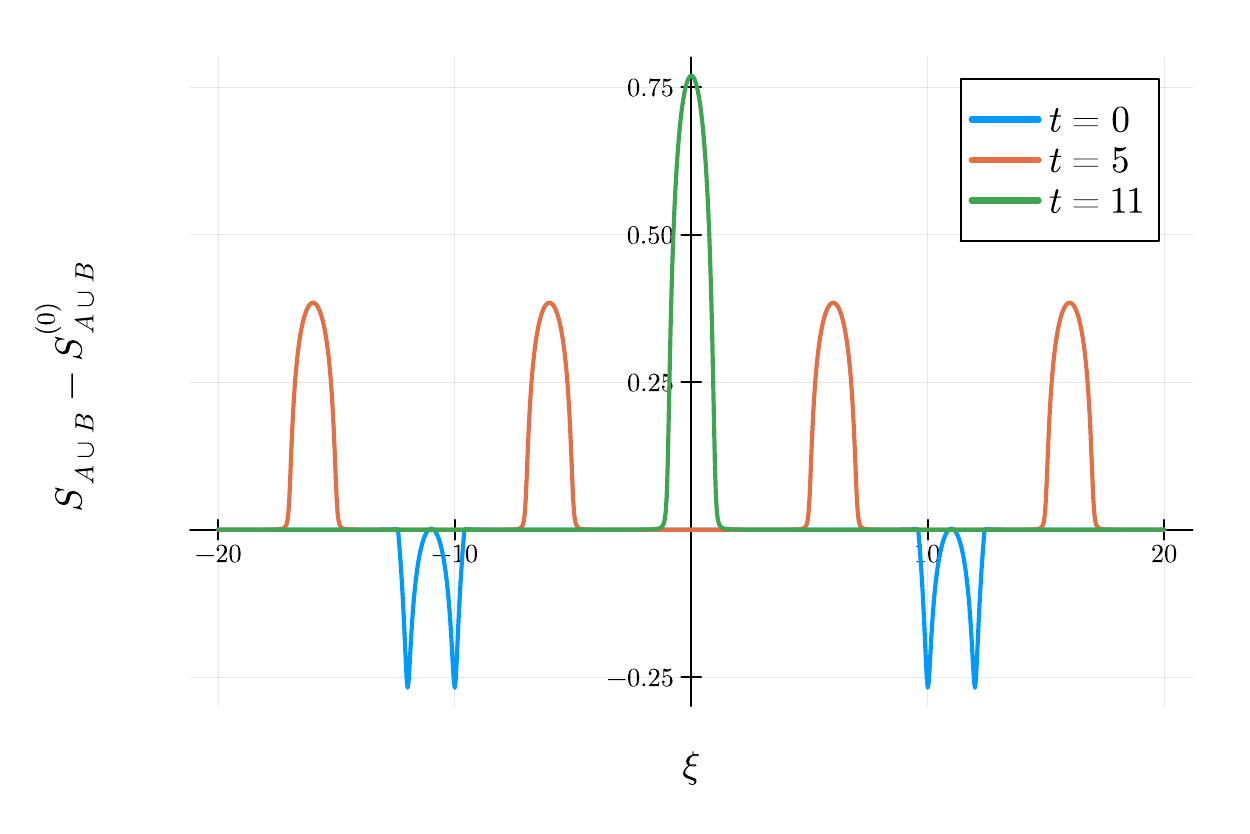}
    \includegraphics[width=0.45\linewidth]{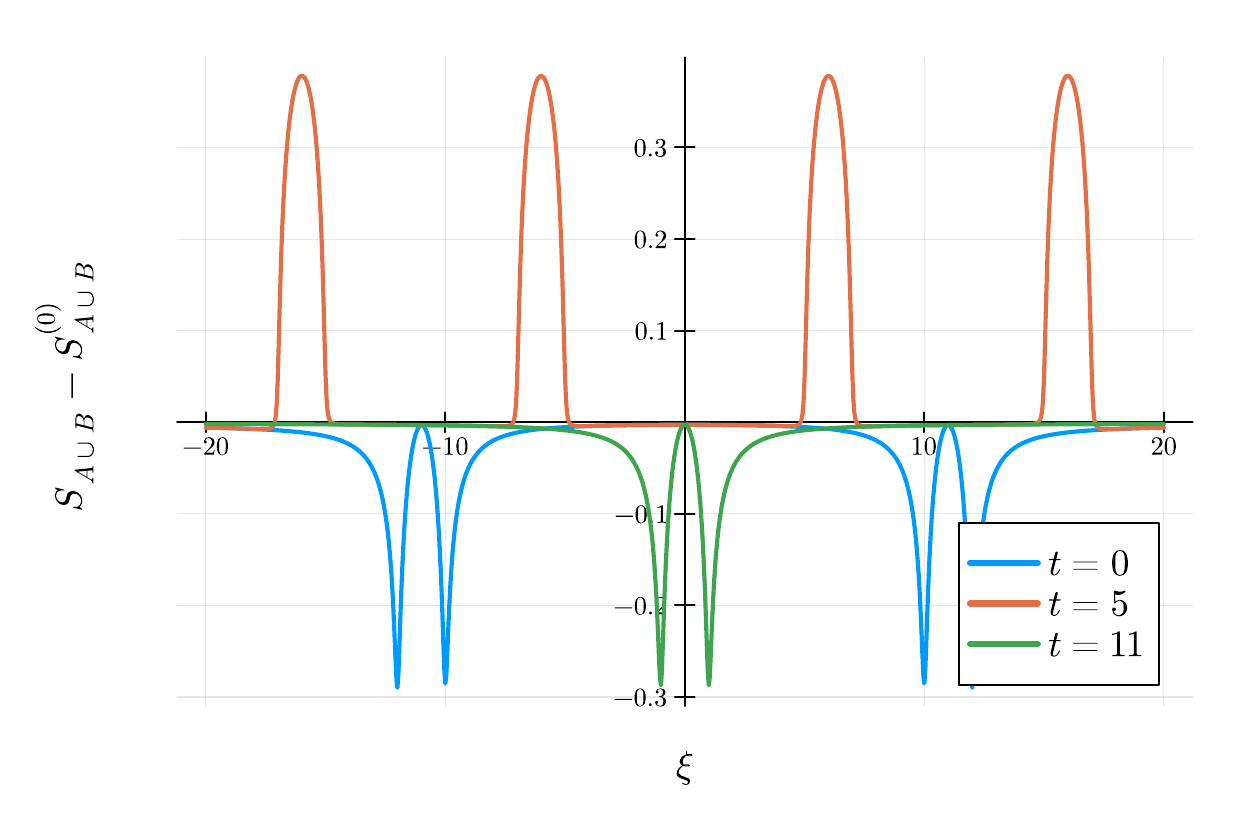}
    \includegraphics[width=0.45\linewidth]{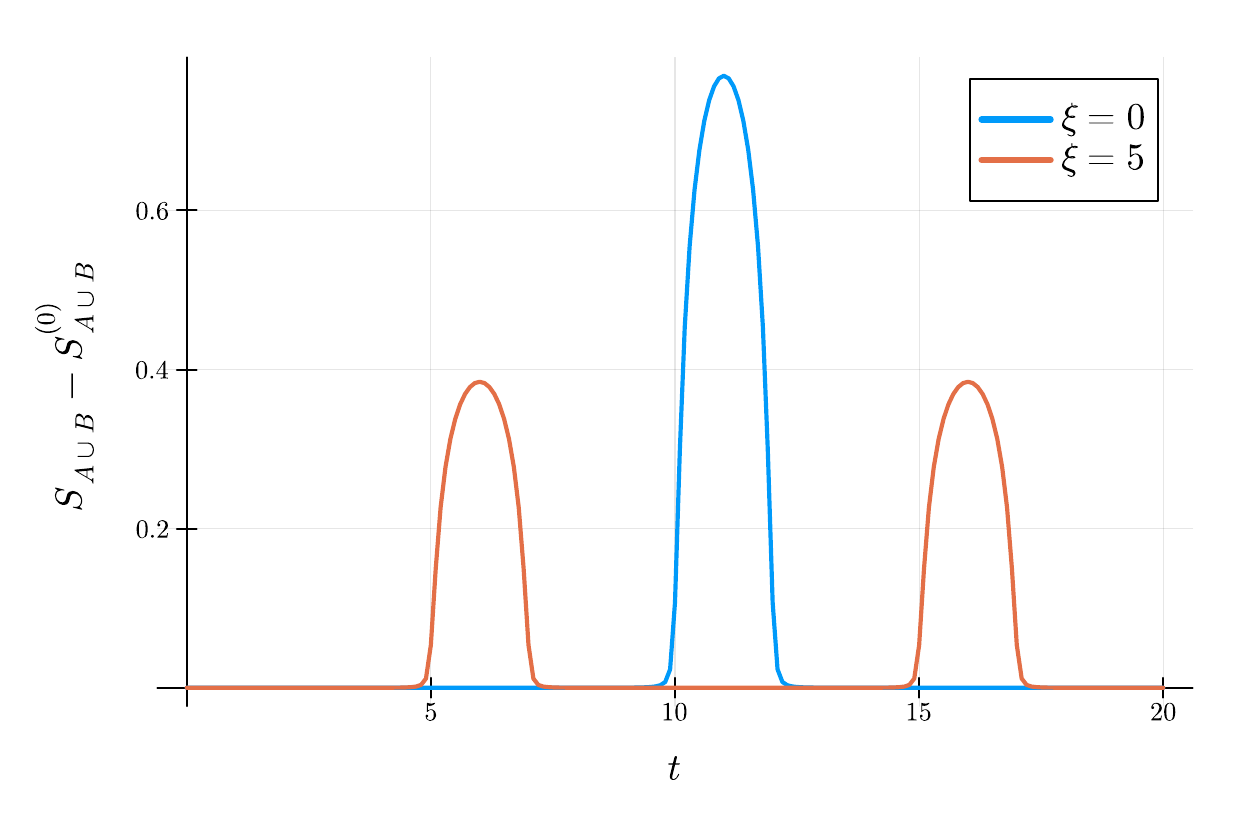}
    \includegraphics[width=0.45\linewidth]{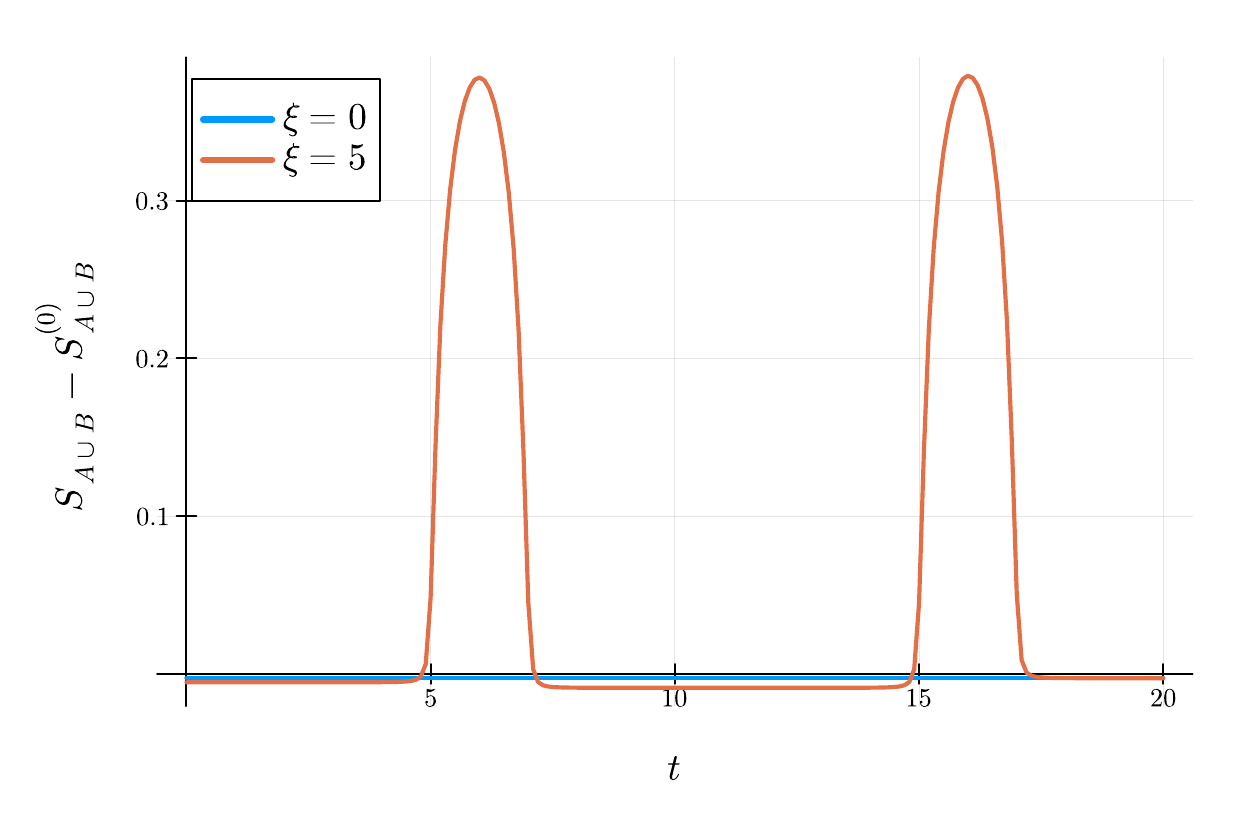}
    \caption{The entanglement entropy \(S_{A(\x) \cup B(\x)}(t) - S_{A(\x) \cup B(\x)}^{(0)}\) in the holographic CFT/the free Dirac fermion CFT for \(A(\x) = [\x - 12, \x - 10], B(\x) = [\x + 10, \x + 12]\) is shown in the left/right column.
    The top/middle/bottom row represents the spatiotemporal/spatial/temporal dependence.
    We chose \(c=1\), \(\ep=1\), \(\d=0.1\), and \(S_{\text{bdy}}=0\).}
    \label{fig:disjEE_L2}
\end{figure}

In Fig.~\ref{fig:disjEE_L10} (\(L=10, D=2\)) and Fig.~\ref{fig:disjEE_L2} (\(L=2, D=20\)), we show plots of entanglement entropy for subsystem \(A(\x) \cup B(\x)\) in the holographic CFT and the free Dirac fermion CFT.
When approximating the entanglement entropy in the holographic CFT using the asymptotic form of two-point functions, we consider all possible ways of contracting eight twist operators, including the mirror images. 
We extend \eqref{S_hol} to two intervals and select the one that yields the minimum contribution.
We obtain the result for the free Dirac fermion CFT by directly evaluating the four-point function of twist operators.

The difference between the results in the two CFTs is particularly pronounced near the region where two inward-propagating waves intersect in Fig.~\ref{fig:disjEE_L2}.
In the context of AdS/BCFT, the results in the holographic CFT are explained by considering the shock wave propagations to the minimal surface \(\Gamma_A\) for \(A\) and \(\Gamma_B\) for \(B\).
This effect is enhanced when each of the two shock waves simultaneously reaches \(\Gamma_A\) and \(\Gamma_B\).
Conversely, in the case of the free Dirac fermion CFT, the results can be explained using the quasi-particle picture. 
The entanglement entropy gets canceled in this region because both quasi-particles of entangled pairs exist within subsystem \(A(\x)\cup B(\x)\). 
This signifies that the quasi-particle picture is not entirely correct for the holographic CFTs, as first pointed out in the context of global quantum quenches \cite{Asplund:2015} (see also \cite{Asplund:2014} for local quenches).

\begin{figure}
    \centering
    \includegraphics[width=0.45\linewidth]{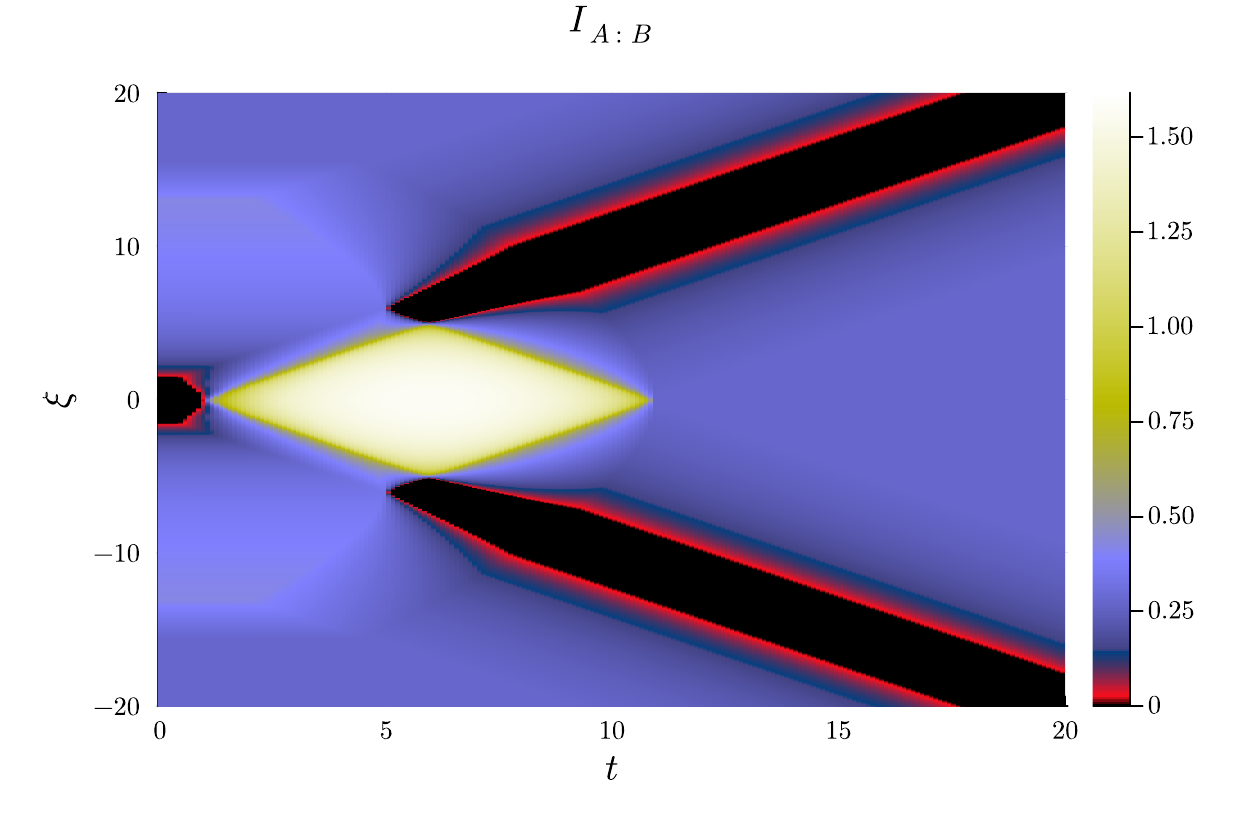}
    \includegraphics[width=0.45\linewidth]{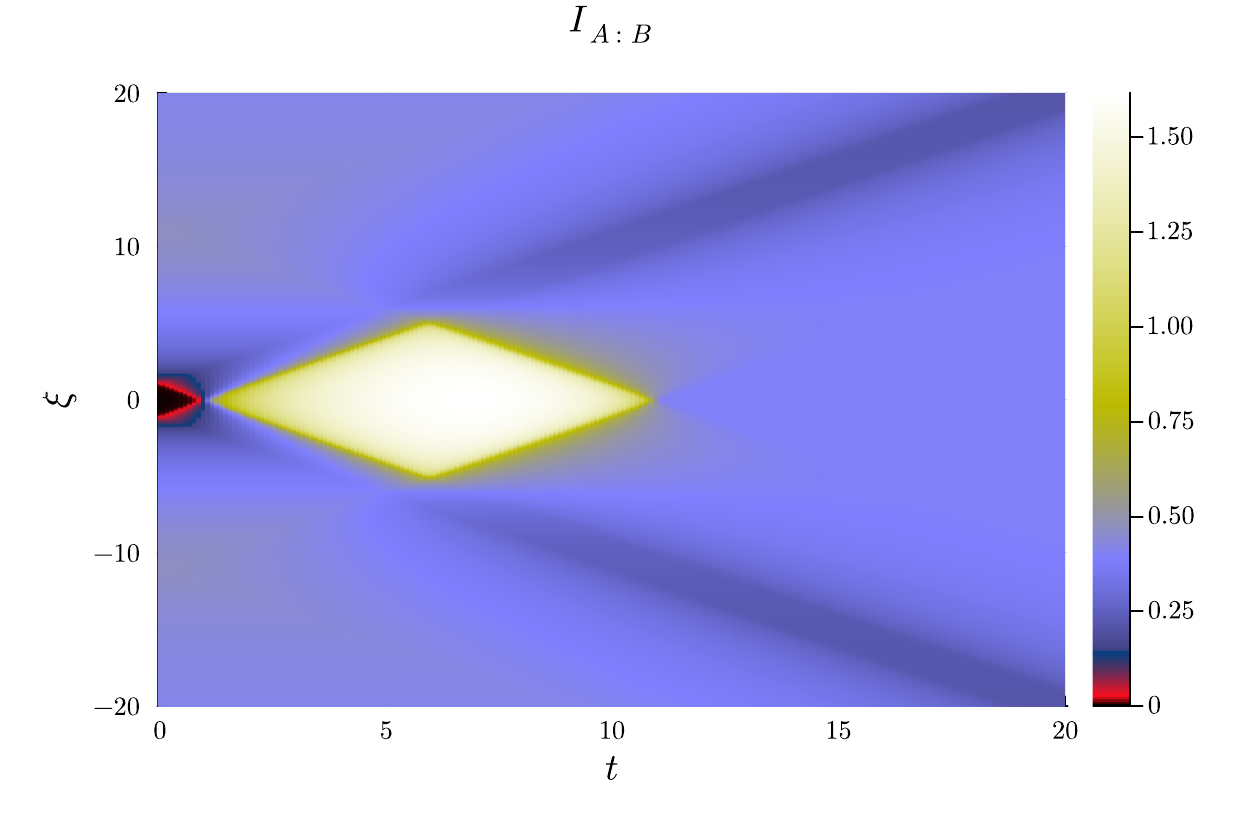}
    \includegraphics[width=0.45\linewidth]{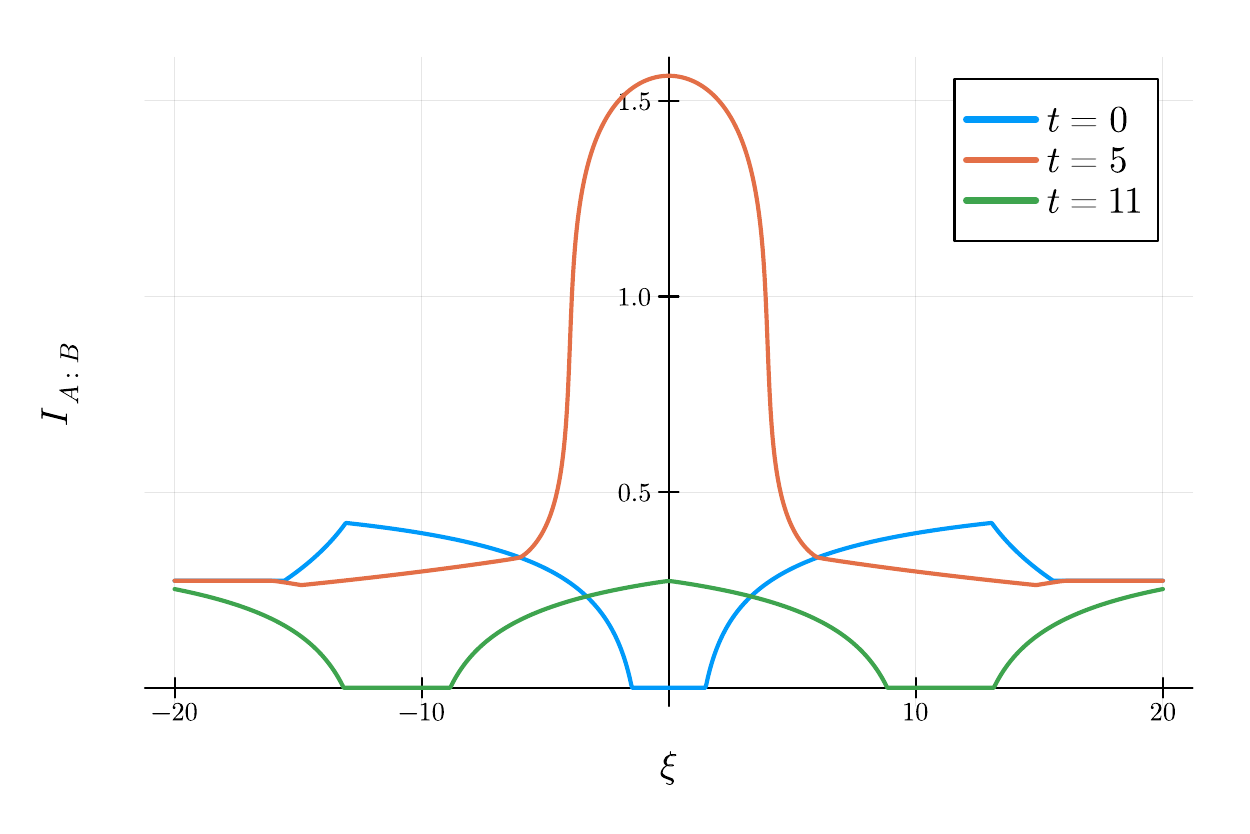}
    \includegraphics[width=0.45\linewidth]{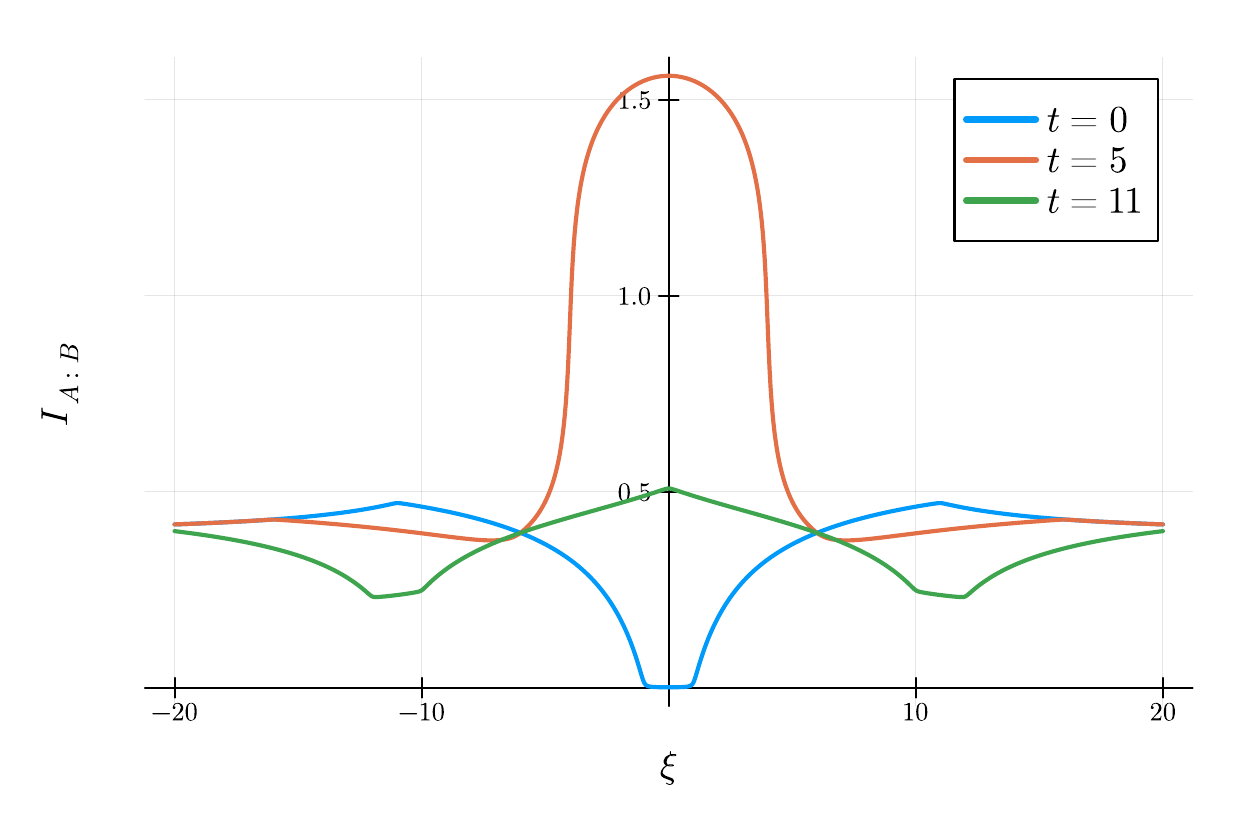}
    \includegraphics[width=0.45\linewidth]{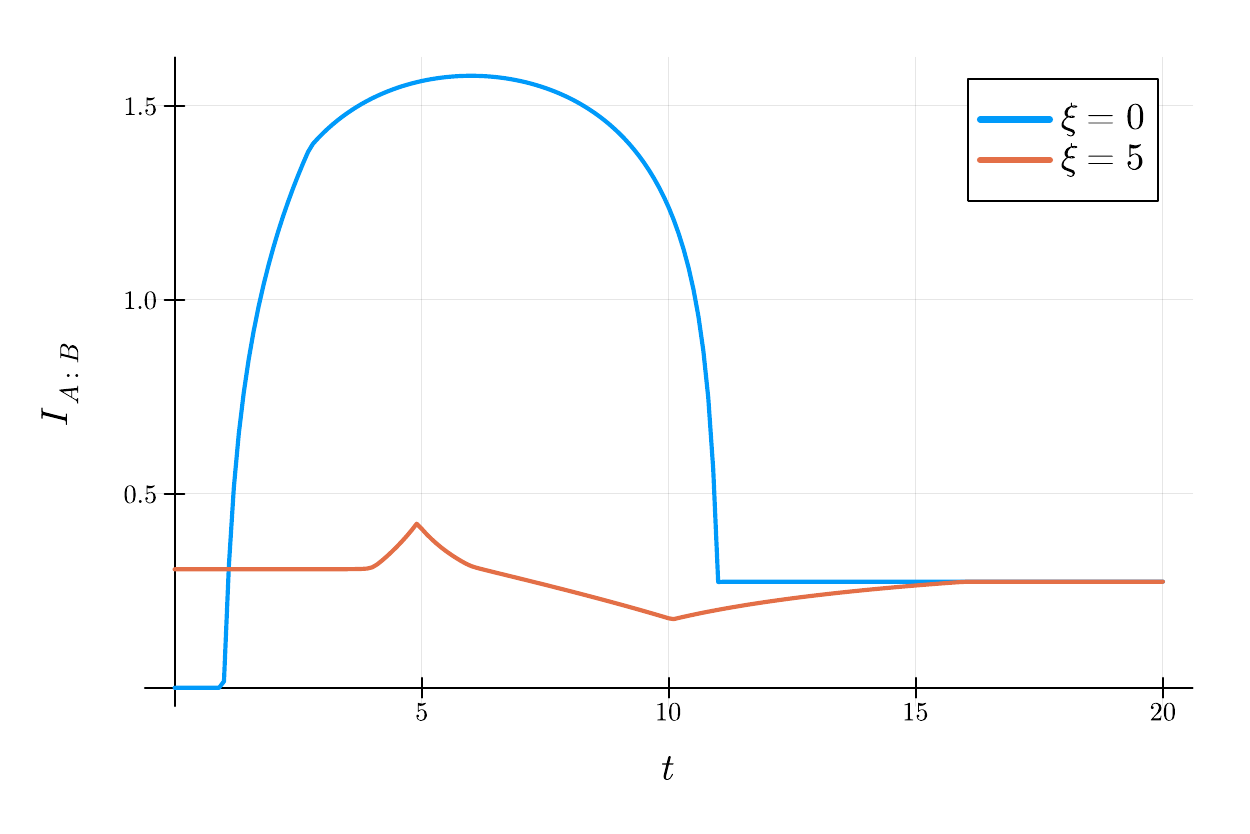}
    \includegraphics[width=0.45\linewidth]{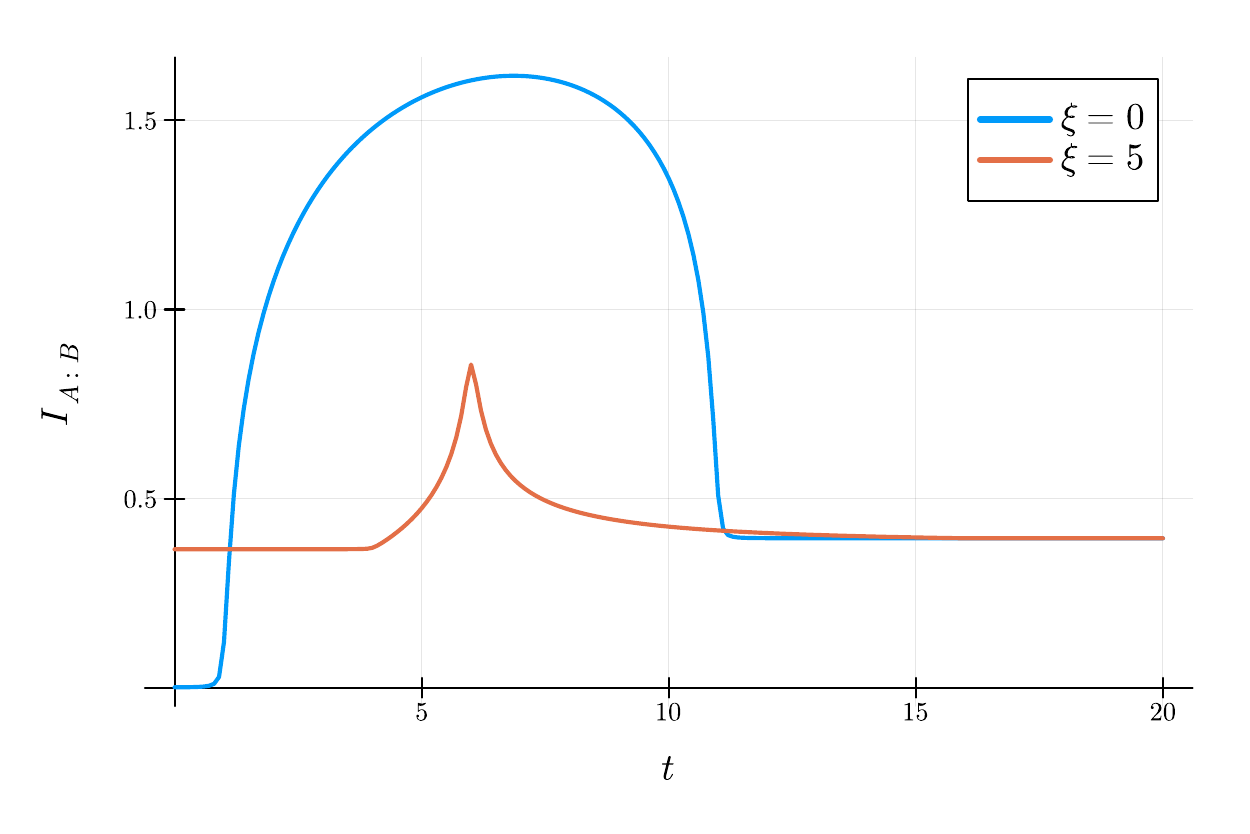}
    \caption{The mutual information \(I_{A(\x):B(\x)}(t)\) in the holographic CFT/the free Dirac fermion CFT for \(A(\x) = [\x - 11, \x - 1], B(\x) = [\x + 1, \x + 11]\) is shown in the left/right column.
    The top/middle/bottom row represents the spatiotemporal/spatial/temporal dependence.
    We chose \(c=1\), \(\ep=1\), \(\d=0.1\), and \(S_{\text{bdy}}=0\).}
    \label{fig:disjMI_L10}
\end{figure}

\begin{figure}
    \centering
    \includegraphics[width=0.45\linewidth]{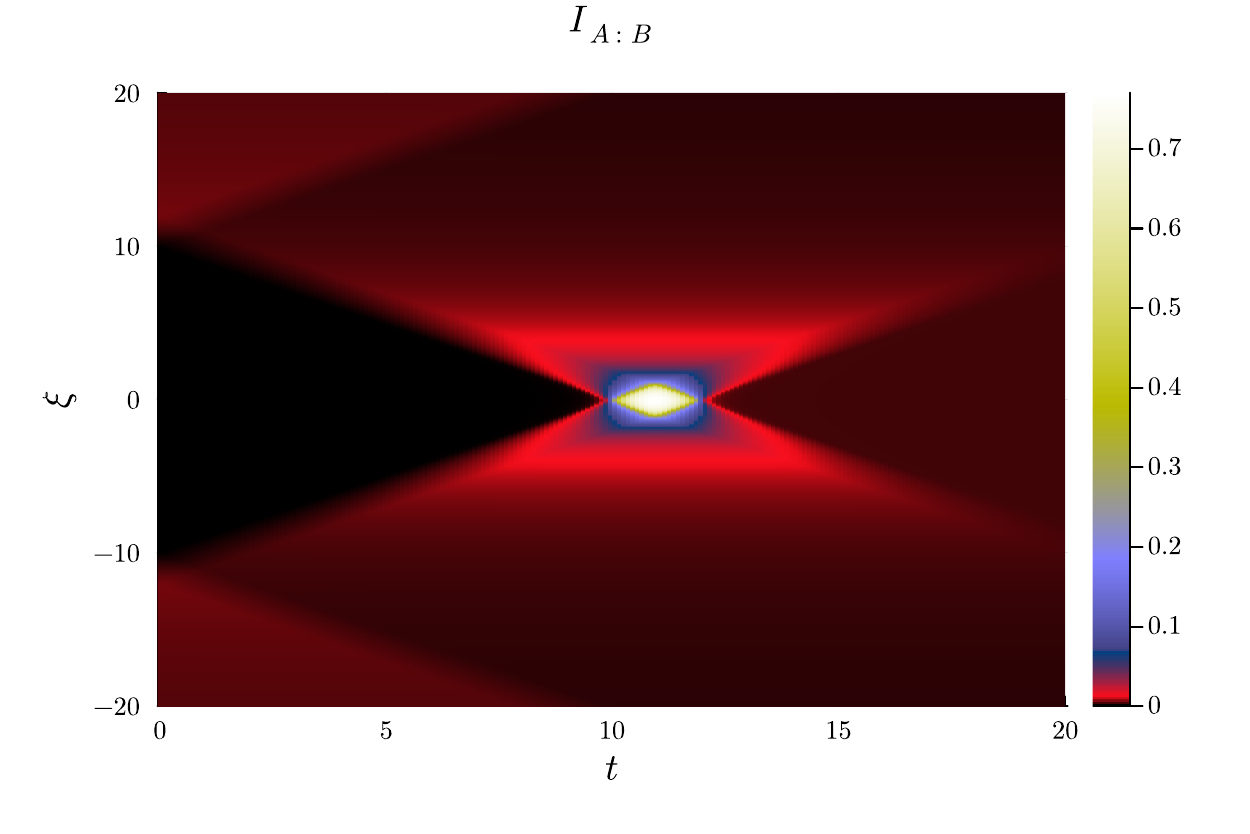}
    \includegraphics[width=0.45\linewidth]{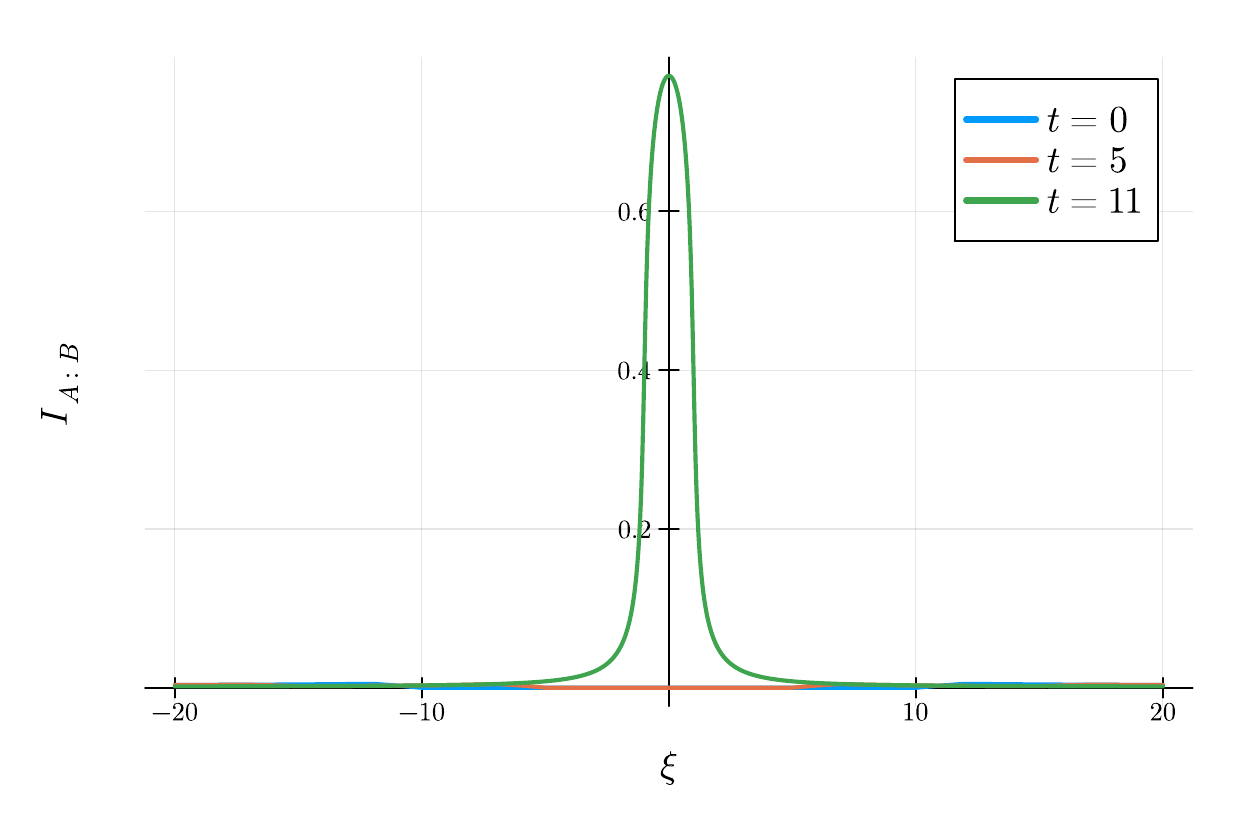}
    \includegraphics[width=0.45\linewidth]{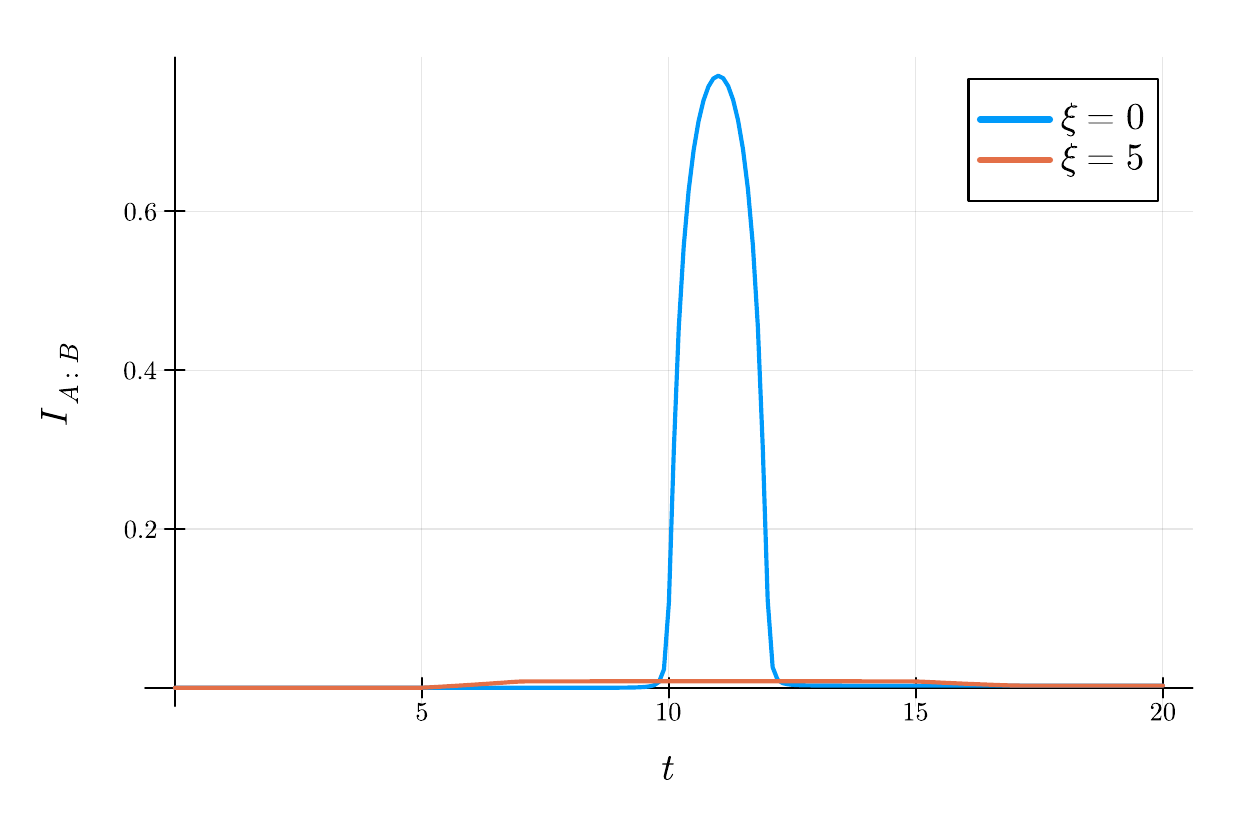}
    \caption{Plots of the mutual information \(I_{A(\x):B(\x)}(t)\) in the free Dirac fermion CFT for \(A(\x) = [\x - 12, \x - 10], B(\x) = [\x + 10, \x + 12]\).
    The left/right/bottom represents the spatiotemporal/spatial/temporal dependence.
    We chose \(c=1\), \(\ep=1\), \(\d=0.1\), and \(S_{\text{bdy}}=0\).
    Note that the mutual information of these subsystems in the holographic CFT is always zero.}
    \label{fig:disjMI_L2}
\end{figure}

We also illustrate the mutual information for the same subsystems in Fig.~\ref{fig:disjMI_L10} (\(L=10, D=2\)) and Fig.~\ref{fig:disjMI_L2} (\(L=2, D=20\)). 
In the holographic CFTs, the mutual information is always zero when \(D\) is larger than the scale of \(L\).

\subsection{Analytical results}
In the following, we present analytical expressions for entanglement entropy in two distinct scenarios, each achieved by taking a specific limit. 
Refer to \cite{Calabrese:2007mtj,Shimaji:2018czt} for the earlier work.

\subsubsection{Entanglement entropy for finite subsystem at \(t=0\)}
Here, we take the limit of \(\delta\to0\) and provide the explicit expressions for the entanglement entropy at \(t=0\) for joint subsystem \(A(\x) = [\x - L / 2, \x + L / 2]\).

For the holographic CFTs, the connected and disconnected contributions are
\begin{align}
    S_{A(\x)}^{\mathrm{con}}(t=0) &\xrightarrow{\d\to0} \begin{cases}\displaystyle
      \frac{c}{3} \log{\frac{L}{\ep}} \quad & (|\x| > L / 2) \\\displaystyle
      \frac{c}{3} \log\left[\frac{2|\x^2-L^2/4|}{\d\e}\right] \quad & (L / 2 > |\x|)
   \end{cases}\label{analytical_Doublet0_con},\\
   S_{A(\x)}^{\mathrm{dis}}(t=0) &\xrightarrow{\d\to0} \frac{c}{12} \log\left[ \frac{2^4(\x^2 - L^2 / 4)^2}{\ep^4} \right] + 2S_{\mathrm{bdy}}\label{analytical_Doublet0_dis}.
\end{align}
For the free Dirac fermion CFTs, we get
\begin{align}
   S_{A(\x)}^{\mathrm{Dir}}(t=0) &\xrightarrow{\d\to0} \begin{cases}\displaystyle
      \frac{1}{12} \log\left[ \frac{L^4(\x^2 - L^2 / 4)^2}{\x^4 \ep^4} \right] \quad & (\x > L / 2) \\\displaystyle
      \frac{1}{12} \log\left[ \frac{2^4(\x^2 - L^2 / 4)^2}{\ep^4} \right] \quad & (L / 2 > \x > - L / 2) \\\displaystyle
      \frac{1}{12} \log\left[ \frac{L^4(\x^2 - L^2 / 4)^2}{\x^4 \ep^4} \right] \quad & (- L / 2 > \x)
   \end{cases}\label{analytical_Doublet0_dir}.
\end{align}
They are consistent with the results in Fig.~\ref{fig:EE} (\(\d = 0.1\)).

\subsubsection{Time dependence for semi-infinite subsystem}
For finite-size subsystems, it is difficult to determine the time dependence analytically.
Therefore, here we consider the subsystem size \(L\) to be infinity.
In other words, we assume a situation where \(w_1=\tau+il\) and \(w_2=\tau+i \infty\) in Fig.~\ref{fig:opposing-slit_geometry}.
In this case, \(w_2\) is no longer a boundary, so we only need to compute a one-point function at \(w_2\).
Then, we get the entanglement entropy:
\begin{align}
    S_{[l, \infty[}(t)= \begin{cases}\displaystyle
       \frac{c}{6}\log{\frac{2l}{\epsilon}} \quad & (l > t) \\\displaystyle
       \frac{c}{6} \log{\frac{2(t^2-l^2)}{\epsilon^2}}+\frac{c}{6}\log{\frac{2\epsilon}{\delta}} \quad & (l<t)
    \end{cases}\label{analytical_Linftyt0_dir}.
\end{align}

\section{Pseudo entropy of JQ state and ground state: single-slit geometry}\label{sec:pejqgs}
Now, we move on to our main example, the single-slit geometry, choosing \(\ket{\psi} = \ket{\JQ(x = 0)}\) and \(\ket{\vp} = \ket{\Omega}\). 
We compute the pseudo entropy of a joining quenched state and the ground state.

\subsection{Pseudo entropy}
\begin{figure}
    \centering
    \includegraphics[width=0.45\linewidth]{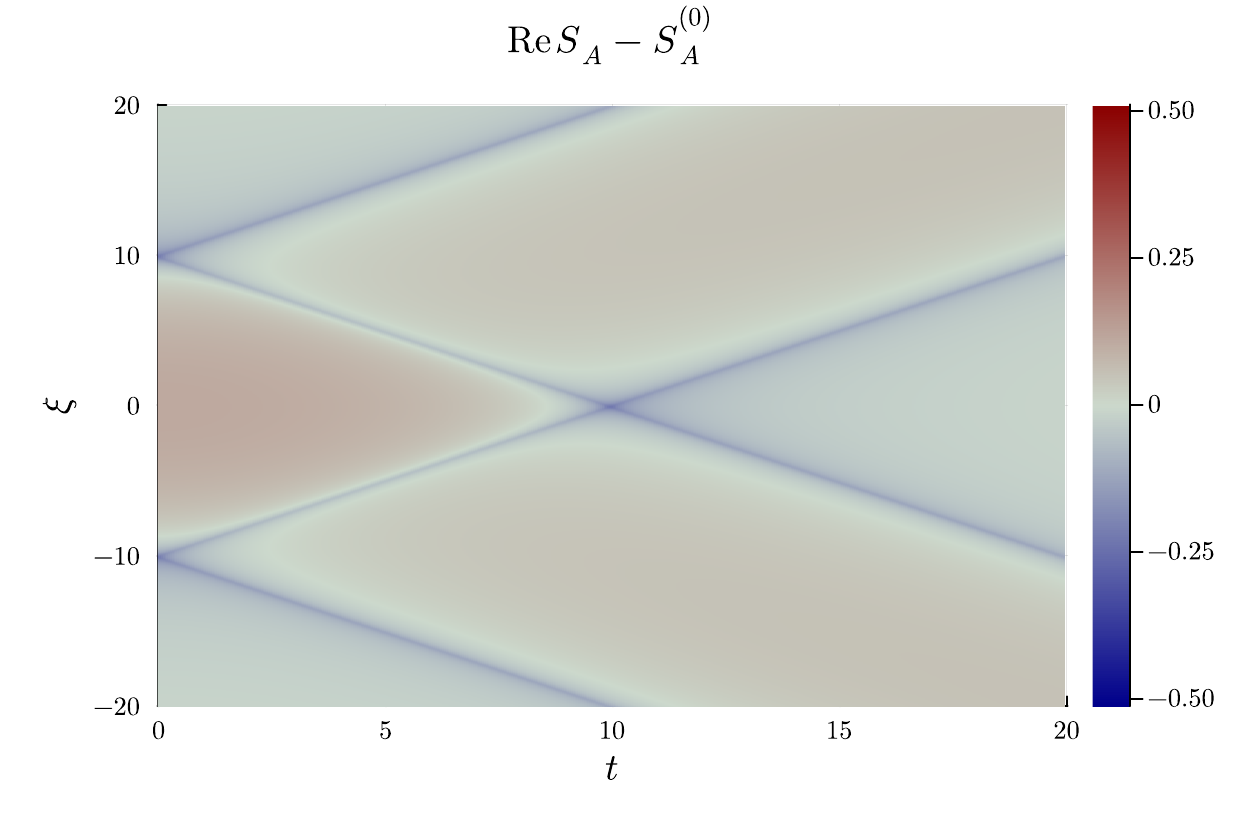}
    \includegraphics[width=0.45\linewidth]{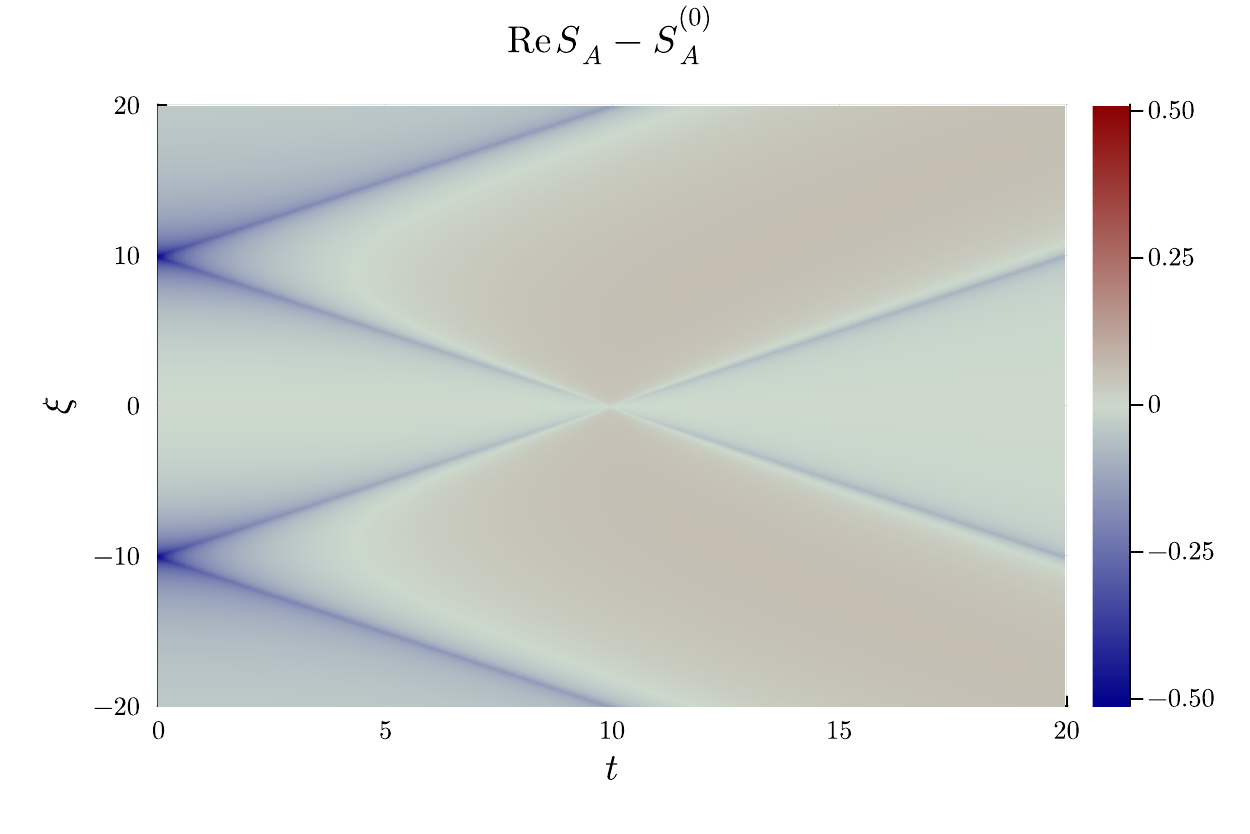}
    \includegraphics[width=0.45\linewidth]{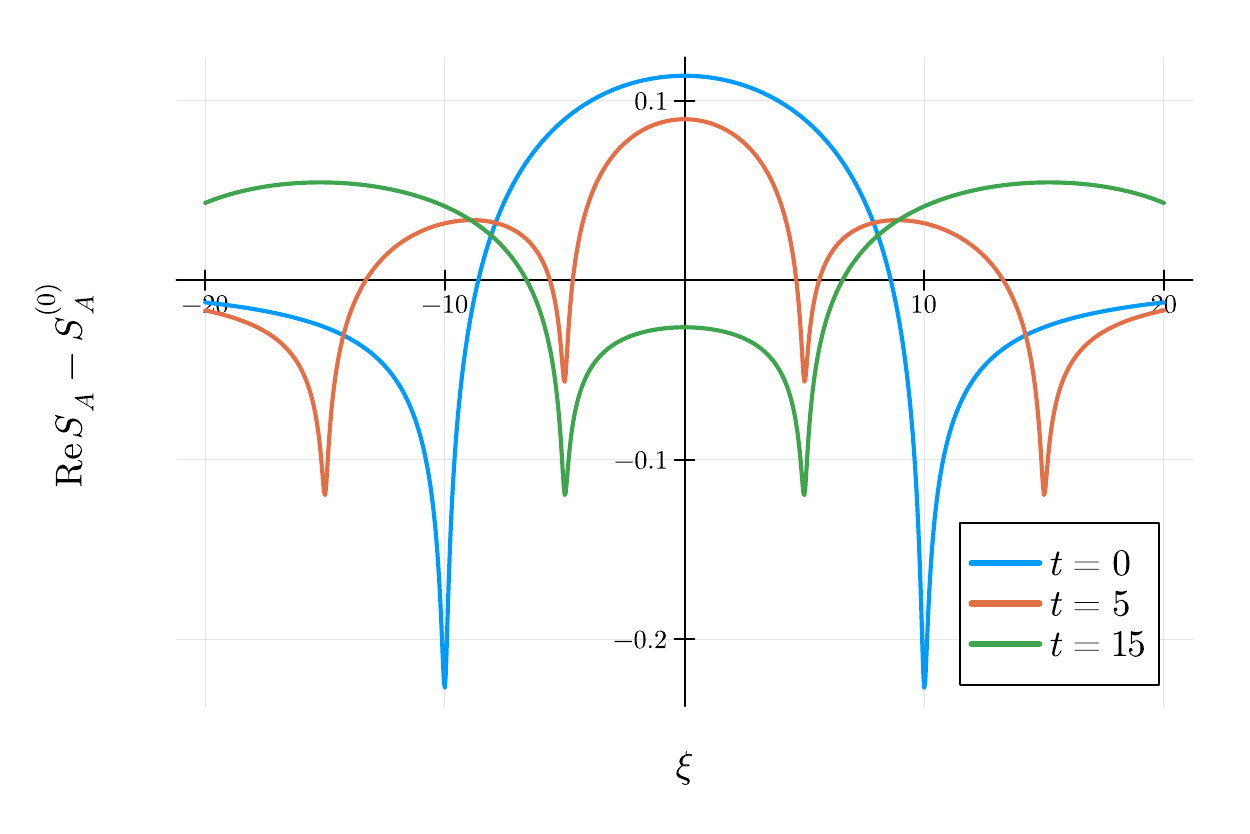}
    \includegraphics[width=0.45\linewidth]{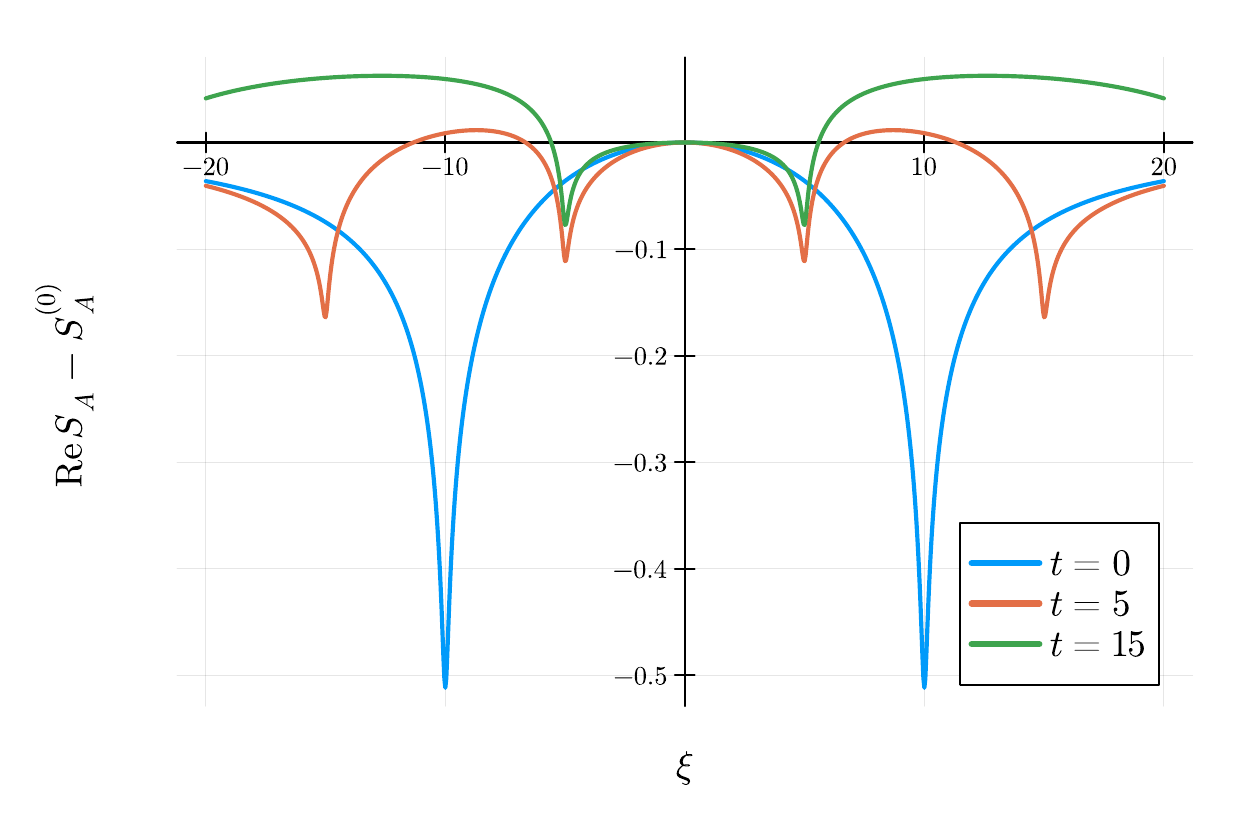}
    \includegraphics[width=0.45\linewidth]{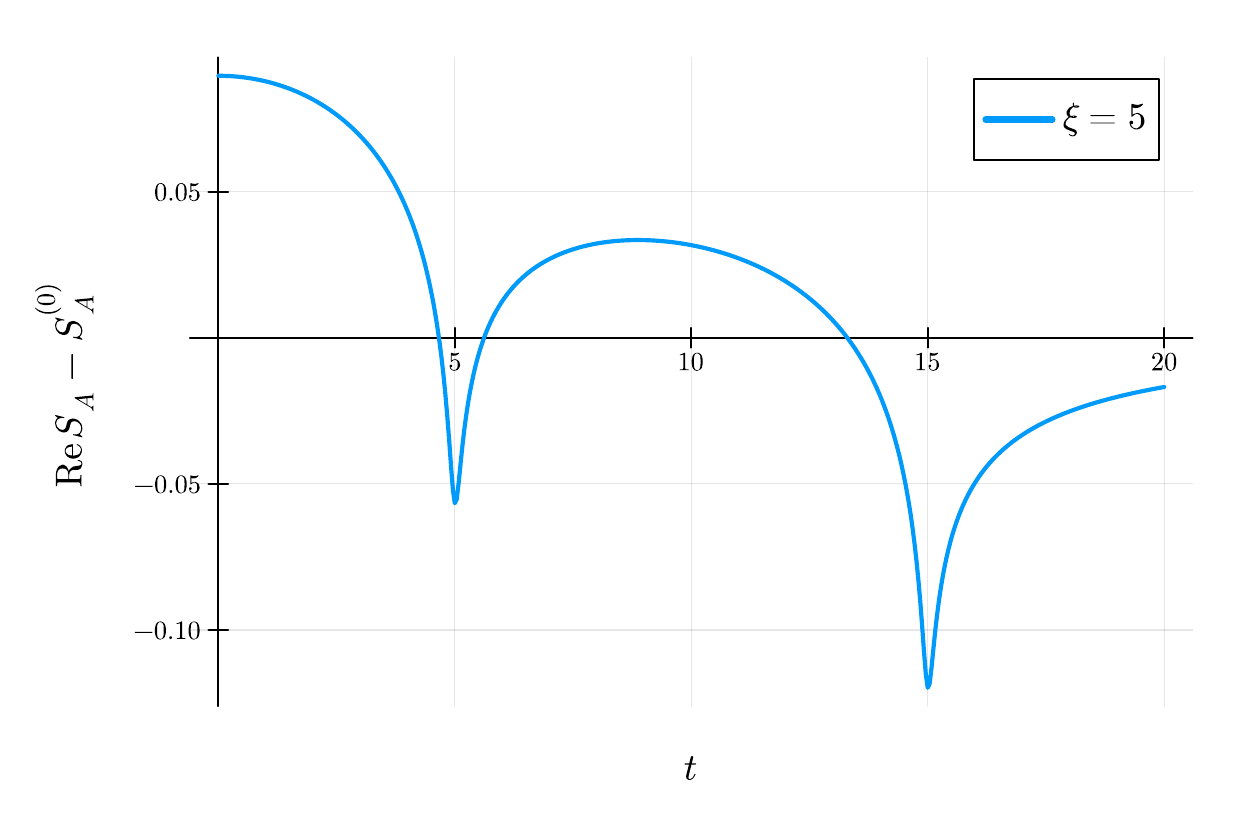}
    \includegraphics[width=0.45\linewidth]{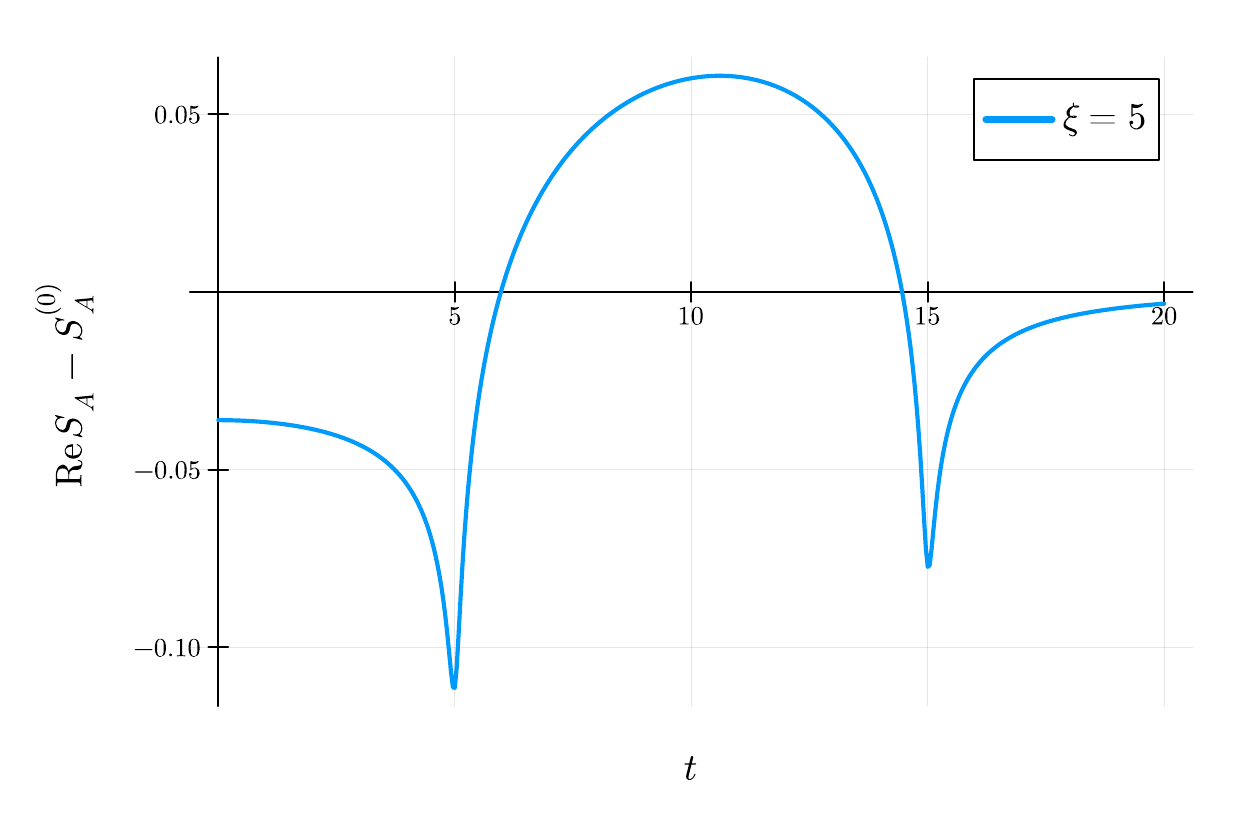}
    \caption{The real part of the pseudo entropy \(S_{A(\x)}^{\JQ|\Omega}(t) - S_{A(\x)}^{(0)}\) in the holographic CFT/the free Dirac fermion CFT for \(A(\x) = [\x - 10, \x + 10]\) is shown in the left/right column.
    The top/middle/bottom row represents the spatiotemporal/spatial/temporal dependence.
    We chose \(c=1\), \(\ep=1\), \(\d=0.1\), and \(S_{\mathrm{bdy}}=0\).}
    \label{fig:rePE_JQGS}
\end{figure}

\begin{figure}
    \centering
    \includegraphics[width=0.45\linewidth]{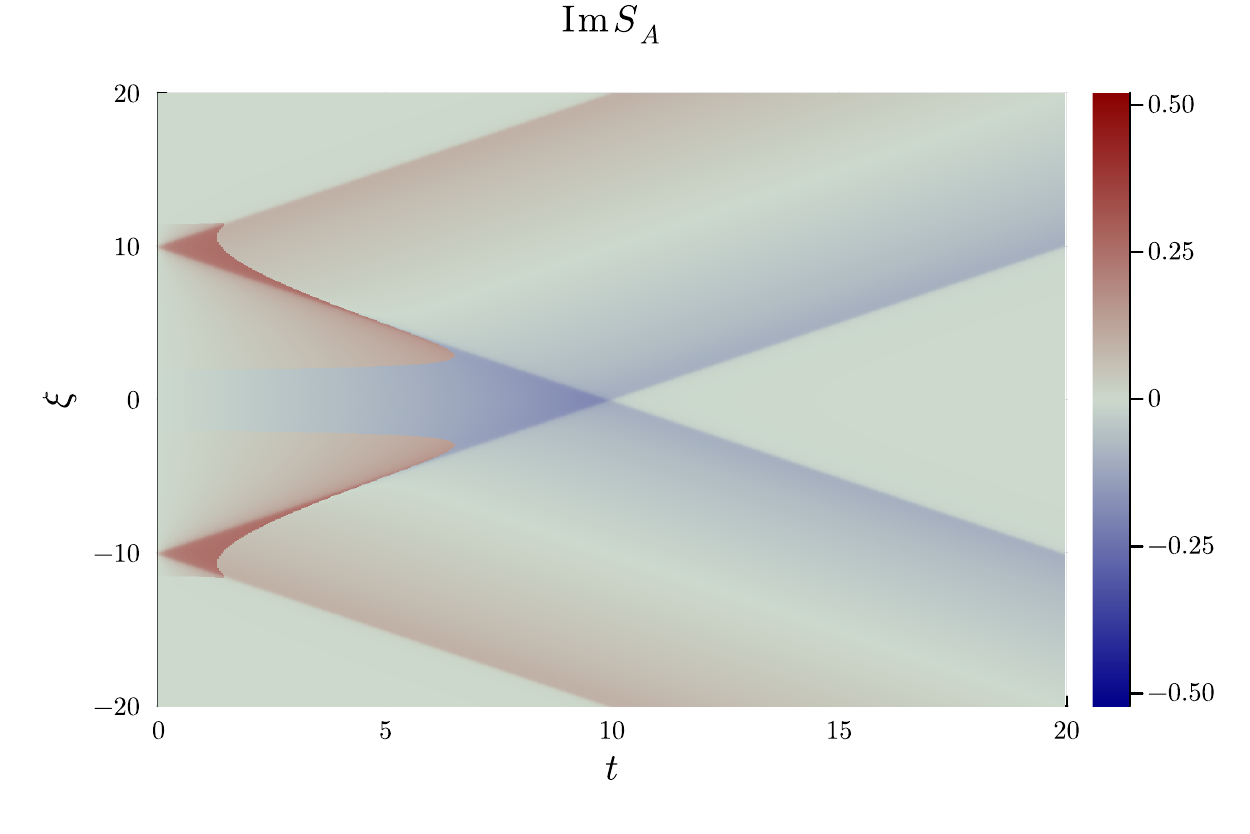}
    \includegraphics[width=0.45\linewidth]{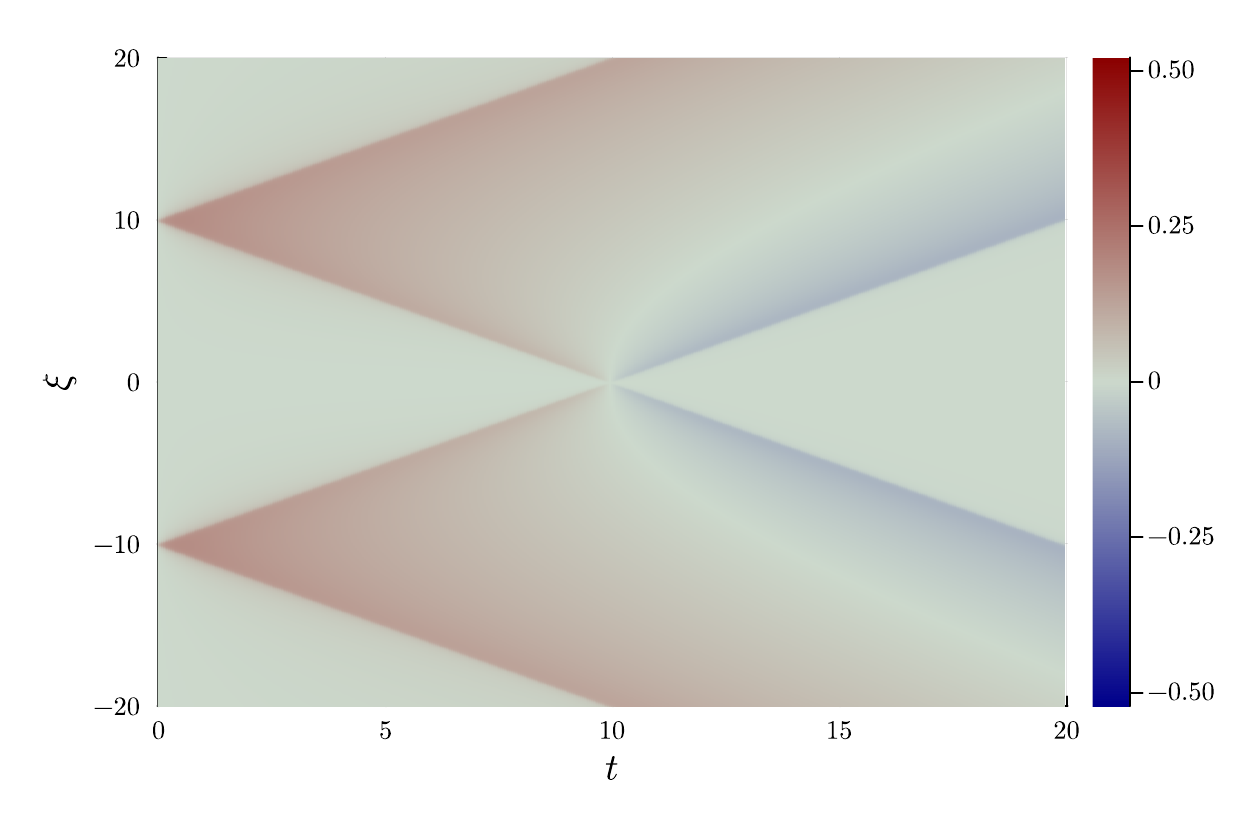}
    \includegraphics[width=0.45\linewidth]{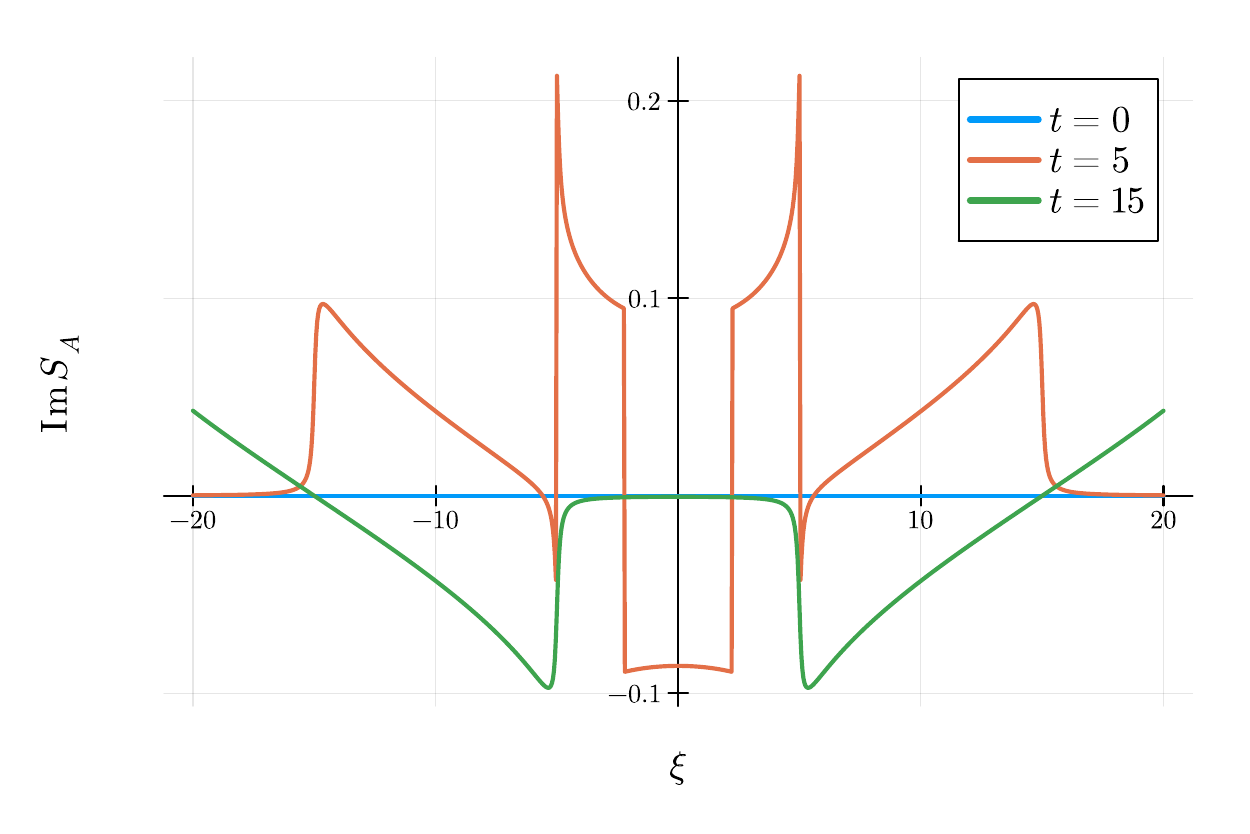}
    \includegraphics[width=0.45\linewidth]{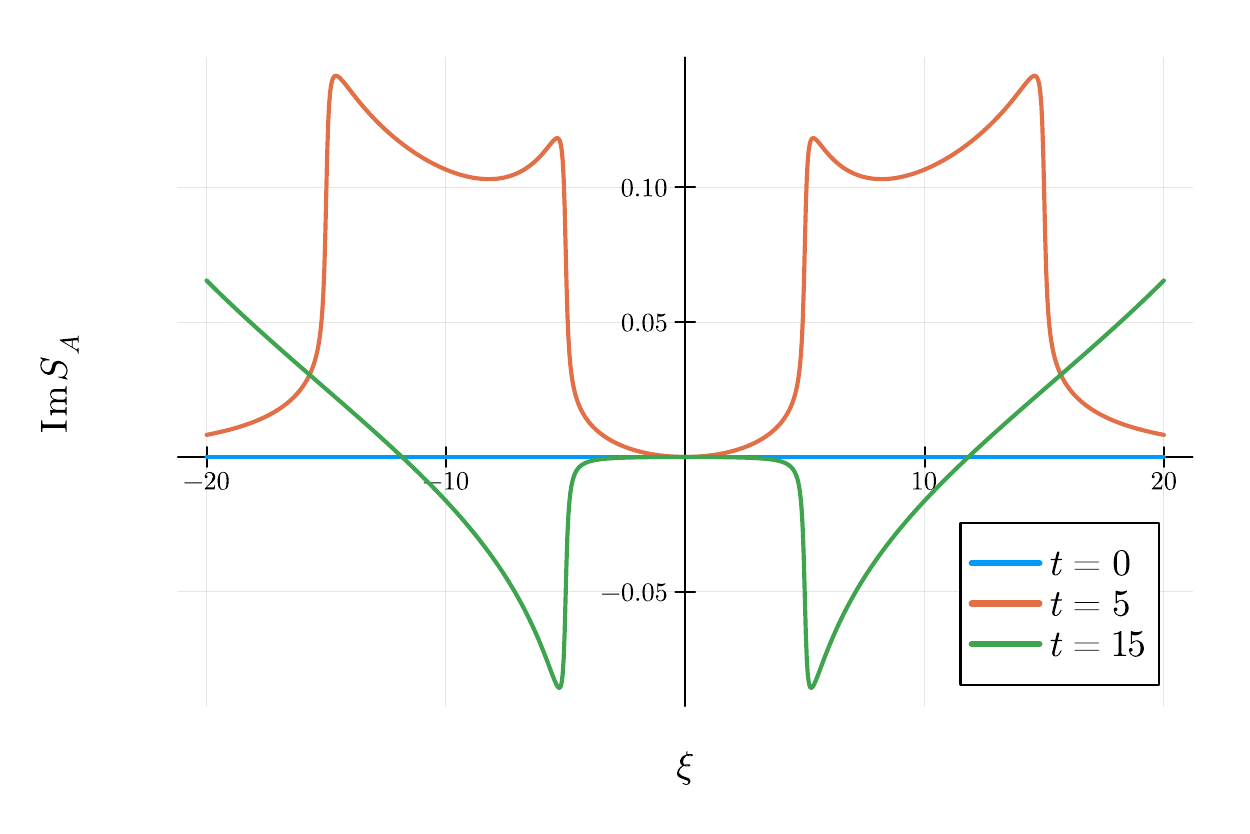}
    \includegraphics[width=0.45\linewidth]{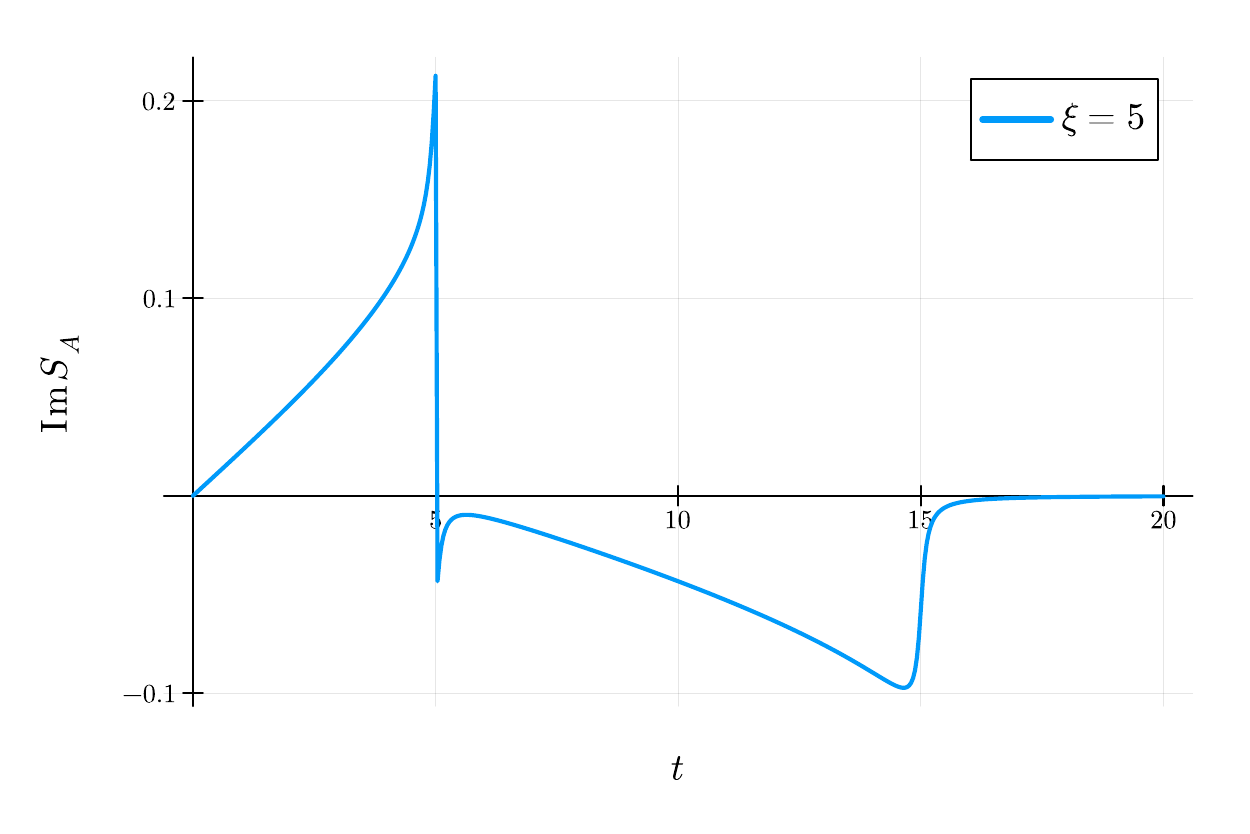}
    \includegraphics[width=0.45\linewidth]{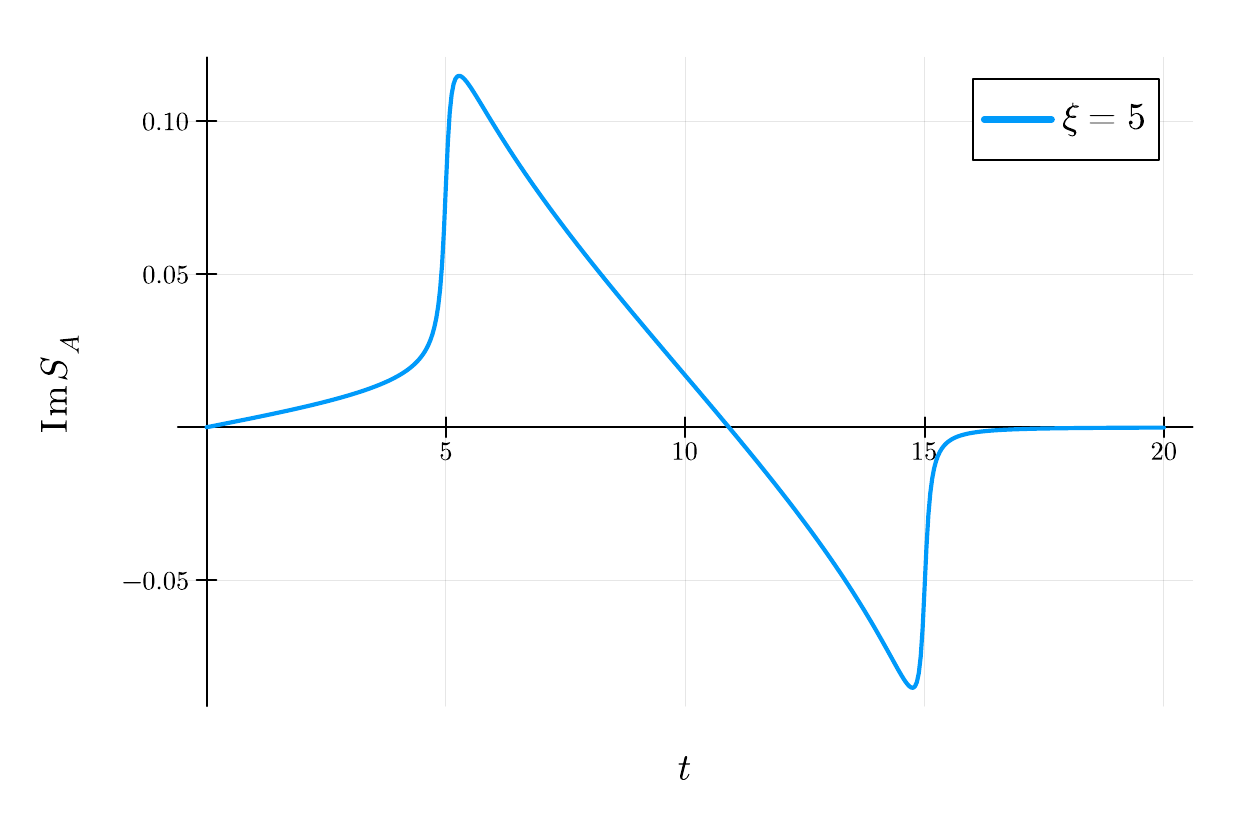}
    \caption{The imaginary part of the pseudo entropy \(S_{A(\x)}^{\JQ|\Omega}(t) - S_{A(\x)}^{(0)}\) in the holographic CFT/the free Dirac fermion CFT for \(A(\x) = [\x - 10, \x + 10]\) is shown in the left/right figure.
    The top/middle/bottom figure represents the spatiotemporal/spatial/temporal dependence.
    We chose \(c=1\), \(\ep=1\), and \(\d=0.1\).}
    \label{fig:imPE_JQGS}
\end{figure}

We take an interval subsystem
\begin{equation}
    A(\xi) = [\x - L / 2, \x + L / 2].  \label{intevgqw}
\end{equation}
In Fig.~\ref{fig:rePE_JQGS} and Fig.~\ref{fig:imPE_JQGS}, we present the pseudo entropy in the holographic CFT, taking the contributions with smaller real parts, alongside the pseudo entropy in the free Dirac fermion CFT.

The imaginary parts of \(S_{A(\x)}^{\JQ|\Omega}(t)\) are non-zero in general due to the Wick rotation, while they are zero at \(t=0\).
In the limit of \(|\x| \to \infty\), the connected contribution converges to the entanglement entropy of the ground state \(S^{(0)}_A\) since the effect of the quench can be neglected.
When the edge of \(A\) coincides with the joining point, the connected contribution takes minima.
The disconnected contribution diverges in the limit of \(|\x|\to\infty\).
This can be explained in the context of AdS/BCFT by considering the minimal geodesic length with the EOW brane, similar to the discussion for entanglement entropy. 

For both the holographic CFT and the free Dirac fermion CFT, we find a dip when the excitations injected at \(x=0\) and \(t=0\) propagate to the two endpoints of \(A\). 
We expect that this is due to the property that the pseudo entropy decreases when we flip quantum entanglement, which is peculiar to the pseudo entropy as first found in \cite{Nakata:2021ubr}. 
This can also be explained by the first law relation of pseudo entropy \cite{Mollabashi:2020yie}, which is very similar to the first law of entanglement entropy \cite{Bhattacharya:2012mi,Blanco:2013joa} that relates energy to pseudo entropy. 
Indeed, the energy density in the joining quench and the single-slit setup, shown in Fig.~\ref{fig:EStensor_JQJQ} and Fig.~\ref{fig:EStensor_JQGS}, also show the dip only in the latter.

\clearpage
\subsection{Entropy excess}
\begin{figure}
    \centering
    \includegraphics[width=0.45\linewidth]{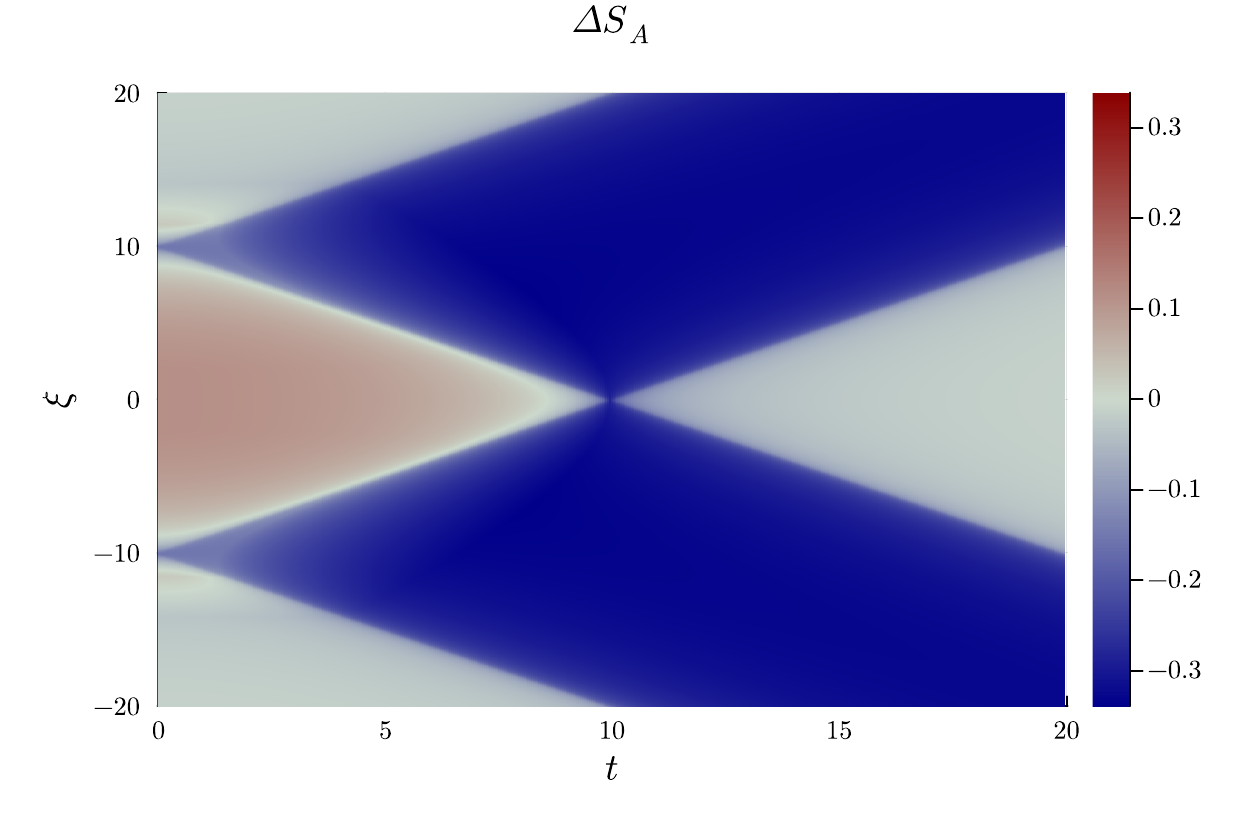}
    \includegraphics[width=0.45\linewidth]{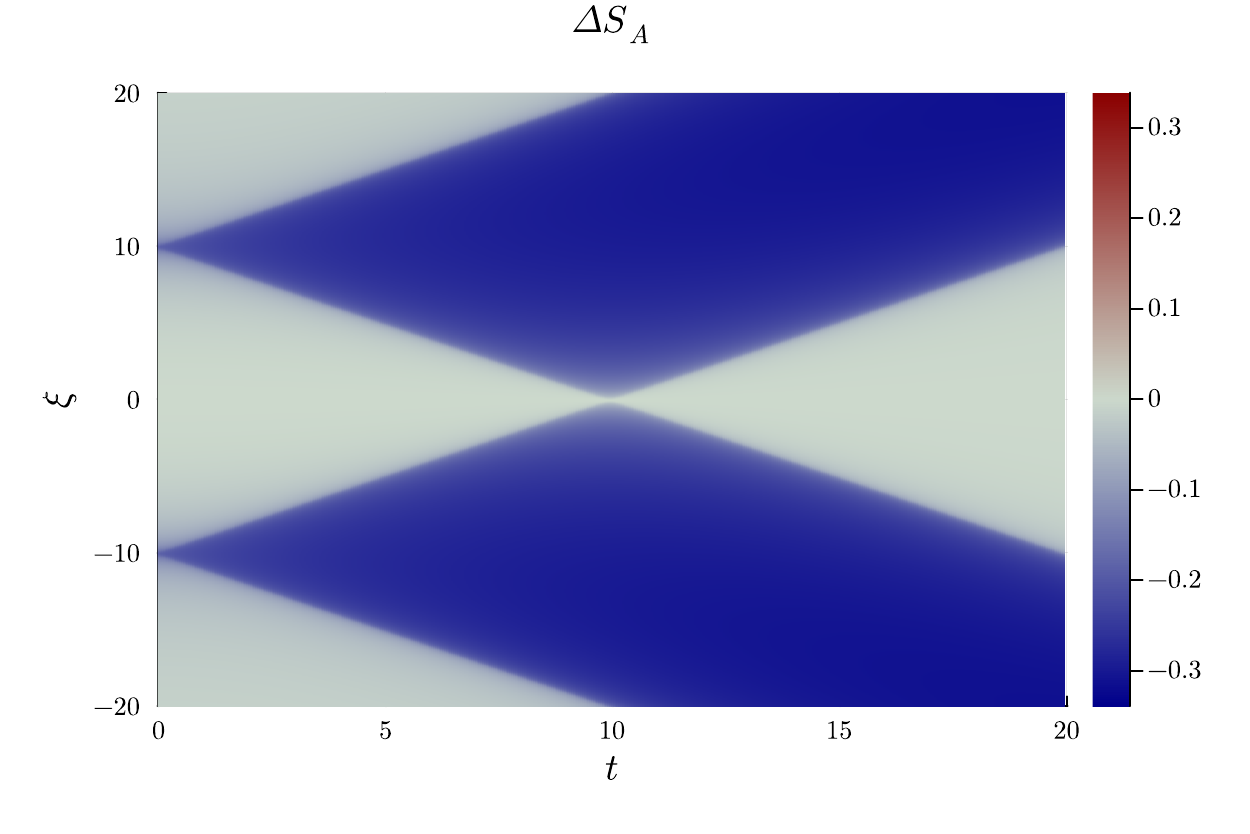}
    \includegraphics[width=0.45\linewidth]{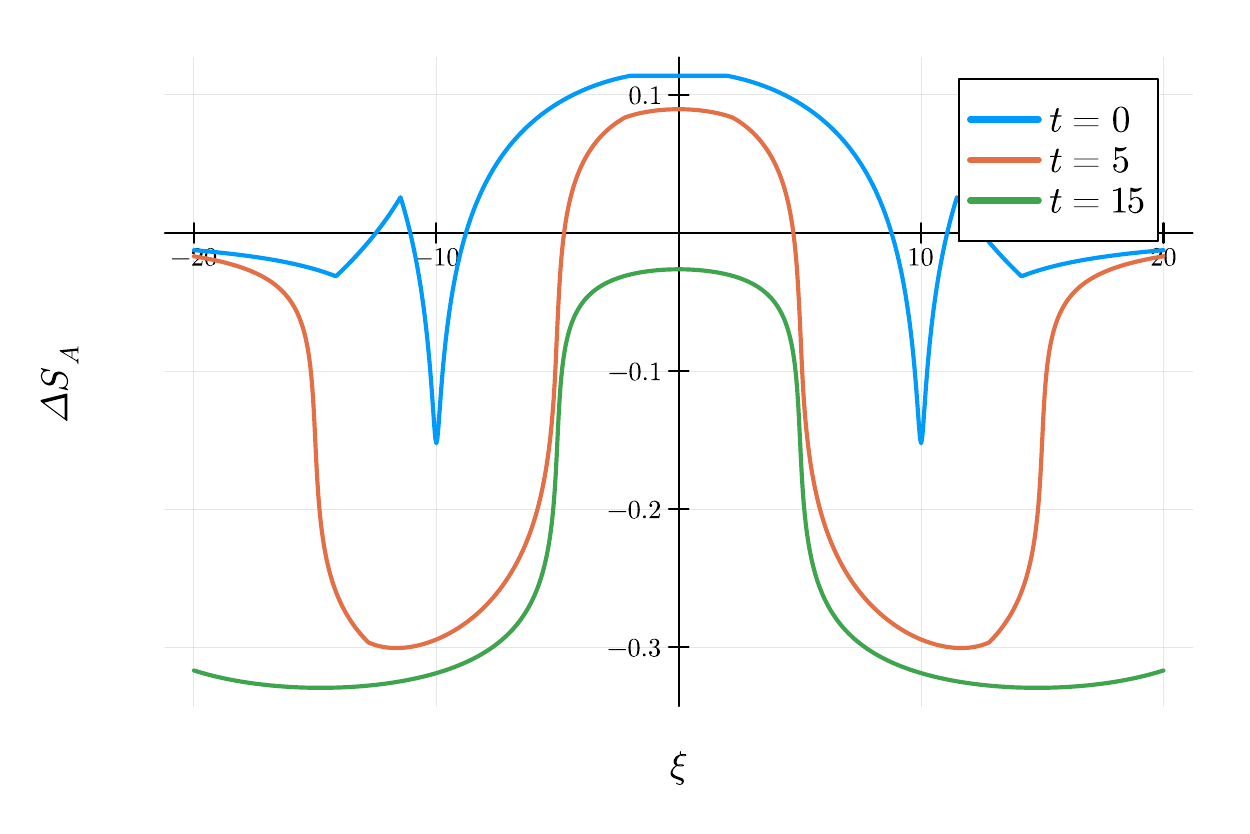}
    \includegraphics[width=0.45\linewidth]{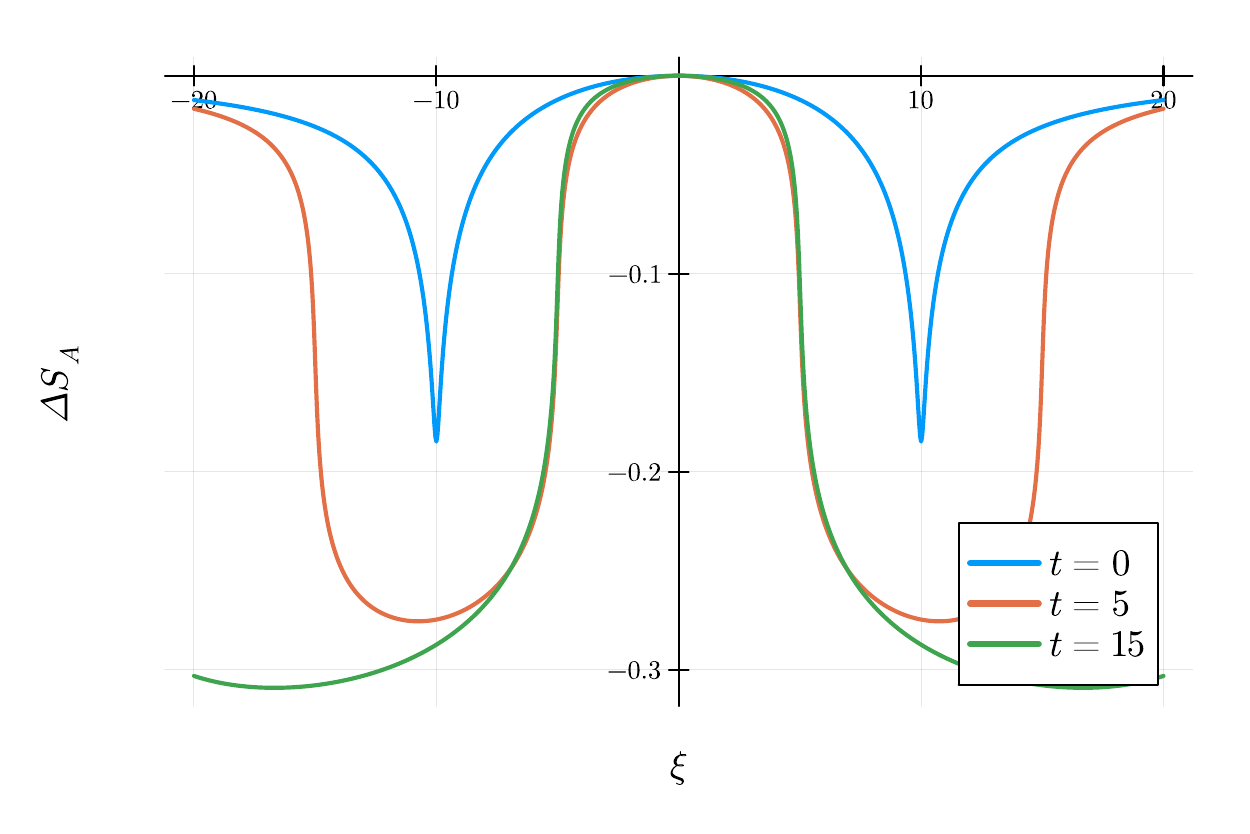}
    \includegraphics[width=0.45\linewidth]{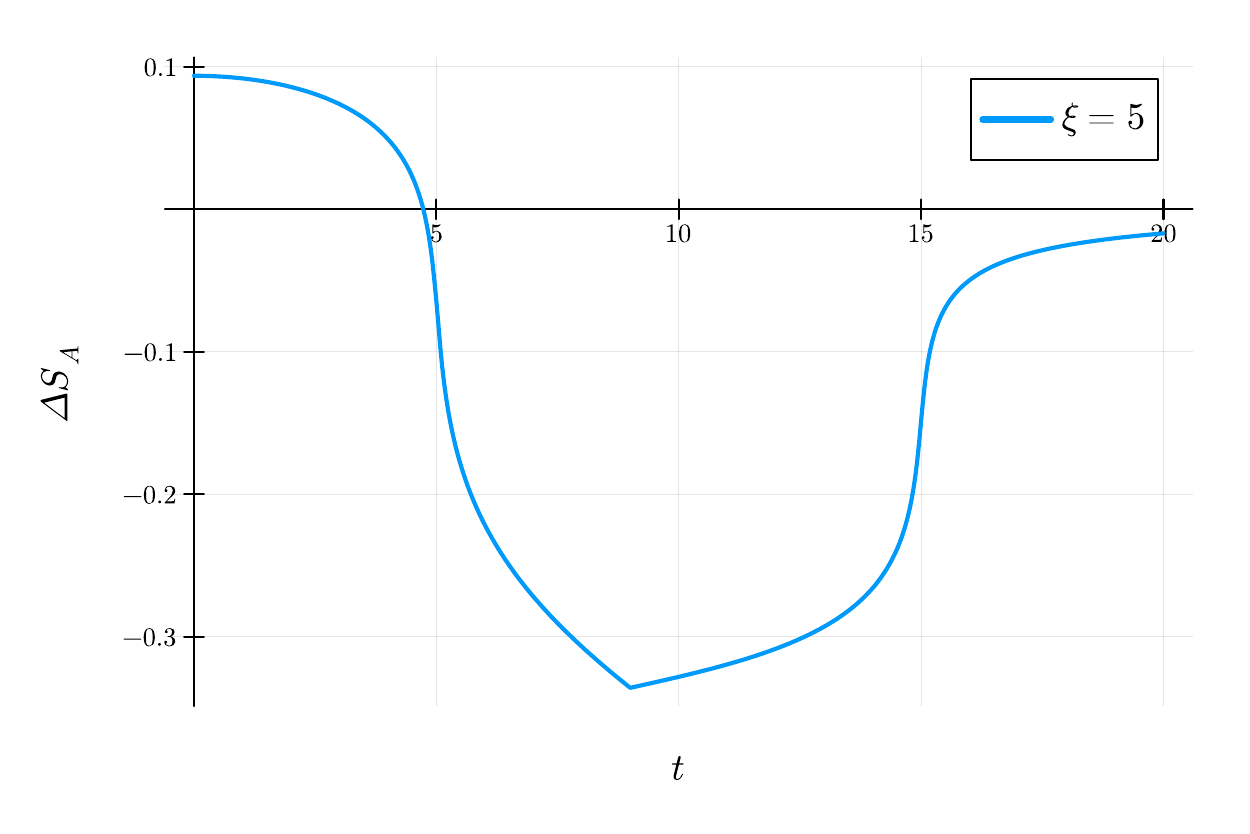}
    \includegraphics[width=0.45\linewidth]{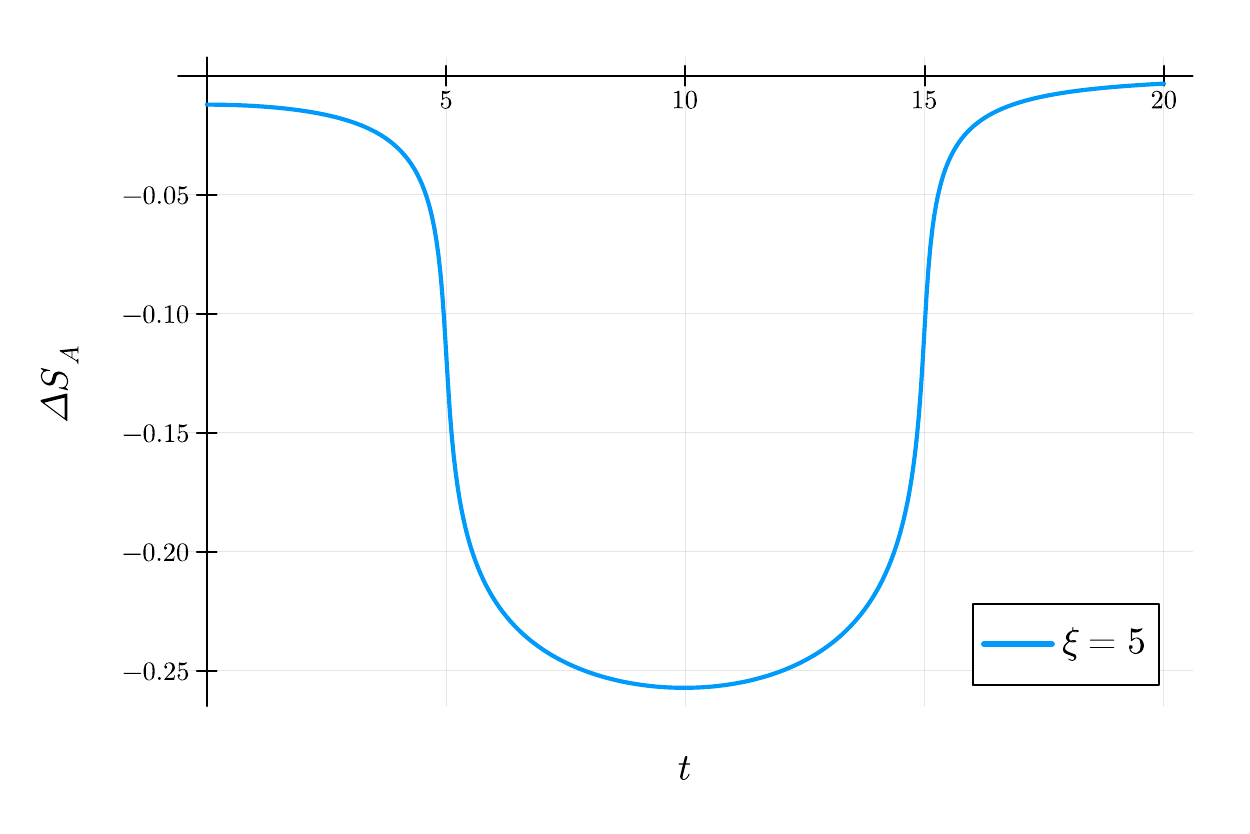}
    \caption{The entropy excess \(\varDelta{S}_{A(\x)}^{\JQ|\Omega}(t)\) in the holographic CFT/the free Dirac fermion CFT for \(A(\x) = [\x - 10, \x + 10]\) is shown in the left/right column.
    The top/middle/bottom row represents the spatiotemporal/spatial/temporal dependence.
    We chose \(c=1\), \(\ep=1\), \(\d=0.1\), and \(S_{\mathrm{bdy}}=0\).}
    \label{fig:deltaS_JQGS}
\end{figure}

The behavior of entropy excess \(\varDelta{S}^{\JQ|\Omega}\) in the holographic CFT and the free Dirac fermion CFT is shown in Fig.~\ref{fig:deltaS_JQGS}. 
Although they exhibit similar relativistic propagation during time evolution, one notable difference is the behavior in the region \(|\xi|<L-t\).
In the holographic CFT, the entropy excess \(\varDelta{S}_A\) can become positive in this region, while in the free Dirac fermions CFT, we always have \(\varDelta{S}_A \leq 0\).
As discussed in section \ref{ssec:PSummary} and Fig.~\ref{fig:DScase}, the absence of multi-partite entanglement leads \(\varDelta{S}_A \leq 0\) as in the Dirac fermion case. 
On the other hand, our results suggest the existence of multi-partite entanglement in holographic CFTs, which agrees with our expectations. 
This is consistent with the fact that the tripartite mutual information vanishes in the free Dirac fermion CFT, while it does not in holographic CFTs \cite{Hayden:2011ag}.

\subsection{Analytical results}
Since the single-slit setup can be examined analytically through a relatively simple conformal map \eqref{comapsin}, below, we also present analytical expressions for pseudo entropy for the interval subsystem \eqref{intevgqw}.

\subsubsection{Pseudo entropy and entropy excess for finite subsystem at \(t=0\)}
\begin{figure}
    \centering
    \includegraphics[width=0.45\linewidth]{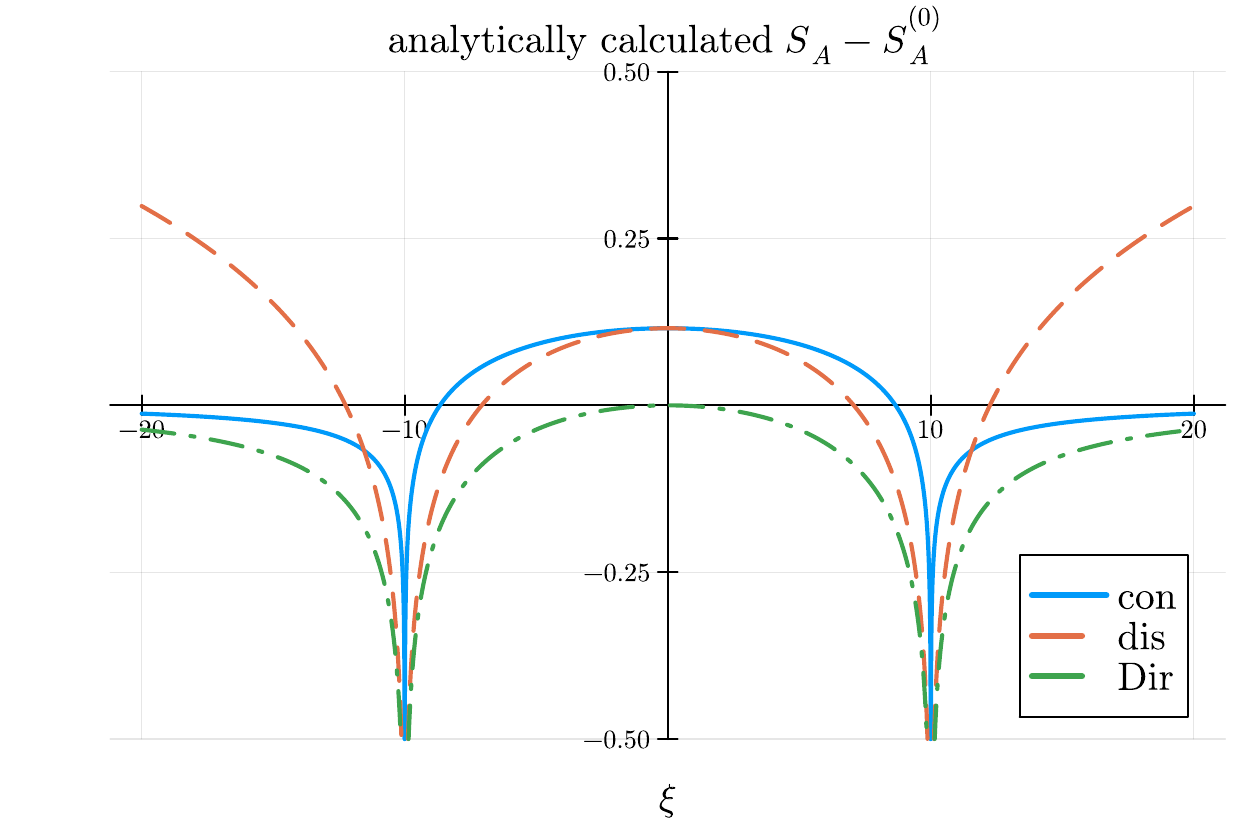}
    \caption{A plot of the analytically calculated pseudo entropy \(S_{A(\x)}^{\JQ|\Omega}(t=0) - S_{A(\x)}^{(0)}\) for \(A(\x) = [\x - 10, \x + 10]\) in the limit of \(\d\to0\).
    The red/blue line illustrates the connected/disconnected contribution in the holographic CFT, while the green line shows the free Dirac fermion CFT result.
    We chose \(c=1\), \(\ep=1\), and \(S_{\mathrm{bdy}}=0\).}
    \label{fig:analytic_PE_JQGS}
\end{figure}

By the same calculation as entanglement entropy, pseudo entropy \(S_{A(\x)}(t=0)\) in the limit of \(\d\to0\) is 
\begin{align}
    &S_{A(\x)}^{\mathrm{con}}(t=0) \notag\\&\quad= \frac{c}{12} \log\left[ \left\{ \left|\x - \frac{L}{2}\right| + \left|\x + \frac{L}{2}\right| - 2\sqrt{\left|\x^2 - \frac{L^2}{4}\right|} \, \Theta\left( \x^2 - \frac{L^2}{4} \right) \right\}^2 \times \frac{16}{\epsilon^4} \, \left| \x^2 - \frac{L^2}{4} \right| \right], \label{analytical_Singlet0_con}\\
    &S_{A(\x)}^{\mathrm{dis}}(t=0) = \frac{c}{12} \log\left[ \frac{64}{\epsilon^4} \left(\x^2 - \frac{L^2}{4}\right)^2  \right] + 2S_\mathrm{bdy}, \label{analytical_Singlet0_dis}\\
    &S_{A(\x)}^{\mathrm{Dir}}(t=0) \notag\\&\quad= \frac{1}{12} \log\left[ \frac{\left\{ \left|\x - \frac{L}{2}\right| + \left|\x + \frac{L}{2}\right| - 2\sqrt{\left|\x^2 - \frac{L^2}{4}\right|} \, \Theta\left( \x^2 - \frac{L^2}{4} \right) \right\}^2}{\left\{ \left|\x - \frac{L}{2}\right| + \left|\x + \frac{L}{2}\right| + 2\sqrt{\left|\x^2 - \frac{L^2}{4}\right|} \, \Theta\left( \frac{L^2}{4} - \x^2 \right) \right\}^2} \times \frac{64}{\epsilon^4} \left(\x^2 - \frac{L^2}{4}\right)^2 \right]. \label{analytical_Singlet0_dir}
\end{align}
The results are plotted in Fig.~\ref{fig:analytic_PE_JQGS}.
They are consistent with the numerical results in Fig.~\ref{fig:rePE_JQGS}.

\begin{figure}
    \centering
    \includegraphics[width=0.45\linewidth]{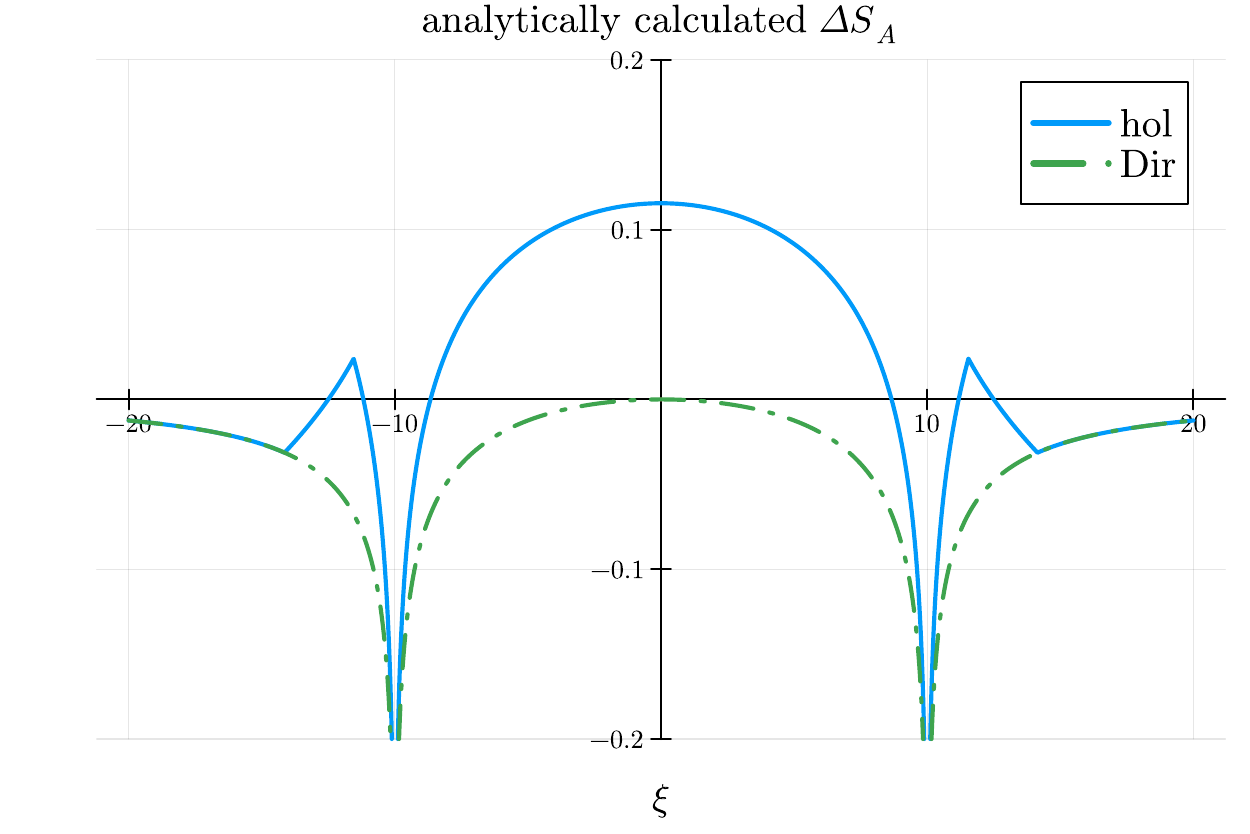}
    \caption{A plot of the analytically calculated entropy excess \(\varDelta{S}_{A(\x)}^{\JQ|\Omega}(t=0)\) for \(A(\x) = [\x - 10, \x + 10]\) in the limit of \(\d\to0\).
    The red/green line illustrates the holographic CFT/the free Dirac fermion CFT result.
    We chose \(c=1\), \(\ep=1\), and \(S_{\mathrm{bdy}}=0\).}
    \label{fig:analytic_deltaS}
\end{figure}

Now we can calculate the entropy excess \(\varDelta{S}_A^{\JQ|\Omega}(t=0)\) in the limit of \(\d\to0\).
The results are plotted in Fig.~\ref{fig:analytic_deltaS}.
They are consistent with the results in Fig.~\ref{fig:deltaS_JQGS} (\(\d = 0.1\)).
Especially for \(\xi = 0\), the pseudo entropy of the joining quenched state and the ground state is evaluated in the $\delta\to 0$ limit as follows:
\begin{align}
    S_{A(\x=0)}^{\JQ|\Omega}(t=0) &= \min\left\{ S_{A(\x)}^{\mathrm{con}}, S_{A(\x)}^{\mathrm{dis}} \right\} \notag\\
    &= \min\left\{\frac{c}{3} \, \log{\frac{L}{\epsilon}} + \frac{c}{6} \log{2}, \frac{c}{3} \, \log{\frac{L}{\epsilon}} + \frac{c}{6} \log{2} + 2S_{\mathrm{bdy}} \right\},
\end{align}
and the entanglement entropy is
\begin{equation}
   S_{A(\x=0)}^{\JQ}(t=0) = \min\left\{\infty, \frac{c}{3} \, \log{\frac{L}{\epsilon}} + 2S_{\mathrm{bdy}} \right\}.
\end{equation}
The entropy excess \(\varDelta{S}_{A(\x=0)}^{\JQ|\Omega}(t=0)\) turns out to be
\begin{align}
   \varDelta{S}_{A(\x=0)}^{\JQ|\Omega}(t=0) &= \begin{cases}\displaystyle
      \frac{c}{6} \log{2} - S_{\mathrm{bdy}} \qquad& (S_{\mathrm{bdy}} \geq 0) \\\displaystyle
      \frac{c}{6} \log{2} + S_{\mathrm{bdy}} \qquad& (S_{\mathrm{bdy}} < 0)
   \end{cases}
\end{align}
and thus, it has an upper bound:
\begin{equation}
   \varDelta{S}_{A(\x=0)}^{\JQ|\Omega}(t=0) \leq \frac{c}{6} \log{2}.
\end{equation}
Recall that entropy excess is related to multi-partite entanglement, as elaborated in section \ref{ssec:PSummary}. 
It is intriguing that, for each cross-section boundary in the AdS$_3$ gravity, the mentioned value is precisely half of the upper bound of a quantity known as the Markov gap \cite{Hayden:2021gno, Liu:2021ctk}, which has also been discussed in the context of multi-partite entanglement. 

\subsubsection{Time dependence for semi-infinite subsystem}
\begin{figure}[t]
    \centering
    \includegraphics[width=4cm]{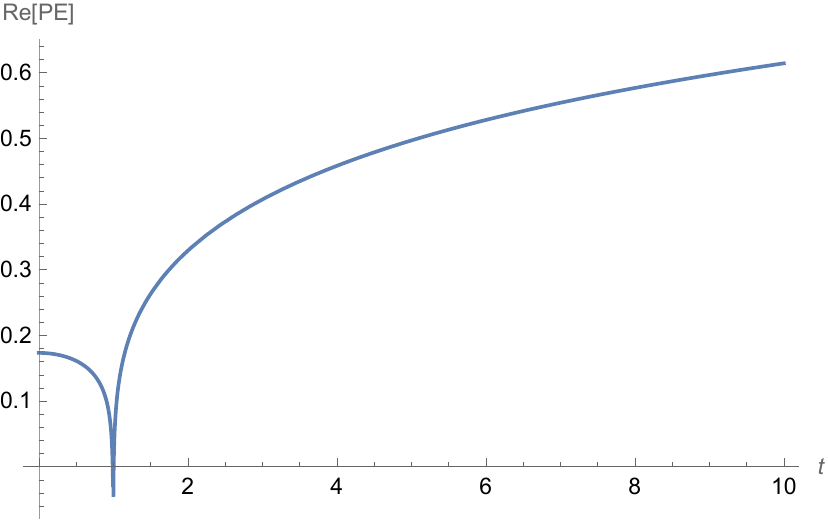}
    \includegraphics[width=4cm]{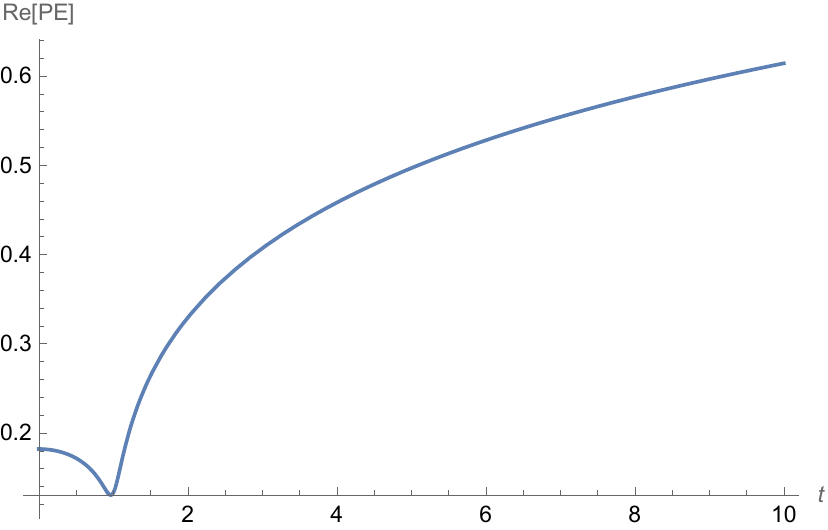}
    \includegraphics[width=4cm]{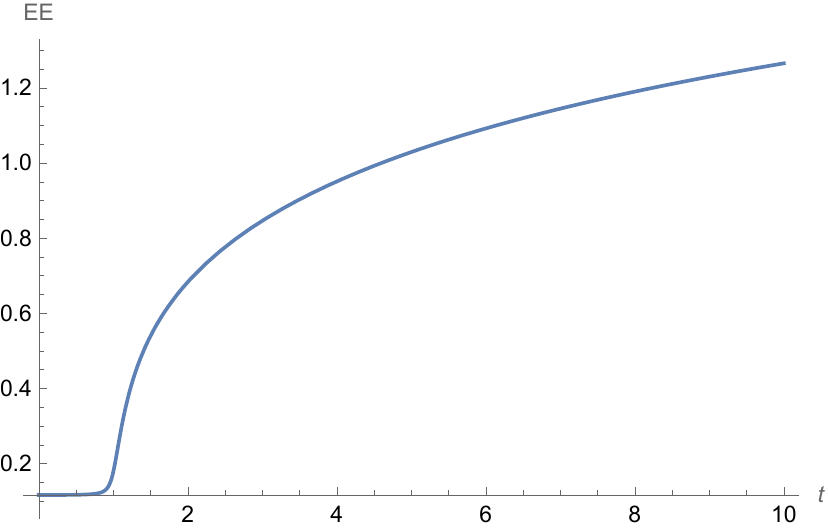}
    \caption{Plots of the real part of pseudo entropy for a semi-infinite subsystem \(A=[l, \infty[\).
    The left shows the analytically calculated pseudo entropy in the limit of \(\d\to0\).
    The middle shows the numerically calculated pseudo entropy with \(\d=0.1\).
    The right shows the numerically calculated entanglement entropy of the joining quench with \(\d=0.1\) for comparison.
    We choose \(c = 1\), \(\epsilon=1\), \(l=1\) and \(S_\mathrm{bdy}=0\) in all cases.}
    \label{fig:analytical_semiinfinity}
\end{figure}

We consider the time evolution of pseudo entropy for a semi-infinite subsystem \(A = [l, \infty[\).
In the limit of \(\d\to0\), we obtain
the following analytical result:
\begin{equation}
    S_{[l, \infty[}^{\mathrm{JQ}|\Omega}(t) = \frac{c}{12}\log{\frac{8\sqrt{l^2-t^2}(it+\sqrt{l^2-t^2})}{\epsilon^2}}.
\end{equation}
The result of pseudo entropy based on the current analytical and previous numerical calculations, alongside that of entanglement entropy in joining local quenches, is plotted in Fig.~\ref{fig:analytical_semiinfinity}.
The numerical and analytical results agree with each other. 
Even though the pseudo entropy exhibits a logarithmic growth \(S_A \sim (c/6) \log{t}\), the coefficient is halved compared with the entanglement entropy in joining local quenches (\(S_A \sim (c/3) \log{t}\)) \cite{Calabrese:2007mtj}. 
Further, the initial behavior (\(0<t<l\)) differs between the two. 
This difference is due to a dip at \(t=l\), which is unique to the pseudo entropy due to the entanglement flipping effect mentioned before. 
A similar effect is observed in the energy stress tensor expectation value on the single-slit setup, as shown in Fig.~\ref{fig:EStensor_JQGS}. 
In this case, the real part exhibits a negative tail around the positive peak, a feature absent in the joining local quenches, as presented in Fig.~\ref{fig:EStensor_JQJQ}.

\section{Pseudo entropy of different JQ states: double-slit geometry}\label{sec:pejqjq}
We analyze the pseudo entropy on the double-slit geometry as a final example. 
This corresponds to the choice of two different joining quenched states 
\(\ket{\psi} = \ket{\JQ(x=x_1)}\) and \(\ket{\vp} = \ket{\JQ(x=x_2)}\). 
For our calculations, we set \(x_1 = - x_2 = 10\) without loss of generality.

\subsection{Pseudo entropy}
\begin{figure}[ttt]
    \centering
    \includegraphics[width=0.45\linewidth]{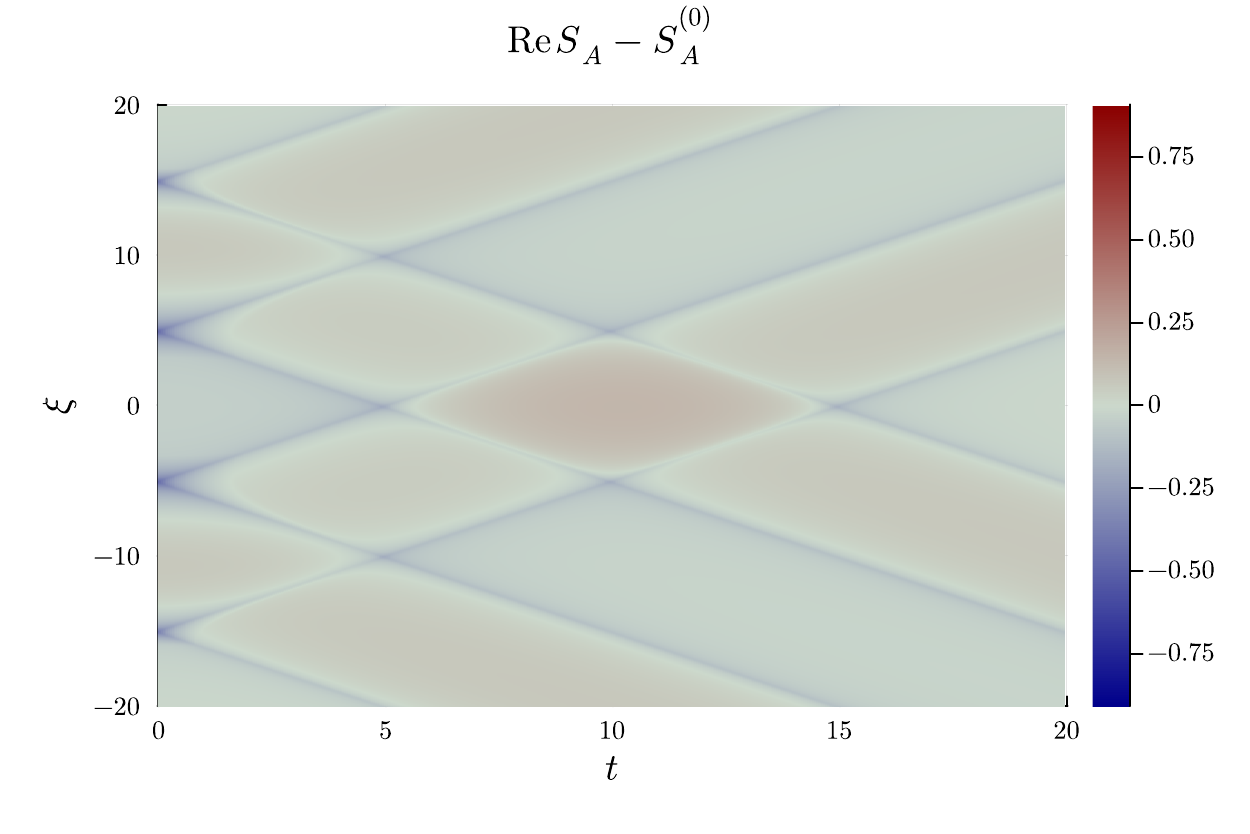}
    \includegraphics[width=0.45\linewidth]{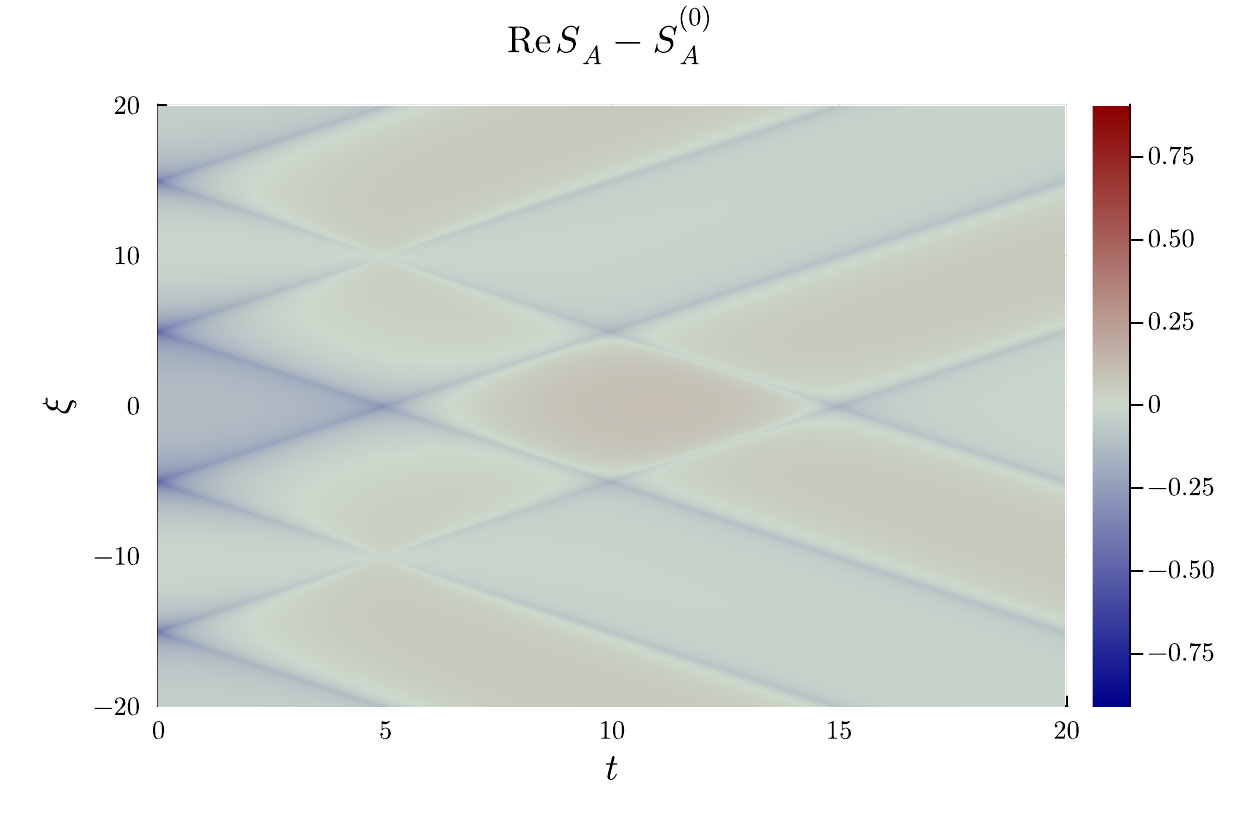}
    \includegraphics[width=0.45\linewidth]{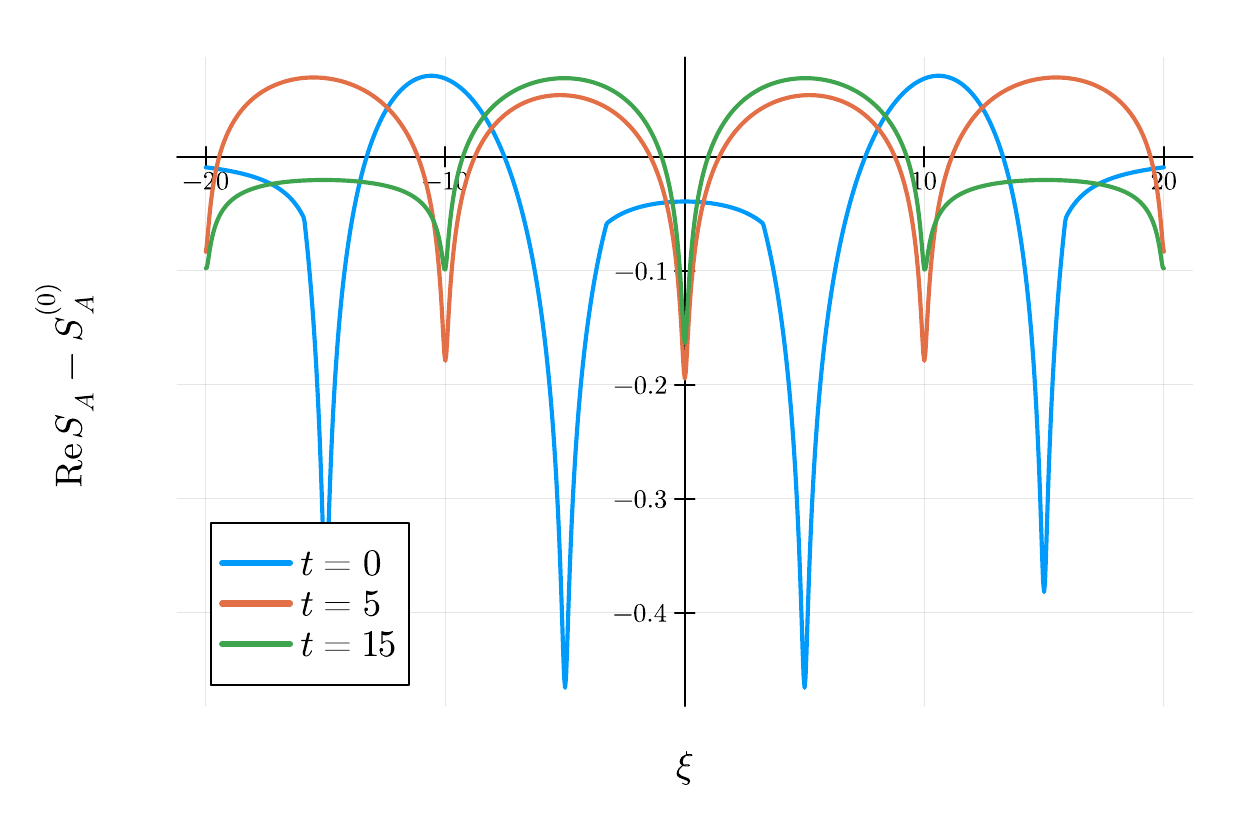}
    \includegraphics[width=0.45\linewidth]{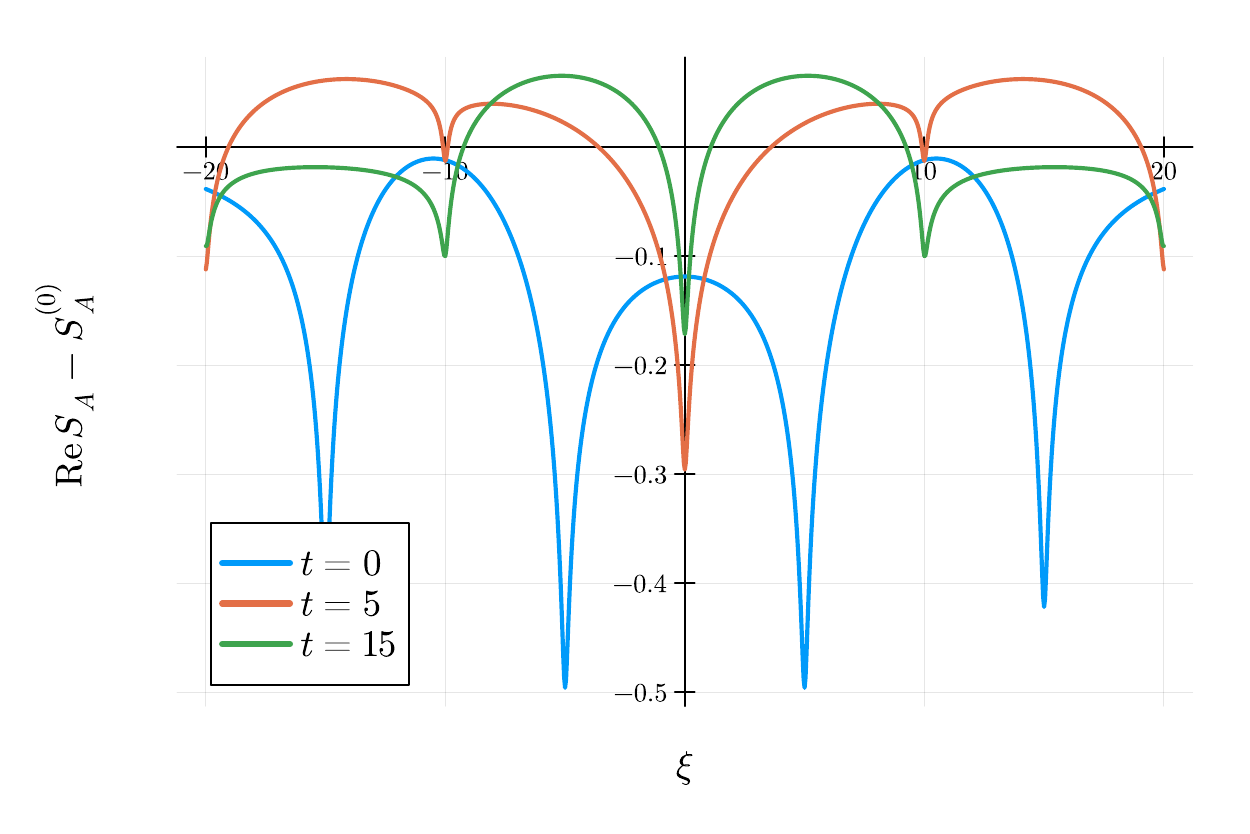}
    \includegraphics[width=0.45\linewidth]{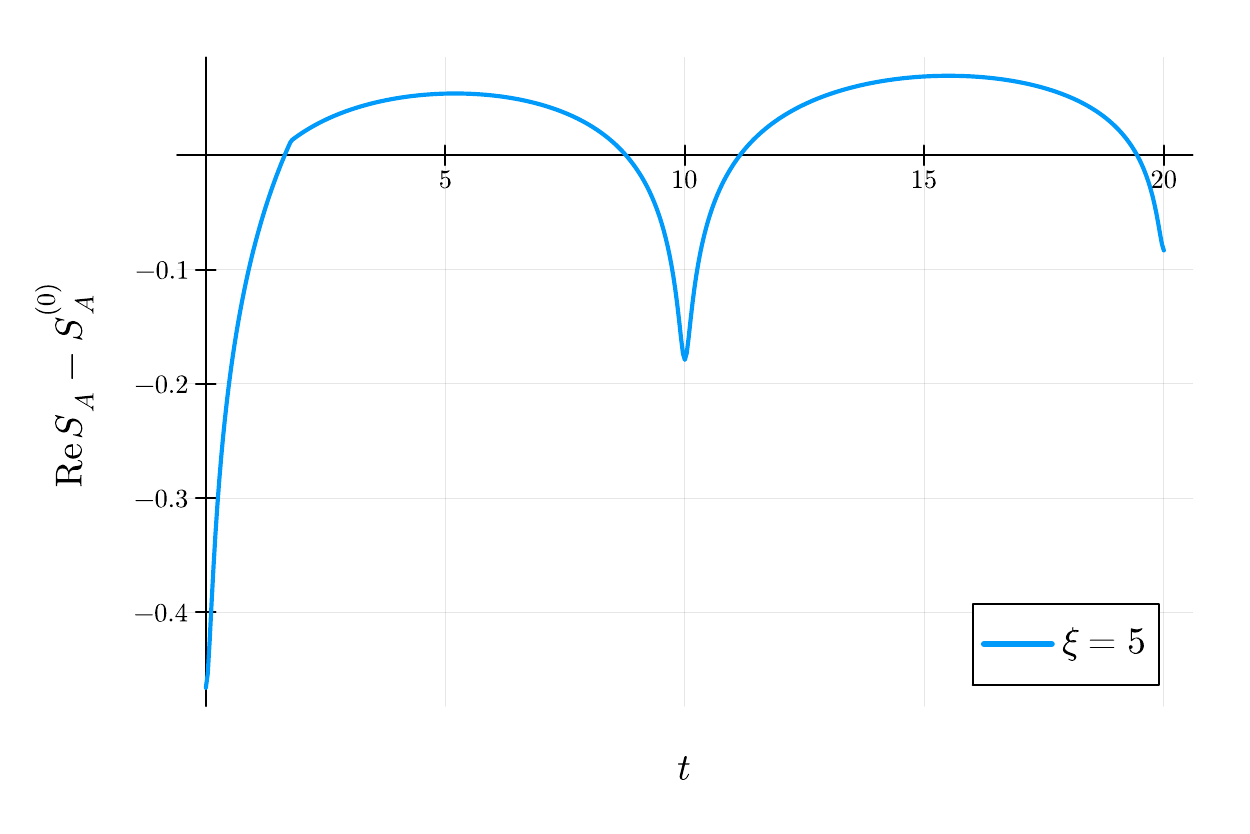}
    \includegraphics[width=0.45\linewidth]{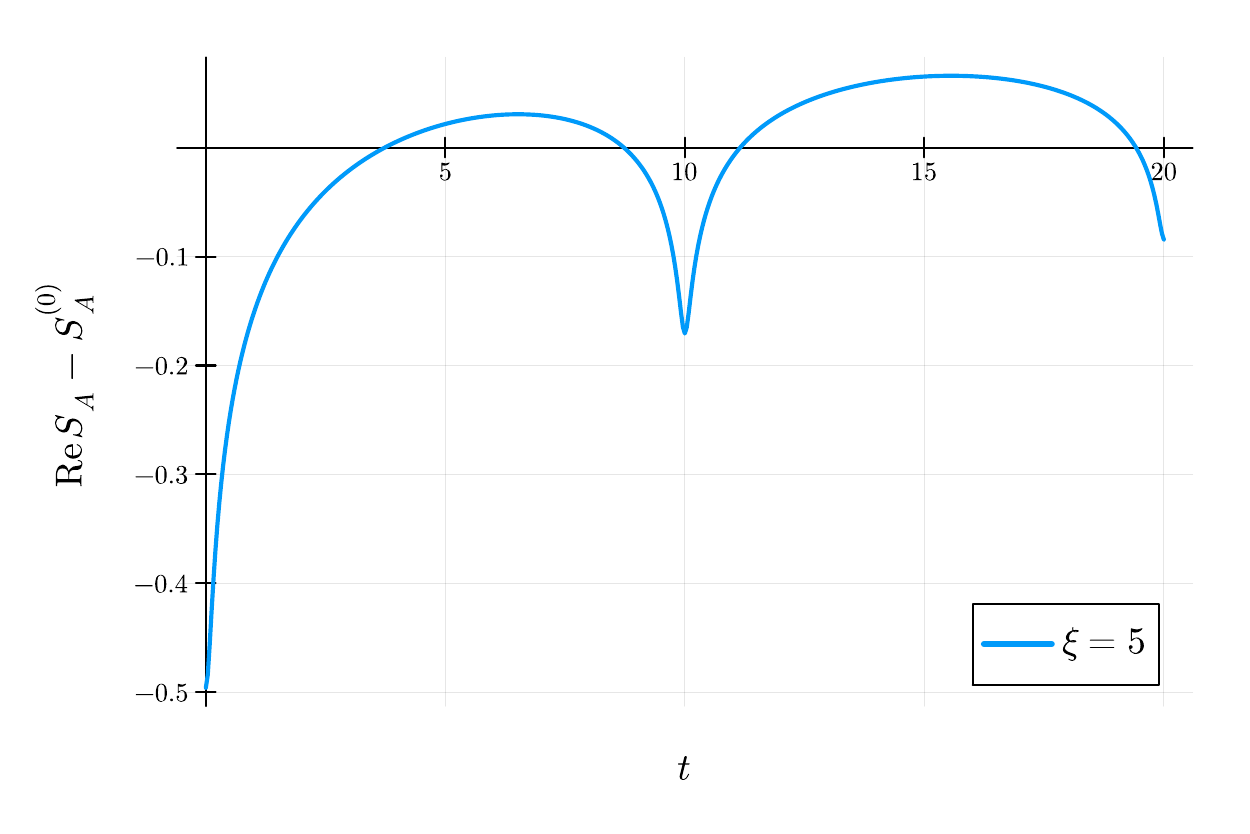}
    \caption{The real part of the pseudo entropy \(S_{A(\x)}^{\JQ_1|\JQ_2}(t) - S_{A(\x)}^{(0)}\) in the holographic CFT/the free Dirac fermion CFT for \(A(\x) = [\x - 5, \x + 5]\) is shown in the left/right column.
    The top/middle/bottom row represents the spatiotemporal/spatial/temporal dependence.
    We chose \(c=1\), \(\ep=1\), \(\d=0.1\), and \(S_{\mathrm{bdy}}=0\).}
    \label{fig:rePE_JQJQ}
\end{figure}

\begin{figure}
    \centering
    \includegraphics[width=0.45\linewidth]{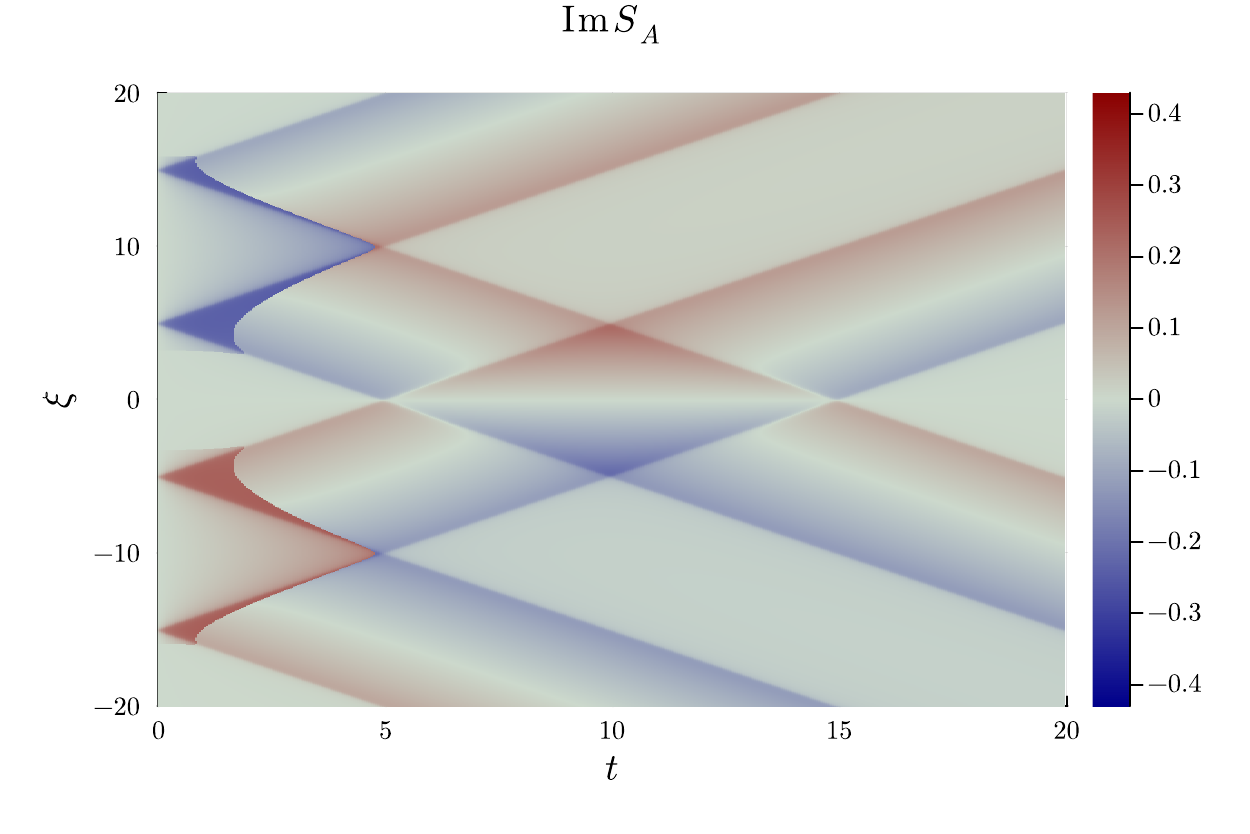}
    \includegraphics[width=0.45\linewidth]{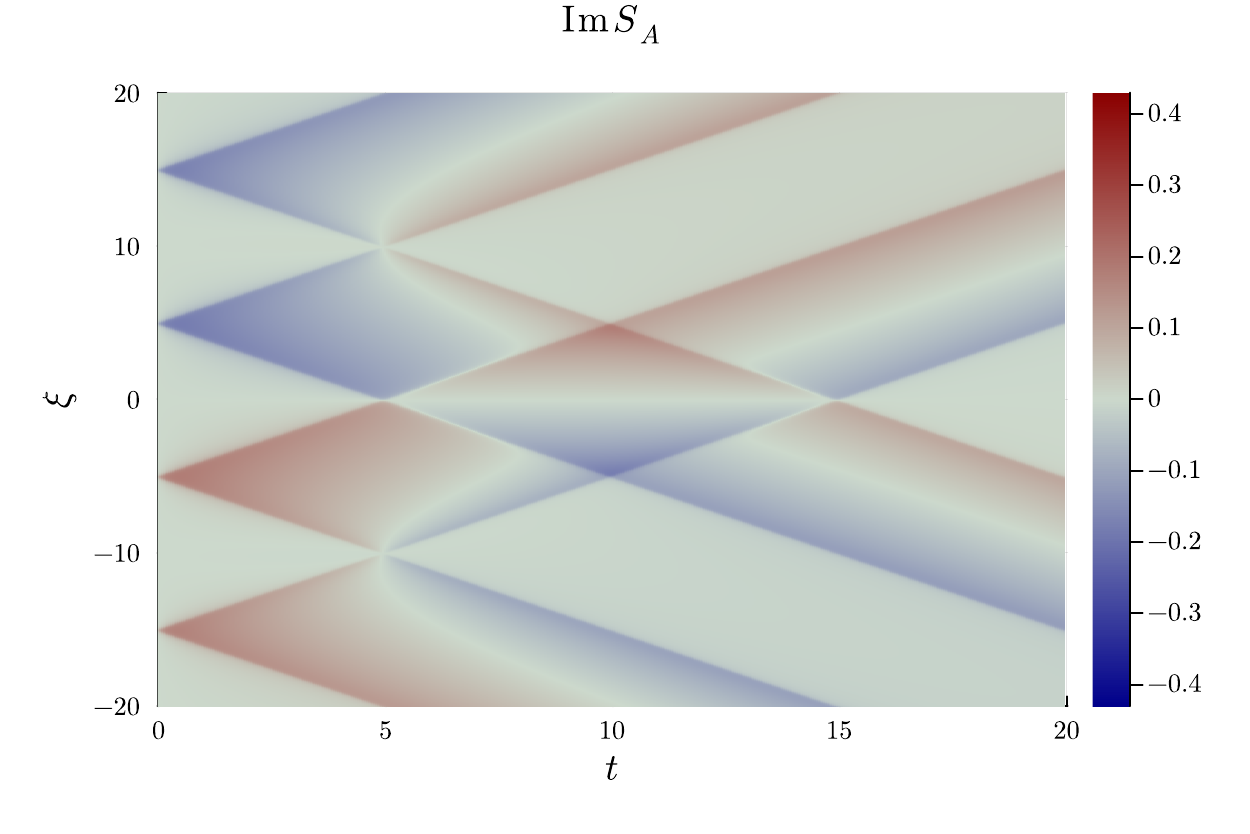}
    \includegraphics[width=0.45\linewidth]{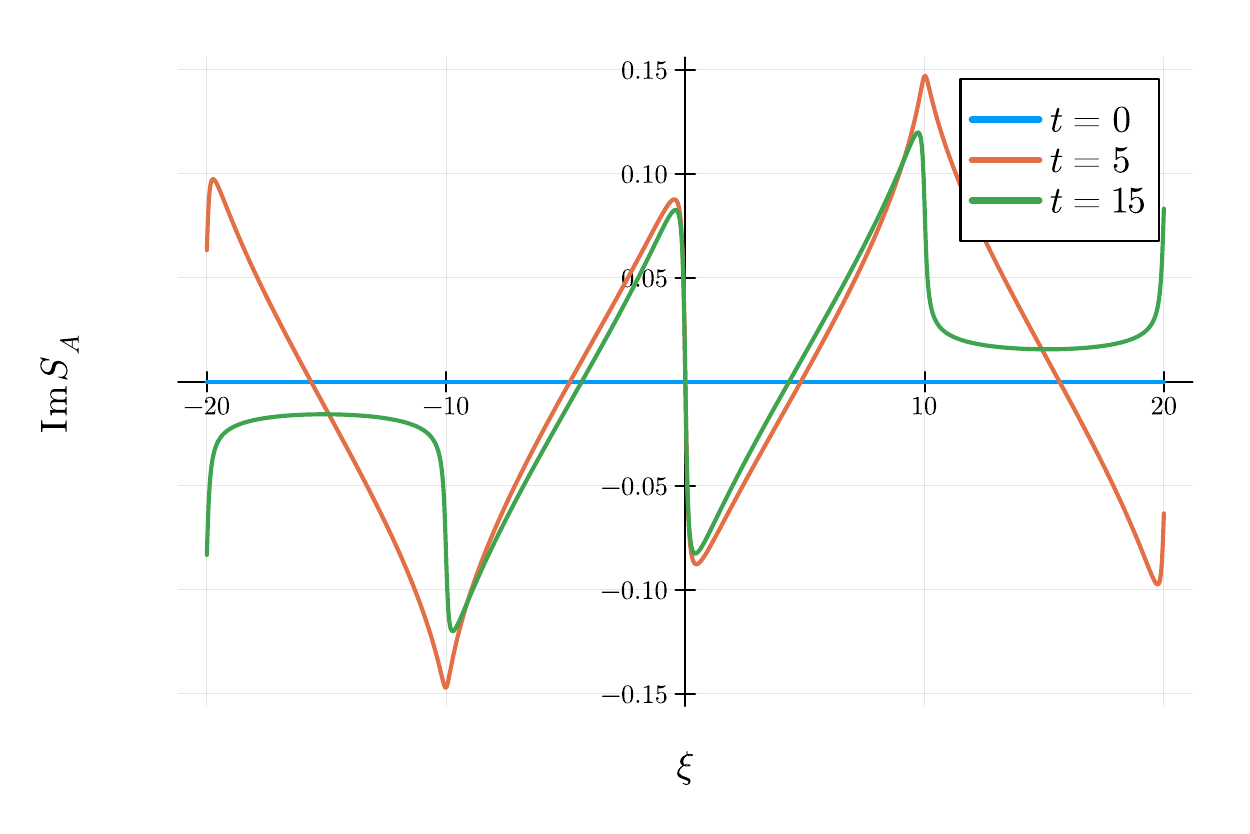}
    \includegraphics[width=0.45\linewidth]{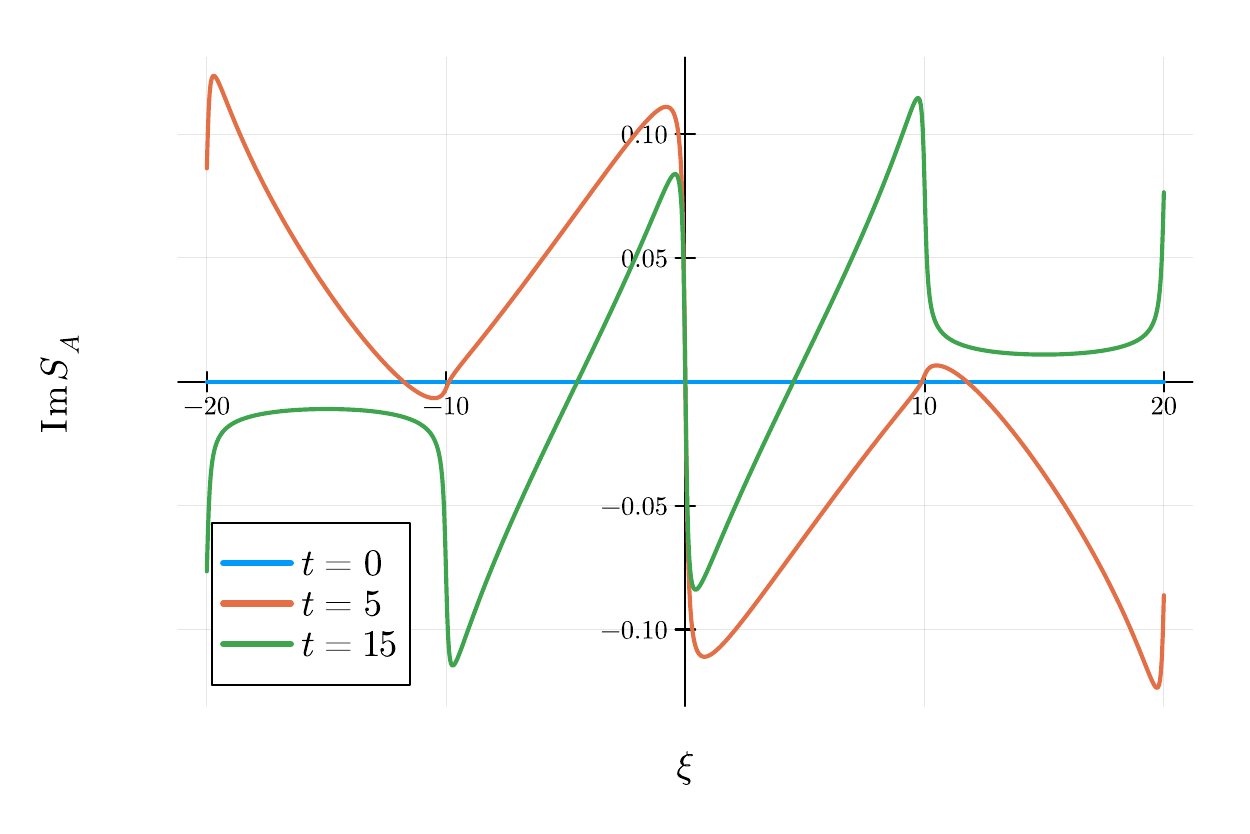}
    \includegraphics[width=0.45\linewidth]{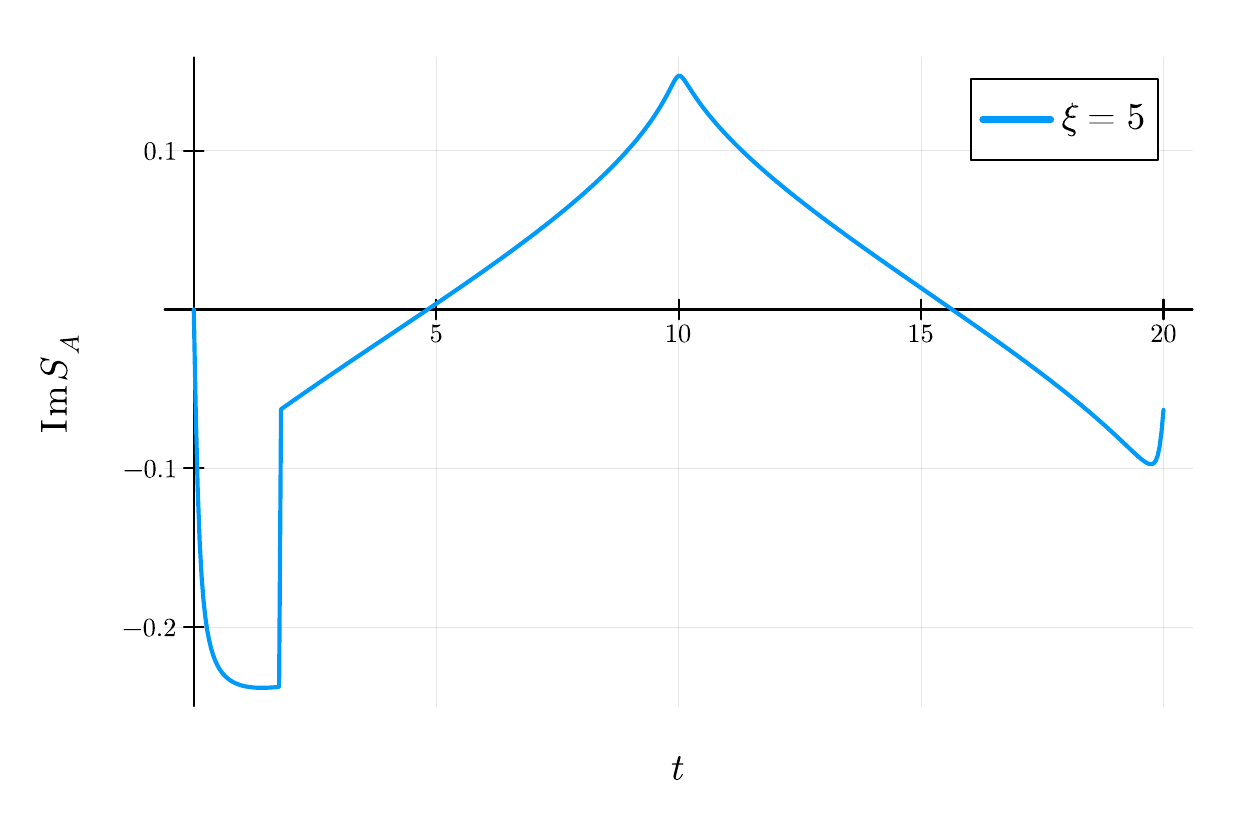}
    \includegraphics[width=0.45\linewidth]{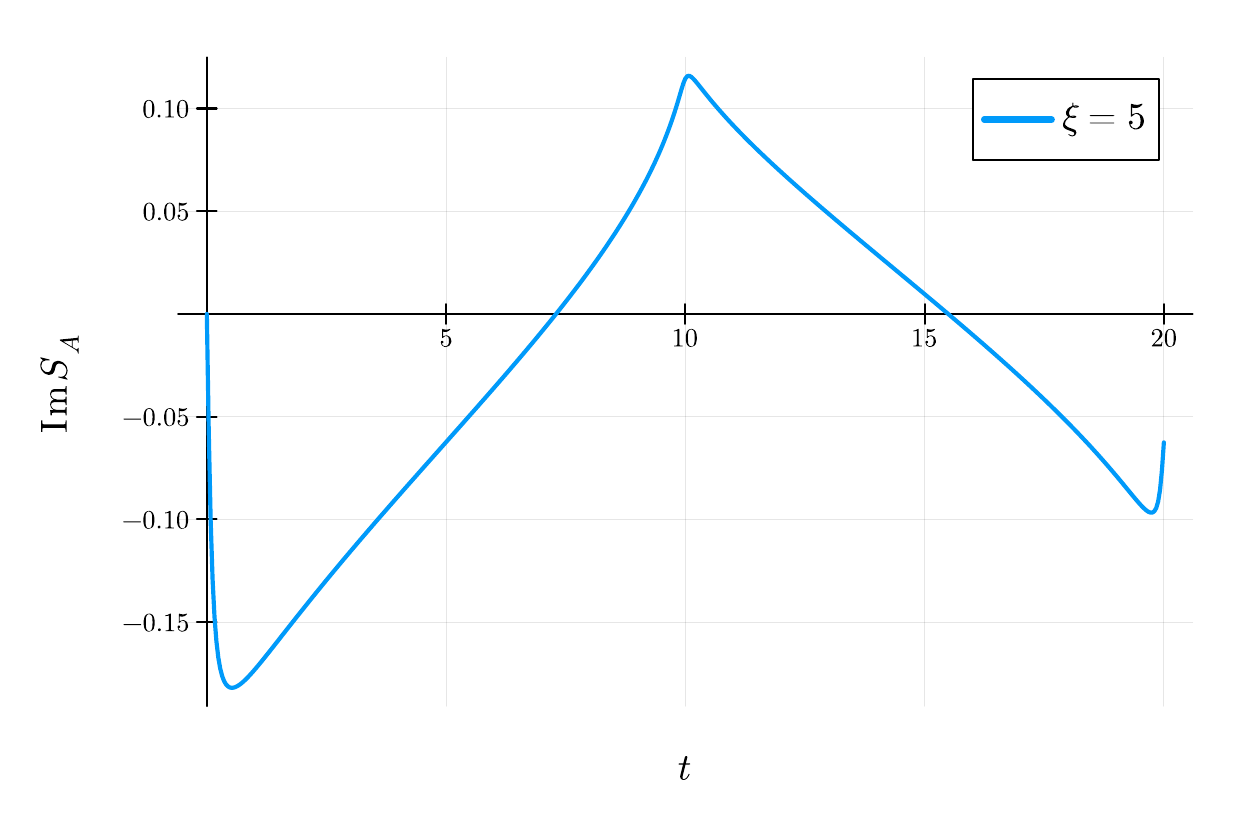}
    \caption{The imaginary part of the pseudo entropy \(S_{A(\x)}^{\JQ_1|\JQ_2}(t) - S_{A(\x)}^{(0)}\) in the holographic CFT/the free Dirac fermion CFT for \(A(\x) = [\x - 5, \x + 5]\) is shown in the left/right column.
    The top/middle/bottom figure represents the spatiotemporal/spatial/temporal dependence.
    We chose \(c=1\), \(\ep=1\), and \(\d=0.1\).}
    \label{fig:imPE_JQJQ}
\end{figure}

We take the interval subsystem \(A\) as defined in \eqref{intevgqw} for the pseudo entropy.
We present plots of the pseudo entropy for both the holographic CFT and the free Dirac fermion CFT in Fig.~\ref{fig:rePE_JQJQ} (real part) and Fig.~\ref{fig:imPE_JQJQ} (imaginary part).
For the holographic CFT, we take the contribution with a smaller real part.

We note that the qualitative behaviors of pseudo entropy are similar between the holographic CFT and the free Dirac fermion CFT. 
In both cases, we observe a dip when the excitations created at the joining points \(x=\pm 10\) at \(t=0\) propagate to the endpoints of the interval \(A\). 
This behavior is once again attributed to the unique property of pseudo entropy when entanglement flippings occur, as we have seen in previous examples.

\subsection{Entropy excess}
\begin{figure}
    \centering
    \includegraphics[width=0.45\linewidth]{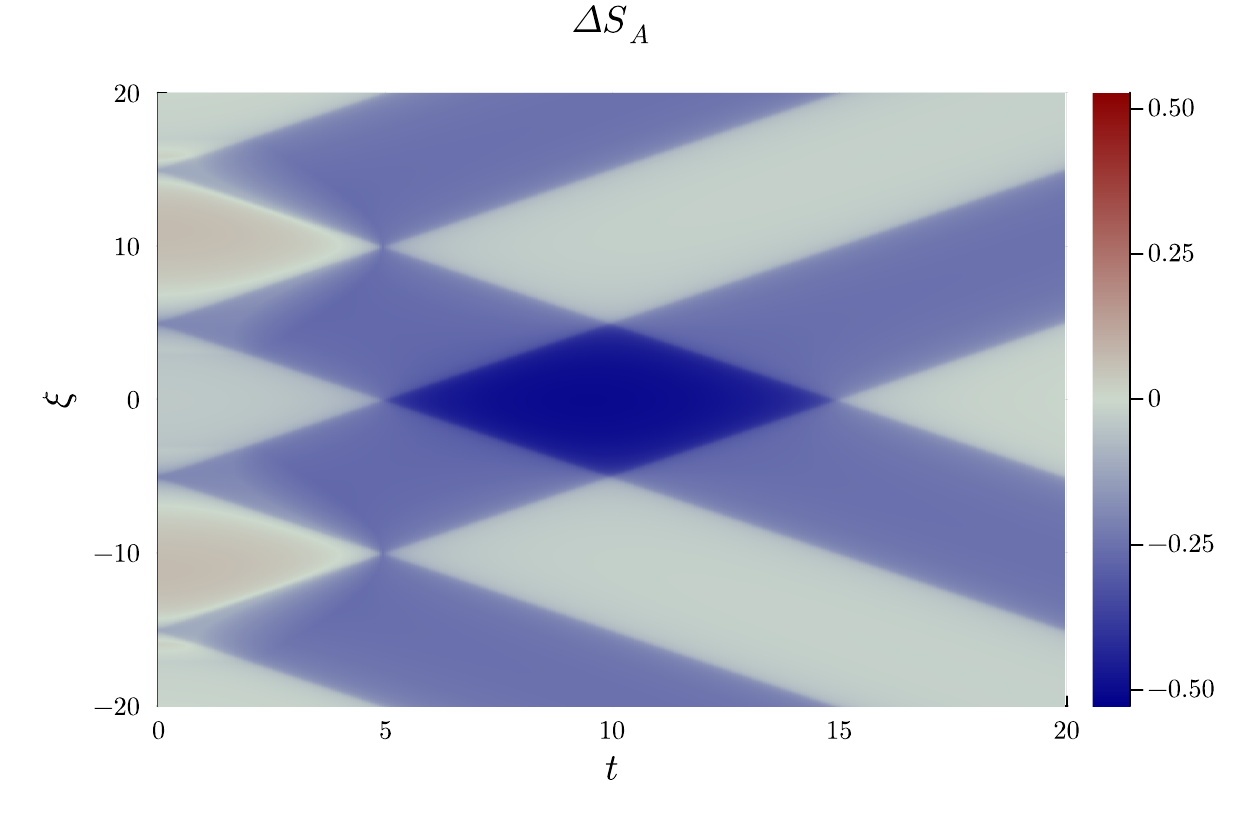}
    \includegraphics[width=0.45\linewidth]{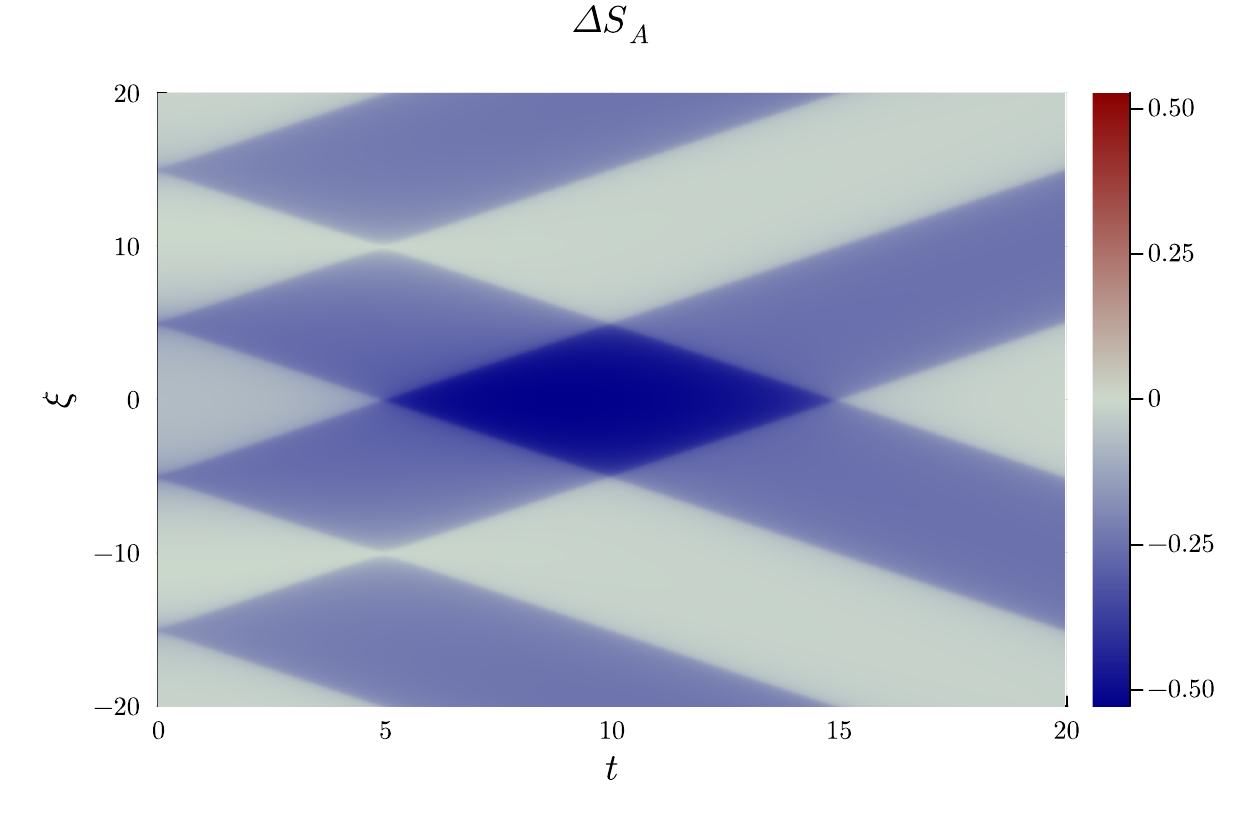}
    \includegraphics[width=0.45\linewidth]{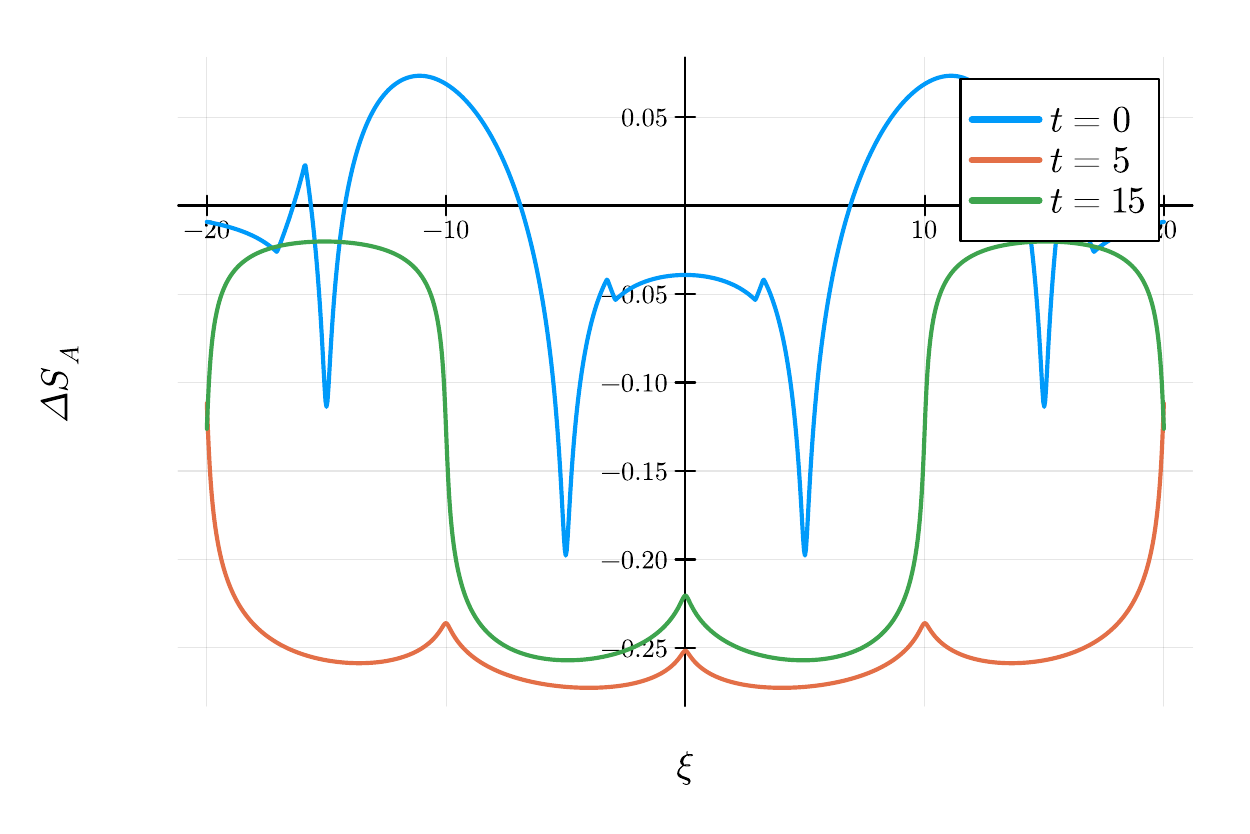}
    \includegraphics[width=0.45\linewidth]{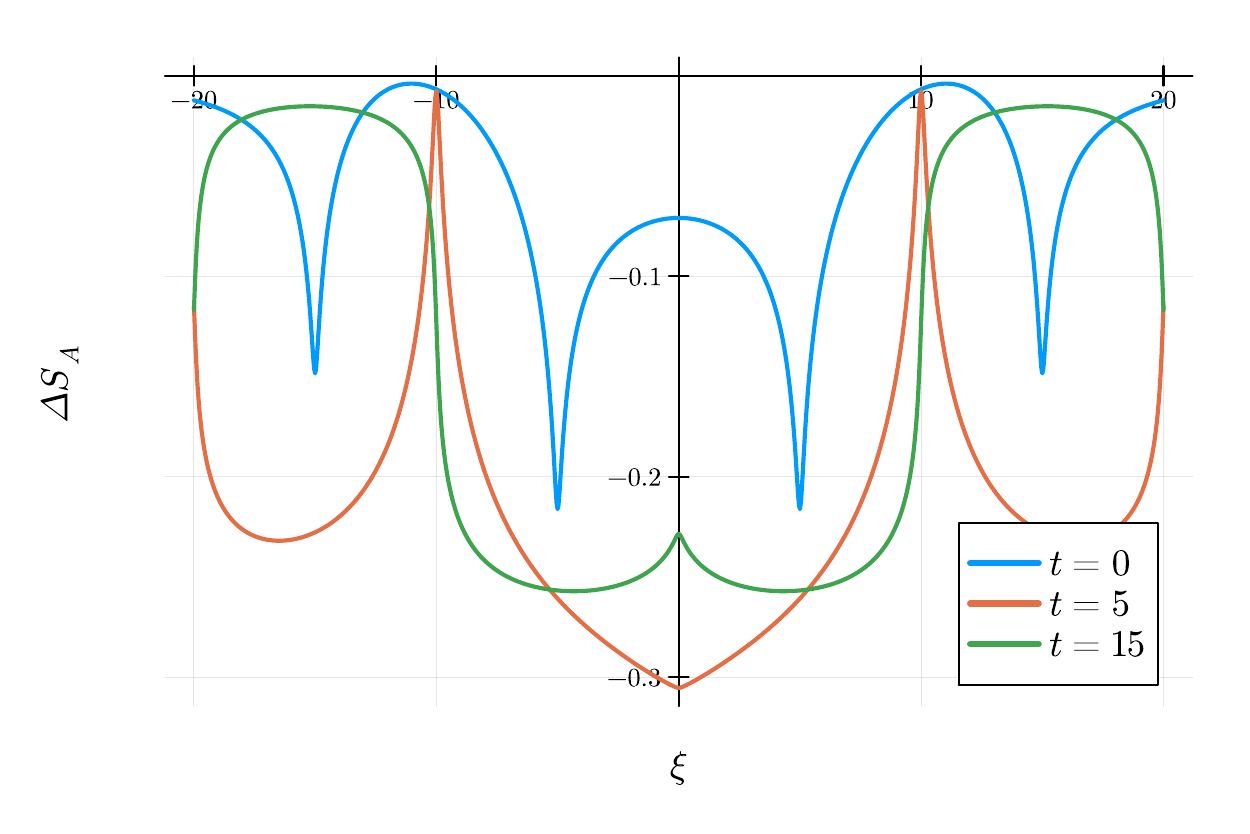}
    \includegraphics[width=0.45\linewidth]{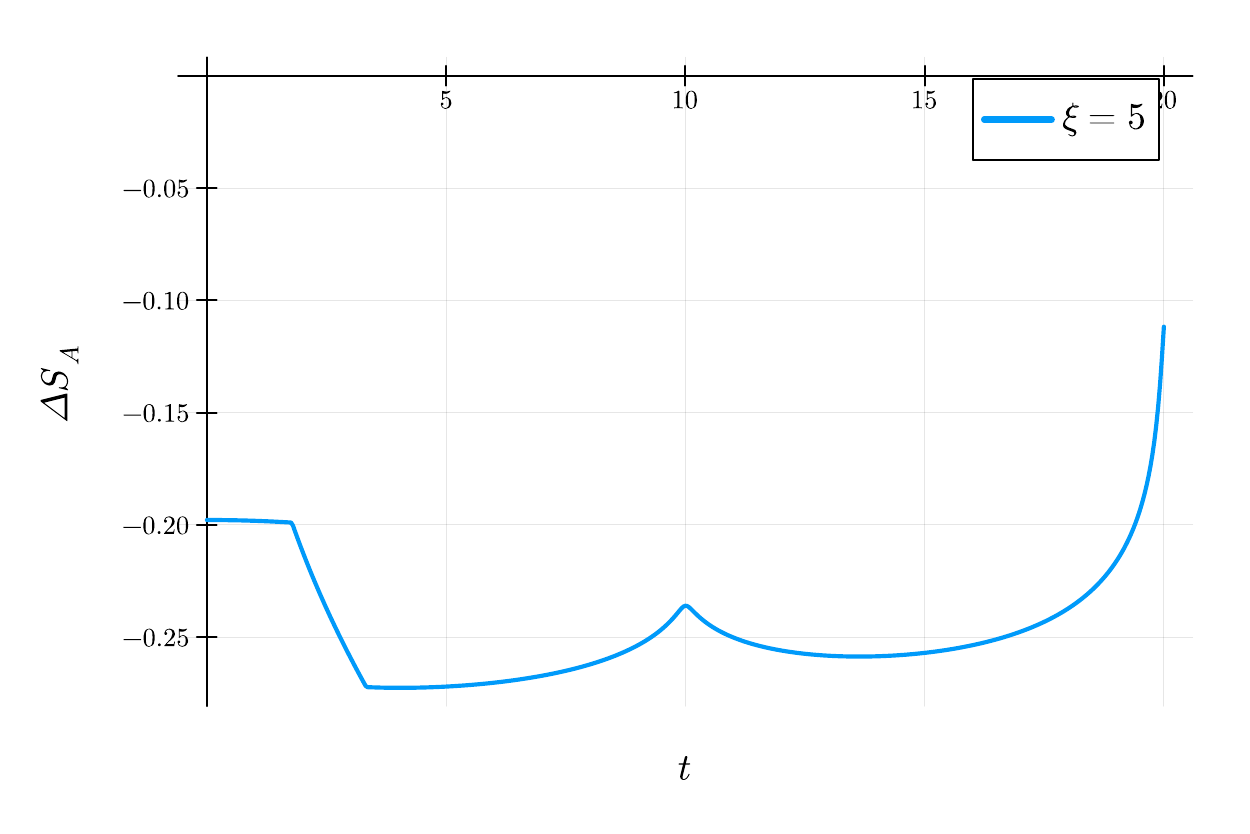}
    \includegraphics[width=0.45\linewidth]{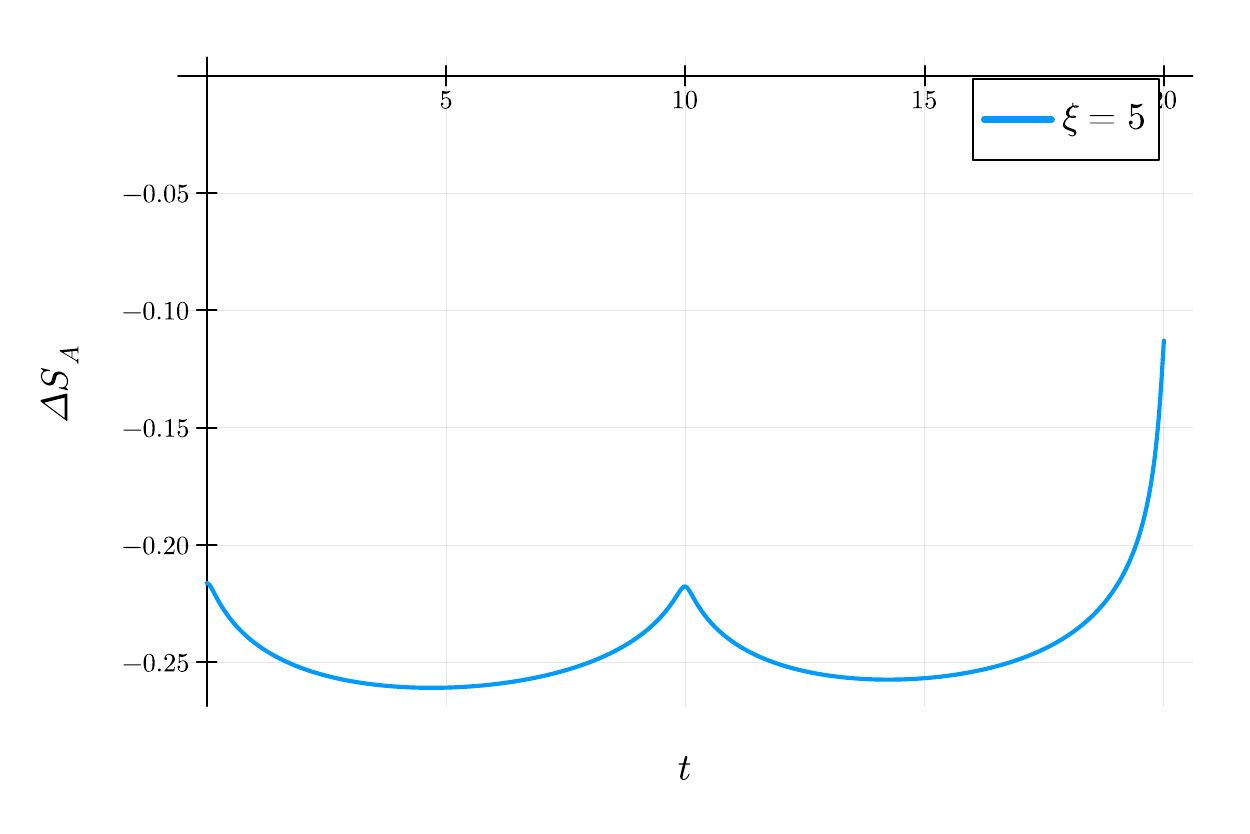}
    \caption{The entropy excess \(\varDelta{S}_{A(\x)}^{\JQ_1|\JQ_2}(t)\) in the holographic CFT/the free Dirac fermion CFT for \(A(\x) = [\x - 5, \x + 5]\) is shown in the left/right column.
    The top/middle/bottom row represents the spatiotemporal/spatial/temporal dependence.
    We chose \(c=1\), \(\ep=1\), \(\d=0.1\), and \(S_{\mathrm{bdy}}=0\).}
    \label{fig:deltaS_JQJQ}
\end{figure}

The entropy excess \(\varDelta{S}_A^{\JQ_1|\JQ_2}\) is shown in Fig.~\ref{fig:deltaS_JQJQ} for both the holographic and free Dirac fermion CFTs. 
In the holographic CFT, we observed that the entropy excess \(\varDelta{S}_A^{\JQ_1|\JQ_2}\) takes positive values when the subsystem \(A\) is affected only by one of the two excitations. 
However, in the free Dirac fermion CFTs, we always find that the entropy excess is non-positive. 
As in the single-slit case, this implies the presence of multi-partite entanglement in the holographic CFT, while it is absent in the free Dirac fermion CFT.

\clearpage
\section{Discussions}\label{sec:discussions}
In this paper, we have investigated the behaviors of pseudo entropy in two-dimensional holographic and free Dirac fermion CFT for (i) the vacuum state and the joining quenched state and for (ii) two different joining quenched states. 
The pseudo entropy of the joining quenched states provides further information that is not accessible through conventional entanglement entropy calculations for single local quenched states alone.

We have already summarized our main results in section \ref{ssec:PSummary}. 
In particular, we found the following two characteristic properties absent in the conventional entanglement entropy under the joining local quenches. 
One is that the time evolution of the pseudo entropy exhibits a dip when the excitations generated by the joining quench propagate to the boundaries of the subsystem \(A\). 
This dip arises due to the suppression of pseudo entropy in entanglement flipping \cite{Nakata:2021ubr}. 
The other is that the entropy excess, defined in \eqref{dPE}, can be positive in the holographic CFTs, while it always remains non-positive in the free Dirac fermion CFTs. 
We noted that \(\varDelta{S}_A>0\) occurs exclusively in joining quenched states with multi-partite entanglement based on simple qubit models. 
This suggests that the holographic CFTs possess multi-partite entanglement in their vacuum states, whereas the free Dirac fermion CFTs do not.

We mention a few interesting future directions as we conclude this paper. 
We currently lack a comprehensive understanding of the imaginary part of pseudo entropy.
In our joining quench calculations, we found many results for the time evolution of the imaginary parts, and it would be interesting to explore its physical and quantum information theoretic implications. 
It is also an intriguing future problem to understand our results in terms of toy models of tensor networks, such as the perfect model \cite{Pastawski:2015qua} and random tensor networks \cite{Hayden:2016cfa}. 
Extending our analysis to calculate pseudo entropy for local quenched states in diverse quantum spin chains can provide a deeper understanding of the physical interpretations and implications of pseudo entropy.

\section*{Acknowledgements}
We are grateful to Pawel Caputa and Shinsei Ryu for useful discussions.
This work is supported by MEXT KAKENHI Grant-in-Aid for Transformative Research Areas (A) through the ``Extreme Universe'' collaboration: Grant Number 21H05187. 
TT is also supported by Inamori Research Institute for Science and by JSPS Grant-in-Aid for Scientific Research (A) No.~21H04469. 
We thank the YITP-ExU long-term workshop ``Quantum Information, Quantum Matter and Quantum Gravity'' (YITP-T-23-01), where a part of this work was completed.

\appendix
\section{Some details of CFT calculations}
\label{ap:formulae}

\subsection{Conformal map for double-slit geometry}
This section derives the conformal map from the double-slit geometry to the upper half-plane.
The conformal map \(F(z)\) from the upper half-plane to the interior of a polygon with \(n\) vertices is given by the Schwarz-Christoffel formula
\begin{equation}
   F(z) = C\int^z \prod_{k=1}^{n-1} (\z - \x_k)^{-\b_k} \dd{\z} + C', \label{SC}
\end{equation}
where \(\b_k\)'s are the exterior angles of each vertex divided by \(\pi\), and \(\x_k\)'s are real numbers mapped to each vertex.

We define \(w_a = - \delta + i x_1\) and \(w_b = \delta + i x_2\) as in Fig.~\ref{fig:double-slit_geometry}.
In the following, we will show that, for given \(x_1\), \(x_2\) and \(\d\), it is possible to construct an appropriate function \(F(z)\) such that \(F(-a) = w_a\) and \(F(b) = w_b\).
In this case, we have \(\x_1 = -a\), \(\x_2 = b\), \(\x_3 = 0\) and \(\b_1 = -1\), \(\b_2 = -1\), \(\b_3 = 2\) for \eqref{SC}. 
Now \eqref{SC} turns to be
\begin{align}
   F(z) &= C \int^z \frac{(\x + a)(\x - b)}{\x^2} \dd{\x} + C' \notag\\
   &= C \left( z + (a-b) \log{z} + \frac{ab}{z} \right) + C''.
\end{align}
By imposing the change on the constants \(C\) and \(C''\), we can take the principal value logarithm
\begin{equation}
   \Log{z} \coloneqq \log{|z|} + i \Arg{z} \quad (-\pi < \Arg{z} \leq \pi)
\end{equation}
as the logarithmic function in the equation, so we obtain a general expression for the conformal transformation
\begin{equation}
   F(z) = C\left( z + (a - b) \Log{z} + \frac{ab}{z} \right) + C''.
\end{equation}
We choose \(a\), \(b\), \(C\) and \(C''\) appropriately as shown below.

For \(x \in \mathbb{R}\setminus\{0\}\), \(F(x)\) is
\begin{align}
   F(x) &= 
   \begin{cases}\displaystyle
      C\left( x + \frac{ab}{x} + (a - b) \log{x} \right) + C'' &\quad (x > 0) \\\displaystyle
      C\left( x + \frac{ab}{x} + (a - b) \log{|x|} + i \pi (a - b) \right) + C'' &\quad (x < 0)
   \end{cases}.
\end{align}
Here \(C''\) corresponds to translation in the \(w\)-plane.

From the constraint that \(F(\mathbb{R})\) is parallel to the real axis, we have
\begin{equation}
   \pa_x \Im{F(x)} = 0 \quad (x \in \mathbb{R}\setminus\{0\}). \label{parallel}
\end{equation}
Setting \(C = R e^{i\theta} \quad (R>0, \theta \in [0, 2\pi[)\), the left hand side of \eqref{parallel} is
\begin{align}
   \pa_x \Im{F(x)} &= \Im{\pa_x F(x)} \notag\\
   &= R\sin{\theta} \left( 1 + \frac{a - b}{x} - \frac{ab}{x^2} \right).
\end{align}
Given that \(F(x) \to +\infty\) as \(x \to +\infty\), we obtain \(\theta = 0\).

From \(F(b) = w_b\) and \(F(-a) = w_a\), we have
\begin{align}
   \begin{cases}
      R\left\{ (a + b) - (b-a) \log{b} \right\} + C'' &= \delta + i x_2 \\
      R\left\{ - (a + b) - (b-a) \log{a} - i\pi(b - a) \right\} + C'' &= - \delta + i x_1
   \end{cases}.
\end{align}
For simplicity, let \(R=1\).
The desired function can be constructed even after making this assumption.
Considering the sum and the difference of the two equations, the conditions now change to
\begin{align}
   \begin{cases}
      - (b - a) \log{ab} - i\pi(b - a) + 2C'' &= i (x_1 + x_2) \\
      2(a + b) - (b - a) \log{(b / a)} + i\pi(b - a) &= 2\delta + i (x_2 - x_1)
   \end{cases}.\label{cond}
\end{align}

Defining \(A \coloneqq (x_2 - x_1) / \pi (>0)\), we obtain \(b = a + A\) from the imaginary part of the second equation.
Substituting it into the real part of the second equation of \eqref{cond}, we obtain
\begin{equation}
   4a + 2A - A \log\left(1 + \frac{A}{a} \right) = 2\delta. \label{tildeF}
\end{equation}
Defining \(K(a)\) as the left-hand side of \eqref{tildeF}, we have
\begin{align}
   \pa_a K(a) &= \frac{(2 + A / a)^2}{1 + A / a} > 0 \quad (\forall a > 0) \\
   \lim_{a \searrow 0} K(a) &= -\infty\\
   \lim_{a \nearrow \infty} K(a) &= \infty,
\end{align}
which shows that for any \(\delta > 0\), the equation \(K(a) = 2\delta\), i.e., \eqref{tildeF}, always has a unique solution \(a=a^* > 0\).

From the first equation of \eqref{cond}, we obtain
\begin{align}
   C'' &= \frac{A}{2} \log{[a^*(a^*+A)]} + i \frac{\pi A}{2} + i \frac{x_1 + x_2}{2} \notag\\
   &= 2a^* + A + A\log{a^*} - \delta + i x_2.
\end{align}

\begin{figure}[t]
   \centering
   \includegraphics[width=5cm]{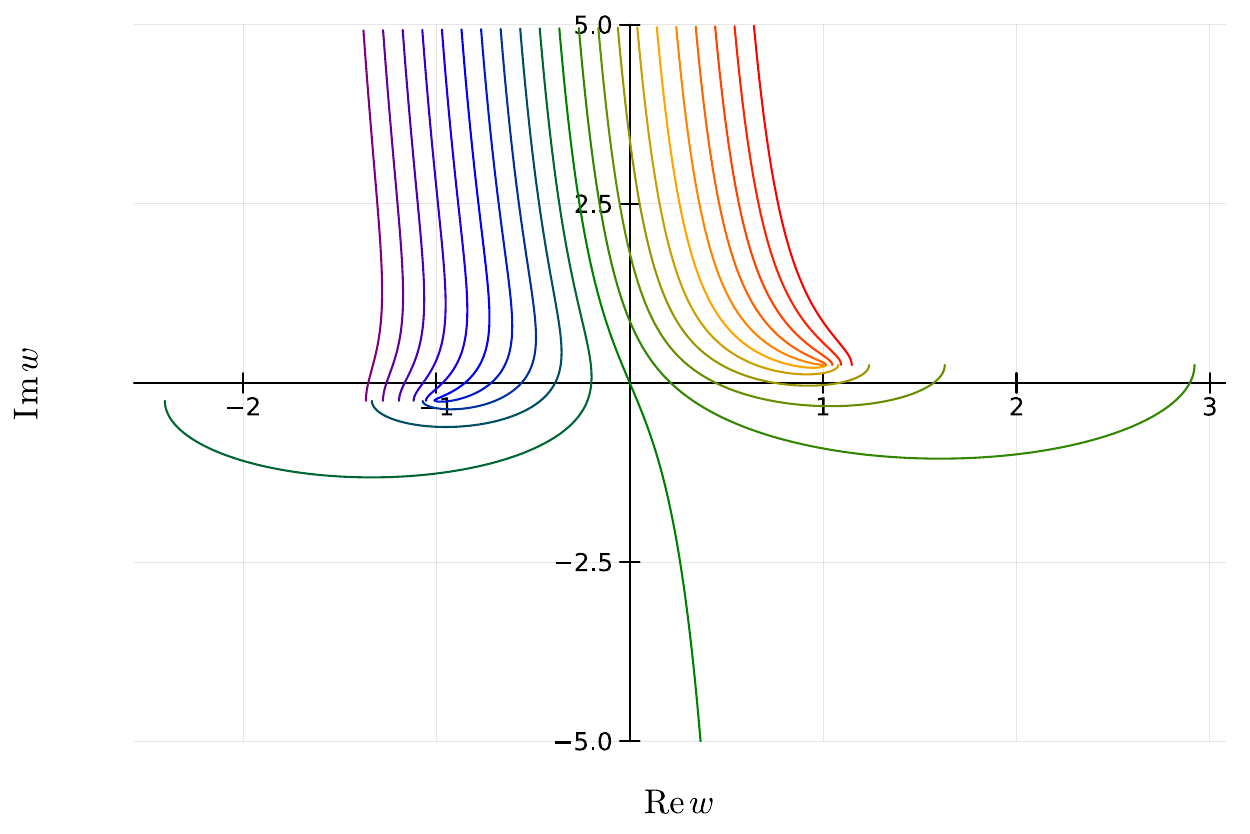}
   \includegraphics[width=5cm]{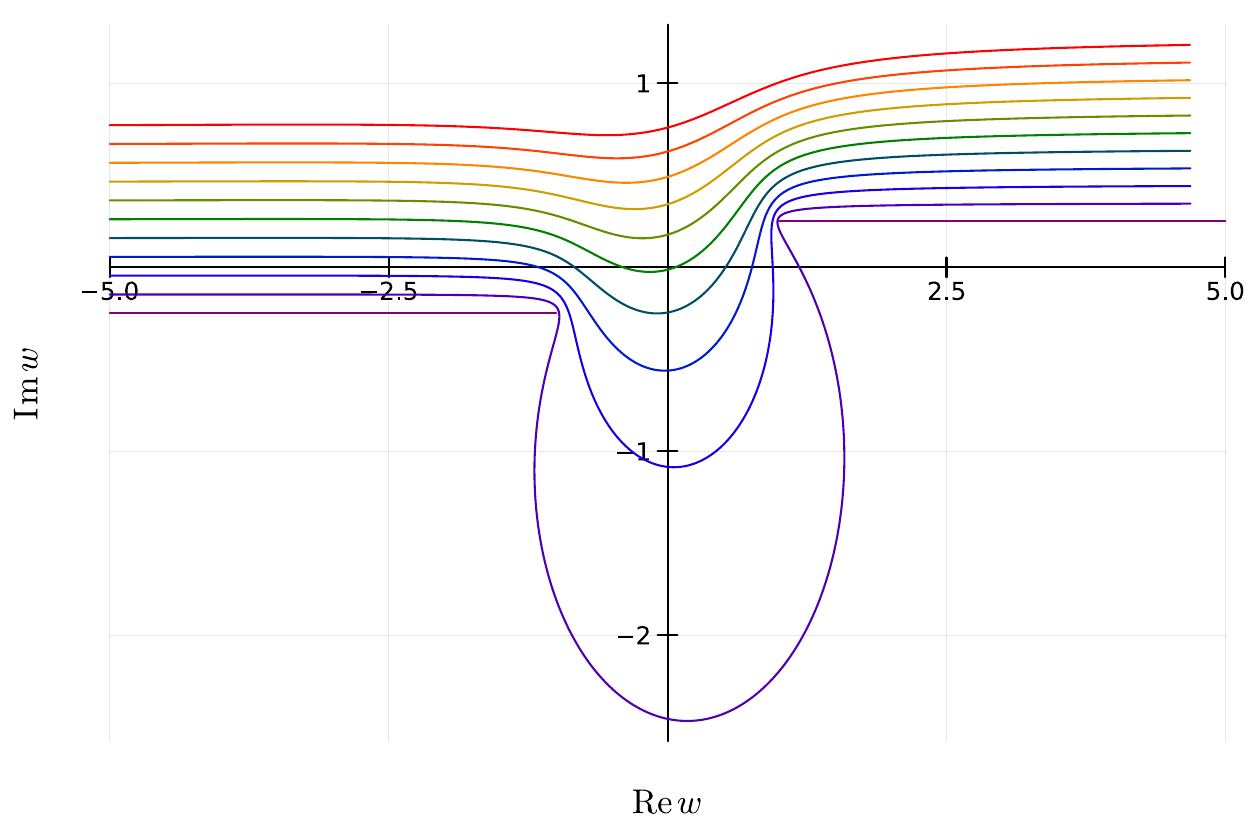}
   \includegraphics[width=5cm]{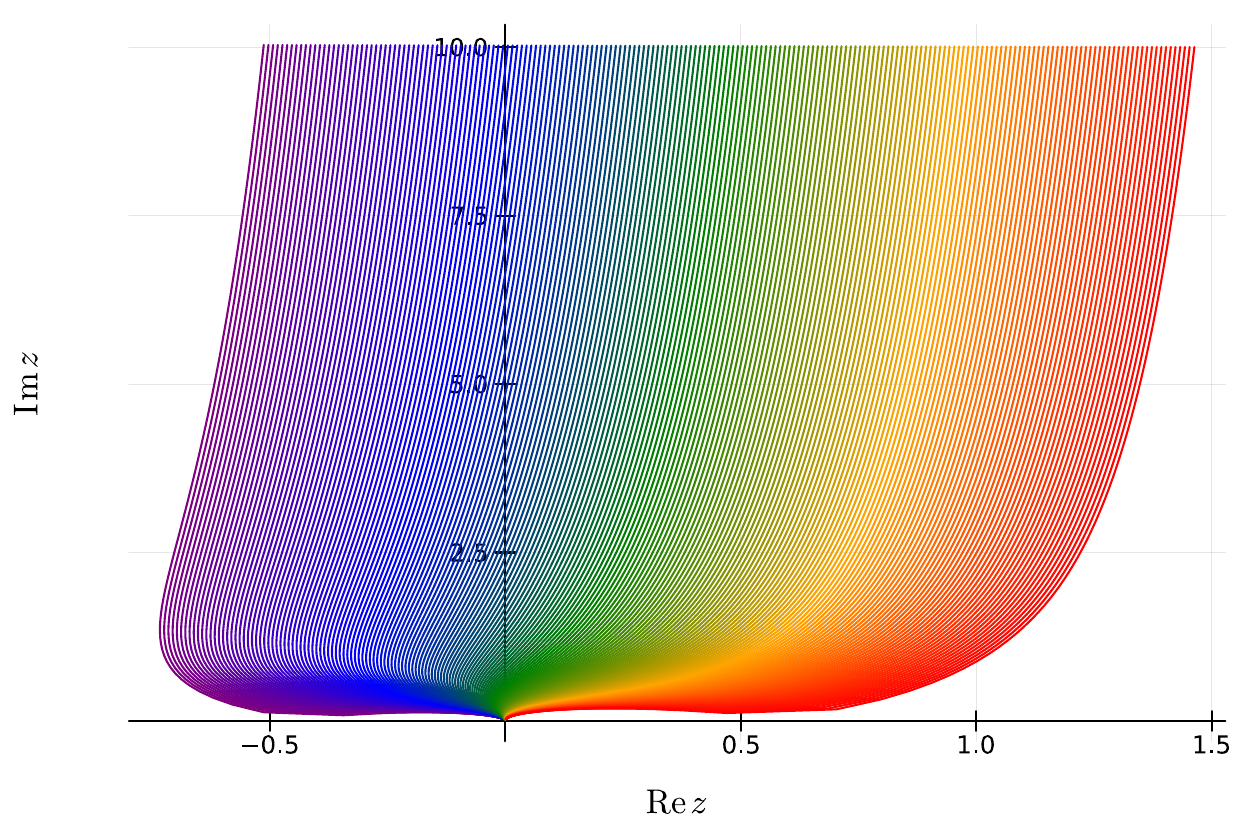}
   \caption{Plots of \(F(z)\) and its inverse \(f(w)\).
   Each line corresponds to the image \(w=F(z)\) for \(z\) with constant real parts (left), the image \(w=F(z)\) for \(z\) with constant imaginary parts (center), and the image \(z=f(w)\) for \(w\) with constant real parts (right).}
   \label{fig:Doubleslitmap}
\end{figure}

Finally, we obtain the expression for the function \(F(z)\) as
\begin{equation}
   F(z) = z - A \Log{z} + \frac{a^*(a^*+A)}{z} + ( 2a^* + A + A\log{a^*} - \delta + i x_2 ),
\end{equation}
where \(A \coloneqq (x_2 - x_1) / \pi\) and \(a=a^*>0\) is the unique solution of
\begin{equation}
   4a + 2A - A \log{\left(1 + \frac{A}{a}\right)} = 2\delta.
\end{equation}
The results are plotted in Fig.~\ref{fig:Doubleslitmap}.

\subsection{Correlation functions in BCFTs}
Correlation functions in BCFTs can be naturally obtained by determining the boundary conditions of BCFTs due to conformal symmetry.

The boundary conditions in BCFTs are obtained from the Ward-Takahashi identity for conformal symmetry.
After a few calculations, we find that for BCFTs to have conformal symmetry, the following boundary conditions are required:
\begin{equation}
\label{boundary_condition}
    T(z)=\bar{T}(\bar{z})\quad\text{for} z=\bar{z}.
\end{equation}
This implies no energy flows across the boundary.
Since both \(T(z)\) and \(\bar{T}(\bar{z})\) are analytic functions and  satisfy the boundary condition \eqref{boundary_condition}, we can analytically connect \(T(z)\) to the lower half-plane:
\begin{equation}
    T(z)=\bar{T}(\bar{z}=z^*)\quad \text{for} z\in \mathrm{LHP}.
\end{equation}
Using this, the Ward-Takahashi identity on UHP is
\begin{align}
    &\langle T(z)O(w,\bar{w})\rangle_{\mathrm{UHP}}\notag\\
    &=\left(\frac{h}{(z-w)^2}+\frac{1}{z-w}\partial_{w}+\frac{\bar{h}}{(z-\bar{w})^2}+\frac{1}{z-\bar{w}}\partial_{\bar{w}}\right)\langle O(w,\bar{w})\rangle_{\mathrm{UHP}}.
\end{align}
This has the same singularity as the Ward-Takahashi identity for \(\langle T(z)O(w,\bar{w})\bar{O}(\bar{w},w)\rangle_{\mathbb{C}}\) using a primary field \(\bar{O}(\bar{w},w)\) with weight \((\bar{h},h)\) in all planes.
From these facts, correlation functions in the upper half-plane are obtained by computing the correlation functions with operators inserted at the points of the mirror map about the boundary.
After inserting the operators, we consider the BCFTs as CFTs in the whole plane with only holomorphic conformal transformation.
This is called the double trick.
Thus, for example, a two-point function in the UHP becomes a four-point function in the entire complex plane, including its mirror image:
\begin{equation}
    \langle O_1(w_1,\bar{w_1})O_2(w_2,\bar{w_2})\rangle_{\mathrm{UHP}}\sim \langle O_1(w_1,\bar{w_1})O_2(w_2,\bar{w_2})\bar{O_1}(\bar{w_1},w_1)\bar{O_2}(\bar{w_2},w_2)\rangle_{\mathbb{C}}
\end{equation}
Note that Virasoro algebra is
\begin{align}
    &L_n \coloneqq \int_{0<\mathrm{arg}z<\pi,|z|=1}\frac{dz}{2\pi i}z^{n+2}T(z)+\int_{-\pi<\mathrm{arg}\bar{z}<0,|\bar{z}|=1}\frac{d\bar{z}}{2\pi i}z^{n+2}\bar{T}(\bar{z})\\
    &[L_n,L_m]=(n-m)L_{n+m}+\delta_{n+m,0}\frac{c}{12}n(n^2-1),
\end{align}
which is only the holomorphic part.

\section{Connected and disconnected contributions}\label{ap:condis}
\begin{figure}[ttt]
    \centering
    \includegraphics[width=6cm]{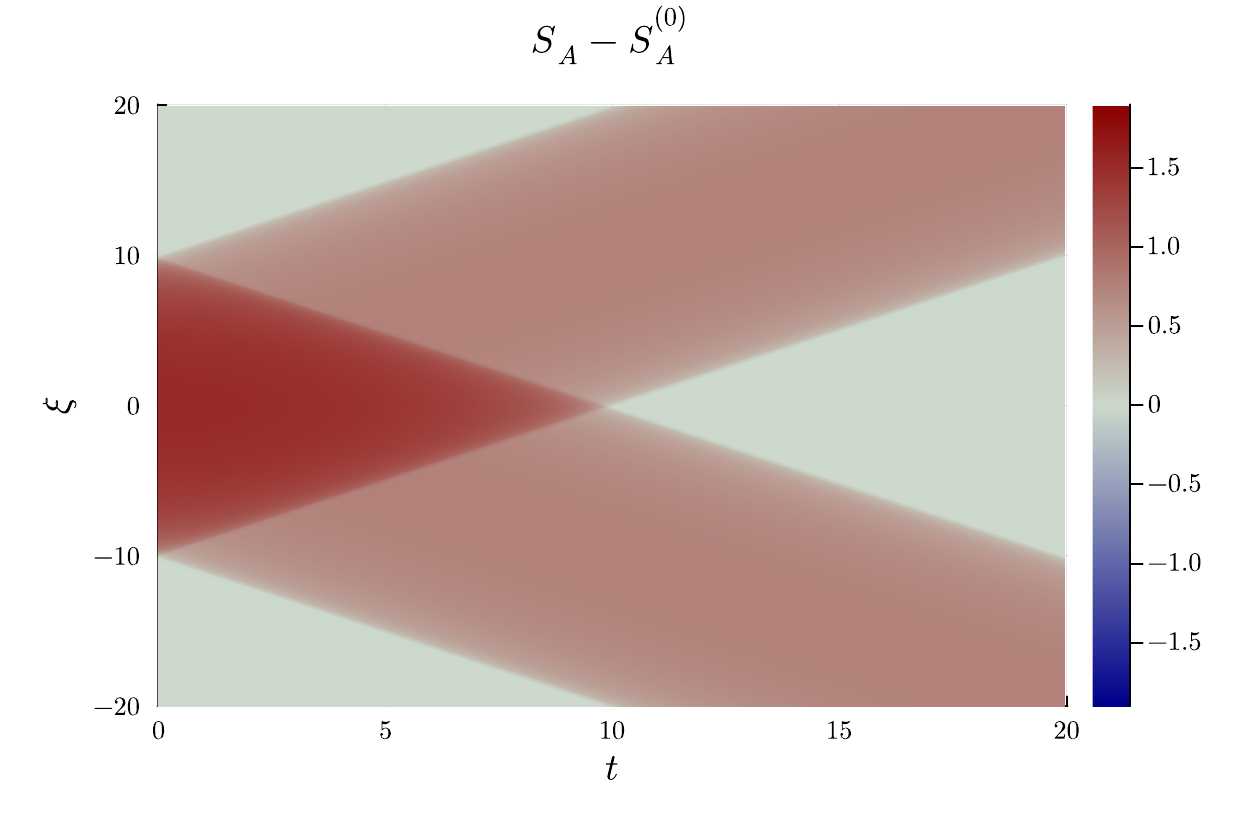}
    \includegraphics[width=6cm]{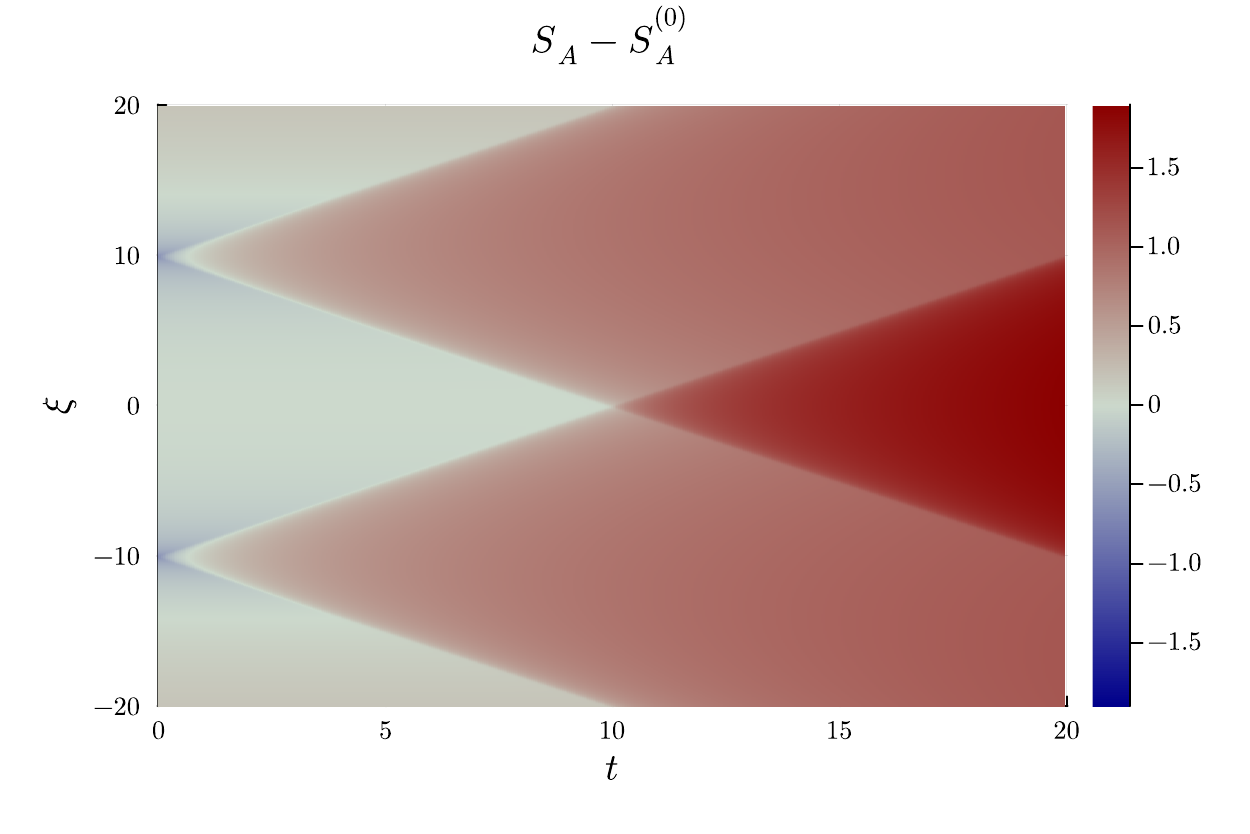}
    \includegraphics[width=6cm]{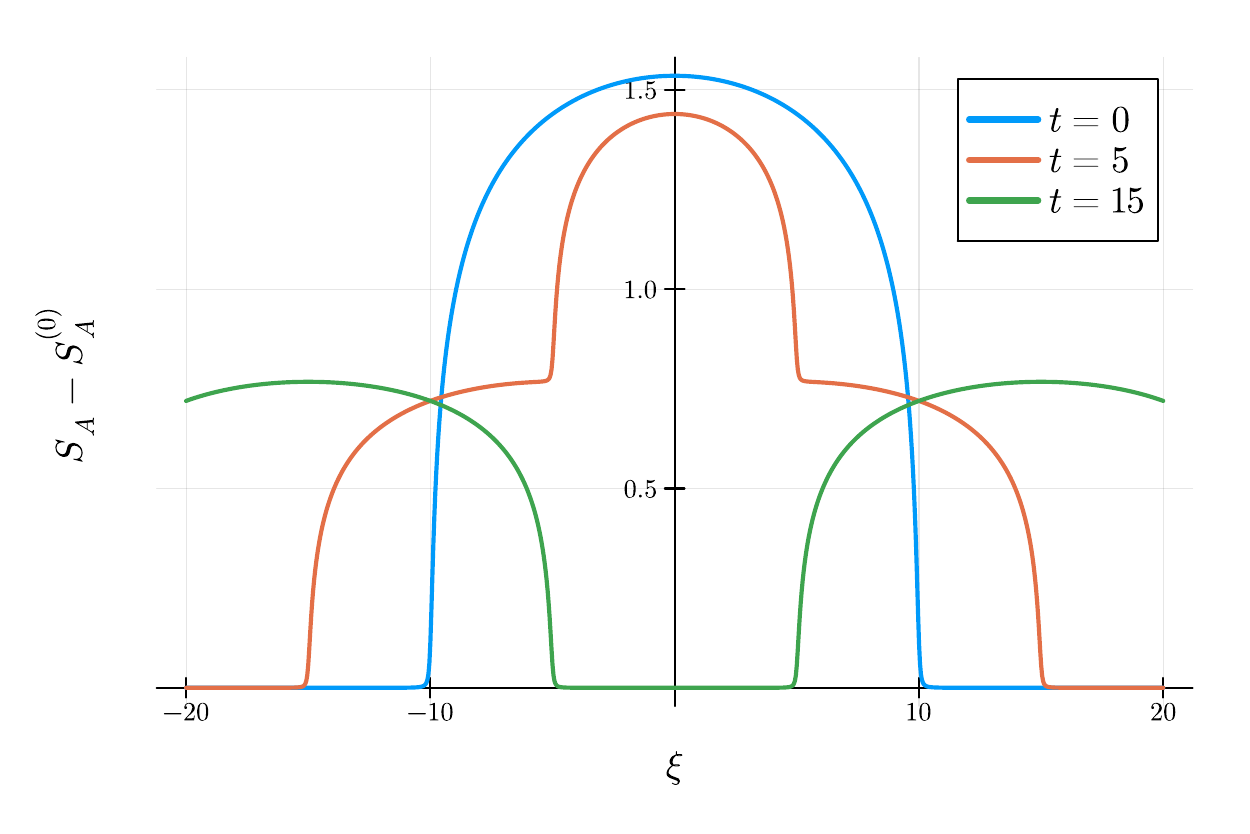}
    \includegraphics[width=6cm]{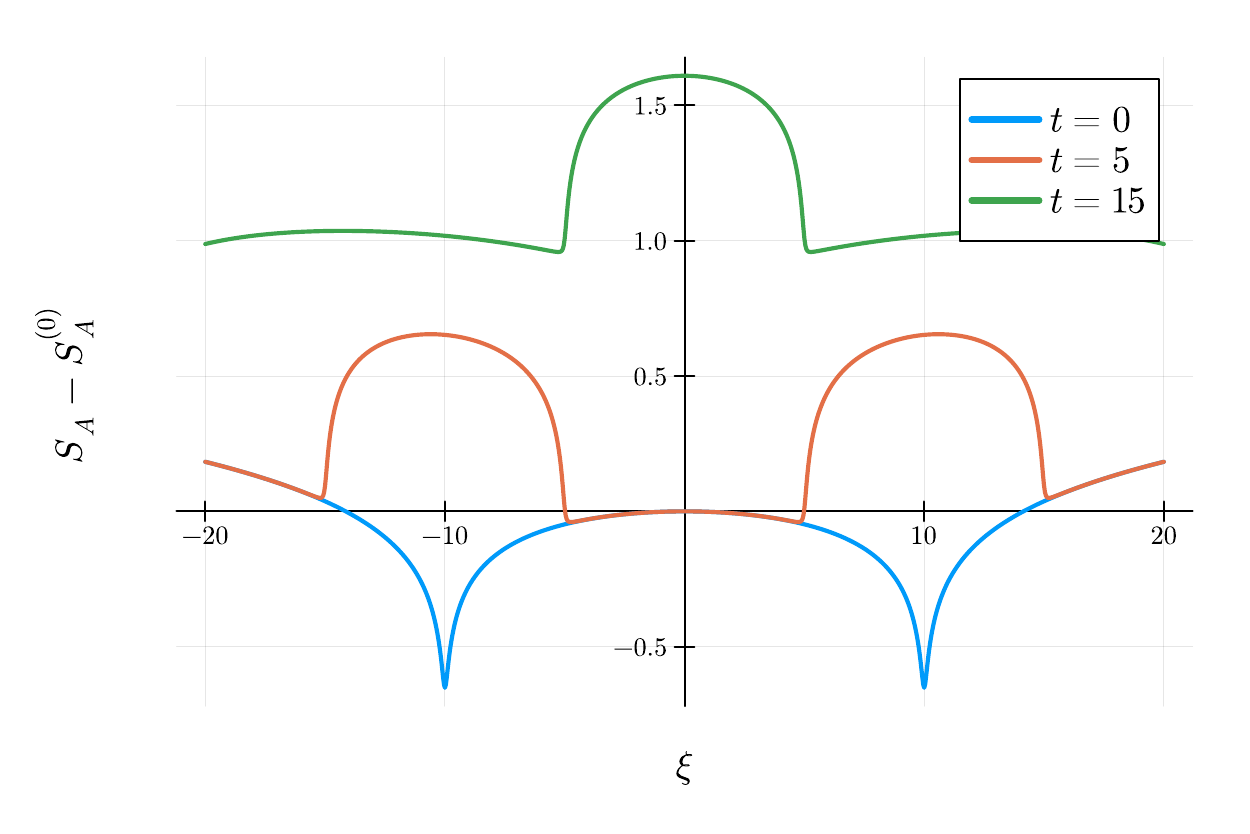}
    \includegraphics[width=6cm]{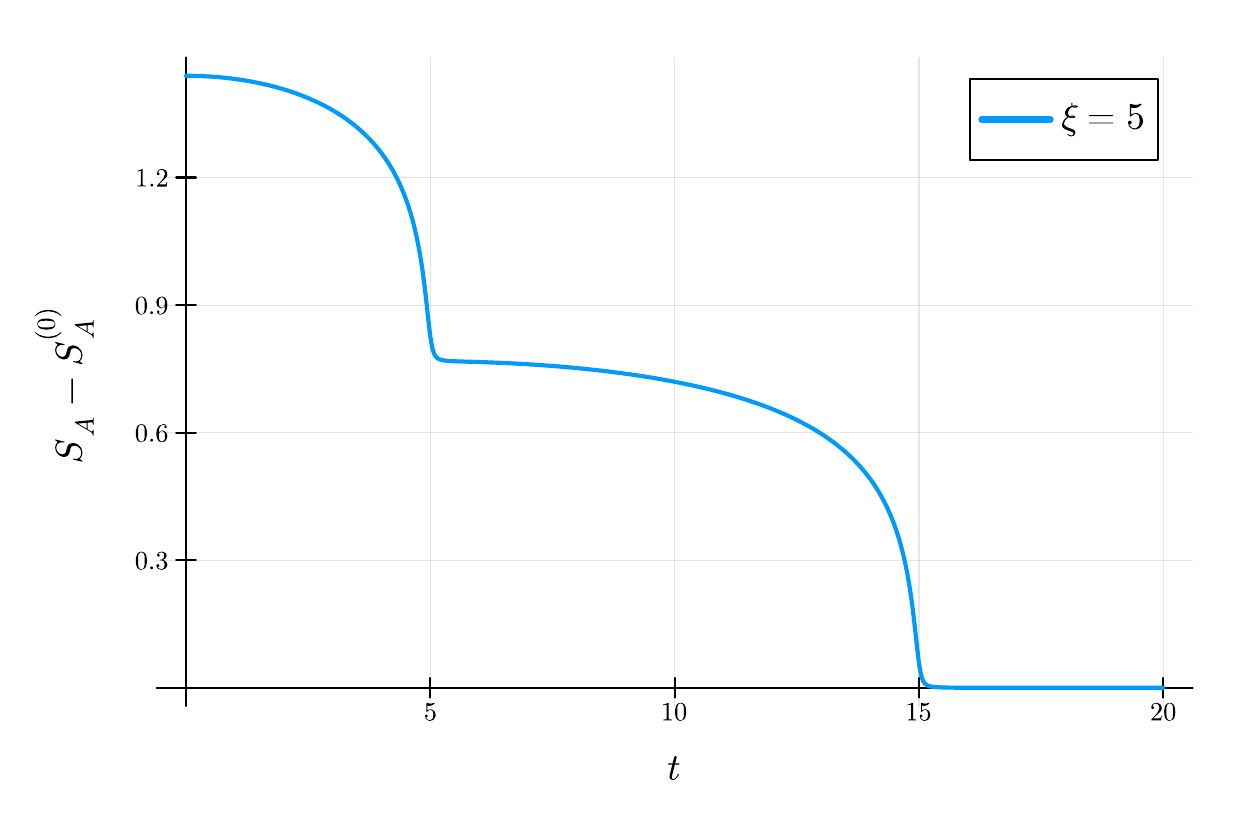}
    \includegraphics[width=6cm]{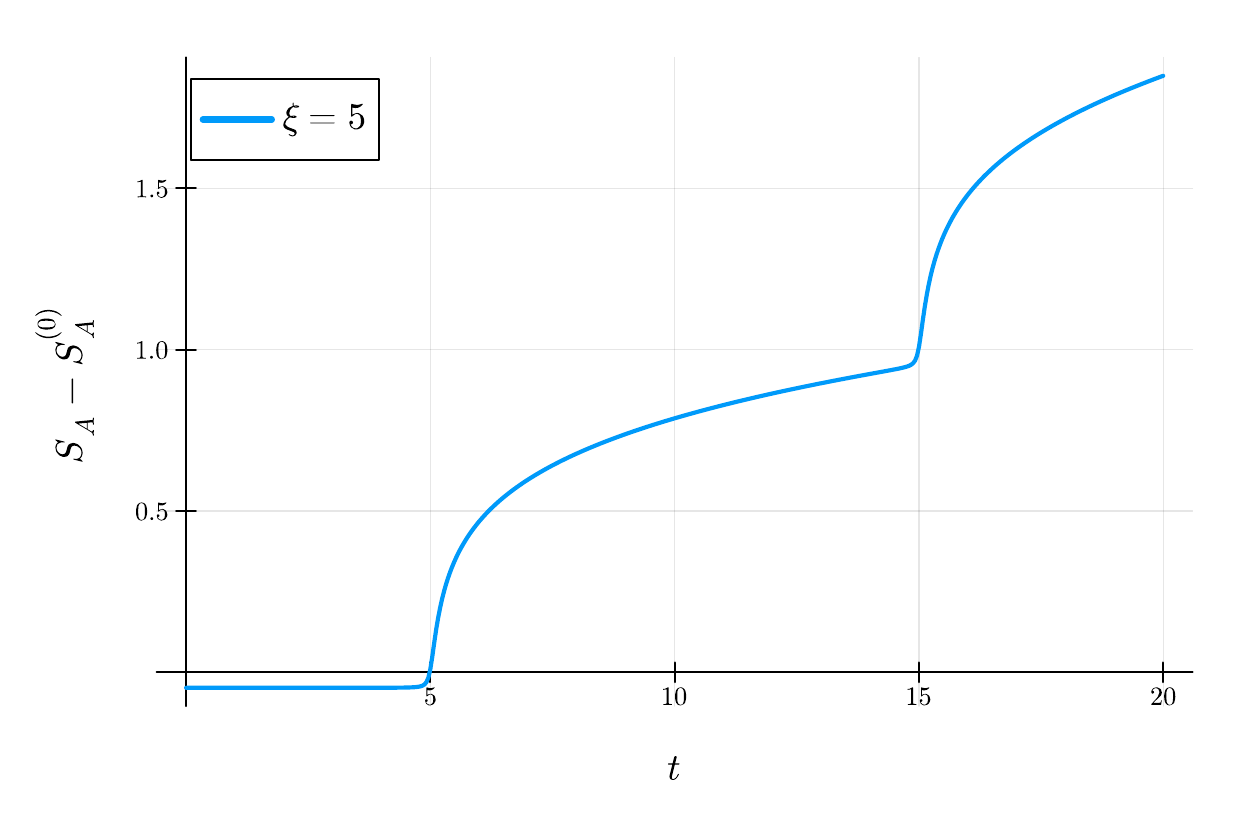}
    \caption{The connected/disconnected contribution to entanglement entropy \(S_{A(\x)}(t) - S_{A(\x)}^{(0)}\) in the holographic CFT for \(A(\x) = [\x - 10, \x + 10]\) is shown in the left/right column.
    The top/middle/bottom row represents the spatiotemporal/spatial/temporal dependence.
    We chose \(c=1\), \(\ep=1\), \(\d=0.1\), and \(S_{\mathrm{bdy}} = 0\).}
    \label{fig:EE_hol}
\end{figure}

\begin{figure}[ttt]
    \centering
    \includegraphics[width=6cm]{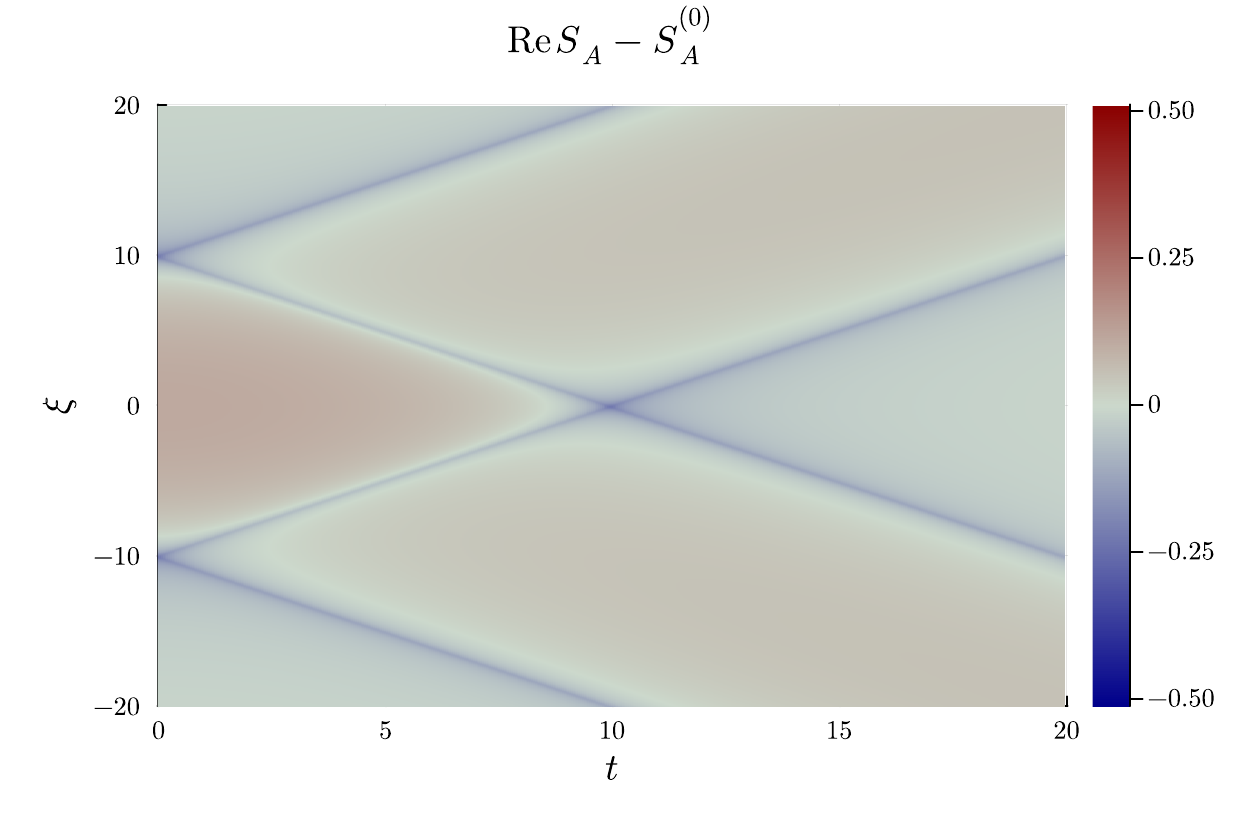}
    \includegraphics[width=6cm]{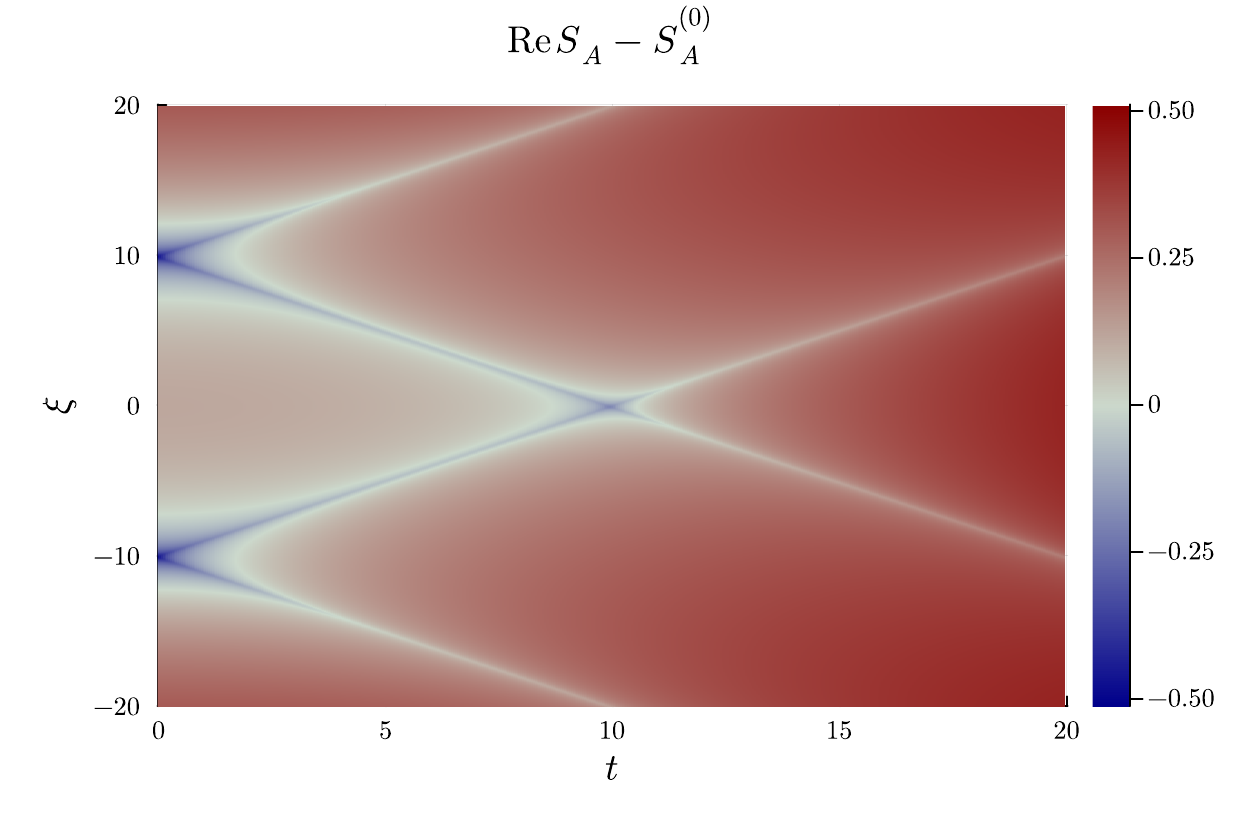}
    \includegraphics[width=6cm]{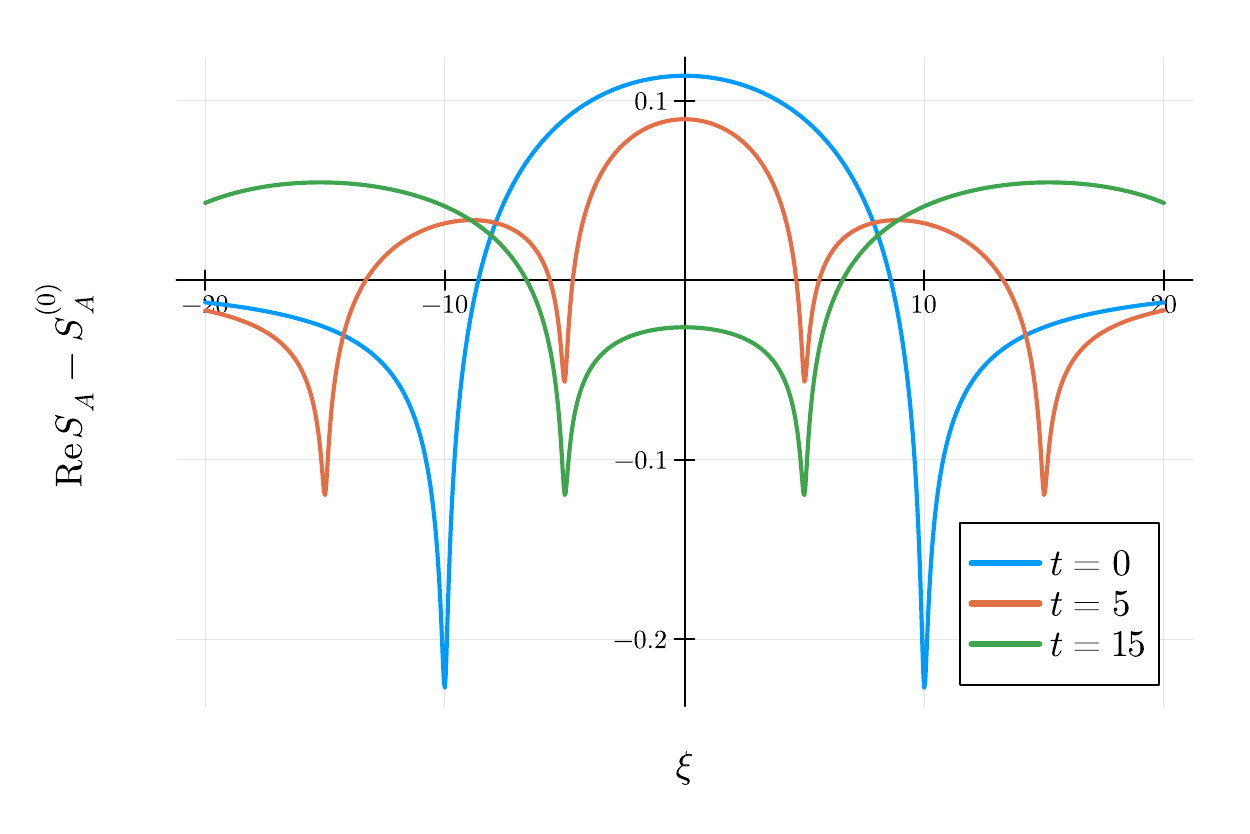}
    \includegraphics[width=6cm]{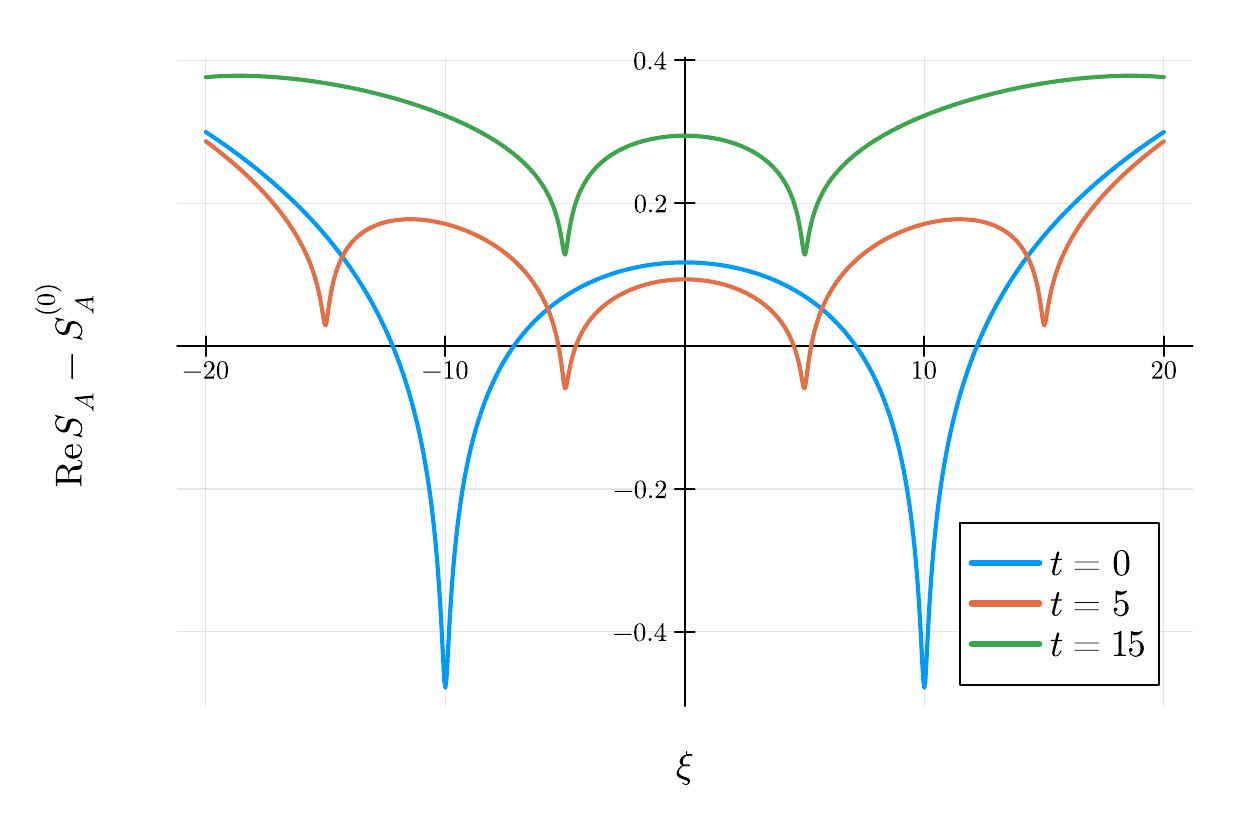}
    \includegraphics[width=6cm]{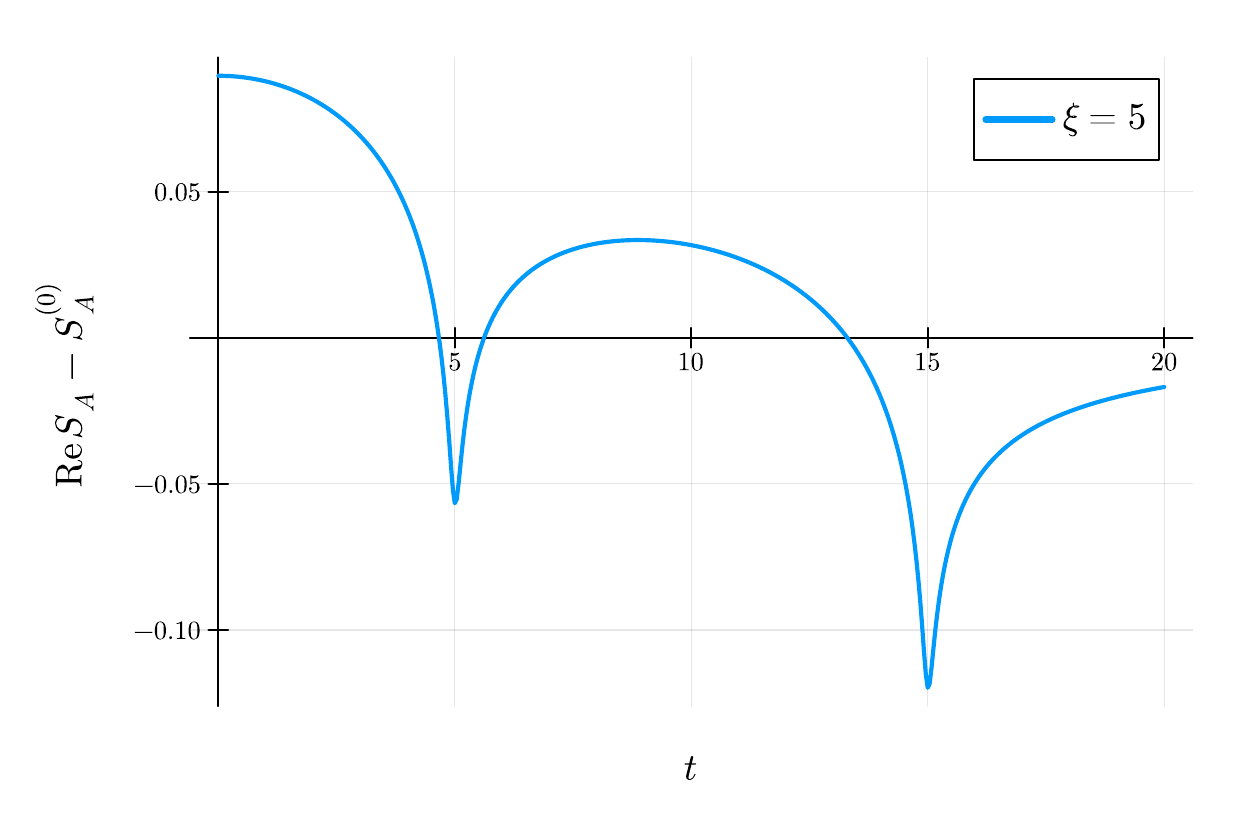}
    \includegraphics[width=6cm]{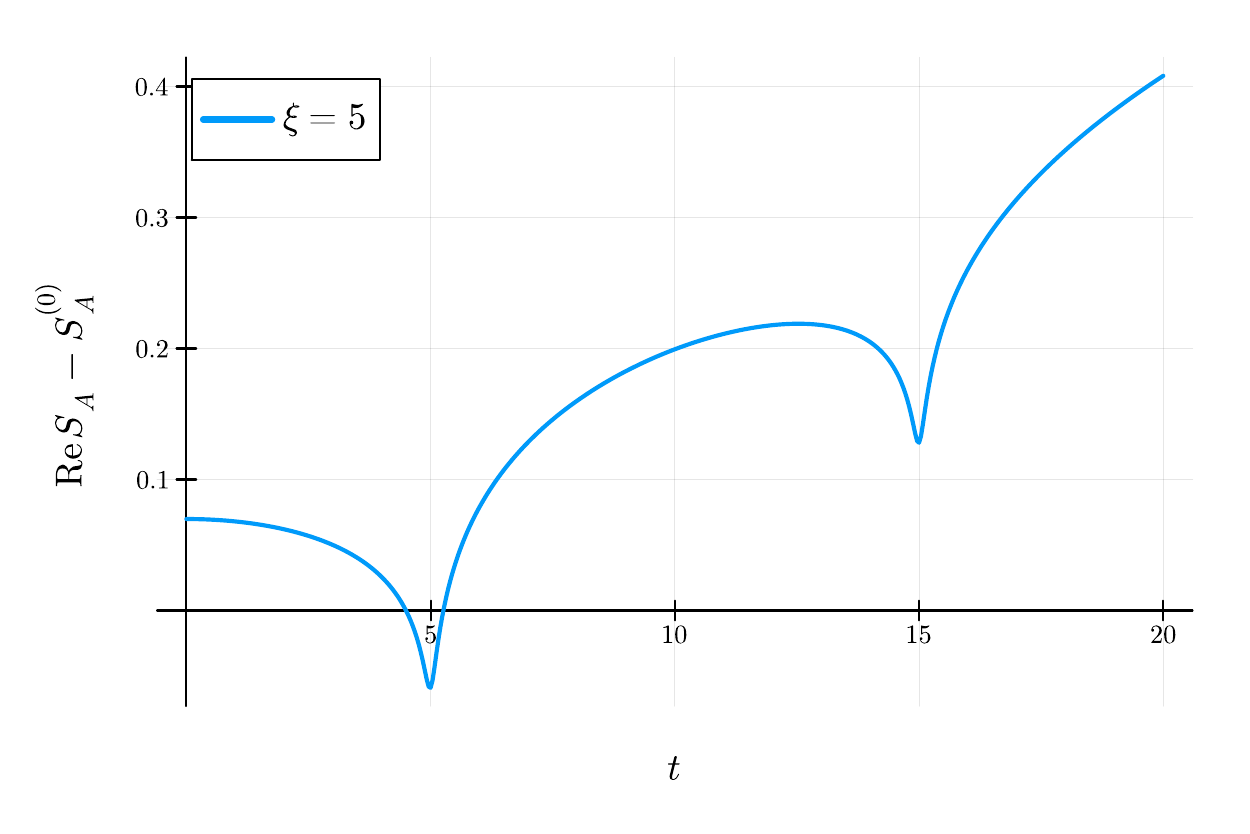}
    \caption{The real part of the connected/disconnected contribution to the pseudo entropy \(S_{A(\x)}^{\JQ|\Omega}(t) - S_{A(\x)}^{(0)}\) in the holographic CFT for \(A(\x) = [\x - 10, \x + 10]\) is shown in the left/right column.
    The top/middle/bottom row represents the spatiotemporal/spatial/temporal dependence.
    We chose \(c=1\), \(\ep=1\), \(\d=0.1\), and \(S_{\mathrm{bdy}}=0\).}
    \label{fig:rePE_JQGS_hol}
\end{figure}

\begin{figure}[ttt]
    \centering
    \includegraphics[width=6cm]{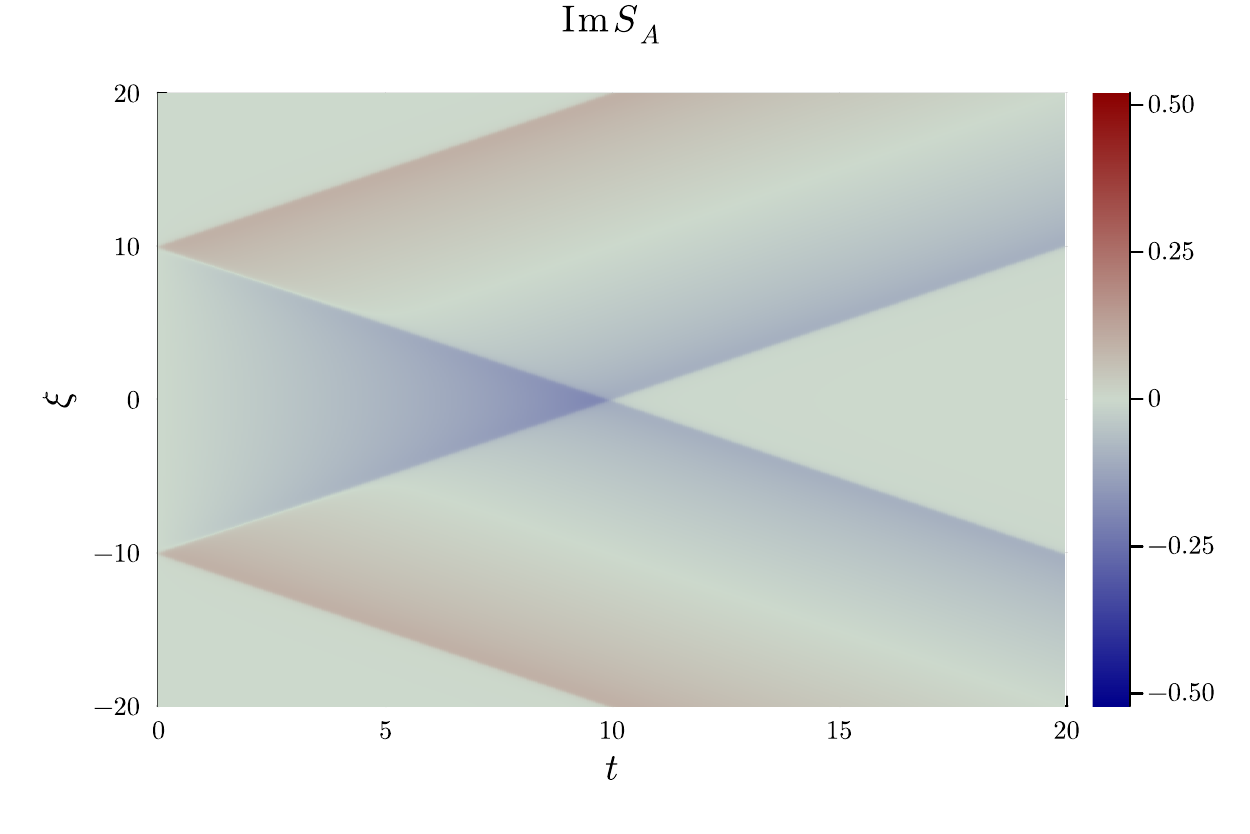}
    \includegraphics[width=6cm]{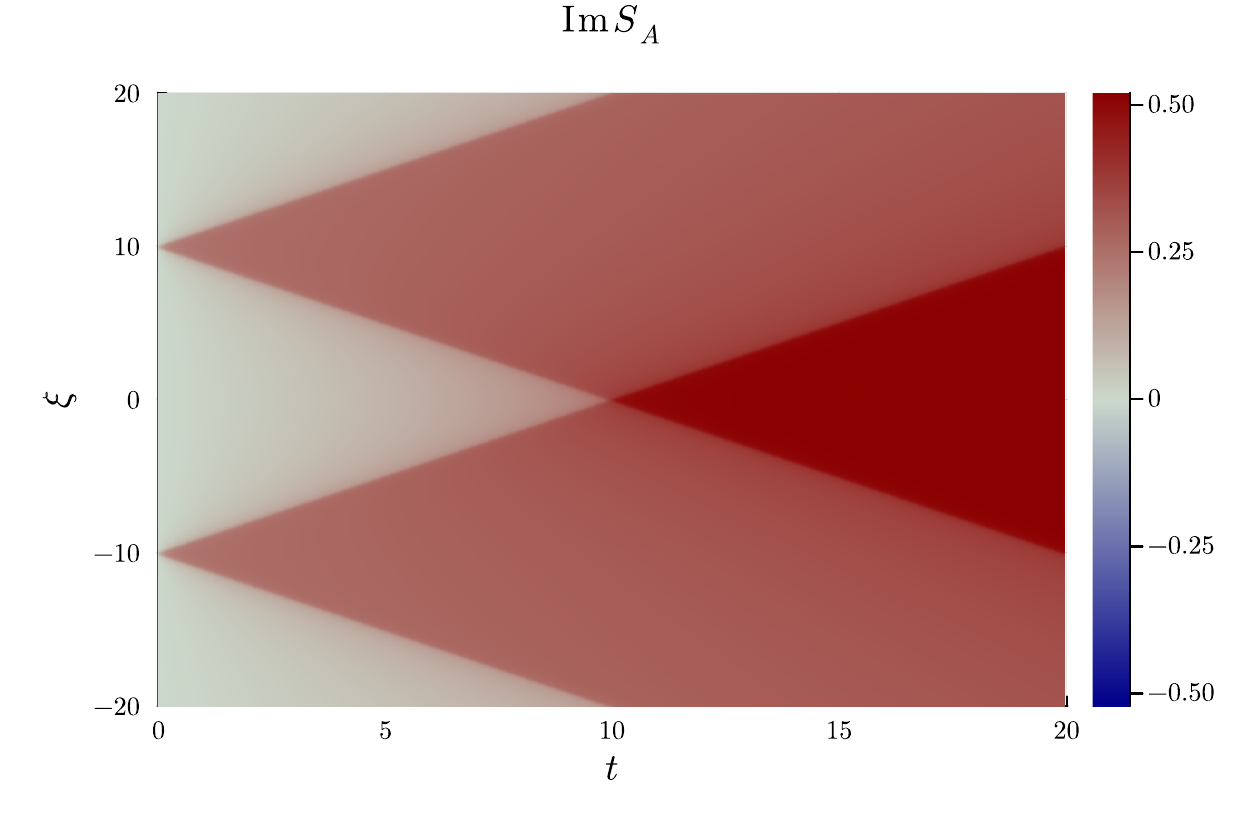}
    \includegraphics[width=6cm]{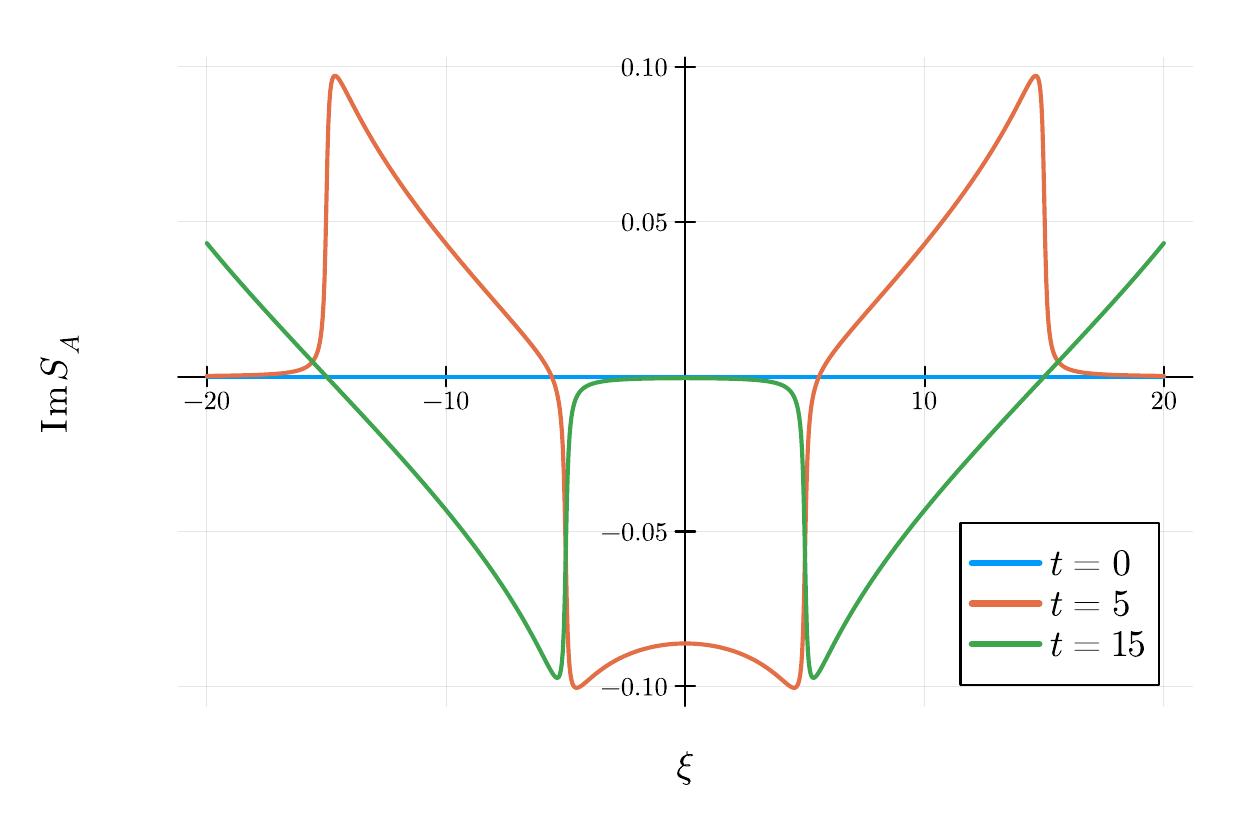}
    \includegraphics[width=6cm]{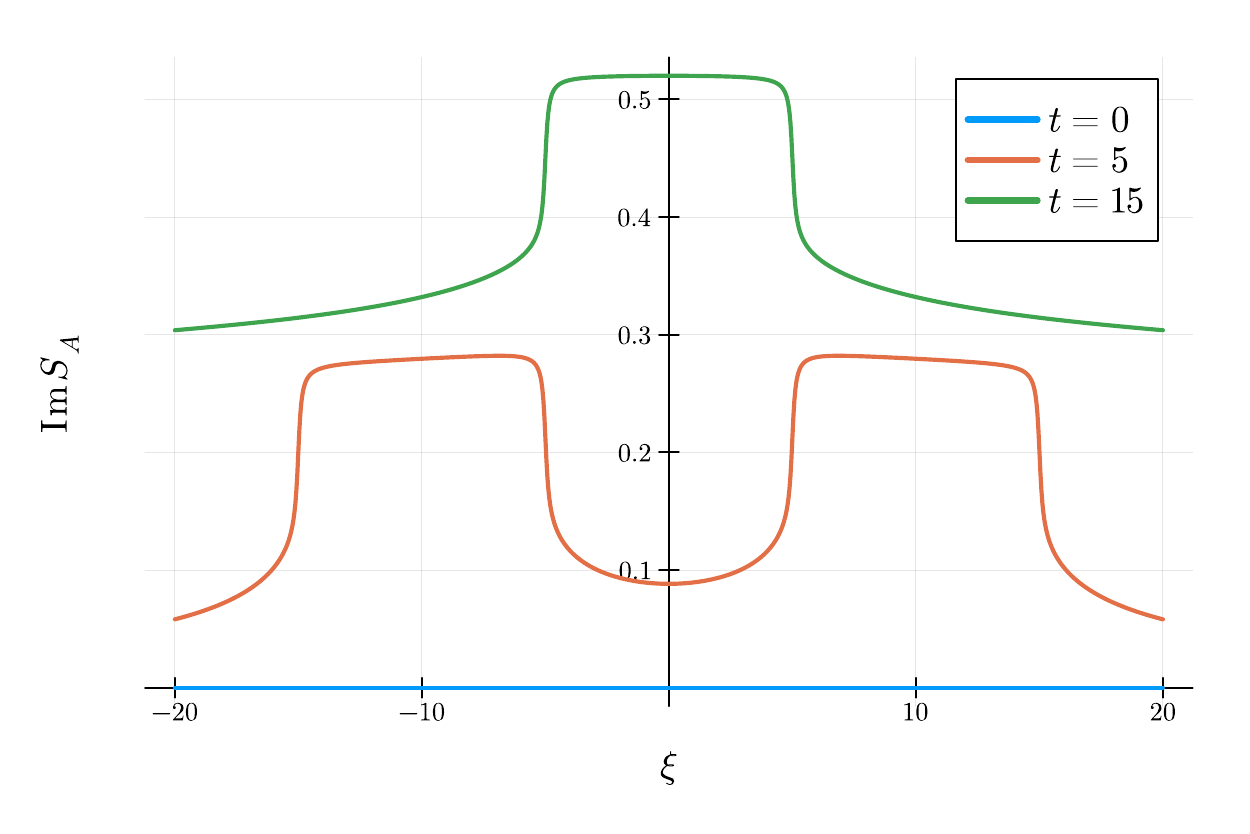}
    \includegraphics[width=6cm]{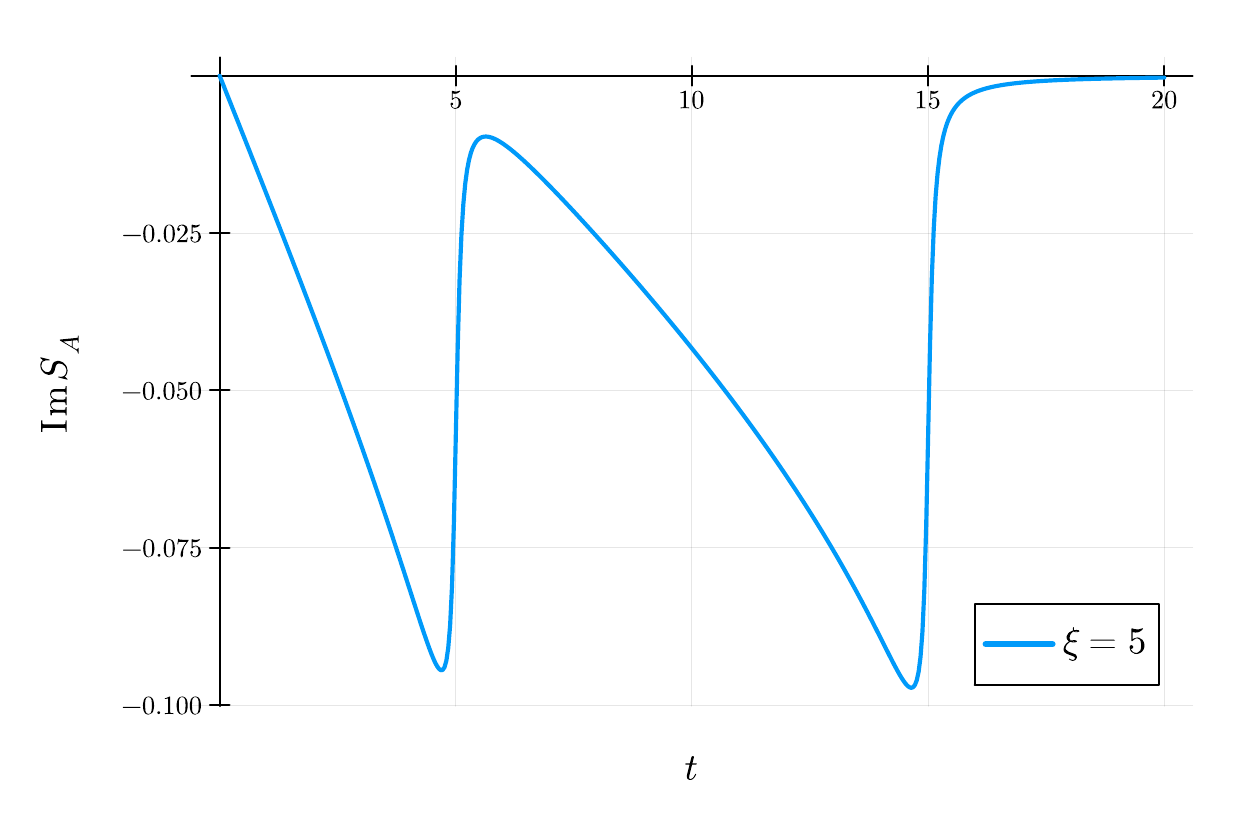}
    \includegraphics[width=6cm]{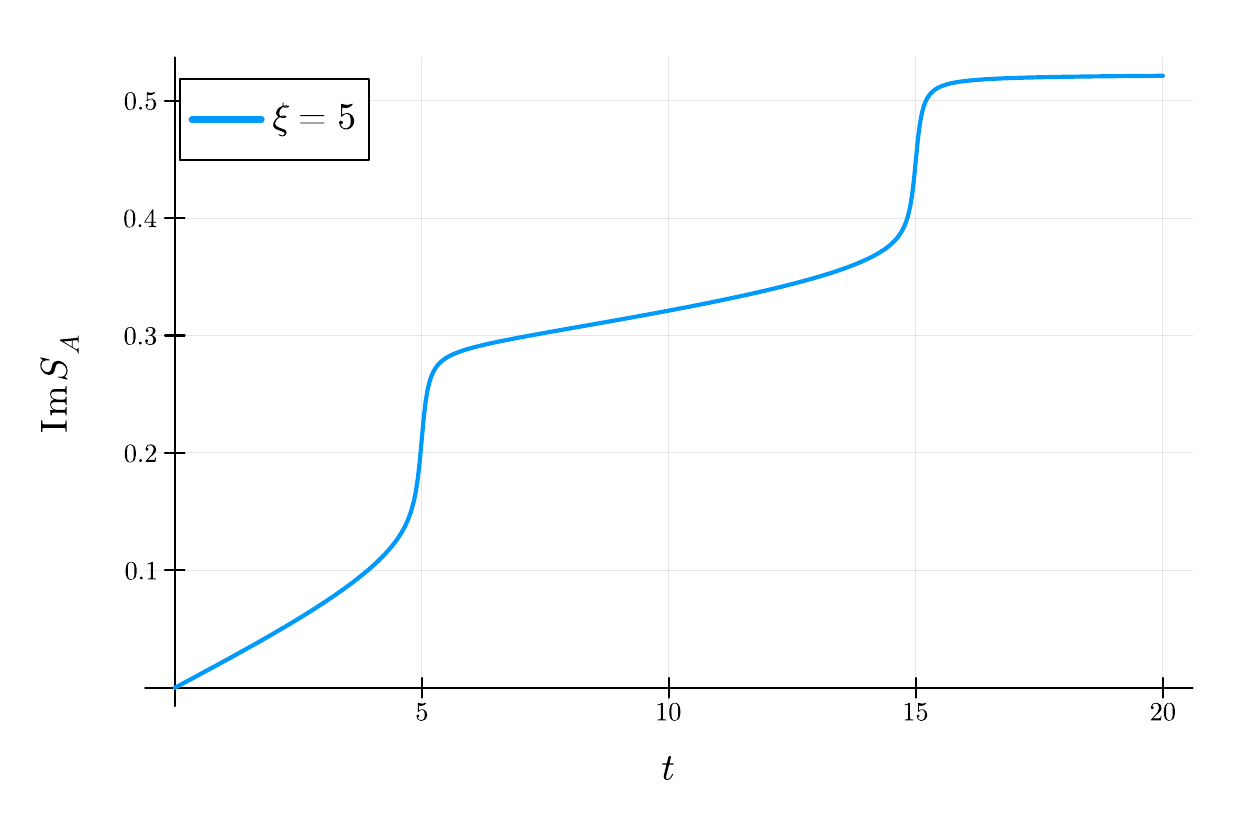}
    \caption{The imaginary part of the connected/disconnected contribution to the pseudo entropy \(S_{A(\x)}^{\JQ|\Omega}(t) - S_{A(\x)}^{(0)}\) in the holographic CFT for \(A(\x) = [\x - 10, \x + 10]\) is shown in the left/right column.
    The top/middle/bottom row represents the spatiotemporal/spatial/temporal dependence.
    We chose \(c=1\), \(\ep=1\), \(\d=0.1\), and \(S_{\mathrm{bdy}}=0\).}
    \label{fig:imPE_JQGS_hol}
\end{figure}

\begin{figure}[ttt]
    \centering
    \includegraphics[width=6cm]{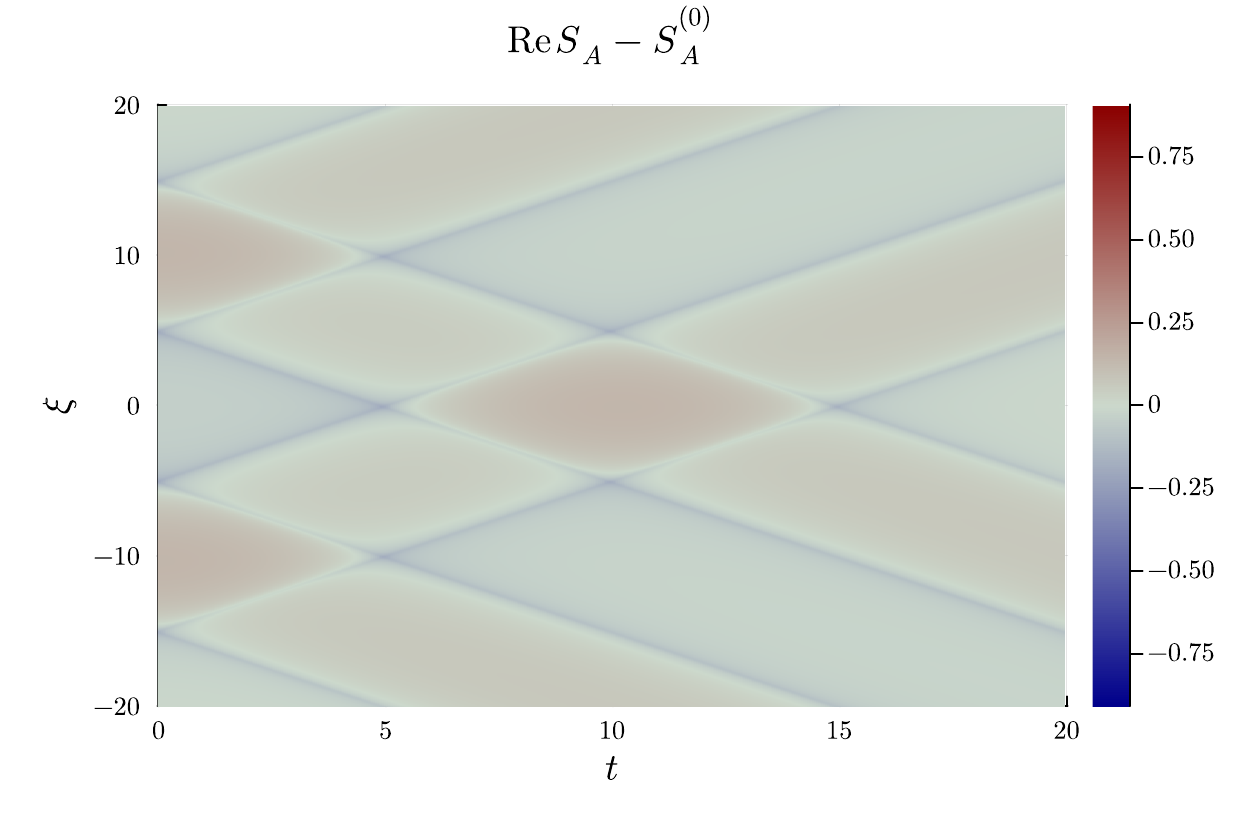}
    \includegraphics[width=6cm]{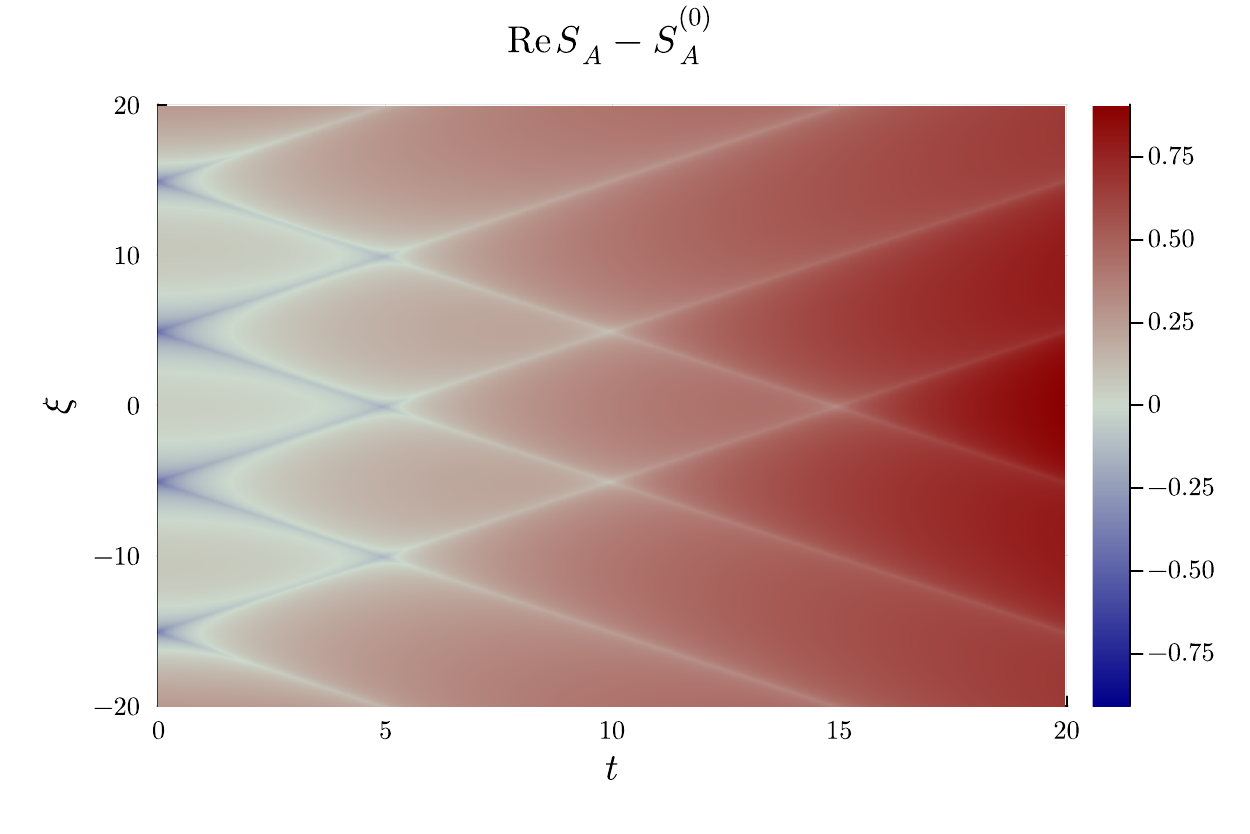}
    \includegraphics[width=6cm]{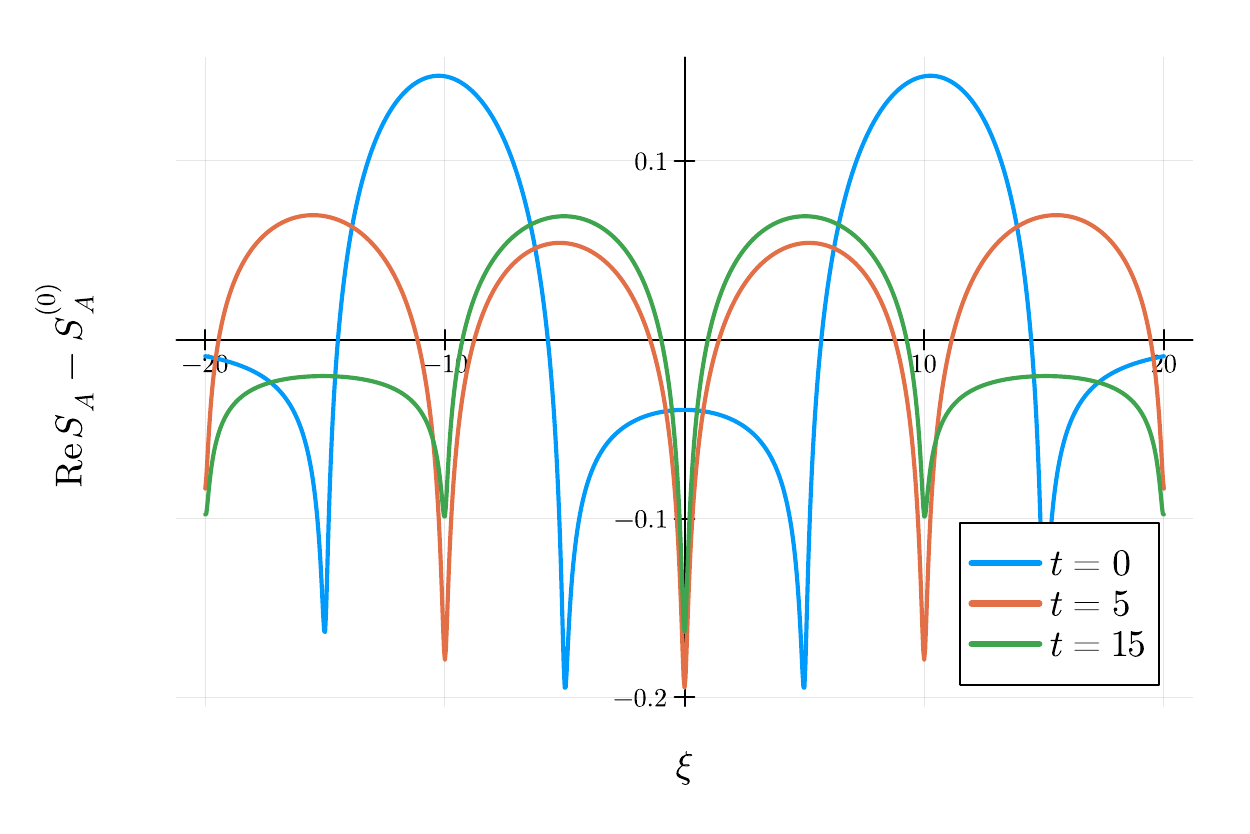}
    \includegraphics[width=6cm]{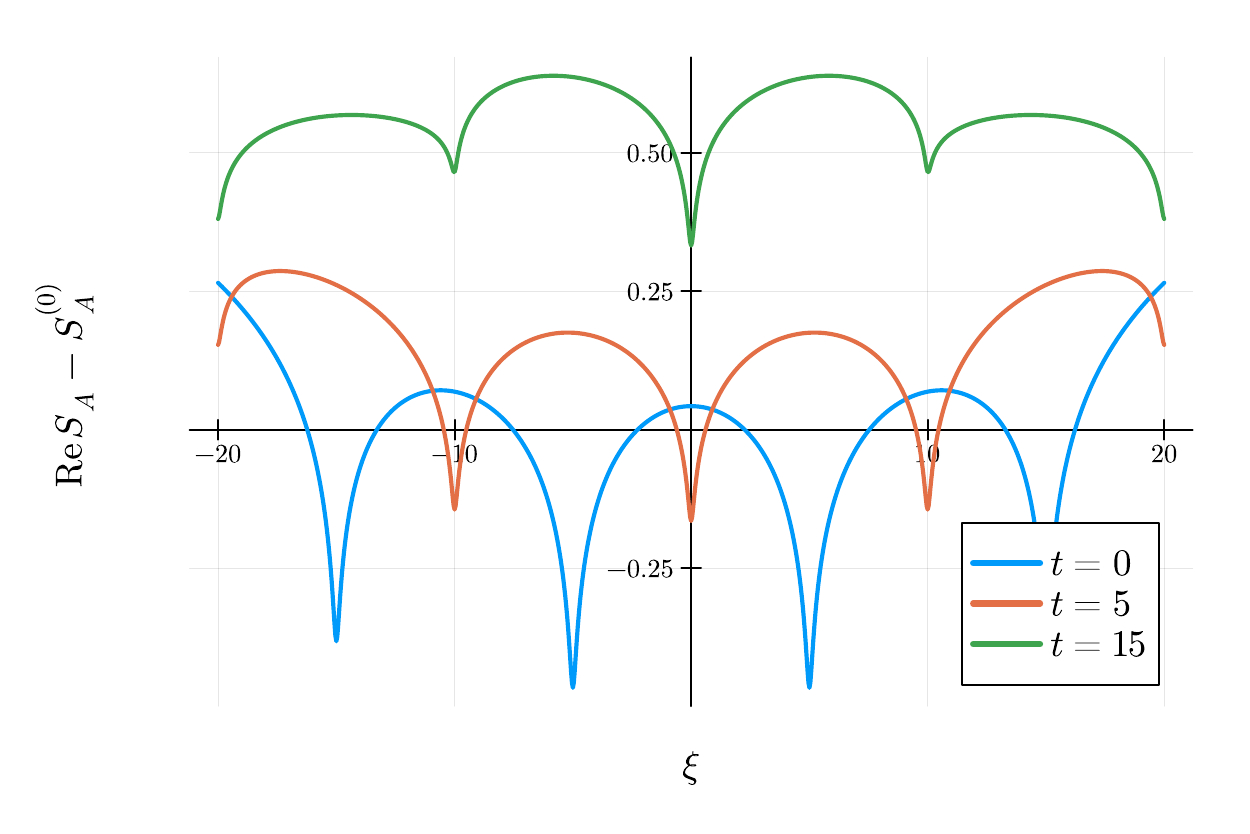}
    \includegraphics[width=6cm]{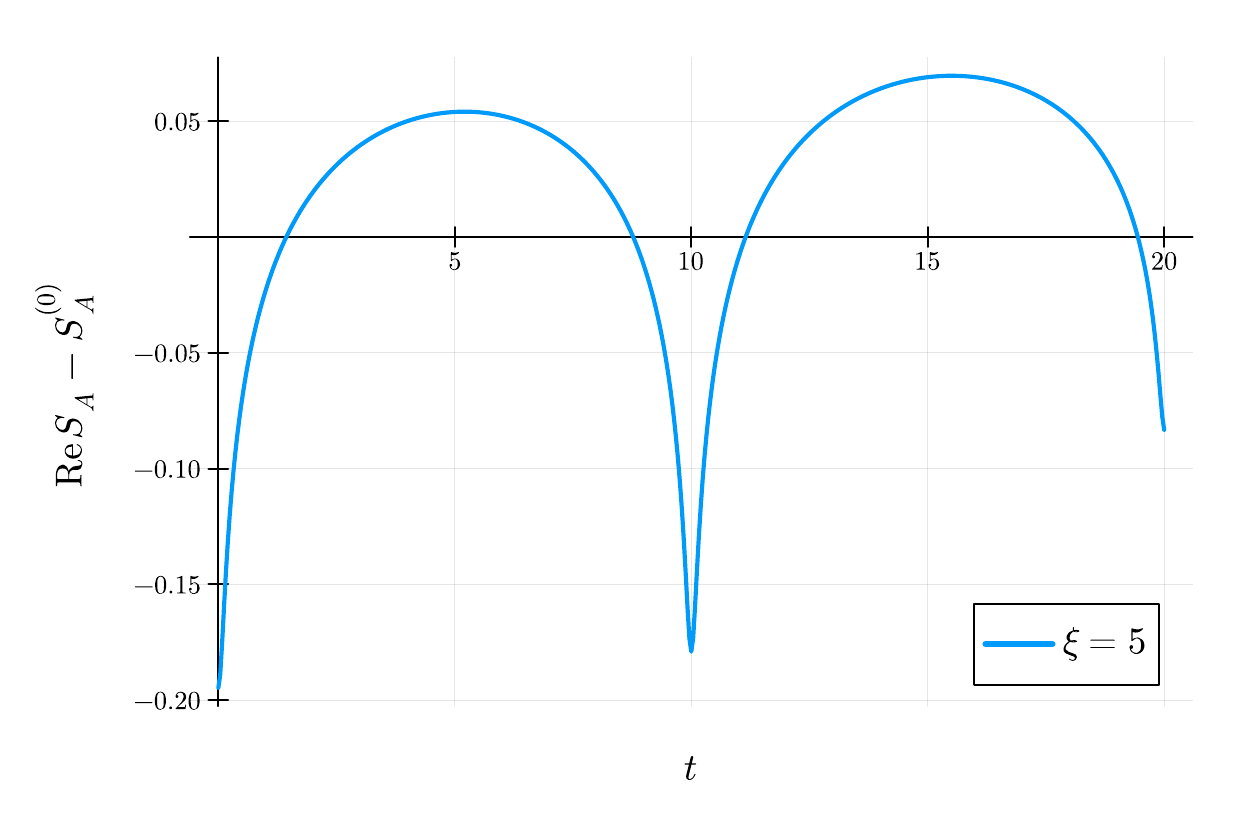}
    \includegraphics[width=6cm]{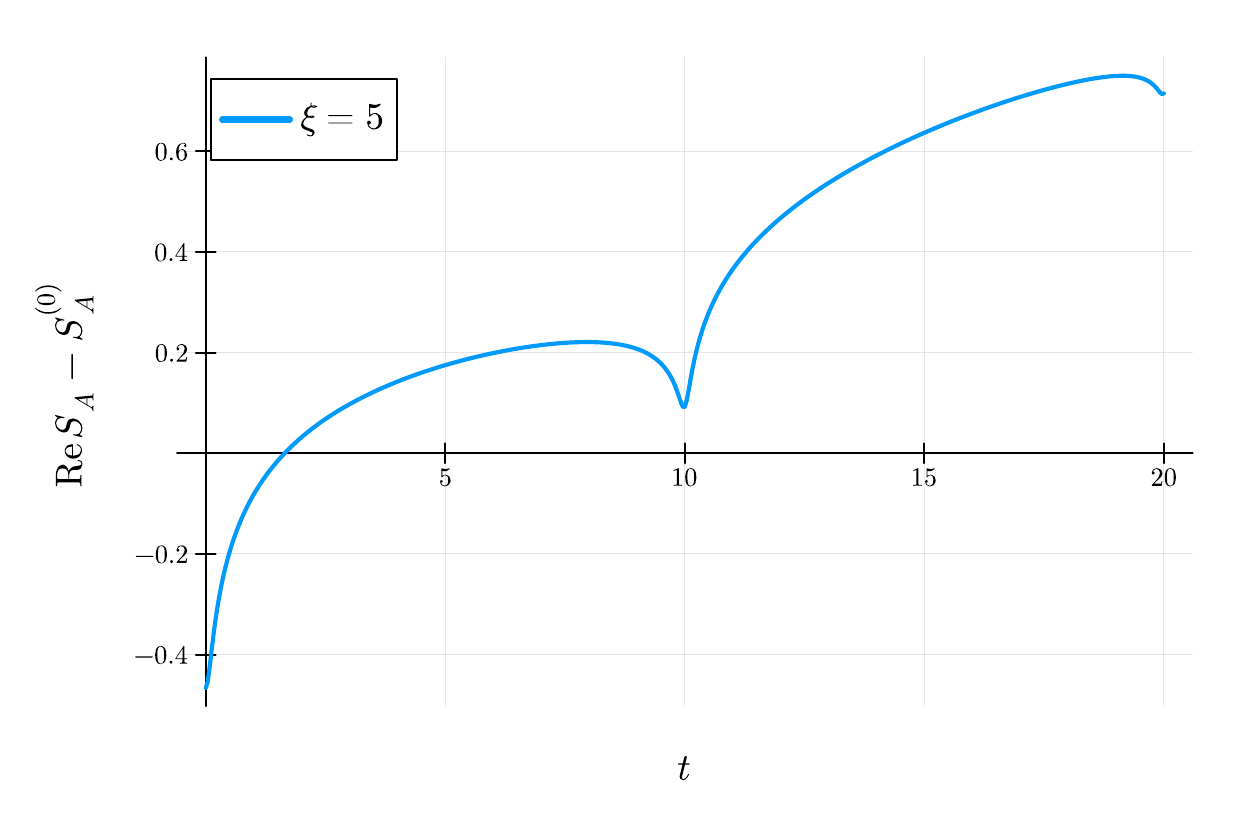}
    \caption{The real part of the connected/disconnected contribution to the pseudo entropy \(S_{A(\x)}^{\JQ_1|\JQ_2}(t) - S_{A(\x)}^{(0)}\) in the holographic CFT for \(A(\x) = [\x - 5, \x + 5]\) is shown in the left/right column.
    The top/middle/bottom row represents the spatiotemporal/spatial/temporal dependence.
    We chose \(c=1\), \(\ep=1\), \(\d=0.1\), and \(S_{\mathrm{bdy}}=0\).}
    \label{fig:rePE_JQJQ_hol}
\end{figure}

\begin{figure}[ttt]
    \centering
    \includegraphics[width=6cm]{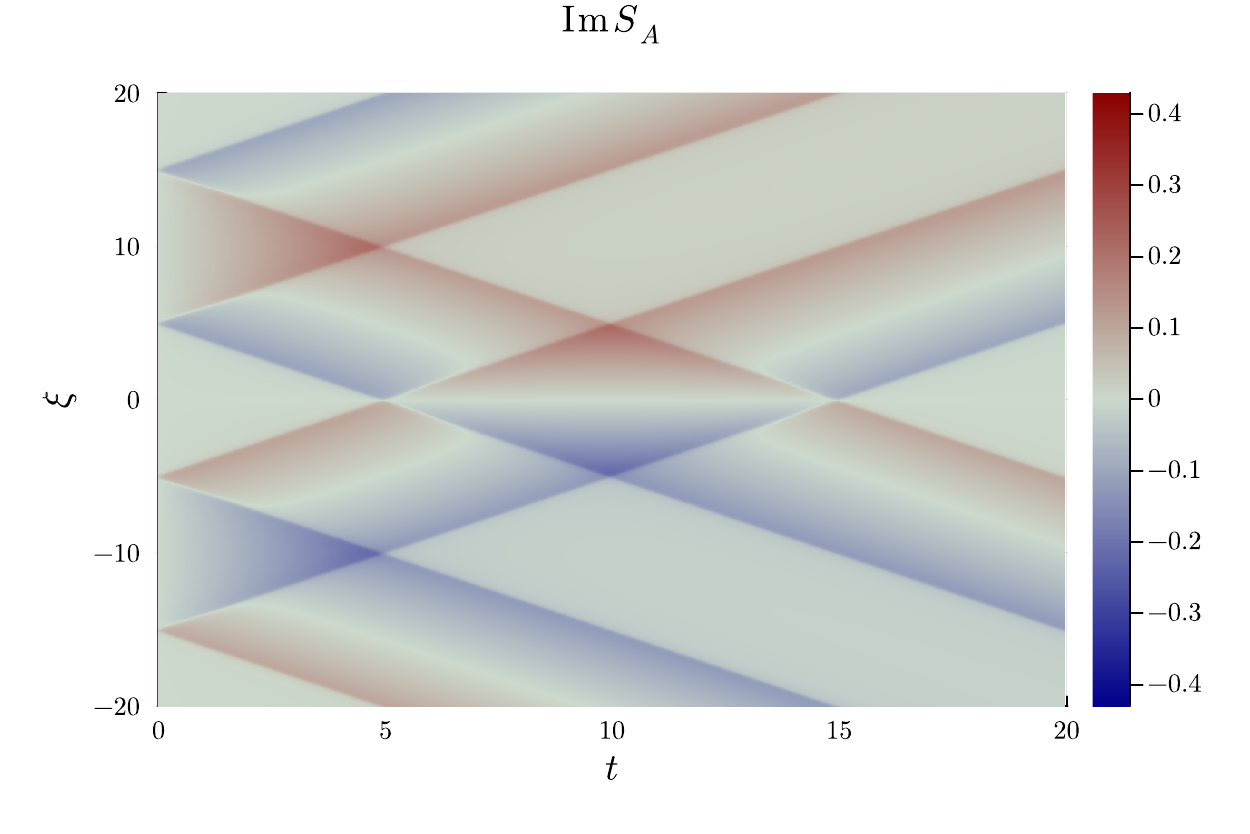}
    \includegraphics[width=6cm]{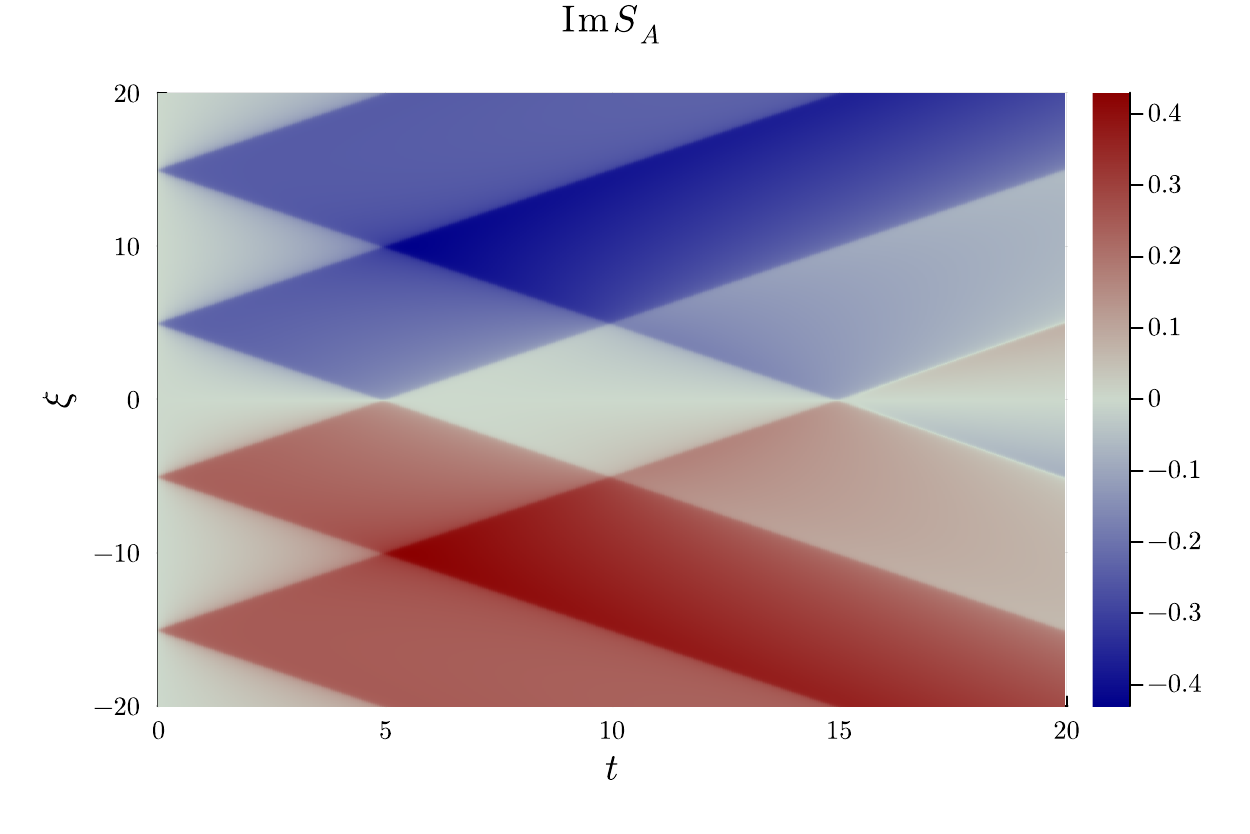}
    \includegraphics[width=6cm]{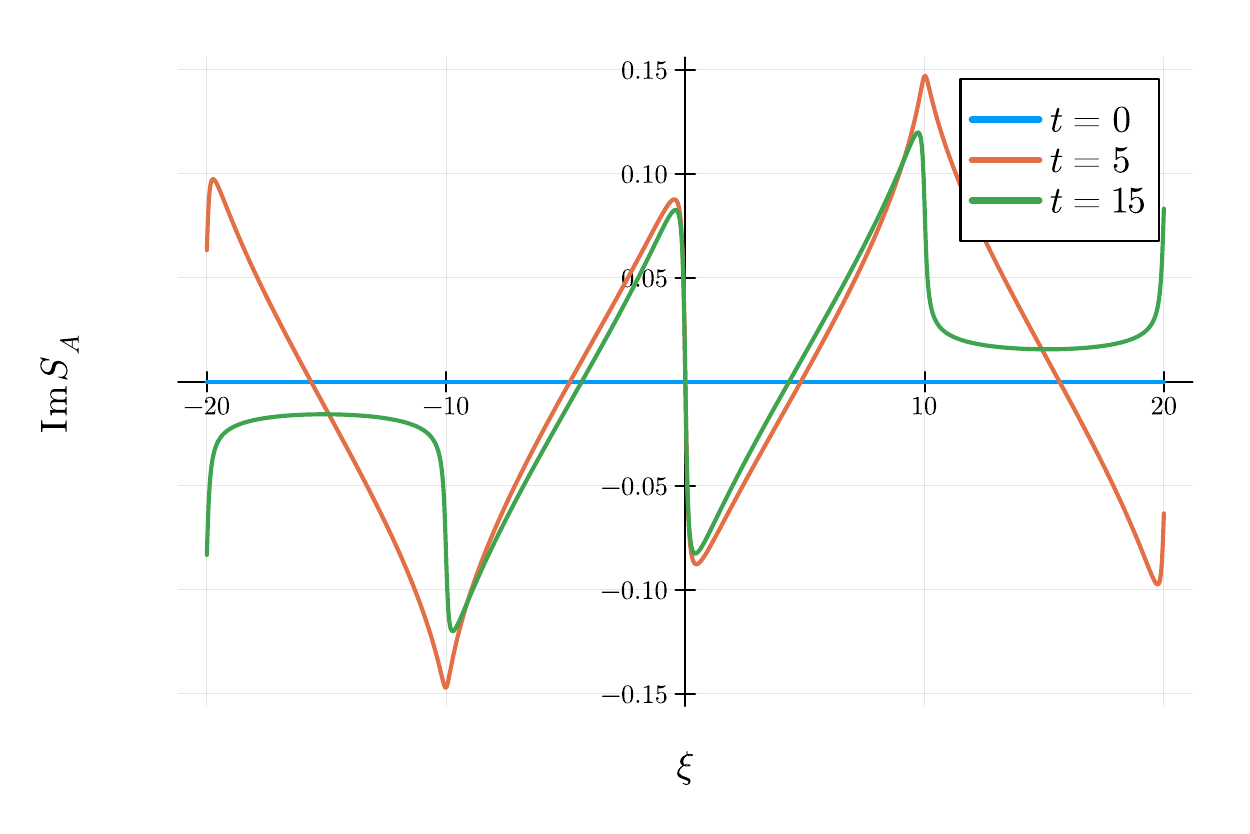}
    \includegraphics[width=6cm]{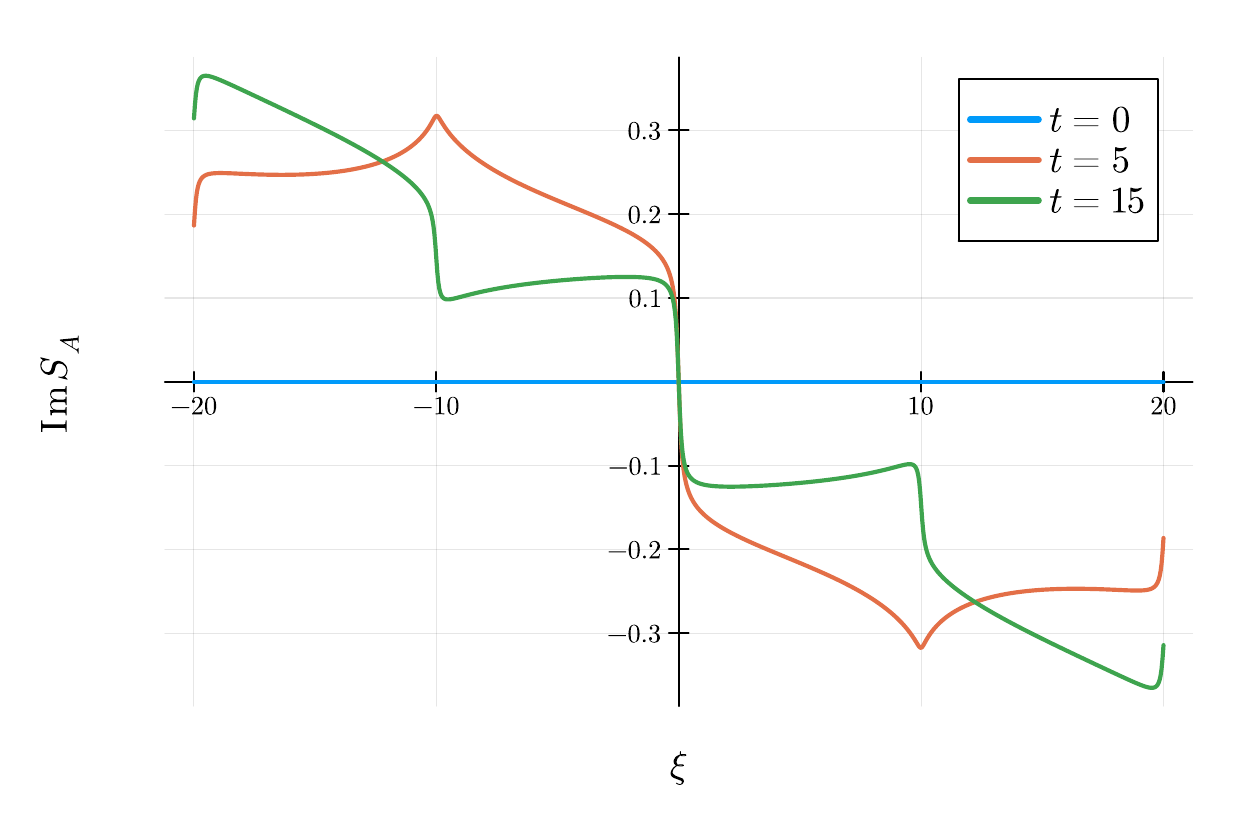}
    \includegraphics[width=6cm]{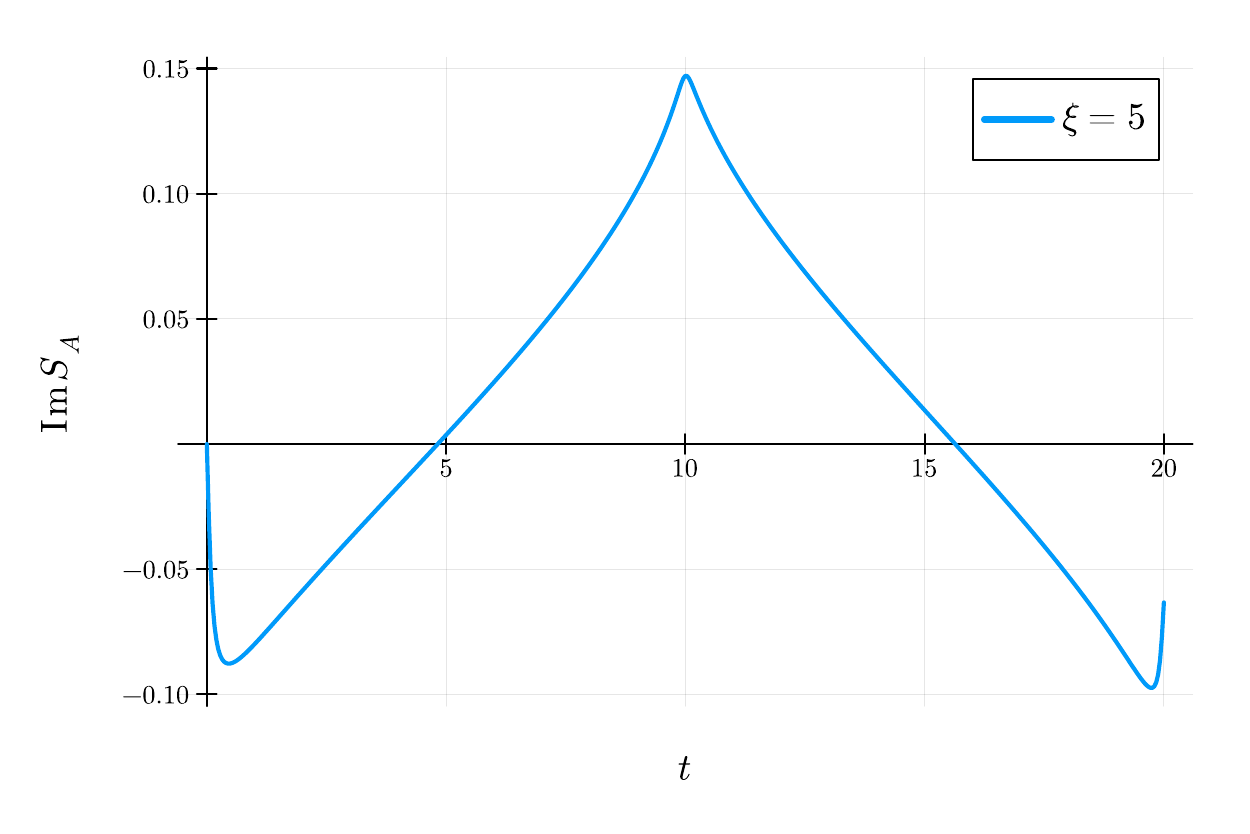}
    \includegraphics[width=6cm]{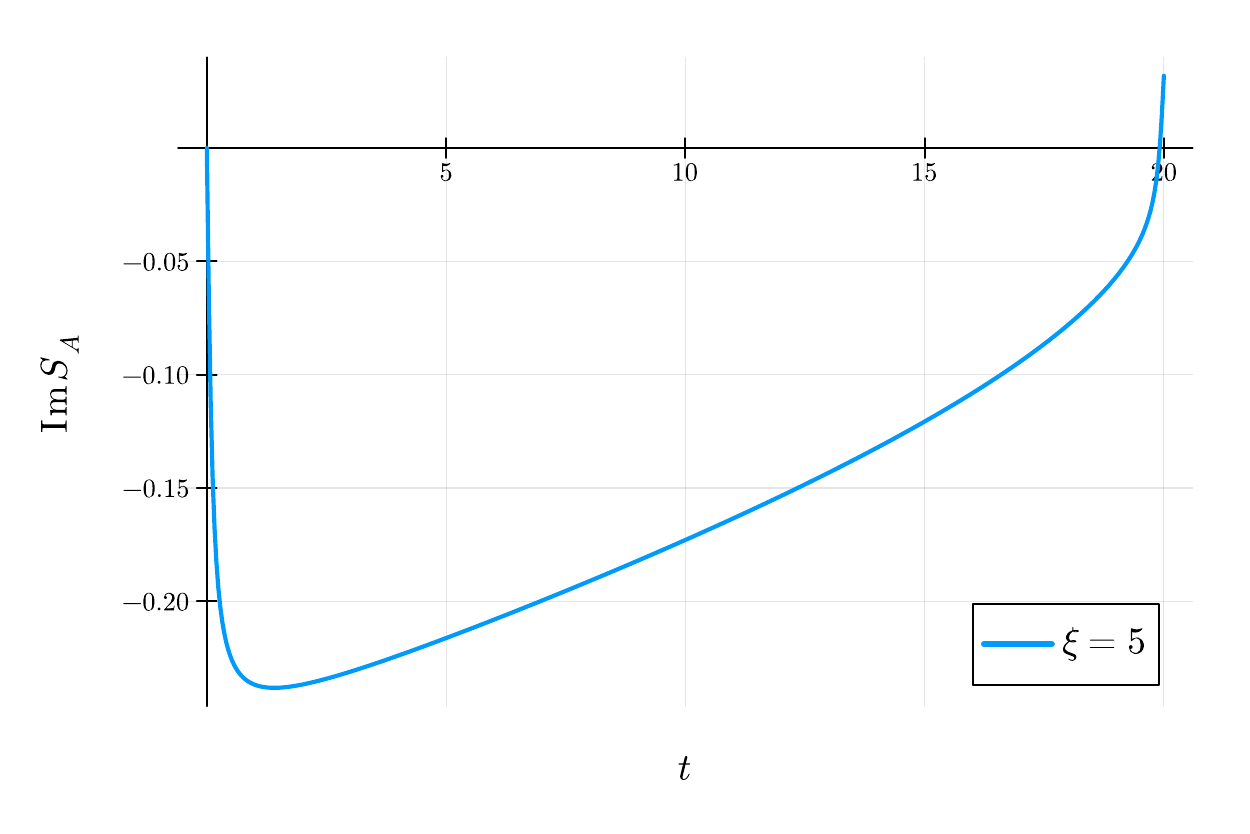}
    \caption{The imaginary part of the connected/disconnected contribution to the pseudo entropy \(S_{A(\x)}^{\JQ_1|\JQ_2}(t) - S_{A(\x)}^{(0)}\) in the holographic CFT for \(A(\x) = [\x - 5, \x + 5]\) is shown in the left/right column.
    The top/middle/bottom row represents the spatiotemporal/spatial/temporal dependence.
    We chose \(c=1\), \(\ep=1\), \(\d=0.1\), and \(S_{\mathrm{bdy}}=0\).}
    \label{fig:imPE_JQJQ_hol}
\end{figure}

For reference, here we present the plots of the real/imaginary part of the connected and disconnected contributions to pseudo entropy in the holographic CFT for the three cases of Table.\ref{tab:def}: 
EE in joining quench (Fig.~\ref{fig:EE_hol}), PE in single-slit (Fig.~\ref{fig:rePE_JQGS_hol}/Fig.~\ref{fig:imPE_JQGS_hol}) and PE in double-slit (Fig.~\ref{fig:rePE_JQJQ_hol}/Fig.~\ref{fig:imPE_JQJQ_hol}). 

\clearpage
\bibliographystyle{JHEP}
\bibliography{PElocalQ}


\end{document}